a/c

# ON THE HEAVY FERMION ROAD

J. Flouquet
Institut de Physique de la Matière Condensée CEA – CNRS
CNRS – BP 166X – 38042 Grenoble cedex France

## 1/ Heavy fermion instabilities

## 2/ Normal phase of Ce compouds

# 6/ The four uranium heavy superconductors : $UPt_3$, $UPd_2Al_3$, $URu_2Si_2$ and $UBe_{13}$

# 7/ Conclusion and perspectives.

# 1/ Heavy fermion instabilities

*Outlooks are :*

- Three singular pressures $P_{KL}$ , $P_C$ and $P_V$ at $T \to 0K$ which are related respectively with the 4f localisation, the magnetic instability and the valence (or orbital) fluctuation.
- The static description of the Doniach Kondo lattice.
- The first glance with a spin fluctuation approach and the non Fermi liquid label.
- The possibility of a Kondo condensate. Over a Kondo coherence length $\ell_{KL} \sim m^*$, the spin memory may be preserved during a long period $\tau_{KL} \approx m^{*2}$.

## *1.1 - Introduction*

The heavy fermion compounds (HFC) belong to the large class of strongly correlated electronic systems (SCES) which covers also 3d intermetallic systems, organic conductors and the high temperature oxyde superconductors. They are also linked to the quantum matter of $^3$He and systems like manganite compounds where the magnetic coupling effects the electronic conduction. Despite three decades of studies, there are still some mysteries concerning the charge and spin dynamics. However major results have been obtained on these specific materials with broad implications in condensed matter physics.

Our article is more a Grenoble laboratory's report on how to track the electronic excitations (charge and spin) and the nature of the ground states than a review which covers all the published works. Furthermore, the approach is that of an experimentalist familiar with low temperature physics i.e the main motivation is to clarify the complex nature of heavy fermions but with references to general basic questions. Most of the figures correspond to data obtained either by us or by collaborators. Of course, references are given on the original discoveries. Our further implication was motivated by the possibilities to add new experimental insights.

For young researchers, a good introduction to the Kondo problem can be found in reference (Anderson 1967), an excellent introduction to unconventional Cooper pairing is of course the review article on $^3$He (Leggett 1975). Unconventional superconductivity is treated theoretically in (Mineev and Samokin 1999 and Gorkov 1987). Reviews devoted to heavy fermion systems can be found in references (Brandt and Moschalkov 1984, Grewe and Steglich 1991, Fulde et al 1988, Springford 1997, Ott 1987, Kuramoto and Kitaoka 2000). A good summary can be found in reference (Heffner and Norman 1996). Extensive discussions on unconventional superconductivity and magnetism have been recently given in the review of Thalmeier and Zwicknagl 2004-a and Thalmeier et al 2004-b. Recent lecture notes have been published (Aliev et al 2001, Coleman 2002). A discussion on singular non Fermi liquid can be found in Varma et al 2002. Up to date points of view on Kondo problems can be found in the Kondo Festschrift edited recently (see Kondo 2005). Our favorite book on HFC is unfortunately up to now only available in Japanese (Ueda et Onuki 1998). Popularization articles are (Hess et al 1993, Cox and Maple 1995, Fisk et al 1998). With the different reviews, the reader will see that the selection of material is often a question of personal tastes. Thalmeier et al (2004-b) have considered that the new ferromagnetic superconductors do not belong to the heavy fermion class. It is a subject extensively discussed in chapter 5.

After discussing the link with the Kondo impurity problem, the relationship with the intermediate valence compounds (IVC), the relevance of spin fluctuations and the key issues in the Kondo lattice (Fermi surface, magnetic and valence instabilities), we will first concentrate on the cerium heavy fermion normal phase properties. We will focus on the appearance of superconductivity in the vicinity of the antiferromagnetic (AF) instability (at a critical pressure $P_C$ ) for 3 dimensional (d) and quasi 2d compounds notably the new 115 series discovered recently in Los Alamos.

The studies of transuranian compounds have been very successful for the understanding of unconventional superconductivity. The discovery of high $T_C$ superconductivity in $PuCoIn_5$ again in Los Alamos illustrates the game between the electronic bandwith and magnetic fluctuations in order to optimise the superconducting critical temperature $T_C$ . The recent observation of the coexistence of superconductivity and ferromagnetism (F) in $UGe_2$ has led to a rush to new examples and also to a revival of theoretical interests on ferromagnetic superconductors. The data on the four archetypal heavy fermion uranium superconductors $UPt_3$ , $UPd_2Al_3$ , $URu_2Si_2$ and $UBe_{13}$ will be examined with special focus on the determination of the superconducting order parameters ($UPt_3$ , $UPd_2Al_3$ ) on the low temperature excitation characteristic of unconventional pairing ($UPt_3$ ) and on the respective temperature (T), magnetic field (H) pressure (P) phase diagrams.

## *1.2 – Localisation, valence and magnetism*

The cerium heavy fermion compounds (HFC) is just at the frontier of classical rare earth intermetallic systems (here the localised 4f electrons and light itinerant electron formed two different baths) (Elliot 1972) and of intermediate valence compounds (IVC) (Wachter 1993 - Newns and Read 1987) where a strong mixing occurs at the Fermi level between the two types of electrons. In the IVC case, the occupancy $n_f$ of the electron in the 4f shell is less than unity and its valence on a given Ce site $v = 4 - n_f$ is clearly intermediate. In HFC, it is the weak departure from $n_f = 1$ which leads to strong low energy magnetic fluctuations at the Fermi level and also to the memory of the local 4f character.

We will first discuss the case of Ce metal where a discontinuity in $n_f$ occurs in pressure and temperature. Our idea is that this discontinuity may be less dramatic and even smooth in HFC but at $T \rightarrow 0K$, it will correspond to a first pressure $P_V$ where the 4f electron looses its local sensitivity to the environment and notably the lifting of its angular momentum degeneracy by the crystal field. Another pressure $P_{KL} < P_V$ will correspond to the critical pressure above which the 4f electron is included in the volume of the Fermi surface. The duality between the localized and itinerant part of the f electron is the core of the heavy fermion problem. In the duality model introduced by Miyake and Kuramoto (1990) and Kuramoto and Miyake (1990), the aim is to use the known results for the single site and to add the renormalization flow for the low temperature regime (see later).

The cerium metal (T, P) phase diagram shows the occurrence of a high temperature trivalent $\gamma$ phase ($n_f = 1$) and of a low temperature $\alpha$ phase ($n_f \sim 0.9$). A first order isostructural line $T_{\gamma\alpha}$ ends up at a critical point around $T_{cr} = 600$ K, $P_{cr} \sim 22$ kbar (Jayaraman 1965). As the intercept $T_{\gamma\alpha}(0) \sim 100$ K at $P = 0$ (figure 1) is high by comparison to any hypothetical magnetic ordering temperature $T_N$, the magnetism was treated crudely (Lavagna et al 1982, Allen and Martin 1982). Recent theoretical developments can be found in Held et al 2001 and in recent publications using a new approach for the band structure (see later).

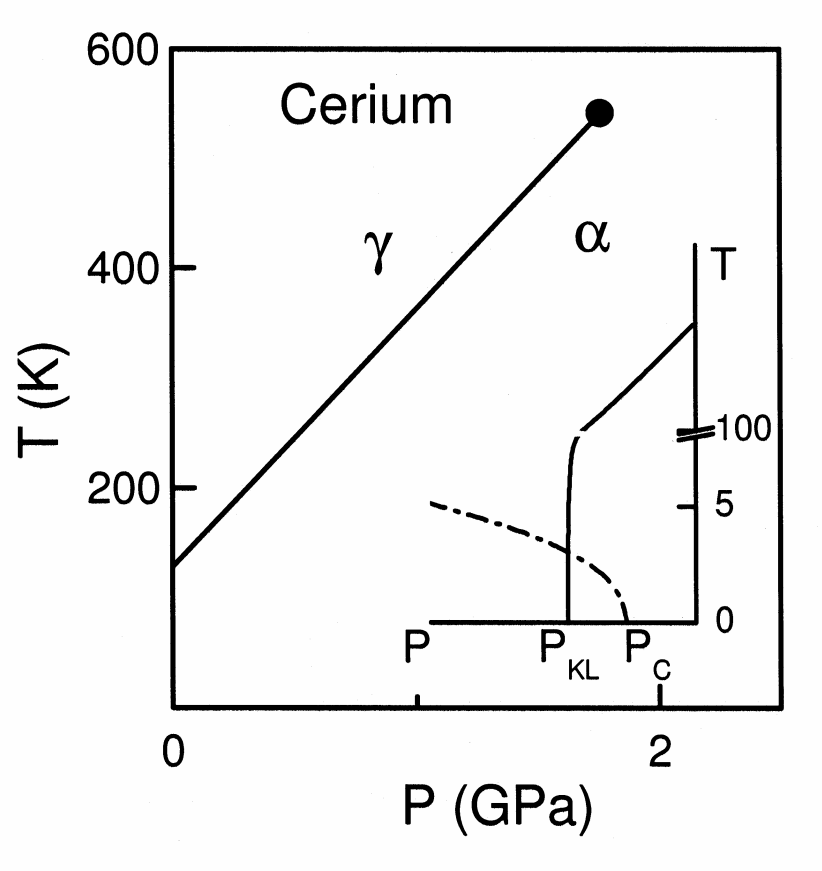

Figure 1 : The simplified (T, P) phase diagram of Ce metal : full line the first order γ - ∝ transition. When it is extended to negative pressure, in the insert, it ends up at $P_{KL}$ for T → 0K. The dashed line represents the hypothetical variation of its magnetic ordering temperature which will collapse at $P_C$ . In HFC, $T_{\gamma\propto}$ (T → 0) will become positive. $T_{cr}$ will collapse to 0 K. The memory of $P_{cr}$ will appear in $P_V$. The cascade of instabilities may be $P_{KL} < P_C < P_V$.

However, if the volume can be expanded to negative pressure the linear extrapolation of $T_{\gamma\alpha}(P)$ to zero occurs at a low negative pressure $P_{KL}$ ~ - 0.3 GPa. Of course at T = 0K, the initial slope $\frac{dT}{dP} = \frac{\Delta V}{\Delta S}$ of the first order transition (ΔS and ΔV respectively entropy and volume jump) must be vertical, according to Clausius Clapeyron relation. At T = 0K no entropy discontinuity can exist but only a volume jump must remain. Near $P_{KL}$ , an important feature is the possibility of a long range magnetic ordering (antiferromagnetic (AF) or ferromagnetic (FM)) for $P < P_C$ . $P_C$ can coincide or not with $P_{KL}$ . In the paramagnetic phase (PM) the 4f electron is found to be itinerant ($P_{KL} < P_C$ ).

In usual HFC, both previous $P_{KL}$ and $P_C$ are positive, the interplay between magnetism and the localisation of the f electron is central. By contrast to the case of Ce metal, there are only few cases of a first order transition between γ and α phase in HFC. That may be due to the presence of other ligand ions in the lattice and consequently to complex electronic structures with a large number of bands. The $P_{KL}$ hypothesis gives the possibility to discuss the localisation of the f electron notably its contribution to the Fermi surface. The pressure $P_C$ marks the disappearance of the long range magnetism. The usual consensus is that $P_C$ is a second order transition at the so-called quantum critical point (QCP). Evidences will be given that a first order transition may occur at $P_C$ . Furthermore if $P_{KL}$ and $P_C$ are two first order singularities, a phase separation may appear between magnetic and paramagnetic phases (see $CeIn_3$).

Also in HFC, the temperature $T_{cr}$ has droped and often vanishes to a negative value but the pressure $P_{cr}$ may be felt with large valence or orbital fluctuations (see $CeCu_2Si_2$, $CeCu_2Ge_2$). Our physical intuition is that it collapses with the pressure $P_V$ where the f electron looses its sensitivity to the crystal field environment since above $P_V$ its angular momentum becomes quenched already by the electronic Kondo coupling. In the case of cerium ions at low temperatures, below $P_V$, the effective spin of the 4f moment is 1/2 while above $P_V$ the full degeneracy J = 5/2 must be taken into account. That will wash out the intersite magnetic coupling and restores a situation of a strong mixing between the electrons without magnetic correlation. This intermediate valence regime corresponds to a Kondo temperature $T_K \geq 100$ K and $n_f < 0.9$ (see below).

Historically, the research on IVC was very active three decades ago. As it involves rather high energy ($T_K > 100$ K), the magnetism was very often ignored. The field of heavy fermion system starts with the discovery of the huge value of the linear temperature coefficient $\gamma \sim 1500$ mJmole $1K^{-2}$ of the specific heat $C = \gamma T$ in $CeAl_3$ (Andres et al 1975). The discovery of the first superconducting HFS ($CeCu_2Si_2$) was reported in 1979 (figure 2) (Steglich 1979). The importance of this observation was boosted by the successive reports of superconductivity in uranium compounds $UBe_{13}$ (Ott et al 1983), $URu_2Si_2$ ( Schlabitz et al 1986) and $UPt_3$ (Stewart 1984). The possible link of superconductivity with the magnetic instability at $T \rightarrow 0K$ i.e. to the critical density or pressure ($P_C$ ) was clear in the pioneering pressure experiments on $CeCu_2Ge_2$ (Jaccard et al 1992) and reinforced by the observation of superconductivity in $CePd_2Si_2$ and $CeIn_3$ (Mathur et al 1998). The direct evidence of heavy quasiparticles was realized in $UPt_3$ (Taillefer and Lonzarich 1988) ; effective masses m* up to 100 $m_o$ was detected ($m_o$ the free electronic mass). It was a major breakthrough as really it demonstrates that heavy fermion particles move on Fermi Surface orbits.

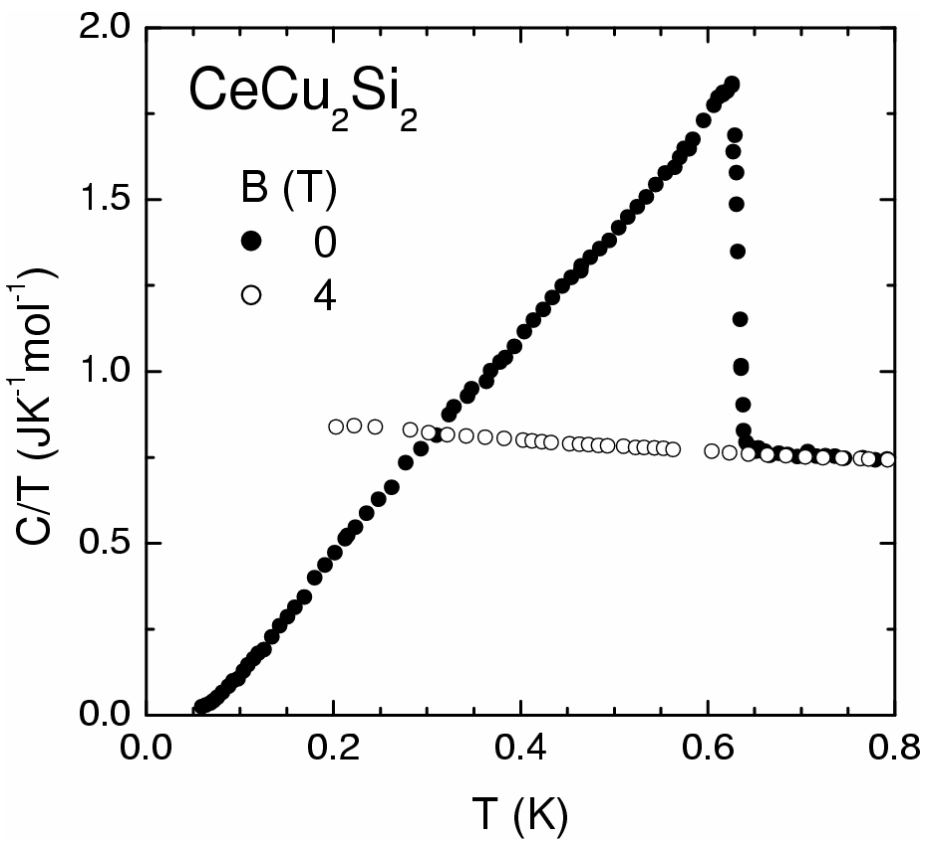

Figure 2 : Thermal variation of C/T of $CeCu_2Si_2$ at (•) H = 0 and (□) H = 4T above $Hc_2(0)$ (Steglich et al 1996, Helfrich 1995) on a superconducting crystal without A phase component. The complicated interplay between A and S phase is discussed in Gegenwart et al 1998.

## *1.3 - From Kondo impurity to Kondo lattice*

HFC are a complex matter where a large effective mass m* appears due to the slow motion of the f electron by hybridisation with the light electron. This magnification is reminiscent of the Kondo effect observed for a single magnetic impurity in a non magnetic normal host. Below a caracteristic temperature $T_K$, a strong coupling occurs leading locally to the disappearance of the magnetism. For $n_f \sim 1$ the coupling of the localized spin S and the spin of the conduction electron s can be reduced to an exchange term (see Blandin 1973).

$$H = -\Gamma \vec{S}.\vec{s}$$

$T_K$ is related to Γ and to the density of states of the light electrons $N(E_F)$ at the Fermi level

$$T_K = \frac{1}{N(E_F)} \exp - \frac{1}{\Gamma N(E_F)}.$$

The negative value of Γ is related to the position $E_0$ of the virtual f level relatively to the Fermi level and its width $\Delta_0$ according to the relation :

$\Gamma N(E_F) = \frac{\Delta_0}{E_0}$ when the on site Coulomb potential goes to infinity.

At low temperature, a Fermi liquid behavior replaces the high temperature Curie like paramagnetic behavior. As T → OK, the specific heat, the susceptibility and the resistivity vary as

$$\gamma = \frac{C}{T} = \frac{1}{T_K}$$

$$\chi \approx \frac{1}{T_K}$$

$$\rho = \rho_0 \left(1 - a\left(\frac{T}{T_K}\right)^2\right)$$

The resistivity reaches a maximum at T = 0 which corresponds to the unitary limit (~ 200 μΩcm per impurity for a metal with one carrier per mole) (see Anderson 1967). The beauty of the single Kondo impurity problem is that, due to the coupling with the Fermi sea of the electronic bath, the entropy can collapse as T → 0K without further coupling with other impurities.

A supplementary effect is that the local interaction mediated by the polarization of the Kondo singlet leads to the famous enhancement (the Wilson ratio R = 2 for S = ½) of the susceptibility χ over C/T with respect to the free electron value $R_0$ (see Hewson 1992, Nozières 1974) :

$$R = 2\,R_0 \text{ for } S = ½ \text{ with } R_0 = \frac{3\,(g\mu_B)^2}{4\pi^2 k_B^2}.$$

For 4f electrons, the spin orbit interaction $\lambda_{SO}\,\vec{L}\,.\,\vec{S}$ between the L = 3 orbital momentum and the spin S governs the formation of the magnetic momentum J carried by the particle. For less than half filled 4f shell, $\lambda_{SO}$ is positive and thus L and S are antiparallel. For the cerium ($4f^1$) case, the J = 5/2 level lies roughly 0.3 eV above J = 7/2 excited level. For the trivalent Yb ions ($4f^{13}$), the spin orbit interaction is negative, the ground state has an angular momentum J = 7/2. Due to the further coupling with the environment, the effective degeneracy of the level changes for $P < P_V$ ; often an effective doublet will be the crystal field ground state. The Coqblin Schrieffer (1969) model has been developed to take into account an orbitally degenerate site. It was used extensively to discuss the competition between Kondo effect (energy $k_B T_K$) and the crystal field splitting $\Delta_F$ (Cornut and Coqblin 1972). If $k_BT_K < \Delta_{CF}$, the low temperature Kondo impurity problem may correspond to the ideal case of a doublet. The proximity to the non magnetic unfilled f level of La explains the importance of hybridisation and thus of a Kondo mechanism. When $k_BT_K > \Delta_{CF}$, only the full 2J+1 = 6 degeneracy ($N_f$) of the 4f level of the trivalent configuration must be considered in all the temperature range. The magnetic interaction drops drastically compared to $k_BT_K$ (Ramakrishan and Sur 1982). The change from $k_BT_K < \Delta_{CF}$ to $k_BT_K > \Delta_{CF}$ is induced at $P = P_V$ under pressure ; as we will show, $T_K$ increases under pressure while the crystal field splitting is weakly pressure dependent (Thompson and Lawrence 1994, Schilling 1979). The relative strength of the Kondo temperature with respect to other energy scales such as the hyperfine coupling (Flouquet 1978), the intersite coupling (Doniach 1977), the pair interaction (Jones et al 1988), or the crystal field splitting is the key parameter to define the ground state properties. By comparison with the cerium impurity, the problem of the 3d Kondo impurity like Mn, Fe or Co is far more complex, as the nature of the magnetism far above $T_K$ involves already the difficult unsolved question of orbital quenching. The experimental paradox is that the study of the rare earth Kondo impurity has been undertaken much later than that on 3d elements and almost at the time when Kondo problem was solved theoretically (Wilson 1975, Nozières 1974, Yamada 1975, see Kondo et al 2005).

Let us stress the feedback of $T_K$ with the valence. The Kondo phenomena is linked with the release of $1-n_f$ from 4f shell to the Fermi sea. In the so called $1/N_f$ expansion, the Kondo energy has been expressed as a function of $n_f$ and $\Delta$ (Hewson 92)

$$k_BT_K = \frac{1-n_f}{n_f}\,\Delta_0\,N_f$$

A large $T_K$ corresponds to a low 4f occupancy. Under P, $n_f$ will decrease and thus $T_K$ will increase. To discuss the strong pressure dependence of $T_K$, the pressure variation of $\Delta$ must be known. It is not so obvious from first principles. In the Anderson Hamiltonian, the width $\Delta_0$ is connected to the mixing potential $V_{df}$ between the 4f and the d light electron and to $N(E_F)$ by the relation

$$\Delta_0 = \Pi V_{df}^2 N(E_F)$$

From high energy spectroscopy (Malterre et al 96), $\Delta$ increases under pressure as $k_BT_F$ the Fermi energy does. So let us assume the proportionality $\Delta = 10^{-2} k_BT_F$. The strong P response of $T_K$ is due to the weakness of $\Delta/k_BT_F \sim 10^{-2}$ by comparison to $Eo/k_BT_F \sim 0.2$. Neglecting the pressure shift of Eo towards the Fermi level, the Kondo formula with $\Delta = 10^{-2} k_BT_F$ leads to a Kondo Grüneisen parameter :

$$\Omega_{T_K} = -\frac{\partial \mathrm{Log} T_K}{\partial \mathrm{Log} V} = 10\ \Omega_{T_F}\ .$$

If the P shift of $E_o$ is taken into account, $\Omega_{T_K}$ will be again enhanced. Neglecting the degeneracy dependence in the expression (2), the physical insight is that the large value of $\Omega_{T_K}$ is linked to the quasitrivalence of the Ce ion ($n_f = 1$). A weak relative variation of $n_f$ magnifies the relative increase of $\partial T_K/T_K$ by $(1 - n_f)^{-1}$. Assuming that $n_f$ varies linearly in the volume according to the Vegard's law with a volume difference of 50% between the $Ce^{3+}$ and $Ce^{4+}$ configuration, $\Omega_{T_K}$ reaches 20 for $n_f = 0.9$. For cerium intermetallic compounds $n_f$ does not drop below 0.8 – 0.85 (Malterre el al 96). As the $4f^0$ and $4f^1$ configurations of the cerium atoms are separated by 2 eV, it will cost too much kinetic energy to drop further $n_f$ . For the Sm, Tm or Yb cases, the separation between the two valence states (2+ or 3+) are far less (100 meV) and thus the valence can vary by one.

Now for a regular array of Ce ions, the Kondo lattice, an extra temperature scale $T_{KL}$ may appear below $T_K$ as the electronic reservoir (carrier number $n_e$) cannot be regarded as an independent infinite bath and furthermore intersite magnetic interactions must be considered. For example neglecting extrasources of light conduction electrons than the release or absorption of an electron on the 4f electron, the valence equilibrium of $Ce^{3+}$ or $Yb^{3+}$

$Ce^{3+} \Leftrightarrow Ce^{4+} + 5d$

$Yb^{2} \Leftrightarrow Yb^{3+} + 5d$

leads already to a quite different Fermi temperature ($T_F$) for the 5d electrons (respectively proportional to $(1\text{-}n_f)^{2/3}$ and $n_f^{2/3}$) and thus to strong differences in $T_{KL}$ for the trivalent configuration if an extrapolation

$k_BT_{KL} \sim (1 - n_f) D(n_f)$ with $D(n_f) \sim 10^{-2} k_BT_F$ is made for the lattice ; $D(n_f)$ represents an unrenormalized 4f band proportional to the Fermi temperature of the 5d electrons. In this naïve frame, the travelling electron seems to be the 5d electron. It gets its effective mass $(1 - n_f)^{-1}$ on jumping from the 4f shell ; its Fermi temperature goes as $(1\text{-}n_f)^{+2/3}$ or $n_f^{+2/3}$ depending the valence equilibrium. The image is similar to the motion of x atoms of $He^3$ (x < 0.05) in $^4He$ liquid medium (see Lounasmaa 1974).

A full understanding of the low energy spectrum of the Kondo lattice even in the PM state has not been given. The difficult controversial point (see Bergmann 1991) is the so called exhaustion principle (Nozières 1985) which points out the impossibility to conserve a rigid picture of the Kondo screening on each site as the required number of available itinerant electron $N(E_F) \times k_BT_K$ is far lower than 1. Within a large $N_f$ approach in the frame work of the socalled slave boson technic, analytical calculations show that $T_{KL} \sim T_K n_e^{1/3}$ at low carrier content $n_e$ (Burdin et al 2000). The ratio $T_{KL}/T_K$ does not depend on the $T_K$ strength. That is not so surprising since magnetic correlations are treated roughly i.e the coherence length is restricted to atomic distances. For experimentalists, the message is that the relation between $\gamma$ and the number of 4f sites and carrier is not trivial when $n_e << 1$. For an intermediate value of $n_e$ (0.5), the differentiation between $T_{KL}$ and $T_K$ becomes difficult. Discussions on band filling effects on Kondo lattice inside a mean field approximation can be found in Coqblin et al 2003, with references to other approaches to Kondo lattice and the periodic Anderson model.

By contrast it was proposed (Nozières 1998) with considerations on phase memory that $T_{KL} \sim T_K^2$ independent of $n_e$. This relation can be found assuming the motion of the quasiparticles on a finite path $\ell_{KL}$ extending far above atomic distancies i.e basically in the vicinity of QCP. To take into account the motion of the heavy quasiparticule (m*) and their strong correlation, the simple step is to introduce an extra correlation length $\ell_{KL}$ . A physical image may be that the quasiparticle of effective mass $m^* \sim T_K^{-1}$ circulated along a Kondo loop of length $\ell_{KL}$ ~ m* leading to a lifetime $\tau_{KL} \sim m^{*2} \sim T_K^{-2}$ (in agreement with Nozières) and even to a small magnetic moment $M_o = \frac{m_0}{m^*} \mu_B$ on each visited site ($m_o$ being the bare electronic mass). For a classical magnetic ordered rare earth compound $\left(\frac{T_K}{T_N} \rightarrow 0\right)$, the length of the magnetic correlation at T = 0K correspond to atomic distancies and the information is carried from site to site by fast light electrons with $m^* = m_o$ . Thus a $10^4$ or $10^6$ longer time constant for HFC ($\tau_{KL} \sim m^{*2}$) will be the results of a slow motion on large distancies.

## *1.4 - The "Doniach model"*

A first discussion on the P collapse of long range magnetism in HFC was given by Donach (Doniach 1977). In the popular Doniach picture, the interplay is between a local Kondo fluctuation given by $k_B T_K$ and the indirect Ruderman Kittel Kasuya Yoshida oscillating interaction (RKKY) $E_{ij}$ between two paramagnetic sites i and j mediated by the conduction electrons (Doniach 1977) :

$$E_{ij} \sim \Gamma^2 N(E_F)$$

As $T_K$ has a strong exponential dependence on Γ and $E_{ij}$ a smoother parabolic dependence, the first idea is that above a critical value of $\Gamma_C$, long range magnetism either ferromagnetic (FM at $T_{Curie}$) or antiferromagnetic (AF at $T_N$ ) will collapse at $P_C$ (figure 3). If the collapse is continuous through a second order transition, $P_C$ corresponds to a quantum critical point. Such a scenario has been discussed (see later) by Hertz (1976) and revisited in the case of spin fluctuation theory (Moriya 1985, 1995 and 2003-a), renormalization group theory (Millis 1993) or in the framework of universality in phase transitions (Continentino 2001).

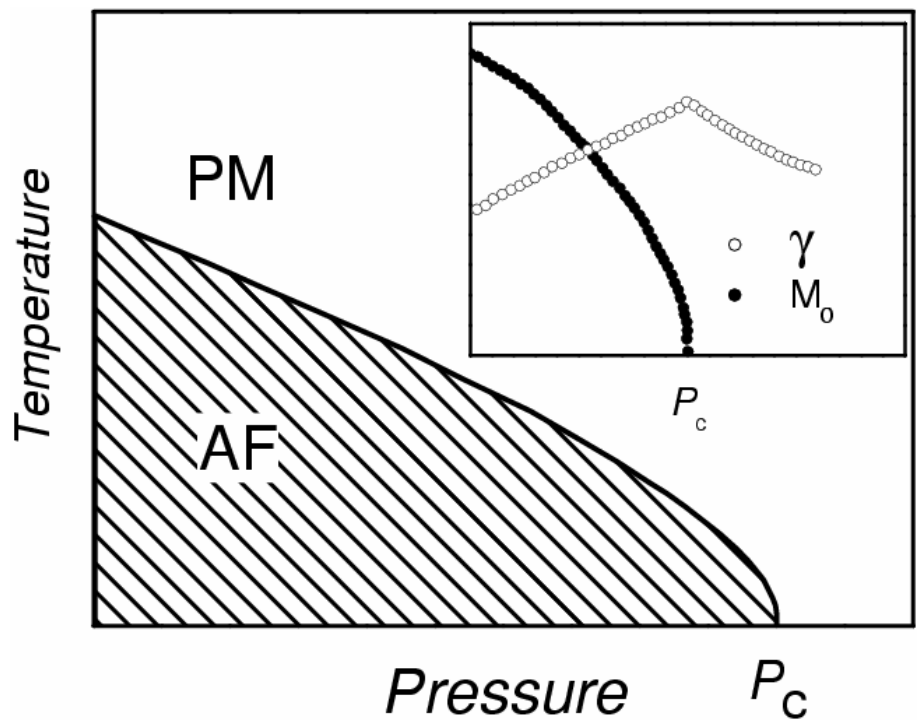

Figure 3 : Phase diagram (T, P at H = 0) (Benoit et al 1979). In this model, the magnetic transition at $T \rightarrow 0K$ is first order with $T < T_K$. In full line dependence of $T_N$ , in (●) of $M_o$ and in (○) of γ.

From a purely phenomenological point of view, the Doniach approach can be related to the discussion of induced magnetism for an array of initially singlet paramagnetic ion caracterized by a deep minimum of the energy $E_i$ at zero local magnetization ($m_i$) ($E_i = a\, m_i^2$) plus a further interaction term b $m_i.m_j$ taking into account first and second neighbour coupling (Benoit et al 1978, 1979). Of course magnetism disappears for a critical value of a/b i.e $T_K$ as shown on figure 3. The problem is quite similar to that discuss for the appearance of magnetism for a singlet crystal field ground state (Wang and Cooper 1968 and 1969).

Assuming that, on the Fermi Sea, the electronic properties can be derived by a lorenztian density of state :

$$\gamma = \frac{k_B\, T_K}{(k_B T_K)^2 + (g\mu_B H_m)^2}$$

where $H_m$ is the molecular field, g the g factor and $\mu_B$ the Bohr magneton ; a maxima of γ i.e. of the effective mass will occur at $P_C$ adjusting the relative variation of a and b with the pressure dependence of $T_K$ and $E_{ij}$ (Benoit et al 1979). The extension of the Doniach model to the magnetic field gives an excellent description of the experiments realized on the magnetic Kondo lattice of $CeAl_2$ (Steglich et al 1977 and Bredl et al 1978) at P = 0.

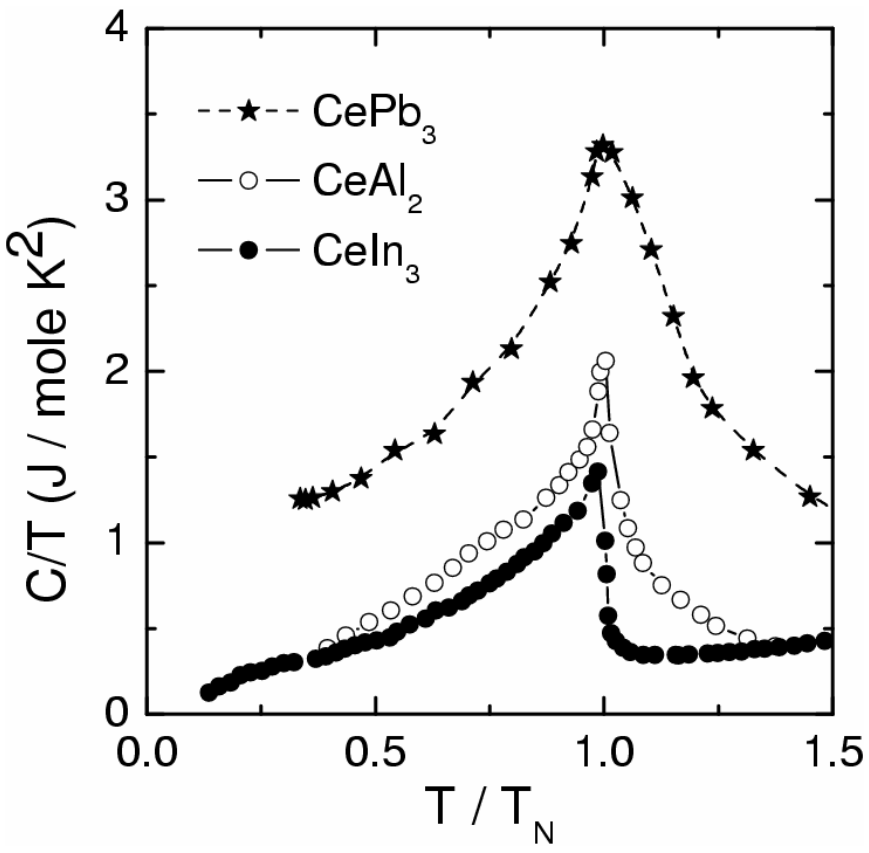

Figure 4 : Specific heat of three magnetic ordered compounds : $CeIn_3$, $CeAl_2$ and $CePb_3$ as a function of $\frac{T}{T_N}$ (Bredl et al 1978, Peyrard 1980, Pietri and Andraka 2000).

Let us look to some magnetic Kondo lattices. Figure (4) represents the variation of C/T in the normalized scale $T/T_N$ for three cubic magnetic ordered Kondo lattice $CePb_3$ , $CeAl_2$ and $CeIn_3$ (Steglich et al 1978, Peyrard 1980, Lin et al 1985, Pietri and Andraka 2000). The compound $CePb_3$ (Vettier et al 1986) as well as $CeAl_2$ (Barbara et al 1979) exhibit incommensurate magnetic Bragg reflections which are stable down to 60 mK. The critical exponent associated with the order parameter M (the sublattice magnetization) is the Wilson value 0.3 for the 3D Heisenberg antiferromagnetic whereas in $CeIn_3$ it is near 0.5 the mean field limit. The trend is that, when the ratio $T_K/T_N$ increases by applying pressure, the magnetic Bragg reflection is commensurate with the same propagation vector (1/2, 1/2 , 1/2) (Morin et al 1988). The refinement of the magnetic structure of $CeAl_2$ on a single crystal (Forgan et al 1990 ) shows that it is not a modulated structure with a single $k_0$ component but a double k structure involving the composition of two helicoidal modes. The Kondo coupling is invoked for the P collapse of $M_o$ . The understanding of the magnetic structure of this SCES is an interesting topic ; an overview of the different facets of exotic structures completely determined by neutron scattering can be found in the review article of Rossat Mignod (1986).

In the two cases of $CeIn_3$ and $CeAl_2$ the estimations $\gamma_{LT}$ and $\gamma_{HT}$ of C/T far below $T_N$ and just above $T_N$ are quite similar while for $CePb_3$ $\gamma_{LT} >> \gamma_{HT}$.

| | $T_N$ in K | $T_K$ in K | $\gamma_{LT}$ $mJmole^{-1}k^{-2}$ | $\gamma_{HT}$ $mJmole^{-1}k^{-2}$ | $M_0^{exp}$ $\mu_B$ / mole | $M_0^{th}$ $\mu_B$ / mole |
|---|---|---|---|---|---|---|
| $CeAl_2$ | 3.8 | 10 | 120 | 150 | 0.53 | 0.89 |
| $CeIn_3$ | 10 | 10 | 140 | 140 | 0.55 | 0.48 |
| $CePb_3$ | 1.2 | 2 | 1300 | 200 | 0.50 | 0.6 |

At least in the two first cases, the establishment of the AF phase seems decoupled from the existence of heavy quasiparticles ($\gamma_{LT} = \gamma_{HT}$). Furthermore, the decoupling looks efficient as an evaluation of $M_0^{th}$ in a classical model with localized moments (susceptibility at $T_N$ equal to $\chi(T_N)$) gives a rather good estimation of the measured value $M_0^{exp}$ according to the relation (Marcenat et al 1988) : $M_0^2 = 2\chi(T_N)\int_0^{T_N} C(T)\,dT$ even for $CePb_3$.

For $CePb_3$, there is a drastic change in C/T through $T_N$. Two components seem to exist : the ordinary magnetic one and the heavy fermion one. Coherence (crossing through $T_N$ ) leads to reach rapidly the low temperature limit of C/T ($\gamma_{LT} > \gamma_{HT}$). In paramagnetic HFC above $P_C$ , the increase of C/T on cooling from $\gamma_{HT}$ to $\gamma_{LT}$ will occur through a continuous slow process over a large temperature range. One may think that $CePb_3$ at P = 0 is near a QCP and thus few kbar will drive the system right to the QCP. That seems supported by an initial P decrease of $T_N$. But the reality is different. At P* = 5 kbar, $T_N$ increases again. The magnetic structure becomes commensurate. The QCP is pushed above 3 GPa (Morin et al 1988, U. Welp 1987-1988). HFC are rather subtle toys with a lot of different possibilities. Notice that even far below $P_C$ > 3 GPa at P = 0, in its AF phase, $CePb_3$ has a $\gamma$ term near 1000 $mJmole^{-1}K^{-2}$.

## *1.5 - Spin fluctuations and the non Fermi properties*

In the almost opposite framework where the f electrons are considered to be completely itinerant ($P_{KL}$ < $P_C$ ) i.e. characterized by an effective Fermi temperature $T^*_F \sim T_K$ it is worthwhile to refer to the results of spin fluctuation theory, developed to understand the magnetism of 3d elements (Moriya 1985, Lonzarich and Taillefer 1985) and recently revisited in connection with heavy fermion, organic and high $T_C$ oxyde compounds.

For simplicity, let us look at the case of the ferromagnetic instability (figure 5) (Moriya 1985, Nozières 1986). For $P < P_C$ the ground state is ferromagnetic. Slowly, on approaching $P_C$, the effective mass m* will be dressed by spin fluctuations. At $P_C$, m* will diverge. In the Hubbard scheme, this will occur when the product UN ($E_F$) = I of the on site coulomb repulsion U by the density of state N ($E_F$) at the Fermi level reaches 1 (m* = log(1-I)) (Moriya 1985). Far below $P_C$ , undamped spin waves can be detected. On approaching $P_C$ , they become overdamped. The regime where Fermi liquid properties can be observed will be pushed to the low temperature $T_I$ which collapses at $P_C$. Above $T_{Curie}$, there is a large regime (III) where the paramagnetism has strong variations in temperature. The uniform susceptibility $\chi_0$ followed a $T^{-4/3}$ law. For the singular pressure $P_C$ , the collective singlet never enters into a low temperature Fermi liquid regime. That leads to the non Fermi liquid (NFL) label (see recent review of Stewart 2001).

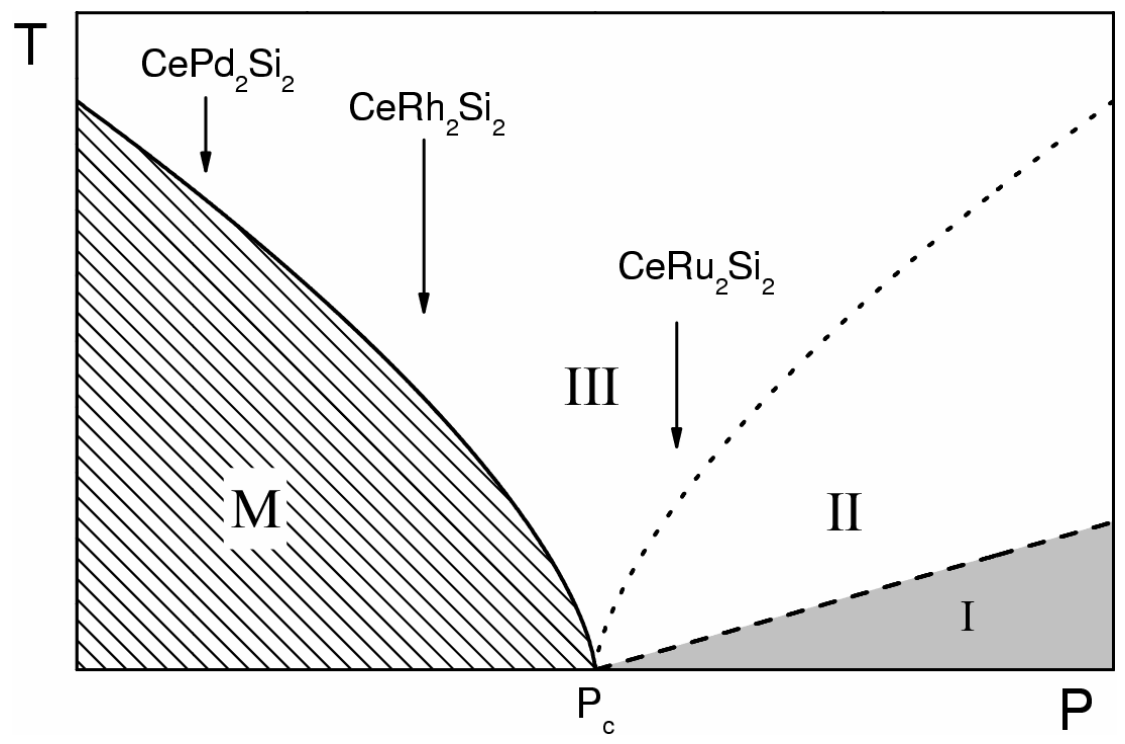

Figure 5 : Magnetic phase diagram predicted for spin fluctuation. In the domain I, Fermi liquid properties will be achieved. In the domain II and III, non Fermi liquid behavior will be found. The location of three HFC described in the text is shown at P = 0. Depending on the nature of the interactions (AF or F), the contour lines I, II, III change. Here the contour is drawn for antiferromagnetism. In case of ferromagnetism, the $T_I$ line starts as $(P - P_C)^{3/2}$.

For $P > P_C$ , the FL regime is reached only at very low temperature below $T_I \sim T_F (1-I)^{3/2}$. Between $T_I < T < T_{II}$, a large crossover regime appears ($T_{II} \sim T_F (1-I)^{3/4}$) before recovering on warming the domain III. As the ratio of $T_{II}/T_I$ diverges as $(1-I)^{-3/4}$ when $I \to 1$, the crossover regime II covers a wide relative temperature range. Near $T_I$, for three dimensional (3d) nearly ferromagnets metals, C/T varies now as LogT in region II ; for a three dimensional nearly antiferromagnet the leading term will go as $\sqrt{T}$ . The interesting concept is that the proximity of quantum criticality will be felt by the electron already at high temperature ($T >> T_I$) via its non Fermi liquid behavior i.e already at $T >> T_I$ (figure 5). The electron "knows" its destiny to be or not to be near $P \sim P_C$ . This statement must be true for simple elements as cerium metal or later ε Fe. The Ce and Yb HFC will give nice illustration on chapter 2 and 3.

Table (1) (Moriya 1987 – 2003a) indicates, for 3d ferromagnetic or antiferromagnetic systems, the respective pressure dependence of $T_I$, $T_{II}$, $T_{II}/T_I$, the ordering temperature $T_{Curie}$ or $T_N$ and their relation with the extrapolated sublattice magnetization $M_o$ at $T \to 0K$ assuming $(1\text{-}I) \sim (P\text{-}P_C)$.

| Table 1 | $T_I$ | $T_{II}$ | $T_{II}/T_I$ | $T_{Curie}$ ou $T_N$ | $T_C$ ($M_0$) |
|---|---|---|---|---|---|
| F | $(P - P_C)^{3/2}$ | $(P - P_C)^{3/4}$ | $(P - P_C)^{-3/4}$ | $(P_C - P)^{3/4}$ | $m_0^{3/2}$ |
| AF | $(P - P_C)$ | $(P - P_C)^{2/3}$ | $(P - P_C)^{-1/3}$ | $(P_C - P)^{2/3}$ | $m_0^{4/3}$ |

For 3d systems the extrapolated value of γ, the susceptibility $\chi_Q$ at the ordering wave vector Q and the average amplitude A of the $T^2$ inelastic term of the resistivity ρ, as P decreases to $P_C$ are given in table 2 (Moriya 2003a) :

| Table 2 | $\gamma$ | $\chi_Q$ | A |
|---|---|---|---|
| F | $\text{Log}(P\text{-}P_C)$ | $(P - P_C)^{-1}$ | $(P\text{-}P_C)^{-1}$ |
| AF | $\gamma 0 - \text{const} \sqrt{(P - P_C)}$ | $(P - P_C)^{-1}$ | $(P - P_C)^{-1/2}$ |

The predictions of their quantum critical behaviors with temperature (plus the nuclear relaxation time $T_1$) for the 3d and 2d case (Moriya 2003-a) are :

| Table 3 | | C/T | $\chi_Q^{-1}$ | $\rho \sim T^n$ | $T_1^{-1}$ |
|---|---|---|---|---|---|
| F | 3d | $- \ell n\, T$ | $T^{4/3} \to CW$ | $T^{5/3}$ | $T\chi$ |
| | 2d | $T^{-1/3}$ | $-T\, \ell n T$ | $T^{4/3}$ | $T\chi^{3/2}$ |
| AF | 3d | $T^{1/2}$ | $T^{3/2} \to CW$ | $T^{3/2}$ | $T\chi_Q^{1/2}$ |
| | 2d | $-\ell n\, T$ | $- T/\ell n T$ | $T$ | $T\chi_Q$ |

The phenomenological model developed by Moriya (self consistent renormalization theory : SCR) allows to classify the different systems with 4 parameters : two dimensionless parameters $Y_0$ and $Y_1$ (Moriya and Takimoto 1995), two characteristic temperatures $T_0$ and $T_A$. $Y_0$ is directly proportional to the inverse of the staggered susceptibility $\chi_Q^{-1}$ at $T \rightarrow 0$ ; $Y_1 = \frac{J_k}{\Delta J_k}$ is the ratio of the exchange at k = Q and its dispersion $\Delta J_K$ in k wavevector space. $T_0$ is related to the frequency response and $T_A \sim \Delta J_K$ is linked to the exchange wavevector dispersion. Figure (6) (Kambe et al 1997) shows the comparison between the experiments and the SF fitting in a reduced temperature scale $t = T/T_0$ for the temperature variation of C/T of three typical HFC : $CeCu_6$, $CeNi_2Ge_2$ and $CeRu_2Si_2$ as well as two cases where $P_C$ is approached by doping $CeCu_{5.9}Au_{0.1}$ and $Ce_{0.925}La_{0.075}Ru_2Si_2$. The corresponding parameters are :

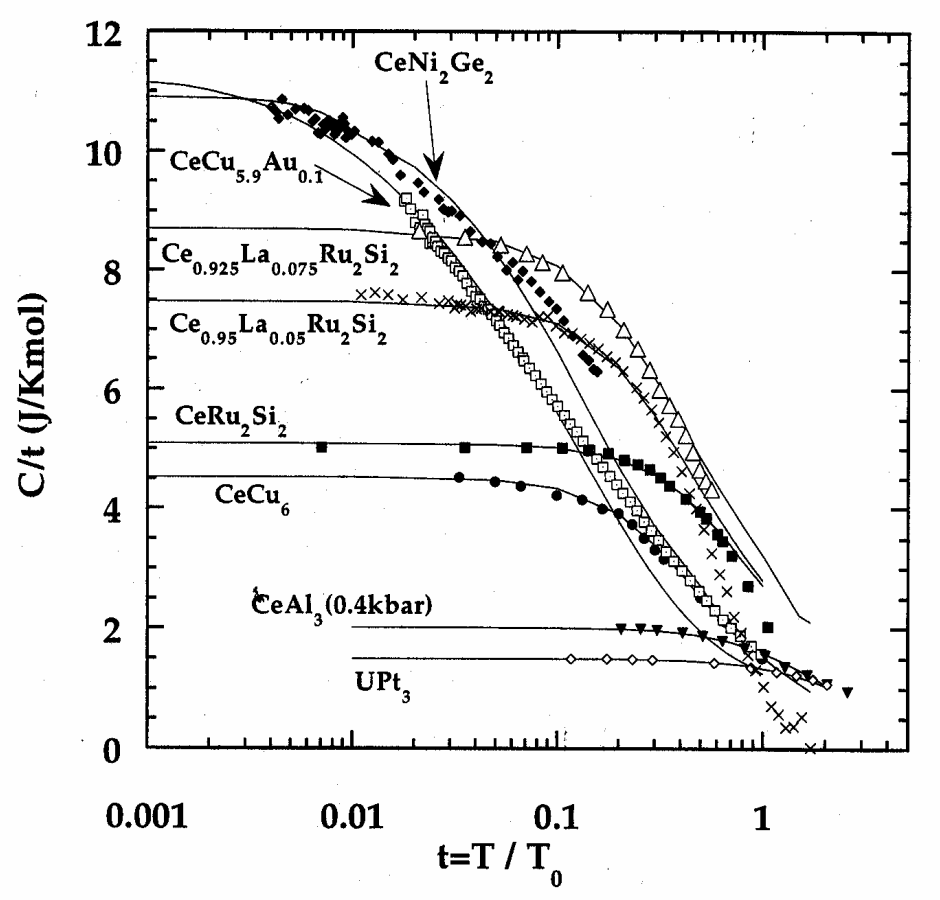

Figure 6 : Comparison between C/t data (symbols) and the SF-model (solid lines) in various heavy electron compounds. Only some experimental points are presented for clarity and a normalized temperature scale $t = T/T_0$ is adopted ($T_0$ for some compounds is presented in table 4). (Kambe et al 1996).

| Table 4 | $Y_0$ | $Y_1$ | $T_0$(K) | $T_A$(K) |
|---|---|---|---|---|
| $CeRu_2Si_2$ | 0.31 | 1.6 | 14.1 | 16 |
| $Ce_{0.925}La_{0.075}Ru_2Si_2$ | 0.05 | 0.77 | 14.7 | 14 |
| $CeCu_6$ | 0.4 | 10 | 3 | 5.5 |
| $CeCu_{5.9}Au_{0.1}$ | 0.003 | 16.7 | 3.4 | 6.7 |
| $CeNi_2Ge_2$ | 0.007 | 7 | 29.7 | 94 |

The coefficient $Y_1$ is far smaller for $CeRu_2Si_2$ (1.6) than in $CeCu_6$ (10) and $CeNi_2Ge_2$ (7) : a more pronounced q structure in $\chi''(q, \omega)$ is observed for $CeRu_2Si_2$ . Rather similar values of $T_0(K)/T_A(K)$ are found for $CeCu_6$ and $CeRu_2Si_2$ . The low temperatures ($T_0$, $T_A$) of $CeCu_6$ combined with the large $Y_1$ parameter explains why $CeCu_6$ may correspond to another situation than $CeRu_2Si_2$ .For $CeNi_2Ge_2$, $T_A(K)/T_0(K) \sim 3$ differs greatly from 1. We will see later that two characteristics energies (4mev and 0.6 mev) characterize the magnetic fluctuations of $CeNi_2Ge_2$.

Let us stress that, by comparison to other SCES systems, the particular interest of heavy fermion systems is that the Néel temperature and the Fermi temperature are comparable and already low (a few Kelvin) far below $P_C$ . Furthermore small shifts in pressure of the bare parameter ($E_0$, $\Gamma$, $N(E_F)$) will be magnified by large shifts of the effective temperatures ($T_N$ , $T_F$, $T_I$, ….) . The corresponding P derivatives $\frac{\partial T_N}{\partial P}, \frac{\partial T_I}{\partial P}, \frac{\partial T_K}{\partial P}$, are enhanced by two order of magnitude (huge Grüneisen parameter). This leads to huge effects on thermal expansion, sound velocity and magnetostriction according to the Maxwell relations. It gives the unique opportunity to scan through $P_C$ by moderate pressure variations (a few GPa) or to sweep through the anomalies of the density of states with a magnetic field H of few tesla. Such conditions are almost unique in condensed matter.

Qualitative agreement with the spin fluctuation theory (SF) is found when other quantities like the dynamical susceptibility or the resistivity are calculated in this framework. The diversity of the HFC in the initial conditions leads to quite different values of $Y_1$ which is related to the problem of the localization of the interaction. In a first approximation, spin fluctuation theory gives a good description of the low energy excitation. Quantitatively, there are discrepancies like the C/T = LogT crossover law observed over a large temperature range (von Löhneysen et al 1994-2000) which is not predicted for 3d itinerant antiferromagnets. Different routes have been proposed : reduction of the dimensionality (3d $\rightarrow$ 2d), occurrence of distributions for the Kondo temperature linked to disorder (Miranda et al 1997, Castro Neto and Jones 2000) or valence fluctuation or drastic change of the Fermi surface's volume (Coleman 1999). For example for $CeCu_{6-x}Au_x$ it was proposed that three dimensional conduction electrons are coupled to two dimensional critical ferromagnetic fluctuations near the quantum critical point $x_c = 0.1$ (Rosch et al 1997). A picture of dynamical heterogeneities (Bernhoeft 2001) invokes different distributions of space and time motions. Of course, broad time and space responses can cover a large diversity of phenomena. This approach can only be a first step to visualize the occurrence of different lengths and times before entering in FL regime.

Let us mention that a simple form has been proposed to describe the temperature variation of the specific heat and the susceptibility of HFC. The high temperature Kondo gas of non interaction Kondo impurities condenses slowly into a heavy electron Kondo liquid of Wilson ratio = 2 with a fraction f. The analysis on the cerium 1.1.5 serie gives that f increases linearly with T on cooling before saturated at a fixed value near 0.9 for $P = P_C$ . The resistivity is dominated by the fraction 1-f of isolated Kondo impurities. For $P << P_C$ , deep inside the AF state f = 1 (Nakatsujii et al 2003, see application to NMR analysis by Curro et al 2004). This continuous two component description assumes that the critical point ($T_{cr}$ , $P_{cr}$ ) will be never achieved ($T_{cr} < 0$). It was stressed that both components have protected behavior according to the "classification" made by Laughlin and Pines (2000) on the "theory of everything".

## *1.6 - Quantum phase transition*

In the Doniach or SF pictures, a second order QCP will mark the transition from long range magnetic ordering to the paramagnetic phase at T = 0K. As it must correspond to the simultaneous collapses of the specific heat and thermal expansion anomalies for $T_N \rightarrow 0$, the initial slope $\frac{\partial T_N}{\partial P}$ can have any finite value. The magnetic coherence length $\xi_m$ will diverge at $T_N$ . For example, in AF-SF theory (Hasegawa and Moriya 1974, Makohsi and Moriya 1975) :

$$\xi_m(T) = \xi_m^o \left[1 - \left(\frac{T}{T_N}\right)^{3/2}\right]^{-1/2}$$

and the magnetic coherence length $\xi_m^o$ diverges on both sides of $P_C$ at T = 0K.

On the AF site, SF theory predicts :

$$\xi_m^o = a\left(\frac{T_0}{T_N}\right)^{-3/4}$$

the jump of the specific heat anomaly collapses with $T_N$ (Zülicke and Millis 1995). In many HFC, the initial slope $\frac{\partial T_N}{\partial P}$ is steep. We will see that the occurrence of a second order QCP is certainly not a general rule. When the transition becomes first order, there will no longer be a divergence of $\xi_m$ ($T_N$ ) but a jump.

The field of quantum phase transition has attracted the interest of both experimentalists and theorists since even if the transition occurs at T = OK, it can govern the physical properties over a wide range of (T, P, H) phase diagram. A quantum phase transition corresponds to competing ground states which may be changed by external variables as pressure or magnetic field. If the crossing corresponds to a second order transition at a quantum critical point ($P_c$ or $H_c$), an universal behavior will appear depending on the initial spin and charge dimensionality.

Quantum fluctuations destroy the long range order at T = 0K in a different way than thermal fluctuations since now statics and dynamics are mixed. Within the renormalization group approach (Hertz76, Millis 93) the effective dimension of the system $d_{eff}$ = d + z couples the geometrical dimension d and the exponent z characteristic of the dynamic (respectively 3 or 2 for ferromagnetic and antiferromagnetic dimension). As $d_{eff}$ = 5 for 3d itinerant antiferromagnet is higher than 4 the upper critical dimension, molecular field treatment (as done in previous SF approaches) may describe the experiments. Two important points must be verified in a microscopic neutron scattering experiment :

- the singular temperature dependence of the static susceptibility $\chi'(k_0,T)$ at the ordered wavevector $k_0$ by comparison to the uniform one,

– the absence of scaling in ω/T of the dynamical susceptibility $\chi''(k,\omega)$ since due to coupling among paromagnons the data must follow a $\omega/T^\beta$ law with β > 1.

As we will see in chapter for $CeCu_{6-x}Au_x$, these SF predictions seem to fail. Previous arguments may be reconsidered. (Coleman 1999) It was suggested a scenario called local quantum criticality. The term local emphasizes that surprisingly $\chi$'(O,T) and $\chi$'($k_0$,T) have the same temperature dependence. This possibility was supported by Si and co-workers (Si et al 2001, 2003a-b) in the treatment of the Kondo lattice for the specific case of 2 d quantum magnetic fluctuations. However the introduction of an extra tridimensionnal fluctuation leads to recover the classical spin fluctuation picture. Numerical identification of a locally quantum critical point is also found for Ising anisotropy and again 2d magnetic correlations. (Grempel and Si 2003 Zhu et al 2003-a). As the phenomena involves all the wavector, it was proposed that it may be associated with a FS melting with the image that the FS is small on the AF magnetic side and large on the PM side with a divergence of the effective mass (Coleman 1999) Quite different conclusions were given by Pankov et al 2004 and Sun and Kotliar (2003). At least, there is yet no consensus on the relevance of quantum local criticality. In the standard SF framework, FS is assumed to be large and conserved in all pressure range ($P \gtrless P_c$ i.e $P_{KL} \rightarrow o$).

The possibility that the main fluctuations responsible for the non Fermi liquid behavior may come from the destruction of the large Fermi surface of the PM state was supported using insights from the theory of deconfined quantum criticality of insulating antiferromagnets (Senthil et al 2004 a and b). In our classification, it corresponds to $P_{KL} = P_c$. The experimental implifications will be a weak moment magnetism due to low energy instability of the small Fermi surface and spinon excitations of the AF state. Spinon extraexcitations must enhance the thermal transport by comparison to the electric one leading to a violation of the Wideman Franz law.

To test these recent theoretical proposals is of course the future experimental challenge. In our view, an important point is the proximity of $P_V$ from $P_C$ which will favor a first order transition. It is even amazing to remark than in the quantum Monte Carlo approach with extended dynamic mean field theory, the finite temperature magnetic transition is first order while the extrapolated zero temperature magnetic transition on the other hand is continuous and locally critical (Zhu et al 2003a). In many cases the critical point is weakly first order with the consequences of large fluctuations but with finite values of the low energy excitations and coherence length at $P_c$. As we will see, often the QCP does not look to be governed only by the spin dynamics but is also associated with an electronic change.

Any first order transition, whatever the discontinuity in volume at T = 0K, may have drastic effects. The universality may be lost since the volume jump (or drop) leads to a non perturbative mechanical work $P\Delta V$ which implies a temperature warming. An interesting discussion is the large P spreading which may occur between moving away from the first order transition $P_C$ and its disappearance at $P_{+C}$. This type of problem was addressed in the lecture notes of Levanyuk (2001) for a structural transformation with the prediction that $P_{+C}$ is the pressure predicted in the frame of Landau Ginzburg theory. But, as he emphasizes, spin matter will offer a large diversity. Depending on microscopic figures (defect, mismatch between them) a large pressure range may occur before the usual behavior after a second order phase transition will be recovered. New magnetic matter may appear as a Griffiths phase or glass state or phase separation with a continuous variation in the mixing between antiferromagnetism and paramagnetism : f = 1 at $P_C$ and 0 at $P_{+C}$. Recent discussions on the quantum phase transition can be found in the review of Vojta (2003) in the recent survey by Continentino (2004-a). A general reference is Sachdev (1999).

It is worthwhile to realize that the discontinuity $\Delta V$ of the volume gives rise to significant mechanical work (W). A ppm or 100 ppm volume variation for a molar volume $V = 40$ cm$^3$/mole leads respectively to $W = 4 \times 10^2$ and $4 \times 10^4$ erg. If this work is absorbed by thermally isolated heavy quasiparticles of $\gamma$ = J mole$^{-1}$K$^{-2}$, that will cause respective warmings to T = 10 mK and 100 mK from T = 0K. Of course, for an isolated system, as the entropy discontinuity between the two phases drops at $T \rightarrow 0$K, the warm up will occur up to recover a significant entropy drop between the two phases. It happens that W on $He^3$ on its melting curve at $P_o \sim 34$ bar and $T \rightarrow 0$K can be quite comparable to that on HFC at its first order transition at $P_C$ . For a HFC case of $\frac{\Delta V}{V} \sim 2.5x10^{-5}$, V = 48 cm$^3$/mole and $P_C$ = 30 kbar, W is the same than that of $He^3$ matter where $\frac{\Delta V}{V}$ = 5%, V = 24 cm$^3$:mole $P_o$ = 34 bar (see Lounasmaa 1988).

Experimentally evidences of first order transition in HFC comes from finite value of the characteristic fluctuation at T = 0K for $P \sim P_C$ (chapter 2) ; from strong departure of the specific anomalies at $T_N$ ($T_N \rightarrow 0$ K) from the molecular field prediction and from the observation of a coexisting pressure range of PM and AF phases (chapter 4). Often, the transition appears as a weakly first order transition. So the expected relative discontinuity in volume at $P_C$ may be near or below $10^{-7}$. To our knowledge, no HFC with a second order ferromagnetic QCP has been found. Theoretical arguments for first order transition at the F instability can be found in Belitz et al (1999) and Chubukov et al (2003). The first evidence on an inhomogeneous medium and thus a phase separation below the ordering temperature in related HFC was given in the NMR work of Thessieu et al (1998, 1999) on MnSi. Such a possibility was rediscovered recently by Doiraud et al (2004), Pfleiderer et al (2004), Yu et al (2004). The occurrence of inhomogeneous intrinsic matter in HFC may belong to a class of phenomena like the geometric order of stripes proposed for high $T_C$ materials (see Zaanen 2001) or the formation of droplets.

## *1.7 - Fermi Surface/Mass enhancement*

In a crystal without Galilean invariance, the FS cannot grow without feeling the Brillouin zone : different bands will occur. Taking into account all possible mobile electrons, the band picture is often that of a spaghetti plate containing more than 10 bands. A large number of orbits need to be detected to discuss the full FS issue. Due to the large number of orbits the distinction between small ($P < P_{KL}$ ) and large Fermi surface is partly misleading. Furthermore the magnetic order creates often new Brillouin zones.

Generally whatever the localisation, the f electron has, the FS are not small with respect to the volume of the Brillouin zone. Far below $P_C$ , one must recover the situation of a normal rare earth compound with the 4 f electron localized and the Fermi sea given by the content of the other electrons. For a well ordered compound like $CeAl_2$ or $CeRu_2Ge_2$ ($P << P_C$ ) their Fermi surface is that calculated for the isostructural lanthanium host i.e. $LaAl_2$, $LaRu_2Ge_2$ ( a small FS in a single band model) (Lonzarich 1988). There are evidences in a system like $CePd_2Si_2$, with $T_K \sim T_N$ at $P = 0$, that the Fermi surface has some itinerant character (Sheikin et al 2003).

On the paramagnetic side, dHvA experiments data are well understood assuming the 4f electron itinerant (chapter 2 and 4). This observation follows either arguments based on the 1D Kondo lattice (Tsunetsugu et al 1988) or the invariance of the Fermi surface volume with the interactions. When the Coulomb repulsion is switched on in the Anderson Hamiltonian (Fazekas 1999). The Fermi surface is predicted to be large. With the new conjectures of local criticality or deconfinement, a new generation of quantum oscillations experiments are underway. To know the present stage, the reader can refer to the works made in Osaka by the group of Onuki (Onuki et al 2003).

On warming one may reach a regime where the 4f electron looses its itineracy before recovering its single impurity $T_K$ behavior. Above $T_K$ the 4f electron plays the role of a paramagnetic Kondo center for the light electronic band of the lanthanium isostructural compound (LaRu2Si2 for $CeRu_2Si_2$ ). Quantum oscillation and high spectroscopy experiments on $CeRu_2Si_2$ ($P > P_{KL}$ ) show the crossing through $T_K$ at P = 0. At $T \rightarrow 0K$ closed to $P_C$, the Fermi surface of $CeRu_2Si_2$ measured by quantum oscillation and derived in band calculations corresponds to an itinerant 4f electron. However on warming (high spectroscopy response), the 4f electrons are localized. (see chapter 2).

A method which takes into account the renormalization was successfully applied to HFC via phenomenological adjustable parameters (Zwicknagl 1992). The FS topology is not affected by strong local correlations ; its contour is already well defined in band calculations with weakly correlated electronic bands. The strong correlations lead to a large effective mass and corresponding anisotropy. The renormalized band approach was successfully applied to $CeRu_2Si_2$ .

Recent progresses have been made in the treatment of the correlation (including even the feedback to the choice on the crystal field arrangement) using the so-called dynamic mean field theory (DMFT) (Georges et al 1996). The combination of DFMT with electronic structure methods is very promising (see Georges 2004). Already application have been made for Ce phase diagram (Anadon et al 2005, Sakai et al 2005). For example, in this last work it has been found that at P = 0 in the $\gamma$ phase, the crystal field splitting due to the hybridisation is 250 K.

A particular situation must appear of course when the number of carriers $n_e$ is just equal to the number of magnetic sites $n_m$ or when the balance between 2 valence configurations releases a light electron. That leads to the prediction of the Kondo insulator (Jullien et al 1979). However in many cases of HFC no insulating phase appears when the magic ratio seems to be achieved as in $CeB_6$ or CeTe. Insulating phases exist in the Sm or Tm chalcogenides in the intermediate valence regime but collapse when the valence approaches three. Kondo insulators found in intermetallic compounds seem to correspond to specific conditions, as found in ordinary metals with an even number of electrons (Takabatake et al 1998, Aeppli and Fisk 1992). We will focus mainly on metallic material with ($n_m$+$n_e$) not integer.

We will take a pedestrian view on considerations on the mass enhancement. The common point with the previous view is that the magnetic intersite interaction in HFC does not modify drastically the mass enhancement originated from Kondo local fluctuations. Different mechanisms are responsible for the dressing of the effective mass m* relative to $m_o$ the bare electron mass. Schematically, one can invoke the renormalization $m_K/m_o$ by strong local fluctuations ($T_K$ or $T_0$ in the Kondo lattice or spin fluctuation approach) and the further renormalisation $m^*/m_K$ due to spin fluctuation or Kondo lattice enhancement. Studies under pressure through $P_C$ and $P_V$ (Brodale et al 1985), under magnetic field (through metamagnetic transition) (Flouquet et al 2002) and also the analysis of the upper critical field $Hc_2$ (T) of the superconducting state (simultaneous fit of the effective mass and the strong coupling constant λ, $\left(\frac{m^*}{m_K} = 1 + \lambda\right)$ can give an evaluation of the respective weights of $m^*/m_K$ and $m_K/m_o$. For HFC cases where AF fluctuations dominate, $m^*/m_K$ seems to be near 2 even at $P_C$ while $m_K/m_o$ may reach 100. Of course, when FM fluctuations play an important role, $m^*/m_K$ can diverge at $P_C$ . We will come back later to the recent claim of the divergence of the effective mass induced by the magnetic field in $YbRh_2Si_2$ (Custers et al 2003).

## *1.8 - Comparison with $^3$He*

The comparison of heavy fermion systems with $^3$He (see Benoit et al 1978 and 1981, Beal Monod and Lawrence 1980 and Leggett 1987) is interesting as it involves similar considerations : the localisation of the $^3$He particle (difference between the solid and liquid phase), the link with the magnetism (AF of the solid phase, paramagnetism of the liquid phase), proximity of the liquid phase from a magnetic instability, tunnelling character of the exchange and multiparticle exchange interaction in the solid and unconventional nature of the p wave superfluidity (see Vollhardt and Wölfle 1990, Leggett 1975). The quantum $He^3$ community likes to represent their phase diagram taking P as the y axis and T as the x axis. The HFC community turns the representation by 90° choosing T and P for the y and x axis.

As experimentalists, we recommend that the HFC physicists turn their heads by 90°C. Then they extend the territory to negative pressure where somewhere a first order phase transition will appear at T = 0K with a horizontal line on the (P, T) frame but vertical line on (T, P) frame. The $UGe_2$, $URu_2Si_2$ , $CeRh_2Si_2$, $CeIn_3$ and $CeRhIn_5$ phase diagrams will be like the localisation phase diagram of $He^3$ and Ce metal. This observation underlines the association of two mechanisms : delocalisation of the particle and drastic change in the spin dynamics. The comparison between HFC and quantum $^3He$ matter was our first motivation to search for the magnetic structure of solid $^3He$ on the melting curve by neutron diffraction (Benoit et al 1985). Attempts are actually made to improve our data which have confirmed the assignment of the up up down down nuclear spin structure of $^3He$ by NMR (Osheroff et al 1980). Results and analysis of excitations in the liquid $^3He$ phase can be found in Fak et al 1998 and Glyde et al 2000.

The strong point of the quantum liquid and solid phase of $He^3$ is first its purity. Secondly the bare parameters such as the bare mass ($m_{He^3}$ nuclei mass), the magnetic moment of the carrier ($\mu_n$ nuclear magnetism) and the experimental condition on its density studies are very well controlled. For the liquid phase, the knowledge of the bare parameters ($m_{He^3}$ , $\mu_n$) allows to derive Landau parameters. Notably $F_1$ and $Z_0$ are connected with the mass enhancement and the reinforcement of the Pauli susceptibility by the spin fluctuation mechanism. As already pointed out, in HFC, different mechanisms are involved in the mass enhancement and even the choice of bare magnetic parameters for the carrier is ambiguous. Basically for $He^3$ there is one type of bare carriers ; in HFC it is a complex two band system.

From the weak pressure variation of the Landau parameter $Z_0$, the low value of the Grüneisen parameter ($\Omega^* = -2$), and the analysis of the neutron scattering experiments, it is well established that liquid $^3He$ is far from a ferromagnetic instability (Anderson and Brinkmann 1975). The role of ferromagnetic spin fluctuations is however crucial in order to explain the stability of the A superfluid phase with respect to the B phase on heating. So HFC are ideal to study the magnetic instabilities. In $^3He$, the transition from the liquid phase to the solid phase corresponds to a huge volume contraction at $T \to 0K$ at the melting pressure P = 34 bar. As there is no hysteresis at this first order transition, the melting $^3He$ curve can be used for very low temperature thermometry.

Of course in HFC, the carrier is not neutral and the motion and pairing of the charge and the spin has its own interest with potential applications. Pumping of itinerant electrons can be induced by pressure or magnetic field ($P_{KL}$ , magnetostriction in $UBe_{13}$). In comparison with high $T_C$ oxyde superconductors, the great advantage is that this low temperature physics deals only with electrons, the normal phase can be studied down to very low temperature without being masked by another superconducting phase transition. Nice scans in (P, T, H) will allow to explore and understand a large diversity of situation. Concerning unconventional superconductivity, the low value of $T_C$ will allow to restore the normal phase with moderate magnetic field as the upper critical field $H_{C2}$ generally does not exceed 10 T.

## *1.9 – Experiments*

### 1.9.1 - Material/Measurements

For an experimentalist, the study of complex materials is rich as it requires the handling of different aspects which require collaborations. As the interplay of the different ingredients ($T_F$, $T_{CF}$ , $T_K$, $T_{KL}$) and their magnetic field and pressure dependence leads to a large variety of situations with critical temperatures ($T_N$ , $T_{Curie}$ , $T_C$ ), crossover temperatures ($T_{KL}$ , and $T_I$), magnetic field (Ha, Hc, $H_M$ , $H_{C_2}$ (T) or pressure instability ($P_{KL}$ , $P_{C,}$ $P_V$ and $P_{-S}$, $P_S$, the two pressures between which superconductivity occurs). The discovery of new materials can be a major breakthrough.

The appearance of new products may open a completely new perspective as happened for the high $T_C$ superconductors, or decisive possibilities to clarify basic issues (see Ce115 and Yb 122 compounds) and also effects which are magnified by the interplay between parameters. For scientists outside the field, HFC may appear as a labyrinth with no exit. They must realize that this physicochemistry spadework is essential to select a clear situation with the possibility to tune later through different phases by application of pressure, uniaxial strain or magnetic field.

Here we have focus on extreme regions of the (T, P, H) phase diagram of HFC. Progress has been made at the frontier of ultimate instrumentation (see Salce et al 2000) to successfully make simultaneous specific heat, resistivity and susceptibility experiments with low temperature tuning of the hydrostatic pressure and also with the increasingly routine P experiments developed more than a decade ago at the high pressure Institute of Troitsk (Eremets 1996). However, often a major development becomes only possible if a new simple detail has been solved. A special attention must be taken on the contact quality in order to be sure that the output thermal power is directly transmitted to the electronic bath. In Grenoble, the boost was given by our Japanese visitor, Y. Okayama (1996), who taught us how to realize tiny beautiful electrical contacts with a microwelding machine on almost any Ce or U intermetallic compound. Sometimes we discuss apparently boring experimental details (pressure gradient, sensitivity to define a characteristic field in quantum critical field problem) in order to motivate young physicists for instrumentation development and for the improvements of accuracy.

Concerning the experimental methods, focus is given on the microscopic probe of neutron scattering done in Grenoble and on NMR performed in Osaka. Inelastic neutron scattering is still the only way to probe the spin dynamics in a wide range of frequency and wavevector. NMR is a "light" probe which gives access to the ultimate low frequency limit at the expense of wavevector integration. It has now been performed with a good accuracy up to $P_C$ in different HFC (chapter 4 – 5).

Thermodynamic measurements such as the specific heat C and the magnetization are fundamental quantities which can be measured with a great accuracy within one percent or better. They are a severe test on any theoretical proposal near the quantum critical point (P or H). Their associated P derivatives (thermal expansion and magnetostriction) or field derivatives (magnetocaloric effect) are very powerfull. They reveal the importance of deformation (compression and shear mode) i.e of density fluctuation. Theoretical calculations at constant volume will miss the main issue. The full detection of dHvA oscillations in $UPt_3$ (Lonzarich and Taillefer 1985) show that the FS topology is rather conventional (given roughly by any band calculation) but with huge renormalized effective masses which establish the validity of the heavy fermion quasiparticle. Few complete FS have been drawn experimentally apart from $CeRu_2Si_2$ and $UPt_3$. A large number of orbits have been determined in the materials addressed in this review (see chapter 2, 4, 6).

As it was underlined, resistivity is certainly the best fast method to detect major breakthroughs in the discovery of superconductivity. Thermal conductivity is a very powerful technique to study unconventional superconductivity : an illustration will be the $UPt_3$ study. The Hall effect is complex in this multiband material. The possibility of a large effect at $P_{KL}$ (at any FS reconstruction) gives now a new experimental fever (Coleman et al 2001), see for the recent applications to V(Cr) (Yeh et al 2002) and to $YbRh_2Si_2$ (Paschen et al 2004) and for the theoretical comments (Norman et al 2003).

If the physicist enjoys measuring a signal which can change its signal amplitude in T, H or P he will certainly select the thermoelectric power $S_{TEP}$ in these narrow renormalized bands (see Jaccard and Flouquet 1985) as well as the Nernst effect. These possibilities were extensively exploited in high $T_C$ materials for the pseudogap issue (Ong et al 2004 and Wang et al 2002). Like in thermal expansion, large variations are expected near $P_{KL}$ or $P_C$ . New experiments are underway notably in Paris (Bel et al 2004-a). As the interplay between FS and magnetic instabilities are still obscure, TEP is an excellent probe to detect subtle effects (see Abrikosov 1988).

The readers can find reviews on other specific techniques on HFC, for the powerful ultrasound probe (Thalmeier and Lüthi 1991) for electrodynamic response of HFC (Degiorgi 1999) for high energy spectroscopy (Malterre et al 1996) and for muon spectroscopy (Amato et al 1997). Few experiments have been performed using tunnelling technique (see $UPd_2Al_3$) but point contact spectroscopy was extensively used (Naidyuk and Yanson 1998). Most of the figures described results obtained in Grenoble, Geneva (D. Jaccard) and Osaka (Y. Kitaoka). Using the internet, readers can recover the important results obtained in the groups of Los Alamos (J. Thompson, J. Sarrao), Tallahassee (Z. Fisk), Zurich (H. R. Ott), Dresden (F. Steglich), Karlsruhe (H. von Löhneysen), Bristol (S. Hayden), London (G. Aeppli), Ames (P. Canfield), Sherbrooke (L. Taillefer), Berkeley (N. Phillips), Wien (E. Bauer), Cambridge (G. Lonzarich), Toronto (S. Julian), Osaka (Onuki), Tokyo (Sakibara, H. Sato), ……

Finally the applied side has not been discussed despite a large possibility for entropy changes in magnetic field, in pressure and of resistivity and thermoelectric power variations in pressure, temperature and magnetic field. At least HFC figures can make very nice exercices as the equivalent of Pomeranchuk effect or fast decompression at constant field in quantum $^3He$, the efficiency of adiabatic demagnetisation or of the Peltier cooling. Our proposal made in March 2004, was applied later by Continentino and Ferreira (2004) and by Continentino et al (2005) to $YbInCu_4$. It is also possible to play with a machine where the vapor or fuel will be the spin. The problem is particularly interesting if there is a (P, H, T) range where a phase separations occurs.

### 1.9.2 – From transport measurements to heavy fermion properties

Resistivity is often the only measurement performed under pressure since it is a sensitive technique (detection of voltage down to $10^{-3}$ nV) with a low dissipative power and the possibility to be coupled thermally with the cold source by excellent electronic contacts. Let us state the related specific problems due to the interplay between charge carriers and scattering.

Since the material is not perfect, residual impurities lead to a residual elastic term $\rho_0$ which is decoupled from inelastic quasiparticle collision at very low temperature. The simple Drude expression for a single type of carrier :

$\rho_0 = \frac{k_F}{n_e^2 \ell}$ shows that $\rho_0$ depends only on the carrier's number ($k_F \sim n_e^{1/3}$) and on the electronic mean free path $\ell$ which does not depend directly on the effective mass but may be enhanced near $P_C$ and $P_V$ $\left( \rho_0 \approx \frac{1}{n_e^{2/3} \ell} \right)$. Often an increase of the impurity scattering is detected near QCP (Flouquet et al 1988, Thessieu et al 1995, Wilhelm et al 2001). The $\rho_0$ enhancement ie a decrease of $\ell$ due to quantum critical fluctuations was calculated by Miyake and Narikiyo (2000) for the spin and Miyake and Maebashi (2002) for the valence. The effect is quite strong near a ferromagnetic (F) QCP and less for an AF QCP. This increase will tend to suppress superconductivity just near $P_C$ for F systems. Critical valence fluctuations near P = $P_V$ also produce a sharp peak of $\rho_0$. Of course if there is a FS reconstruction (for example at $P_{KL}$ ), $\rho_0$ will be a basic probe to detect a change of the carrier. Finally, as the impurity sites may correspond to a so called Kondo hole i.e a deformed Cerium site near a lattice imperfection (intersticial, dislocation, stacking fault, …), they may also give a temperature dependant term $\rho_{imp}$ (T) to the resistivity. Before claiming that any T variation of $\rho$ is an intrinsic property, proof must be given that it does not depend on $\rho_0$.

The low energy excitations of the dressed particles appear in the temperature contribution of the resistivity. In SF, below $T_I$, the $AT^2$ law is one of the signature of the Fermi liquid regime. The behaviors reported in table 3 for the SF predictions of ρ(T) at the QCP assume an average scattering over all the Fermi surface i.e basically one type of carrier but different scattering processes. Using the parameter ($Y_0$, $Y_1$, $T_0$, $T_A$) extracted from the specific heat and inelastic neutron scattering data in the $CeRu_2Si_2$ , the calculated SF contribution ($\rho_{th}$) with the hypothesis of an average on relaxation times is greater than that measured $\rho < \rho_{th}$. An extra source of electronic conductivity i.e from different types of carriers may occur. It was first assumed that a source of the new conduction channel might come from an impurity band (Kambe and Flouquet 1997).

Another hypothesis is to assume that on the Fermi surface, the hot spots (corresponding to a momentum $k = k_0$ transfer on the Fermi surface) and the cold spots (insensitive to AF SF) will give two different channels of conduction. Such a model was developed by Rosch (1999) for 3d HFC. Right at the AF QCP, the electronic conductivity σ is written as :

$$\sigma = \sigma_{hot} + \sigma_{cold} = \frac{\sqrt{t}}{x_{imp} + t} + \frac{1 - \sqrt{t}}{x_{imp} + t^2}$$

The impurities are represented by a reduced parameter $x_{imp}$ inversely proportional to $\ell^{-1}$ $x_{imp}^{-1} \sim \ell$, and the temperature by its normalized value $t = \frac{T}{T_I}$. The fraction of the FS considered as hot spots is $\sqrt{t}$ ; its spin fluctuation rate is linear in T. The second term describes the cold regions where FL quasi particle scattering proportional to $T^2$ occurs. In the clean limit $x_{imp} < t$, the cold quasi particle term dominates $\Delta\rho \sim T^2$, in the dirty limit $x_{imp} > t$ $\Delta\rho \sim T^{3/2}$. For a value of x = 0.01, a linear T term appears in an intermediate range of temperature.

As shown by Rosch (2000), a departure from $P_C$ (i.e the appearance of a finite coherence length) restores the observed situation with a minimum n = 3/2 for the inelastic SF term ($\rho \sim T^n$) at $P_C$ . In the specific case of HFC, the k structure of the AF correlation emerges on top of a large continuous signature of the local fluctuations. The recovery of the $T^2$ term may be faster than in the SF model where all the weight of the fluctuation is at $k \sim k_0$. The effect of dimensionality can modify the present scenario : 2d fluctuations lead to n = 1 at the QCP as well as valence fluctuations (see $CeCu_2Si_2$ chapter 4). In some HFC, different AF wavevectors ($k_0$, $k_1$, $k_3$ for $CeRu_2Si_2$ ) may have different instability points. However in many cases, n reaches a minimum at $P_C$ which may be a simple test to check for a critical pressure.

A scaling of A with $\gamma^2$ contrary to the SF predictions, the so called Kadowaki – Woods (1972) relation ($A/\gamma^2 = 10^{-5}$ μΩ cm mol$^2$ K$^2$ mJ$^{-2}$) is often observed in HFC due to the broad wavevector response by contrast to the strong frequency one. Theoretical discussions on the validity of this assumption can be found in references Miyake et al (1989) and Takimoto and Moriya (1996). A recent discussion on the Kadowaki Woods relation is given in reference Tsuji et al (2003) with an analysis for its deviation in Yb based intermediate valence systems.

It was stressed that another interesting ratio is the Seebeck coefficient ($S_{TEP}/T$ extrapolated at $T \rightarrow OK$) by the corresponding term γ of the specific heat ($C = \gamma T$) (Sakurai 1994 and 2001 and Behnia et al 2004). A log-log plot of $S_{TEP}/T$ versus γ for different SCES going from cuprate, organic conductor, HFC and even simple metal (Behnia et al 2004) gives results aligned mainly on the same line with the ratio :

$$S_{TEP}/T\gamma \sim (N_{av}e)^{-1}$$

where $N_{av}$ is the Avogrado number. Deviations may indicate :

- a large difference between the number of itinerant carrier and the heat carrier generally assigned to the f electron (1 per formula unit in the case of Ce),
- different channels of diffusion (case of IVC),
- specific cancellation of $S_{TEP}/T$ in compensated metals where hole like and electron like bands can give difference in the sign of the thermoelectric power. However it is remarkable for the Ce HFC that, whatever is $T_K$ at very low temperature and also, whatever is the band structure and notably the degree of compensation between holes and electrons, $S_{TEP}$ is positive. The current flow from a free electron gas state to a heavy fermion phase gives the remarkable result of $S_{TEP} \sim C$ as $T \rightarrow OK$ as pointed out by Sakurai (1994, 2001). This very low temperature behavior is quite different from that observed at intermediate temperature ($T \sim T_K$) where $S_{TEP}$ seems to be sensitive to the energy derivative of the density of states (see Jaccard and Flouquet 1985). Experimental references on the thermoelectric power of cerium and Ytterbium intermetallics can be found in Zlatic et al (2003). Calculation of the TEP, specific heat and susceptibility on the based of mixed valence systems or of high $T_K$ assumption using $\frac{1}{N_f}$ expansion can be found in Bickers et al (1985) and Houghton et al (1987) At the limit of large degeneracy the proportionality $S_{TEP}/T\gamma \sim (N_A e)^{-1}$ is recovered. It was recently shown that the quasi-universal ratio of the Seebeck coefficient to the specific heat term $\gamma$ can be understood on the basis of the Fermi liquid description of strongly correlated metals (Miyake and Kohno 2005 ). The Kadowaki-Woods rule for resistivity and its equivalent for the thermoelectric power are not golden rule. At least, they emphasize that the strong renormalization is due to the strong frequency dependence of the response.

Finally it is interesting to read old papers published seven decades ago on electrical conductivity and thermoelectricity to elucidate the properties of transition metals taking into account two bands with light and heavy carrier (Mott 1935, Baber 1937 and Wilson 1938).

# 2/ Cerium Normal phase properties

*The Outlook are :*

- HFC magnetism is furtive : its ground state cannot be predicted from first principles.
- The $CeRu_2Si_2$ Kondo lattice ($P_C$ < $P_V$) shows that :
  The collapse of the long range magnetism may be not associated with a divergence of the coherence length. Tiny moment is an intrinsic property which can be tuned by P or H. The transition from a weakly polarized paramagnetic phase to a strongly polarized paramagnetic phase is driven by the Kondo collapse at the pseudo-metamagnetic transition,
- Experiments on $CeCu_6$ were the first support of local quantum criticality.
- Just above $P_C$ , $CeNi_2Ge_2$ looks as a nearly antiferromagnetic metal … but a complete study is yet not achieved.
- The hole Kondo lattice of Yb HFC appears different from electron analog of Ce HFC.

## *2.1 - Magnetic furtivity of $CeAl_3$*

To illustrate the difficulty to predict ab initio the ground state of HFC, the story of $CeAl_3$ is briefly summarized. In the Ce, Al series, the cerium systems appears down to low temperature (10 K) as a "normal" rare earth ion sensitive to the crystal field splitting. However tiny differences (huge volume and anisotropy sensitivity) will lead to drastic changes in the ground state. Figure (7) represents the magnetic entropy of three Ce, Al compounds : $CeAl_2$, $CeAl_3$ and $Ce_3Al_{11}$ (Steglich et al 1977, Bredl et al 1978, Peyrard 1980,

Flouquet et al 1982). Already at T = 10 K, the magnetic entropy of a free doublet (RLog2) is recovered. Thus down to 10 K, the magnetic properties of these compounds appear similar. It will be very difficult to predict their low temperature destinies. The drop of entropy for $CeAl_2$ at $T_N$ = 3.8 K and for $Ce_3Al_{11}$ at $T_{Curie}$ = 6.2 K and $T_N$ = 3.2 K marks the entrance in magnetic phases. As for $CeAl_3$ (Andres et al 1975), there is no trace of magnetic ordering, it was admitted during more than a decade that $CeAl_3$ ends up in a Pauli paramagnetism (PM). The maxima of C/T for T ~ 350 mK was taken as an evidence of the entrance in a low temperature correlated Fermi liquid regime. Furthermore, the combination of thermal expansion and specific heat shows, that the huge C/T value and the concomitant large negative thermal expansion $\left(\frac{\partial V}{\partial T} = \frac{-\partial S}{\partial P}\right)$ (as for liquid $^3$He) is not a continuation of the single impurity Kondo picture. As $T_K$ increases under pressure, the classical dilatation of a metallic solid on heating is expected and thus a positive thermal expansion for a single Kondo impurity. At least, it was obvious that a characteristic temperature other than $T_K$ occurs (Ribault et al 1979, Flouquet et al 1982).

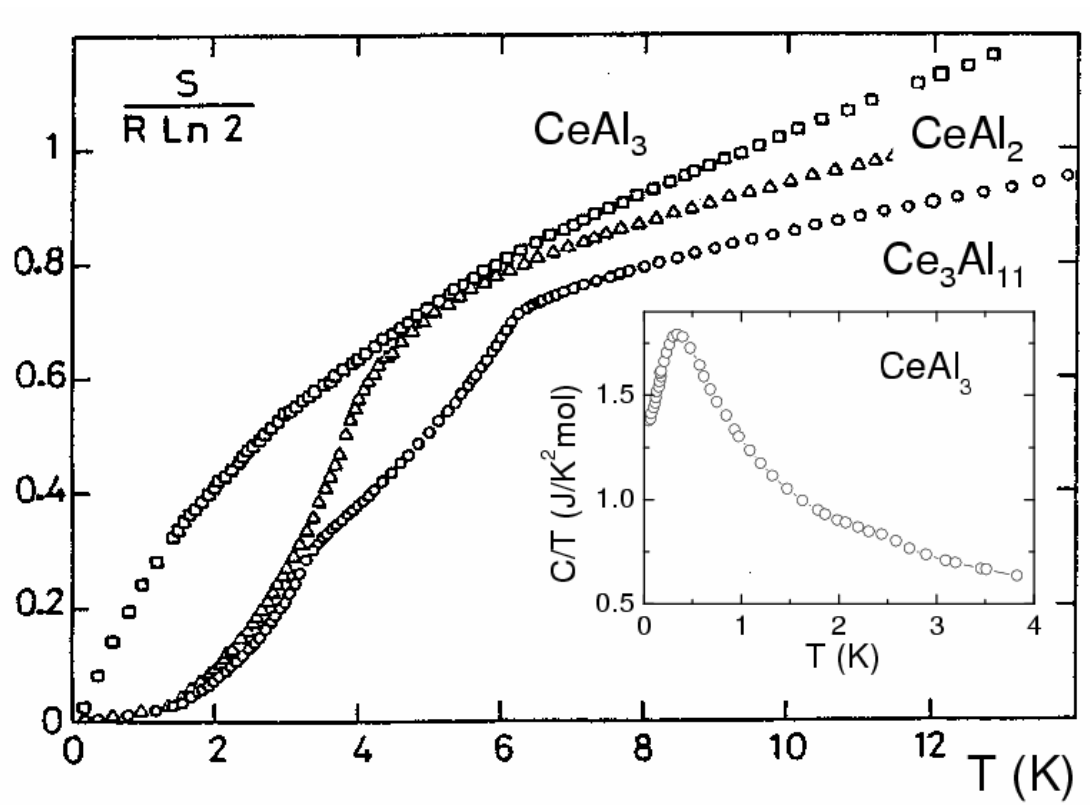

Figure 7 : Entropy of $CeAl_3$ (□), $CeAl_2$ (△), and $Ce_3Al_{11}$ (○) measured on polycristal (Flouquet et al 1982). Temperature insert variation of C/T of $CeAl_3$ .

Experimentally, the main difficulty with $CeAl_3$ is that the compound is not a "line" compound i.e. it is formed in a peritectic solid reaction. All data were taken on polycristals. The growth of tiny single crystals of this hexagonal lattice shows clearly the occurrence of AF ordering with pronounced AF anomalies in specific heat and resistivity (Jaccard et al 1987-1988, Lapertot et al 1993). The data on polycristals appear a result of a broad distribution of the Néel temperature due to its unusual high sensitivity to pressure and uniaxial stress. At P = 0, $CeAl_3$ may be on the AF boundary with $P_C$ ~2 kbar. The negative value of the thermal expansion can be well explained as, on the AF verge, on approaching $P_C$, the Kondo like picture and the spin fluctuation approach predict an increase of $\gamma$ with pressure. Through the Maxwell relation $\left(\frac{\partial V}{\partial T} = -\frac{\partial S}{\partial P}\right)$ that will lead to a negative thermal expansion.

In this situation closed to $P_C$ at ambient pressure, the difference on the sample preparation is directly linked to the fact that the compound $CeAl_3$ appears as a solid phase only below 1135°C i.e its peritectoid formation. Starting with the magic composition 1-3 of $CeAl_3$ , the first solid created phases will be $CeAl_2$ and $Ce_3Al_{11}$ on cooling from T ~ 1500°C. The realisation of nice polycristals with a single $CeAl_3$ phase is achieved by a slow interdiffusion of $CeAl_2$ and $Ce_3Al_{11}$ at T = 850°C during a week ; residual resistivity down to µΩcm was achieved. Using a highly non equilibrium process with a start on the $Ce_3Al_{11}$ side, Lapertot et al (1993) succeed to produce large separate millimetric grains of $CeAl_3$ and $Ce_3Al_{11}$. In Geneva, the growth of single crystals was successful by a mineralisation just below the temperature 1135° C of the peritectoid landing (Jaccard et al 87-88). In both cases, the $CeAl_3$ crystals are extracted among a mixture of $Ce_3Al_{11}$ or $CeAl_2$ agregates. Whatever is the process, single crystals are characterized by residual resistivity one order of magnitude higher than that of polycrystal. Obviously that leads to enhance the disorder by respect to the polycrystal material with the consequence of the appearance of clear magnetic transitions. Weak perturbations (not observable by relative changes in the lattice parameters down to $10^{-4}$) cause the material to select either a pure AF ordering or a dominant PM phase. Whatever the sample elaboration (polycrystal – single crystal), an extra pressure of P = 0.2 GPa or a magnetic field of H = 2T pushes the system into an ordinary FL-PM state. (Flouquet et al 1988, Cibin 1990).

## *2.2 - The Kondo lattice $CeRu_2Si_2$ : P, T phase diagram*

$CeRu_2Si_2$ is a key system for the understanding of HFC (Besnus et al 1985) as the tetragonal lattice has an axial symmetry and the local $\chi_L$ magnetic susceptibility shows an Ising behavior (Flouquet et al 2002). The crystal field ground state of $CeRu_2Si_2$ is almost a pure ±5/2 doublet with respective g factor along the c and a axis $g_{//} = 5g_J$ and $g_\perp = 0$. The growth of large excellent crystals has allowed a large diversity of microscopic studies in different laboratories notably extensive high energy spectroscopy, neutron scattering, NMR and quantum oscillation experiments. The possibility to achieve high electronic mean free paths of $\ell$ ~1000 Å gives the opportunity of a full determination of the Fermi surface in de Haas van Alphen measurements. The high quality of the shiny surface is a guarantee of excellent robust metallic connections which allow nice thermal and pressure studies.

At P = 0, the pure compound $CeRu_2Si_2$ can be located few tenths of GPa above $P_C$ = - 0.3 GPa. Positive pressure experiments move the system away from $P_C$ . Negative pressure experiments has been realized artificially by expanding the lattice by doping. A dilution of the Ce ions by $x_c$ = 7.5 % Lanthanium ions drives the system to $P_C$ ~ - 0.3 GPa. For simplicity, an unique variable P will be used. The negative pressure has been calibrated knowing the change of the lattice parameters and the compressibility.

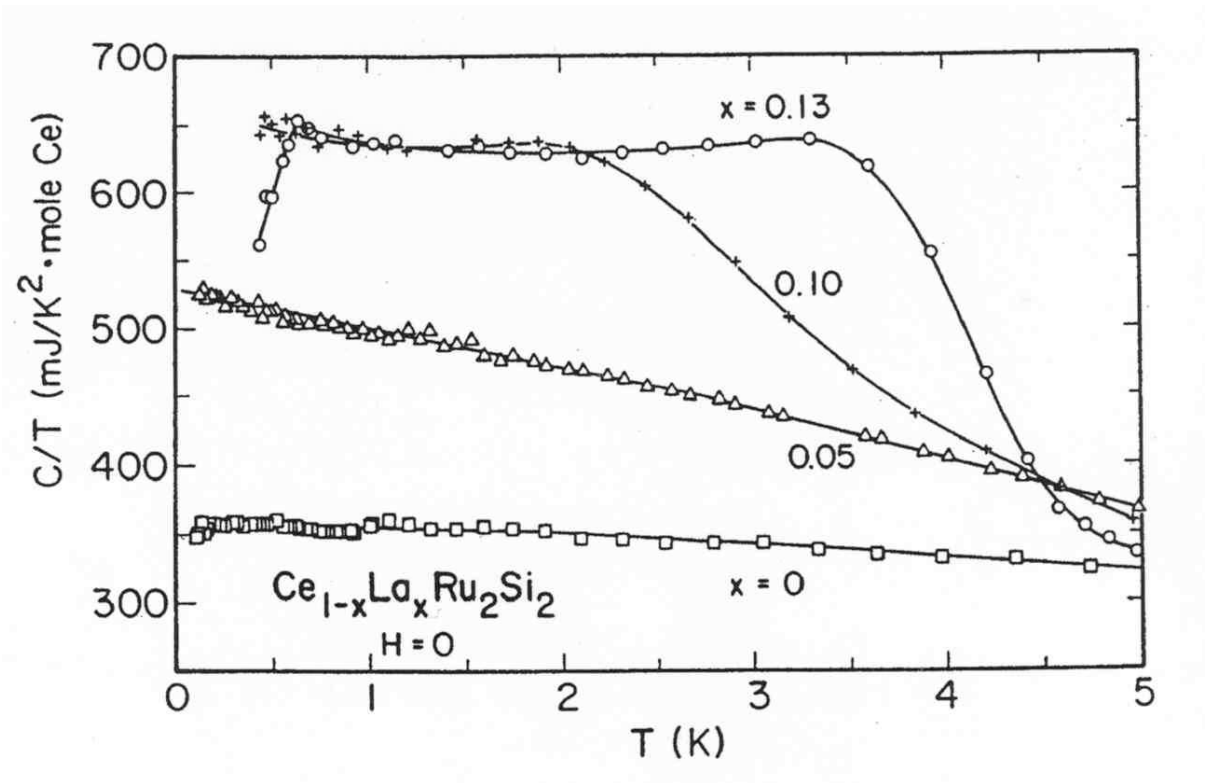

Figure 8 : Variation of C/T as a function of T for $Ce_{1-x}La_xRu_2Si_2$ for x = 0, 0.05, 0.10 and 0.13 (Fisher 1991).

The NFL behavior of $CeRu_2Si_2$ can be recognized by the slow increase of C/T on cooling before reaching the Fermi liquid regime (figure 8) (Fisher et al 1991). Neutron scattering experiments were successfully explained in the framework of spin fluctuation theory (Raymond 1999-a). New data of Kadowaki et al (2004) confirm the validity of SF. A scan in wavevector at constant energy transfer (figure 9) shows a large local response i.e. invariant in wavevector and superimposed AF correlations caracterized by different vectors $k_0$, $k_1$, $k_2$ (Rossat Mignot et al 1988, Regnault et al 1988). For a negative pressure, below $P_C$ , long range magnetism appears at the wavevector $k_0$.

*Normal phase of Ce compounds*

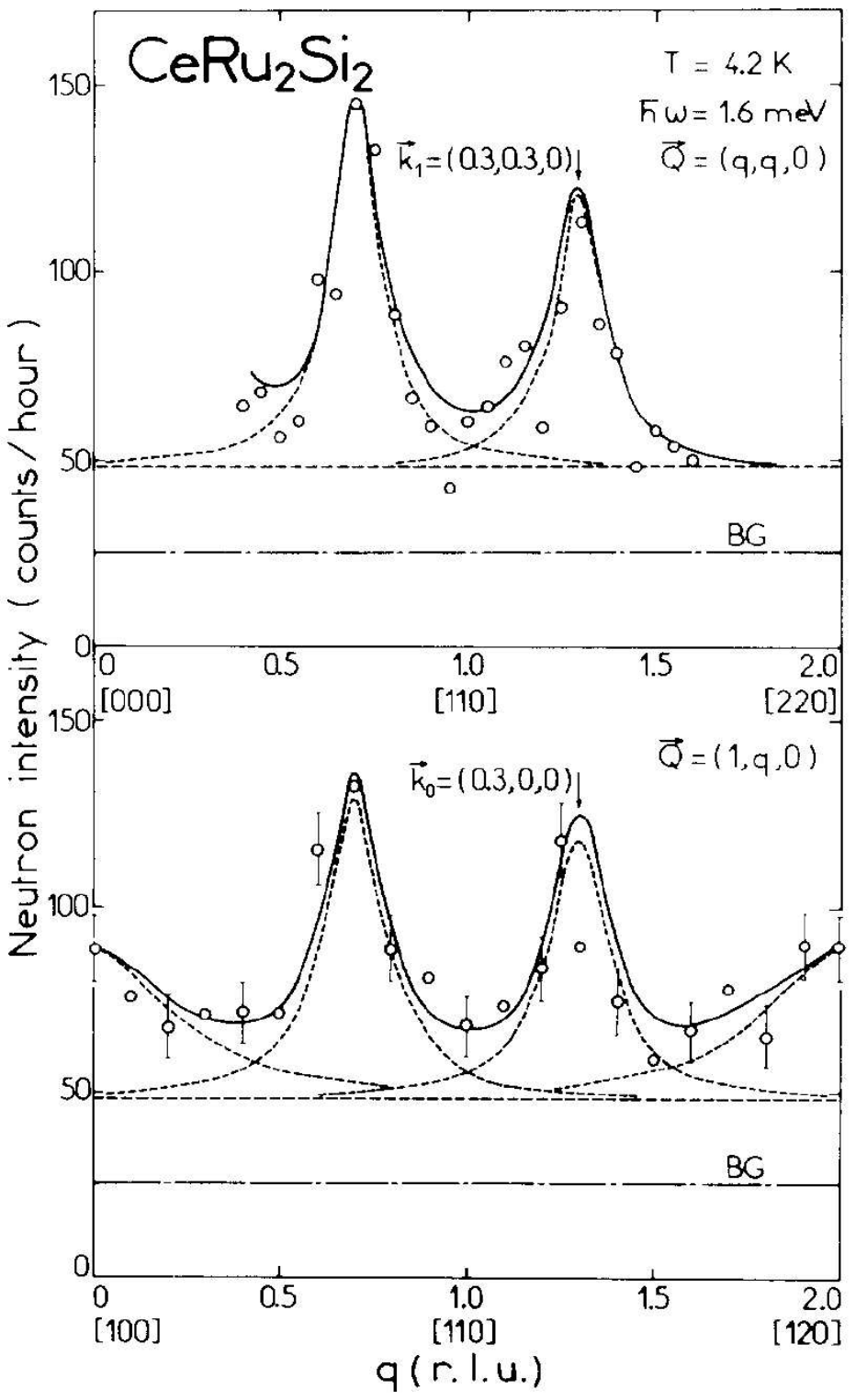

Figure 9 : q-scans performed at a finite energy transfer $\hbar\omega$ = 1.6 meV along the directions [110] at T = 4.2 K for $CeRu_2Si_2$ , showing the incommensurate wave vector $k_0$ = (0.3, 0, 0) and $k_1$ = (0.3, 0.3, 0) (Rossat-Mignod et al 1988).

A new generation of inelastic neutron scattering experiments (Raymond et al 2001 and Knafo et al 2004) was performed recently, notably on $Ce_{0.87}La_{0.13}Ru_2Si_2$ (i.e 0.3 GPa below $P_C$ at P = 0) by a pressure tuning through $P_C$ and on $Ce_{0.925}La_{0.075}Ru_2Si_2$ at the critical concentration $x_c$ but only at P = 0. In this last study, the imaginary part of the susceptibility $\chi''$ was almost continuously analysed in frequency ω at different T for the AF wavevector $k_o$ and for $k_s$, a wavevector characteristic of the local response.

In a phase transition, the frequency enters in the scaling function :

$$\chi'' \sim T^{-\alpha} f\left(\frac{\omega}{T^{\beta}}\right)$$

As pointed out previously, when the dangerous irrelevant interaction must be considered, $\omega \xi_m^z = \frac{\omega}{T^\beta}$ with an exponent β > 1 (Continentino 2001, Sachdev 1999). If the fit is made over a large temperature window from 1K to 100 K, the respective value of α and β are found for the AF wavevector $k_o$ (α = 1, β = 0.8) and for $k_s$ a wavevector far from magnetic instabilities (α = 1, β = 0.60) at least down to the temperature where a saturation occurs. In any temperature range, the inelastic response is well described by a Lorentzian form

$$\chi'' = \frac{A}{\Gamma\left(1+\left(\frac{\omega}{\Gamma}\right)^2\right)}$$

but the linewidth $\Gamma_{k_o}$ at $k_o$ stays finite (~ 0.2 meV) below $T_{k_o}$ = 3K while the linewidth $\Gamma_{k_s}$ (~ 1.5 meV) at $k_s$ saturates below $T_{k_s}$ = 17 K (figure 10). So, contrary to the SF prediction, ($\Gamma_{k_o}$ ~ $Y_0$, $Y_0 \to 0$ at $P_C$ ), $\Gamma_{k_o}$ does not collapse. The so called magnetic QCP may not exist. A finite coherence length seems to occur at $P_C$ (figure 10). This residual value may be linked to the observation that a tiny sublattice magnetization ($M_o$ ) is observed on cooling (Raymond et al 1997) at $k_o$ ($M_o = 0.02\mu_B$ , $T_N$ ~ 2K). As a concentration gradient can occur in the crystal, it was suggested that it may lead to "residual" long range magnetic order which will disappear under pressure. However the quasicoincidence in temperature (T ~ 2K), where the inelastic linewidth $\Gamma_{k_s}$ saturates and the signal elastic magnetic diffraction appears, favors an homogeneous picture. This is reinforced by the fact that similar saturation of $\Gamma_{k_o}$ right below $P_C$ was found at low temperature in the pressure measurements on $Ce_{0.87}La_{0.13}Ru_2Si_2$ (Raymond et al 2001).

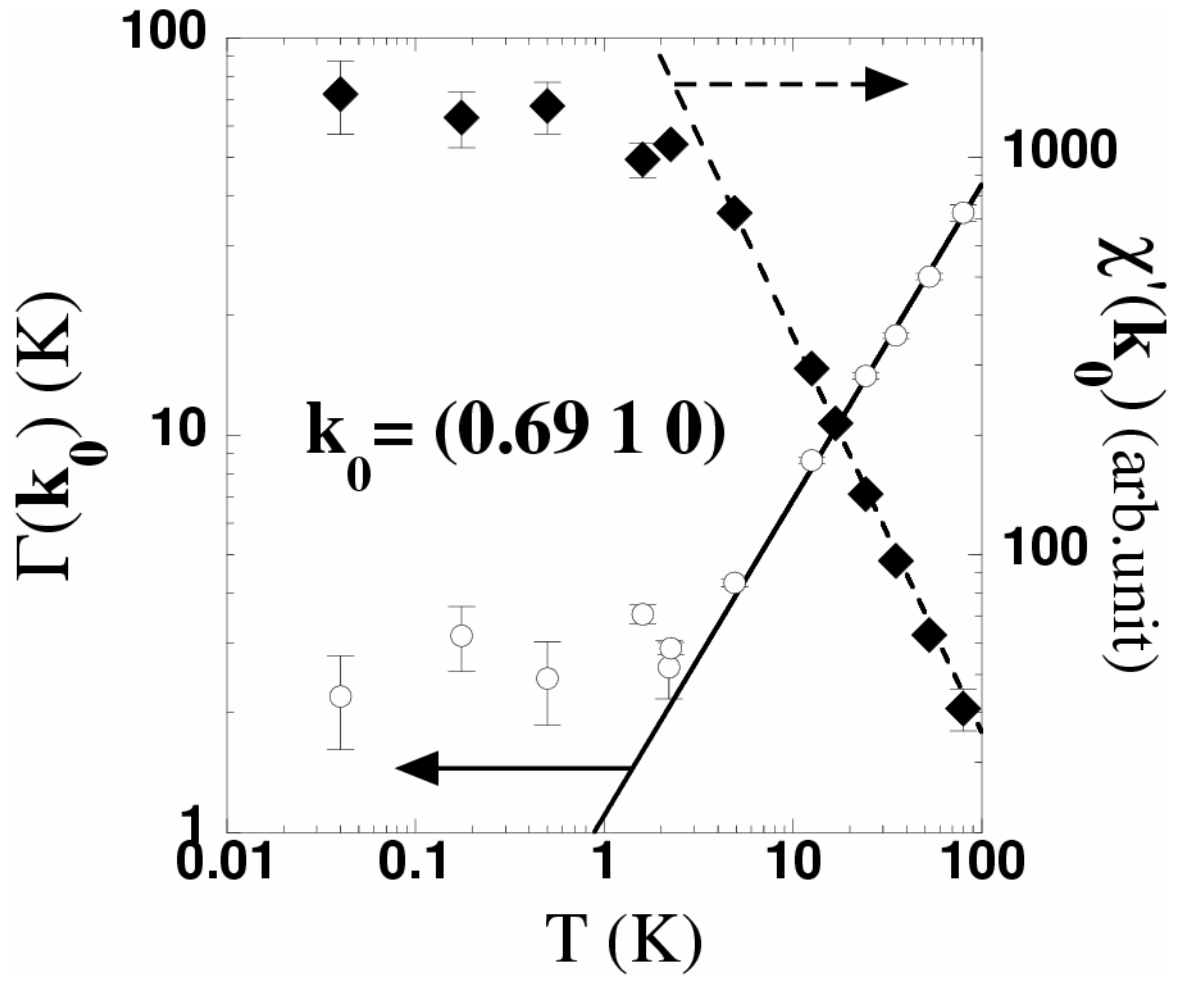

Figure 10 : Temperature variation of the quasielastic energy width $\Gamma_{k0}$ and of the static susceptibility $\chi'(k_0)$. The wavevector $k_0$ is characteristic of AF correlations. The dashed and full lines are fits in $T^{-1}$ and $T^{0.8}$ (Knafo et al 2004).

The fact that small moment antiferromagnetism (SMAF) (tiny $M_o$ )may be the signature of a new heavy fermion matter (Kondo condensate) seems supported that, even in excellent single crystal of $CeRu_2Si_2$, tiny ordered moments $M_o \sim 10^{-3}\mu_B$ have been detected (figure 11) below 2K on top of dynamic spin fluctuations involving may be two components in this heavy fermion matter (Amato et al 1993). Migration of one electron from one constituant to the other produce a thermoelectric power $S_{TEP}$ (insert of figure 11) (Amato et al 1989). As explained by D.O. Edwards for the mechanism of a $^3$He – $^4$He dilution refrigerator (Edwards 1970), the migration of electron from one component to another can produce a Peltier cooling power.

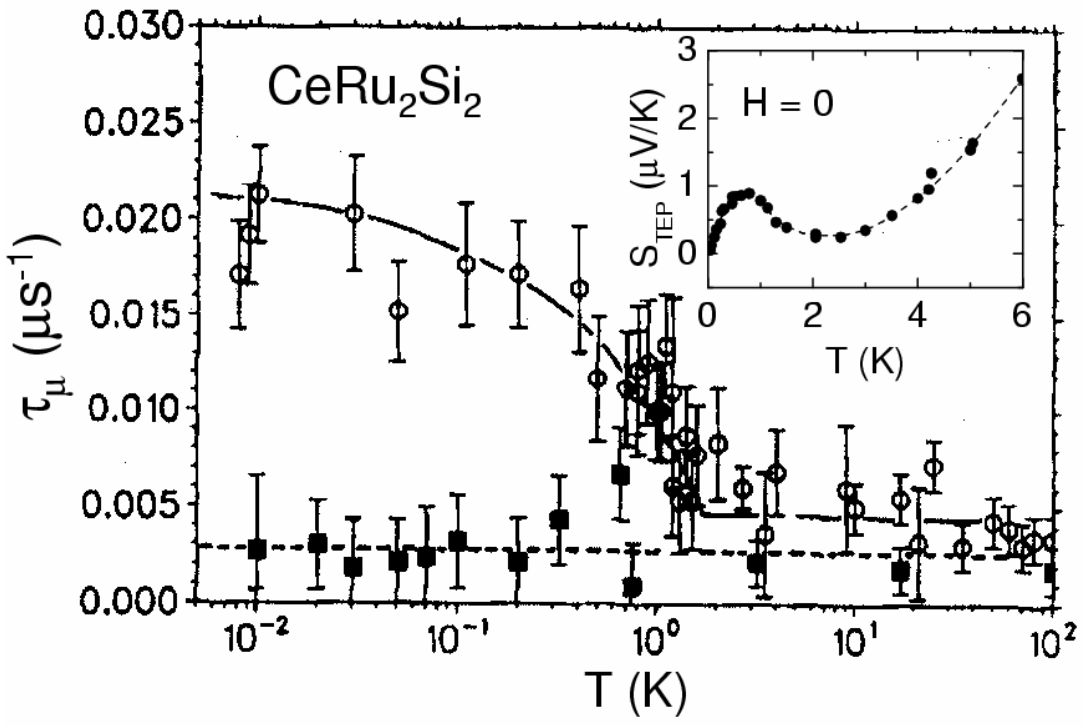

Figure 11 : Evidence for static ultrasmall moment magnetism in $CeRu_2Si_2$ (Amato et al 1993). Relaxation rate $1/\tau\mu$ of the $\mu+$ polarisation measured in zero-field (0) and longitudinal

applied field (■), $H_{ext}$ = 6 kOe). The initial μ+ polarization is along the $\hat{c}$-axis. The lines are to guide the eye. In insert, the temperature dependence of the TEP measured in the basal plane at H = 0 (Amato et al 1989).

Even if a relative good agreement has been obtained with the SF approach, there are still uncertainties on the situation at $P_C$ : order of the transition, role of disorder, Fermi surface formation. On warming above $T_0$ or $T_K$, the derived β exponent < 1 is not predicted for a quantum phase transition. This behavior is due to the fact that $T_K$ (~ 20 K) occurs just in the temperature window of the fit (3-100K). Taking into account that one order of temperature lower than $T_K$ or $T_{KL}$ is necessary to feel a simple regime ($T_F$/10 for a free electron gas), it is not so surprising that the $T^\beta$ scaling reflects the single impurity dynamics and the slow formation of the Fermi surface.

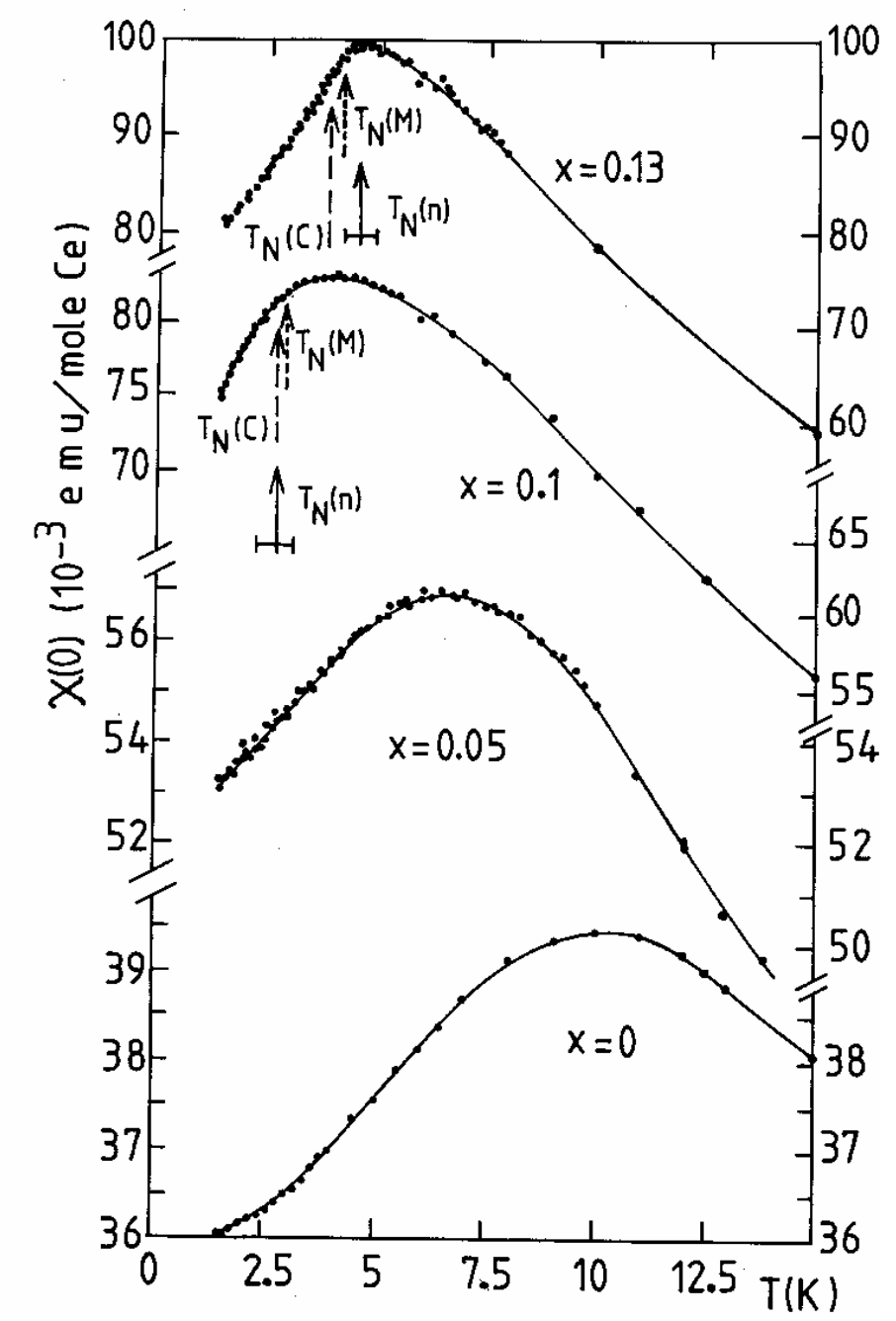

Figure 12 : Susceptibility of $Ce_{1-x}La_xRu_2Si_2$ for x = 0.13, 0.10 on the AF side and x = 0.05 x = 0 on the Pa side. In term of pressure by reference to $CeRu_2Si_2$ at P = 0, $P_C$ = - 0.3 GPa, xc = 0.075. $T_N$ (M, n , C ) indicate the Néel temperature determined by magnetization , neutron scattering and specific heat. (Fisher et al 1991)

The analysis of $CeRu_2Si_2$ neutron scattering data may require to introduce at low temperature an inelasticity ($\omega_0$) for the frequency at least to explain consistently the dynamical and static response. As Γ increases strongly on heating, the inelasticity is smeared out already above 20 K. Macroscopic evidence of a pseudogap i.e. a dip in the renormalized density state near the Fermi level appears clearly by the occurrence of a susceptibility maximum $\chi_M$ at $T_M$ identified often as the Kondo lattice temperature (figure 12) and the concomitant metamagnetic phenomena. For the magnetically ordered compound ($P < P_C$ ), $T_M$ is just above $T_N$. When $T_N$ vanishes at $P_C$ , the maximum $\chi_M$ subsists as a signature of the interplay between AF exchange coupling and Kondo like fluctuations. The behavior is opposite to that of a spin ½ Kondo impurity where the susceptibility continuously increases on cooling. Translated into the density of states a pseudogap is required. In good agreement with this picture, the Zeeman splitting of the spin up and spin down band can induce a pseudo-metamagnetism with a continuous jump of the magnetization.

To demonstrate the complexity of this heavy fermion matter, the simple visualization is via old fashioned thermodynamic language (Grüneisen 1912, Peyrard 1980, Benoit et al 1981, Takke et al 1981). In the same spirit as the Clapeyron (or Ehrenfest relations) of P, T, dependence on discontinuities in the entropy (or specific heat) and volume (or thermal expansion) for first (or second order) transitions, the Grüneisen parameter defined as the ratio of $\alpha$ over C at each temperature :

$$\Omega^*(T) = \frac{\alpha}{C}\frac{V_0}{\kappa}$$

(where $\alpha$, $V_0$ and $\kappa$ are respectively the volume thermal expansion, the molar volume and the isothermal compressibility) is a excellent probe towards a single parameter scaling. It will be reduced to a constant $\Omega^*(0)$ independent of the temperature only if the free energy F can be expressed by a single parameter T* i.e.

$$F(T) = T\Phi\left(\frac{T}{T^*}\right) \; with \; \Omega^*(0) = -\frac{\partial \log T^*}{\partial \log V}$$

Figure 13 represents the temperature variation of the Grüneisen parameter of $CeRu_2Si_2$ at P = 0 i.e. roughly 3 kbar above the critical pressure $P_C$ = - 3kbar. The two singular points are : (i) the huge extrapolated value of $\Omega^*(0)$ = +190 and (ii) the slow entrance into a simple regime (T ~ 1K) where $\alpha$ and C reach their proportionality. At low pressure close to $P_C$ ,

$\Omega^*(P, T = 0K)$ decreases strongly with P and then reaches a plateau $\Omega^*(P_V) \sim 80$ at $P_V \sim 3.5$ GPa before decreasing again above 5 GPa (Payer et al 1993, Flouquet et at 2004 ).

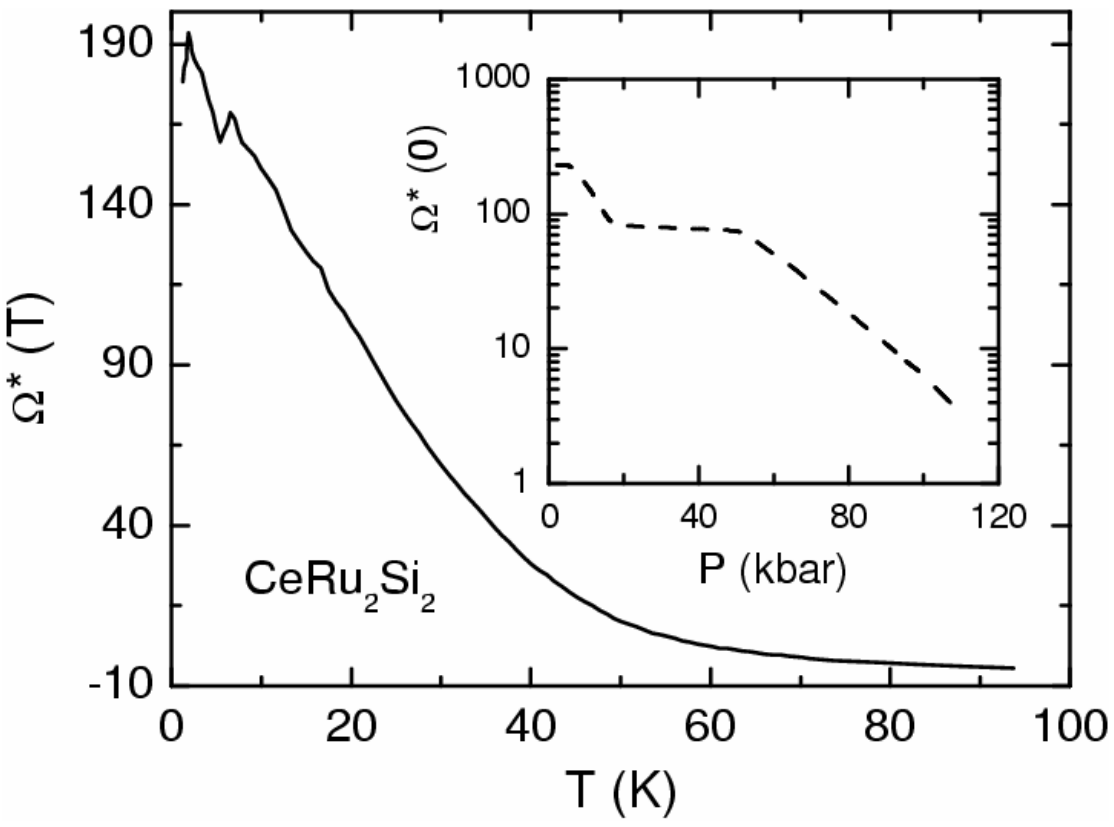

Figure 13 : Heavy fermion Grüneisen parameter ($\Omega^*$ (T)) of $CeRu_2Si_2$ , the phonon contribution has been substracted (Lacerda et al 1989). The insert : pressure dependence of $\Omega^*(0)$ (Flouquet et al 2004).

Extensive studies of $CeRu_2Si_2$ were performed by dHvA experiments (Aoki H. et al 1993 et 1995, Julian et al 1994). Quantitatively below the metamagnetic field $H_M$ i.e. in the PM phase, the data are well understood by assuming itinerant 4f electrons. For the main hole orbit $\psi$ detected below $H_M$ (figure 14), the dHvA signal can be only observed for H close to the basal plane (100) axis. Since its effective mass reaches 120 $m_o$ very large magnetic fields and very low temperature are needed for its detection. As $H_M \rightarrow \infty$ in the basal plane, there is no field limitation to observe the low field PM phase. On warming above $T_M$ , photoemission spectroscopy (Denlinger et al 2000) is well explained assuming the 4f electron localized i.e excluded from the FS which corresponds to $LaRu_2Si_2$ Fermi surface. Qualitative arguments can be found in (Fulde 1994).

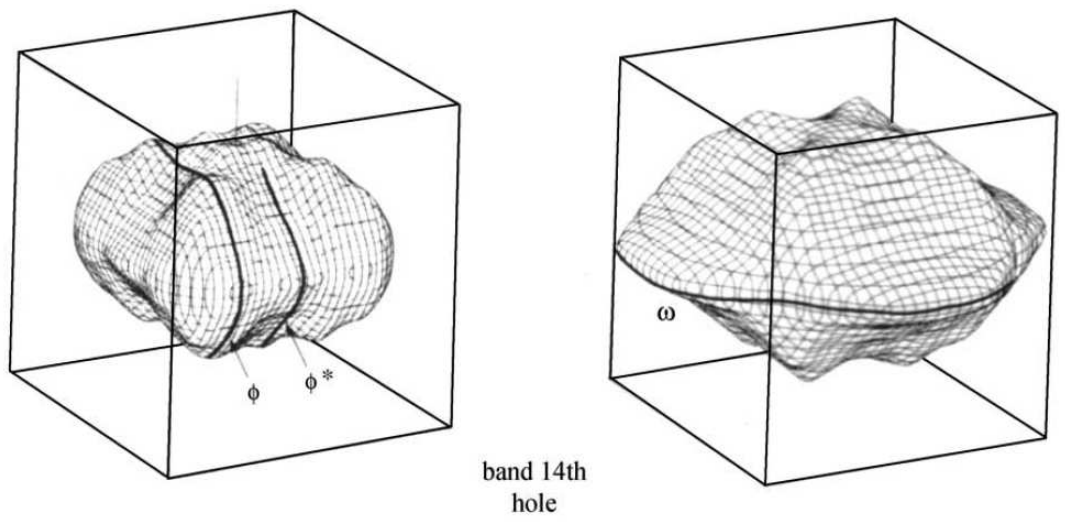

Figure 14 : Hole Fermi surface of $CeRu_2Si_2$ on both side of $H_M$ (Aoki et al 1995).

## *2.3 – The Kondo lattice $CeRu_2Si_2$ : (H, T) phase diagram*

The application of a magnetic field along the easy c axis leads to switch (through a drastic crossover at $H = H_M$) from a low field paramagnetic phase (PM), dominated by AF correlations and large local fluctuations, to a highly polarized state (PP), dominated by the low wavevector q (ferromagnetic) excitation and the surviving local fluctuations (Haen et al 1987) (figure 15 a-b). Neutron scattering experiments show the continuous spread of the AF response with the same characteristic energy $\omega_{AF}$ ~1.6 meV up to $H_M$ (Raymond et al 1999-a) and the emergence just in the vicinity of $H_M$ of an inelastic ferromagnetic signal at far lower energy transfer $\omega_F$ ~ 0.4 meV than $\omega_{AF}$ (Flouquet et al 2003, see also Sato et al 2004).

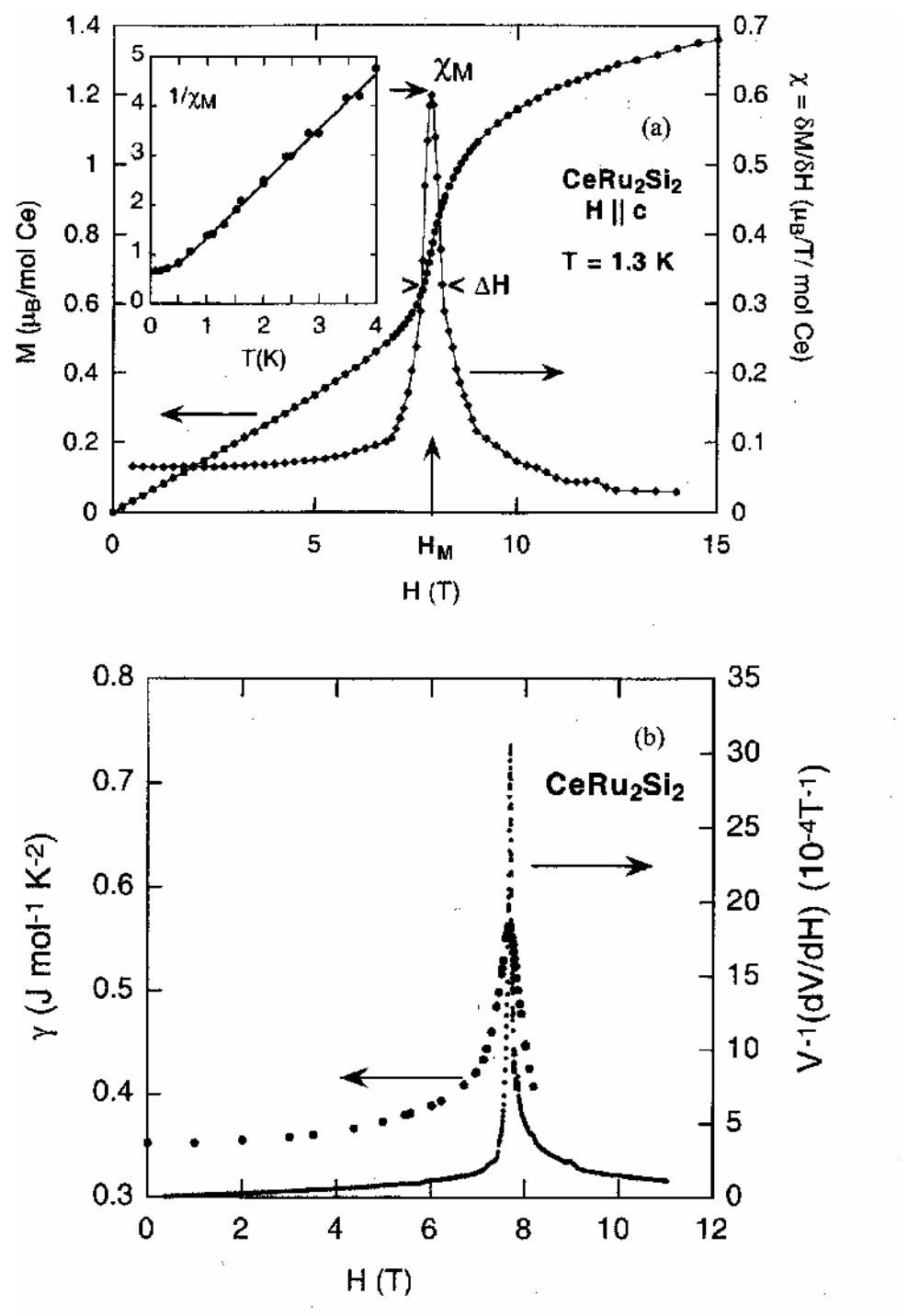

Figure 15 : In (a) low temperature magnetization M(H) of $CeRu_2Si_2$ , the insert is the temperature dependence of the differential susceptibility $\chi_M$ at $H_M$ . In (b) field variation of $\gamma = (C/T)$ $T \to 0$ and of the derivative of the magnetostriction ($1/V \frac{dV}{dH}$) (Flouquet et al 2002).

A skilful mechanism, characteristic of the Kondo lattice $CeRu_2Si_2$, controls the dominant magnetic interaction. Quite remarkably, the field and temperature response is well reproduced by a simple form of the thermodynamic quantities such as the entropy.

$$S = S\left(\frac{T}{T^*}, \frac{H}{H_M}\right)$$

with equal Grüneisen parameters $\Omega_{T^*} = \Omega_{H_M} = +190$ (Lacerda et al 1989). With this hypothesis one can predict the field variation of $\gamma_H$ knowing the field variation of the linear thermal expansion $\alpha_v = a_H T$. A satisfactory agreement is found between this scaling derivation of $\gamma_H$ and the bare determination either by specific heat or by magnetization (Paulsen et al 1990). Using the determination of $a_H/\gamma_H$, the corresponding field variation of $\Omega_H$ shows a positive and negative divergence through $H_M$ (figure 16) (Holtmeier 1994). It has been recently pointed out that a hyperbolic divergence of $\Omega_H = \frac{1}{H - H_M}$ is associated to a field quantum phase transition (Zhu et al 2003b). The phenomenon is not driven by a sole change of the AF phase transition but involves a concomitant transfer to ferromagnetic fluctuations. The driving mechanism is the field shift of the Fermi level in the pseudogap.

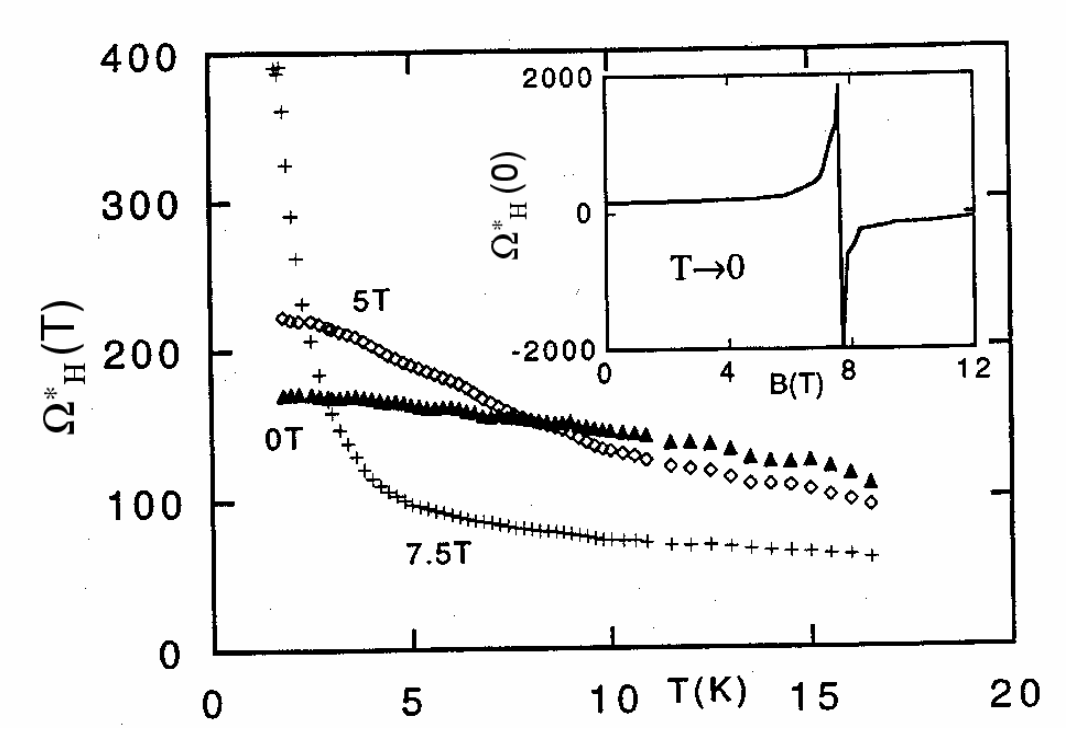

Figure 16 : $\Omega^*_H(T)$ Grüneisen parameter at constant magnetic field of $CeRu_2Si_2$ for H < $H_M$ . In insert, $\Omega^*_H$ (0) through $H_M$ (Holtmeier 1994).

The Ising spin character of the bare local magnetic ion plays a crucial role in the sharpness of this electronic substructure and thus for the pseudo-metamagnetism. The pseudogap shape of the density of states of $CeRu_2Si_2$ is the manifestation of the strong anisotropic hybridisation induced by the $|\pm 5/2>$ Ising crystal field ground state (Hanzawa et al 1987) and by the periodicity of the lattice (Evans 1992) In a Fermi liquid theory based on the periodic Anderson model (Ikeda 1997, Ikeda and Miyake 1997), the density of states of the quasiparticle bands has a singularity in $\sqrt{\varepsilon}$ of the energy ($\varepsilon$) in agreement with the observed T and H dependence of the specific heat (Aoki et al 1998). Satoh and Okhawa (2001) introduced a pseudogap as an input parameter in the Anderson lattice. The magnetic exchange interaction J (k) between the quasiparticules, caused by the virtual exchange of pair excitations of quasiparticles, depends on the structure of the density of states. The field sweep in the pseudogap produces a change of sign of J(k) at $H_M$ . As the volume dependence of $J_k$ (M) mimics that of $T_{K,}$ $J_k$ (M, x) scales with $T_K$ whatever its sign. Despite the H switch from AF to F, a scaling can occur through $H_M$ (the real mechanism comes from the Kondo lattice).

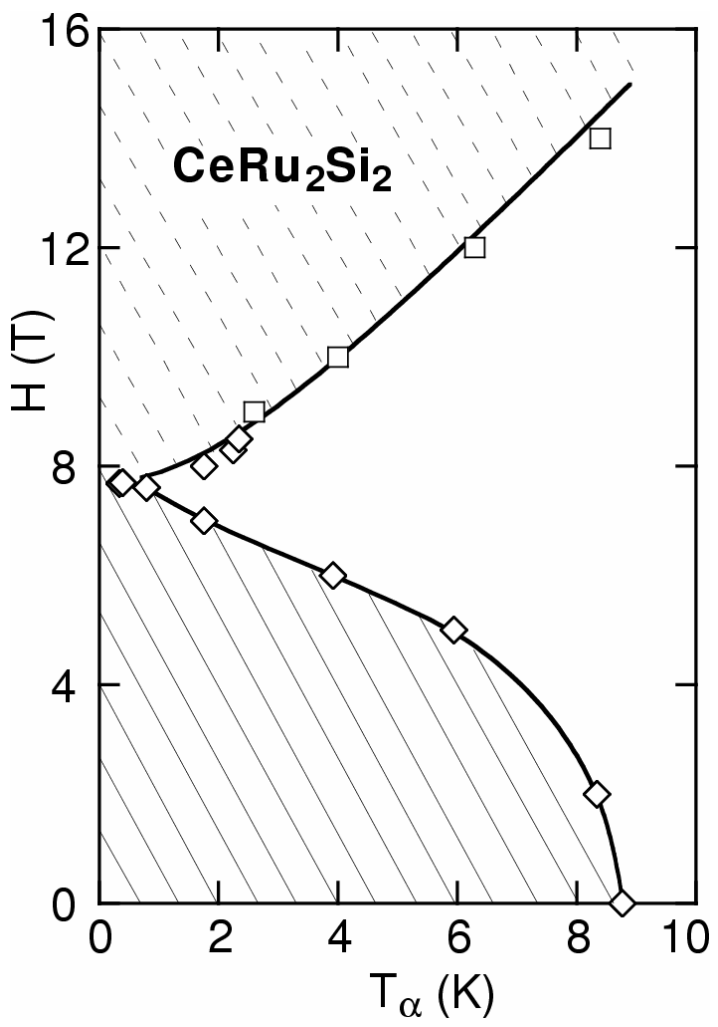

Figure 17 : The crossover phase pseudodiagram $T_\alpha$(H) derived from the thermal expansion measurements (◇). The high-field data (□) are the temperature of the C/T maxima. (Lacerda et al 1989)

Reminiscent of the α - γ collapse of the cerium metal, the associated spectacular lattice softening (~ 30%) (Kouroudis et al 1987) and the large volume expansion (~$10^{-3}$) illustrate the interplay (see below) between magnetic, electronic and lattice instabilities in the vicinity of this critical end point. By the sensitive technique of thermal expansion, it was possible to match the crossover line $T_\propto$ between PM, PP and the uncorrelated paramagnetic state (figure 17). Below $T_\propto$, $\propto_H$ ~ $a_H$ T, $T_\propto$ (H) defines the singular contour where this low temperature electronic property is recovered. Up to H = $H_M$, the thermoelectric power (TEP) shows a maximum at $T_{TEP}^{Max}$ = 0.50 K with roughly the same initial positive slope : $\left(\frac{\partial S_{TEP}}{\partial T}\right)$ (Amato et al 1989). While above $H_M$, $T_{TEP}^{Max}$ increases rapidly with the field. The quasi invariance of $T_{TEP}^{Max}$ below $H_M$ coincides also with the weak H variation of the temperature $T_A$ ~ 0.3 K below which $AT^2$ is obeyed in the resistivity despite changes in the amplitude of A by 80 % (Kambe et al 1995). Future experiments may demonstrate if $M_o$ collapses above the metamagnetic transition i.e entering in the homogeneous polarized paramagnetic phase. The invariance of $T_A$ and $T_{TEP}^{Max}$ for $H < H_M$ may be the fingermarks of the occurrence of weakly antiferromagnetism.

At P = 0, i.e for $P = P_C + \varepsilon$ , the thermal expansion is zero at $H_M$. The effective mass cannot diverge but reaches a pronounced maxima. For AF/SF as well as for ferromagnetic fluctuations in finite magnetic field, no divergence of the effective mass is expected. At least with the spin dynamics, only a drastic decrease of m* will appear above $H_M$ . In the Doniach Kondo picture, no divergency will be expected as it will occur for $T_K \rightarrow 0$ but here classical magnetism will lead to $\frac{m^*}{m_o} \rightarrow 1$.

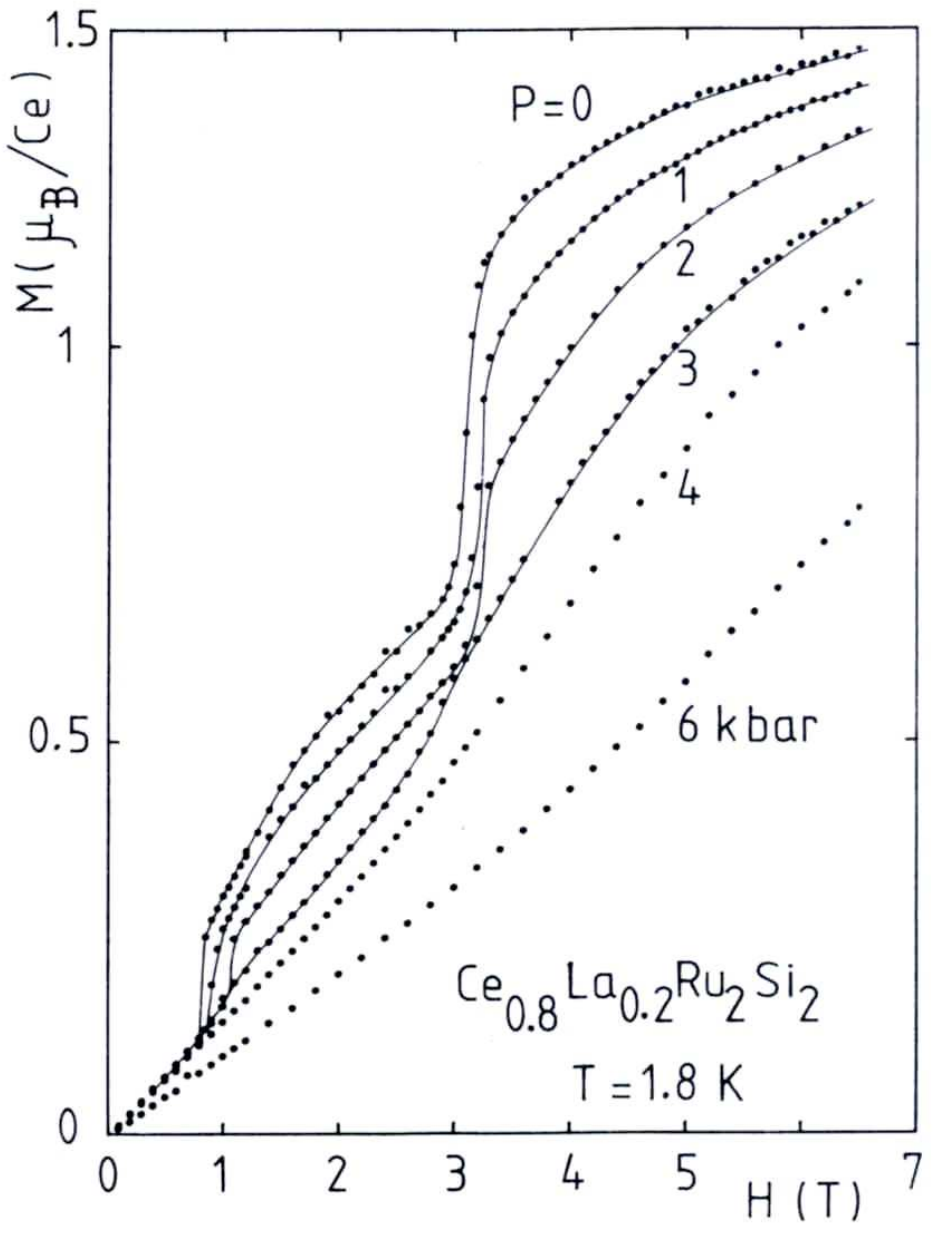

Figure 18 : Magnetization vs. H (//c) of $Ce_{0.8}La_{0.2}Ru_2Si_2$ at 1.8 K at different pressures indicated in kbar (0.1 Gpa). At $P_C$ , $H_M$ will be the extension of $H_C$ which ends up at the critical field $H_M$ ($P_C$ ) at $P_C$ . $T_N$ decreases strongly with the pressure. (Haen et al 1996).

For a volume expansion with x = 0.20 La subsitution, a 6 kbar negative pressure by comparison to $CeRu_2Si_2$ , the AF ordered phase at P = 0 shows two successive first order metamagnetic transitions at Ha and Hc (figure 18). Their weak initial pressure dependence corresponds to the observation that the magnetization jump occurs at the critical values $M_a$ and $M_c$ independent of the pressure (Haen et al 1987, 1996). The pseudometamagnetism at $H_M$ emerges on warming above $H_c$. As P approaches $P_C$ ~ + 3 kar, its differentiation from $H_c$ (T = 0) becomes less pronounced. In the ordered phase, the initial field variation of the magnetization i.e. here the Pauli susceptibility appears quasi-independent of the pressure. Constant $M_a$ and $M_c$ corresponds to fixed values of $H_a$ and $H_c$. In the PM state, the Pauli susceptibility $\chi_0$ becomes now strongly pressure dependent and correlatively $H_M$ since the product $\chi_0$ $H_M$ is invariant. The discontinuous disappearance of $H_C$ and Ha at $P_C$ has not been observed of course under ideal conditions as the lanthanum doping has introduced disorders. The sound idea is that $H_C$ reaches its critical point at $P_C$ and $H_M$ is the crossover continuation of the metamagnetic field $H_C$ above $P_C$ . A schematic picture of the pressure variation of $H_C$ and $H_M$ in $CeRu_2Si_2$ is drawn figure 19. It will be underlined later that, for $CeNi_2Ge_2$ and also for $YbRh_2Si_2$, the lost of the strong Ising character leads to a quite different situation with the evidence of a decoupling Kondo field $H_K$. In $CeRu_2Si_2$ , $T_A \sim T_O$ thus $H_K \sim H_M$ .Of course, the metamagnetic field $H_M$ changes in temperature. Its collapse on warming near 40 K corresponds to the disappearance of an inflection point in the magnetization, of the maximum of the magnetoresisitivy and of the temperature window where AF correlations are no any more visible in neutron scattering.

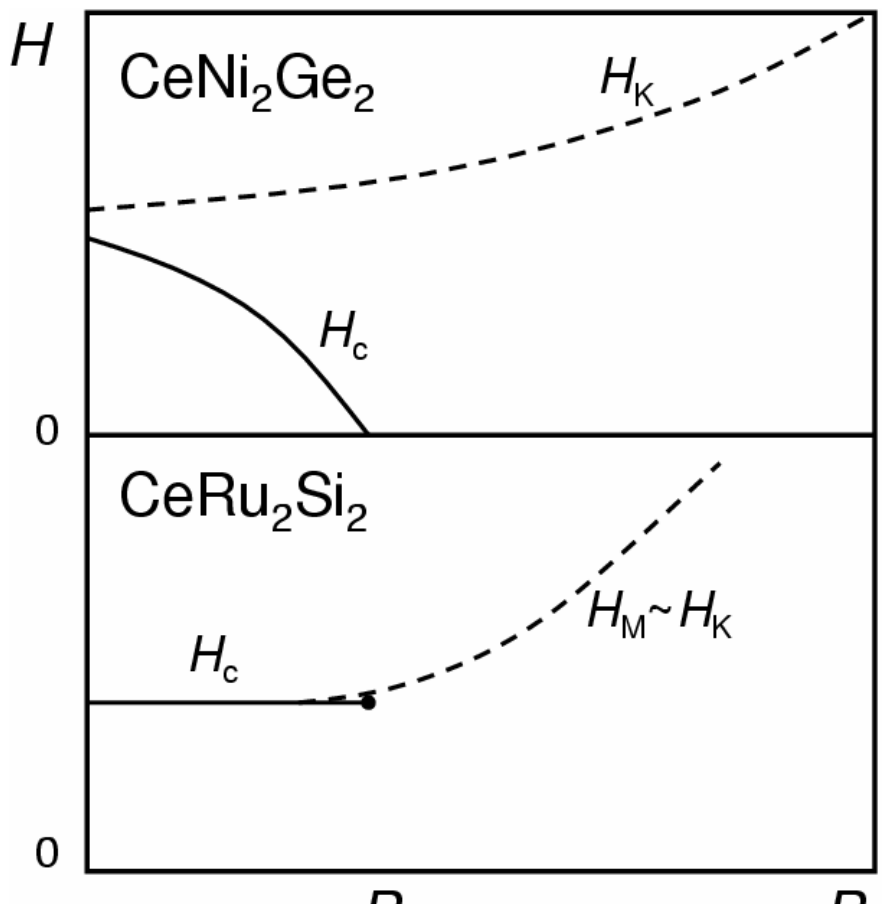

Figure 19 : Schematic pressure dependence of the critical field $H_C$ and the crossover field $H_M$ or $H_K$ for the two cases with $H_C$ ending at a critical point $P_C$ ($CeRu_2Si_2$ ) or $H_C$ collapsing with $T_N$ ($CeNi_2Ge_2$, $YbRh_2Si_2$ ).

An intriguing question is the FS evolution (figure 14). Above $H_M$ there is no trace of the heavy (Ψ) orbit ($m^* \sim 120\ m_o$), a new orbit (w) is detected now for H close to the (0, 0, 1) axis with rather moderate effective mass. As such an orbit is predicted in the $LaRu_2Si_2$ band calculation, it was first claimed that the magnetic field leads through $H_M$ to a localisation of the 4f electron (Aoki H. 1993, 1995). However important orbits are still missing in dHvA experiments as the measured FS is too small to explain the remaining large contribution of the electronic specific heat. Attempts to detect a carrier change at $H_M$ by Hall effect and tranverse magnetoresistivity failed to detect any variation (Kambe et al 1996). It may happen that the FS volume does not change but drastic modification occurs in the Kondo lattice. At least, from the analysis of the dHvA amplitude, the effective masses of spin up and down electrons are different above $H_M$ . A simple physical idea is that the minority spin bands get a high mass as, on travelling, they feel the repulsion of the majority spin electrons. Below $H_M$ , the spin up $n_\uparrow$ and spin down $n_\downarrow$ carriers can be regarded as undistinguished and their effective mass equal $m_\uparrow^* = m_\downarrow^*$. In the polarized frame, drastic change will occur. Then the current flow is not given by a single type of quasiparticles but by two types. It will be now the sum of the current of each spin entity. Running with their respective masses which changes with magnetic field. For H → ∞, the spin down carrier must become immobile $m_\downarrow^* \to \infty$, the spin up carrier completely undressed $m_\uparrow^* = m_0$.

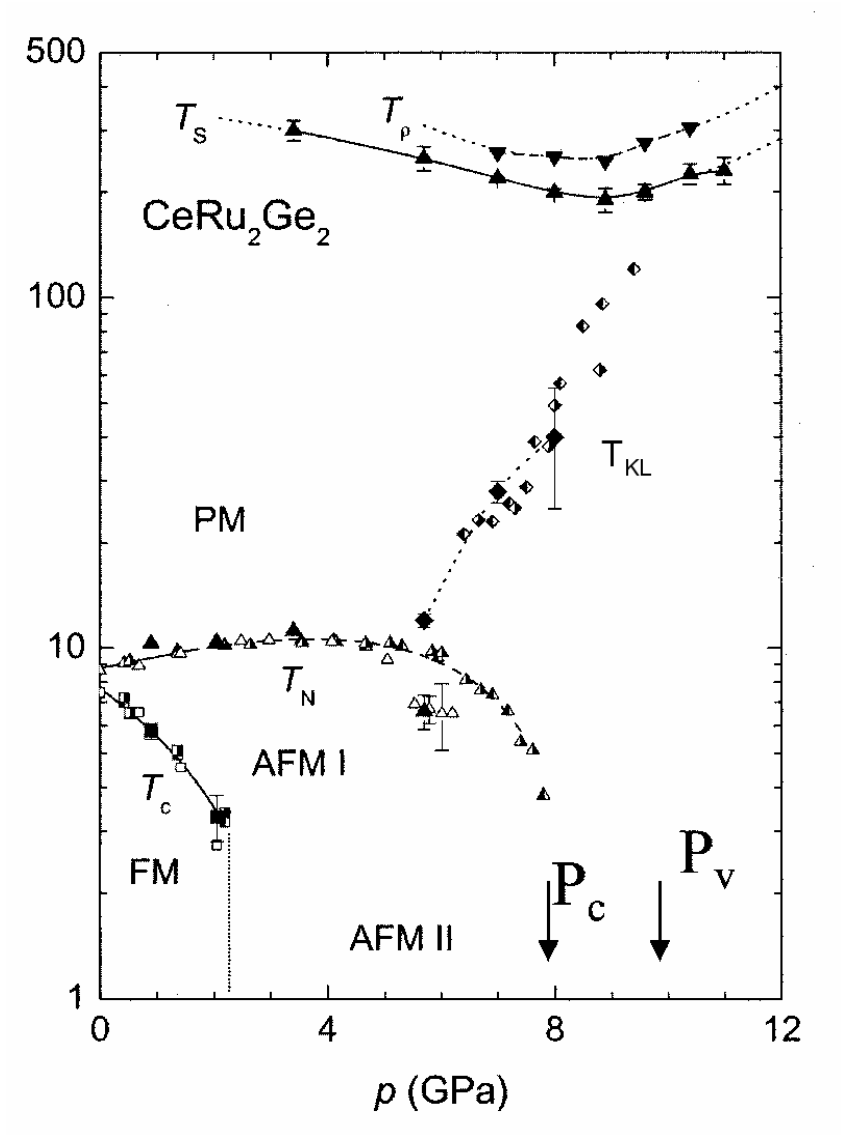

Figure 20 : (T, p) phase diagram of $CeRu_2Ge_2$ obtained from electrical resistivity (half open symbols) calorimetric (open symbols) and the combined ρ(T) and $S_{TEP}$ (bold symbols) measurements. At low pressure a paramagnetic (PM) to antiferromagnetic (AFM I) phase transition occurs at $T_N$ and a subsequent transition into a ferromagnetic phase (FM) takes place at $T_C$ . The FM ground state is suppressed at 2.3 GPa. The combination of all data suggests that long range magnetic order is suppressed at a critical pressure $P_C \sim 8.0$ GPa. $T_\rho$ and $T_S$ is the position of a maximum in ρ(T). The half filled diamonds indicate $T_{KL}$ α $1/\sqrt{A}$. (Wilhelm and Jaccard 2004).

A worthwhile reference will be to compare the FS of the polarized state of $CeRu_2Si_2$ versus that of the ferromagnetic $CeRu_2Ge_2$ (King and Lonzarich 1991 – Ikezawa et al 1997). $CeRu_2Ge_2$ represents the situation where the lattice of $CeRu_2Si_2$ would be expanded by a virtual negative pressure near 8GPa. For a positive pressure of 8 GPa, the electronic properties of $CeRu_2Ge_2$ reproduce the same behavior as $CeRu_2Si_2$ at P = 0. As shown in the (T, P) phase diagram (figure 20) (Wilhelm and Jaccard 2004), at P = 0, $CeRu_2Ge_2$ presents two successive magnetic states on cooling (Raymond et al 1999-b). The first AF phase has the same incommensurate propagation vector $k_0$ as that found previously while the low temperature phase is ferromagnetic. This last phase disappears rapidly under pressure at P = P*. The Curie temperature $T_{Curie}$ does not collapse to zero but merges at a finite temperature of 1.6 K. Above 3 GPa, only the AF order survives. In the low temperature F phase, FS measurements show that at P =0, the f electrons appear localized, i.e. the orbits are those found in band calculations for $LaRu_2Si_2$. Thus, the transition of F to AF ground states at P* appears discontinuous. It may coincide with a discontinuous changes in FS in contrast with the previous case of $CeRu_2Si_2$ through $H_M$ at P = 0. To clarify the situation above $H_M$ is still an experimental challenge ; progress needs to be made in band calculations taking the magnetic field into account since the powerful method of quantum oscillation requires strong magnetic fields which lead to finite polarization ($M > 0.1\ \mu_B$). One can see on the phase diagram that $T_K$ will reach the maxima of the resistivity $T_{Max}$ for $P = P_V \sim 10$ GPa i.e 2 GPa above $P_C$ . This observation agrees with the conclusion made for $CeRu_2Si_2$ that also $P_C < P_V$.

The absence of unconventional superconductivity in $CeRu_2Si_2$ associated with AF fluctuation can be due to the Ising character of the magnetism which precludes favourable transverse fluctuations (Monthoux and Lonzarich 2001) (see chapter 3). As we will see for $CeRh_2Si_2$, the superconducting dome may be also restricted in a narrow pressure range at $P_C$ . Already, at P = 0, superconductivity might be lost. One may speculate whether superconductivity will occur also for $P \sim P_V$ since valence fluctuations may be an efficient mechanism (see chapter 4). The superconductivity of $CeRu_2Si_2$ (or $CeRu_2Ge_2$ ) is still an open experimental problem as the results were obtained only in Bridgman cell with weakly hydrostatic solid pressure transmitted medium. A new generation of measurements must be performed.

If superconductivity occurs at $P_C$ , the interesting situation is if the upper critical field $H_{C_2}$ (0) becomes larger than $H_M$ . The switch in the magnetic interactions may induce change in the Cooper pairing and in the nature of the different transitions. Quite generally, an interplay of a previous (H, T) phase diagram as observed in $CeRu_2Si_2$ and other phase diagrams as superconductivity or long range magnetism may lead to novel phases. Such a case may happen for the superconductor $CeCoIn_5$ and the hidden order phase of $URu_2Si_2$ (chapter 4 and 6). Let us compare now $CeRu_2Si_2$ with three examples of highly documented NFL behavior : $CeCu_6$ , $CeNi_2Ge_2$ and $YbRh_2Si_2$ .

## *2.4 – $CeCu_6$ , $CeNi_2Ge_2$ : local criticality versus spin fluctuation*

The $CeCu_6$ familly (doping with Au, Ag) (von Löhneysen et al 1994, 1996, 2000) was extensively and carefully studied in the past since it was the first huge HFC ($\gamma$ = 1500 $mJmole^{-1}K^{-2}$) (Onuki and Komatsubara 1987, Amato et al 1987) where large single crystals can be obtained by contrast to $CeAl_3$. By comparison with $CeRu_2Si_2$ , its crystal structure basically orthorhombic is far less symmetric and also highest residual resistivity was realized ($\rho_0$ ~ 10 µΩcm). At low temperature $CeCu_6$ becomes even monoclinic. The respective anisotropies of the susceptibility along the a, b and c axes are $\chi_c/\chi_a$ ~ 5, $\chi_a/\chi_b$ = 2 while in $CeRu_2Si_2$ $\chi_c/\chi_a$ =15. FS measurements are still incomplete with the detection of few orbits (Reinders et al 1987). In $CeCu_6$ , the neutron inelastic spectrum shows less pronounced peaks in the wavevector response than in $CeRu_2Si_2$ (Aeppli et al 1986, Rossat Mignot et al 1988). Special focus was given to the critical doping $x_c$ = 0.1 of the $CeCu_{6-x}Au_x$ serie. Carefull thermodynamic measurements (Löhneysen 1994) points out NFL laws with unexpected temperature variation for 3d itinerant AF. For example, the specific heat follows a TLog T law over 2 decades of temperature ; the uniform static susceptibility has a strong increase in $T^{-0.75}$ on cooling. From neutron scattering date, it was claimed that the main effect is due to 2d magnetic fluctuations (as maxima of intensity at a given frequency occurs on wavevector rods (Stockert et al 1998). Ignoring the spreadout of the rods, this statement is given for the validity of a 2d treatement of the magnetic fluctuations (see Si et al 2001). In contrast to the previous case of $Ce_{1-x}La_xRu_2Si_2$, a unique ω/T scaling describes $\chi''(q, \omega)$ with $\alpha$ = 0.75, $\beta$ = 1 whatever the wave vector (Schröder et al 1998). The same $T^{+0.75}$ variation of the inverse susceptibility $\chi^{-1}(q)$ is found at all q (figure 21) with no sign of a saturation in temperature. As such invariance of $\alpha$ contradicts a SF description, a new picture called Fermi liquid destroying spin localizing transition ie with $P_{KL} = P_C$ has been suggested (Coleman 1999) with the intuition that m* may diverge on a single point $P_C$. The stimulating remark was that, at $P_C$, Fermi surface stability must be reconsidered. As we discuss chapter 1, this scenario seems supported by the theoretical developments (Si et al 2001 – 2003).

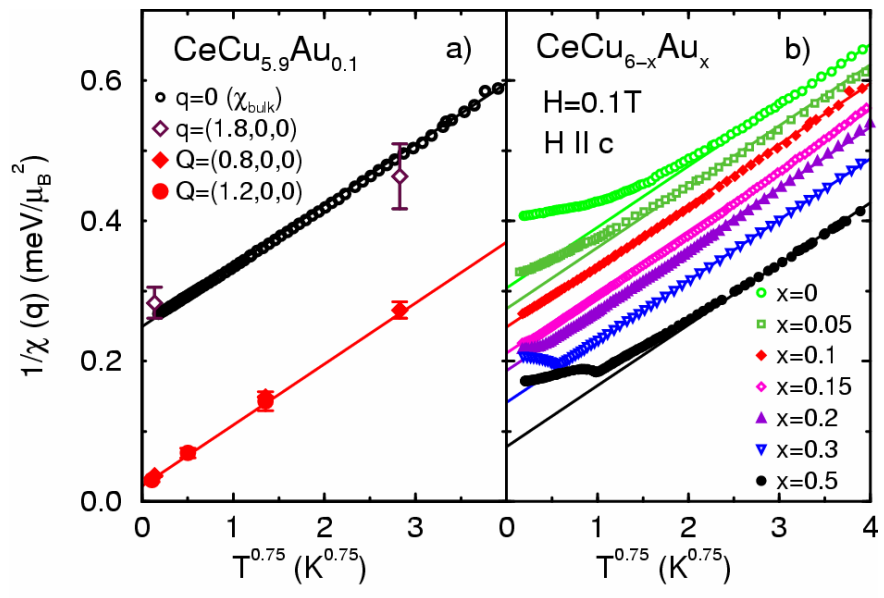

Figure 21 : Results on $CeCu_{6-x}Au_x$. In a), inverse of the static susceptibility $\chi_q$ measured at different wave vectors in $CeCu_{5.9}Au_{0.1}$ as a function of $T^{0.75}$. (Schröder et al 2000). In b), inverse of the uniform susceptibility $\chi q = 0$ for different x as a function of $T^{0.75}$. The proximity of $x_c = 0.1$ from QCP to a large T range where a NFL behavior in $T^{0.75}$ is observed.

To compare different HFC is not a easy task as the different ingredients ($\Delta_{CF}$, $T_K$, $T_{KL}$, $n_e$, $P_C$, ….) may give quite different temperature ranges for the observation of non Fermi liquid behaviors (domain I, II, III of figure 5). Furthermore the other ions (the ligands) play a role in the bandstructure. On table 5, typical parameters of the pure lattice $CeCu_6$ and $CeRu_2Si_2$ are shown. Both of them are located few tenth of GPa above $P_C$ respectively near –0.4 and –0.3 GPa. In the table 5, $\gamma$ is the extrapolated value of C/T at T $\rightarrow$ 0K, $T_A$ is the temperature below which $AT^2$ law in resistivity is observed, $T_M$ is the temperature where the susceptibility reaches its maximum, $T_{corr}^{(a)}$ the temperature characteristic of the magnetic correlations (Jacoud 1989) and $T_{corr}^{(b)}$ the temperature below which the usual positive magnetoresistivity of metals appears (values can be found in the references). A very low temperature needs to be achieved in $CeCu_6$ for entering in the FL regime. The drastic contrast between $CeCu_6$ and $CeRu_2Si_2$ is on $T_{corr}^{(a)}$ and $T_{corr}^{(b)}$ which respectively involves the onset of magnetic correlations and the formation of the Fermi surface. The low $T_{corr}^{(b)}$ value of $CeCu_6$ suggests a slow construction of the Fermi surface for the $CeCu_{6-x}Au_x$ serie. The open problem is if the very low temperature regime is different from that of $CeRu_2Si_2$ .

| Table 5 | $\gamma$ in mJ mole$^{-1}$K$^{-2}$ | $T_A$ in K | $T_M$ in K | $T_{corr}^{(a)}$ in K | $T_{corr}^{(b)}$ in K |
|---|---|---|---|---|---|
| $CeCu_6$ | 1500 | 0.1 | 0.8 | 5 | 0.15 |
| $CeRu_2Si_2$ | 360 | 0.5 | 10 | 40 | 70 |

As indicated in chapter 1, the specific heat and resistivity data indicates that $CeNi_2Ge_2$ may be located close to $P_C$ ($Y_0$ = 0.007). By contrast to $CeRu_2Si_2$ , in $CeNi_2Ge_2$, the ratio $\chi_c/\chi_a$ is relatively weak ($\chi_c/\chi_a$ ~ 1.4 at T = 4.2 K) (Fukuhara et al 1996). $CeNi_2Ge_2$ may appear ideal to study a 3d QCP as, at P = 0, it is very close to the QCP ($Y_0$ = 0.007). Neutron scattering experiments have been performed recently with mono isotopic Ni in order to avoid the large incoherent elastic scattering of natural Ni (Kadowaki et al 2003). This allows to detect the low energy excitations at $\omega$ ~ 0.6 meV around the AF vectors (½, ½, 0) and (0, 0, ¾). This energy range is lower than the previous excitations at 4 meV centered at incommensurate wavevector (0.23, 0.23, 0.5) (Fak et al 2000). The magnetic fluctuation becomes independent of the temperature below 2K. A crude fit of the specific heat can be obtained with this new low energy of 0.6 meV in the SF framework. But here again the low caracteristic energy is finite despite that $Y_o$ is considered near zero. That may support the previous statement on the $CeRu_2Si_2$ series of a non divergence of the coherence length at $P_C$ or push any QCP in $CeNi_2Ge_2$ to a deeper negative pressure than predicted.

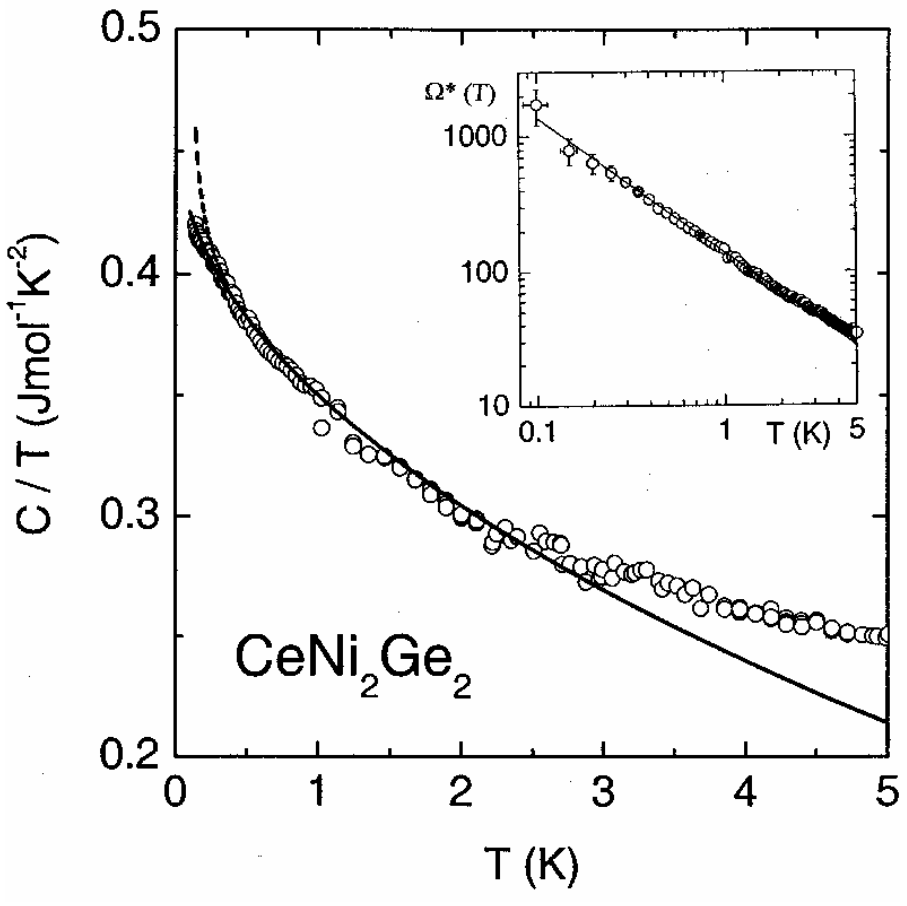

Figure 22 : Temperature variation of C/T and of $\Omega$* of $CeNi_2Ge_2$ (Küchler et al 2003). From the raw data (dashed line at low T) a nuclear contribution has been substracted giving the low T open circle. The solid line is a fit with SCR-SF i.e assuming C/T varies as C/T = $\gamma_0$ - C$\sqrt{T}$.

Recently new thermodynamic measurements (notably coupled specific heat and thermal expansion) were realized on $CeNi_2Ge_2$ down to 50 mK by Küchler et al (2003) (figure 22). Now in agreement with AF spin fluctuation theory at QCP, the Grüneisen parameter diverges at very low temperature ($\Omega^*$ = 70 at T = 1K goes up to 1000 at T = 0.1 K). The observed temperature variation of C and $\alpha$ is predicted by 3d AF spin fluctuation ($C/T = \gamma - \alpha \sqrt{T}$ , $\gamma(Y_0) = \gamma_{0-a} \sqrt{Y_0}$, $A(Y_0) \sim Y_0^{1/2}$ and $\Omega(Y_0) \sim Y_0^{-1}$ neglecting the pressure dependence of $\gamma_0$) (Moriya and Takimoto 1995). Thus thermodynamic measurements confirm the location of $CeNi_2Ge_2$ almost right at QCP. The finite value of $\Gamma_0$ stays an enigma. The metallurgy of $CeNi_2Ge_2$ is quite sensitive to the chemical composition (see Chichorek et al 2003). Appearance of superconductivity at $P_C$ or $P_V$ is still unclear (Grosche et al 2001, Braithwaite et al 2000). Thus, differences may exist between, large and small crystals as shown from studies on $CePd_2Si_2$. Thus, the location right at $P_C$ may be not so obvious. A complet view of $CeNi_2Ge_2$ is still not achieved.

Thermal conductivity experiments (Kambe et al 1999) have been performed on $CeNi_2Ge_2$ to test if the Wiedeman-Franz law $\frac{\rho K}{T} = L_0$ is well obeyed close to T $\rightarrow$ 0K, at least close to $P_C$ on the PM side. An excellent agreement was found. Of course, the effect of the proximity to the magnetic instability appears in the thermal response of both $\rho$ and K quantities. At least in this study, the charge carriers are also the heat carrier : the Wiedeman-Franz law is verified. Now, the remained question is if the Wiedeman Franz law will be also obeyed on approaching $P_C$ from the AF side.

Pseudometamagnetism in $CeNi_2Ge_2$ has been found at $H_K$ = 42T but the microscopic origin of the cross-over field is different than that of $CeRu_2Si_2$ where an equal balance exist between the intersite and Kondo coupling ($T_0 \sim T_A$ table chapter 1) (Fukuhara et al 1996). In $CeNi_2Ge_2$, $T_0$ and $T_A$ are quite different, weak magnetic fields restore rapidly Fermi liquid behavior with a concomitant strong H decrease of $\gamma$ (Gegenwart et al 1999). Thus bare AF correlations may strongly collapse with the magnetic field or interfer rapidly with ferromagnetic fluctuations. However high magnetic fields at H > $H_K$ are required to wipe out the Kondo lattice gap i.e for the breaking of the local Kondo fluctuations. To summarize for the $CeNi_2Ge_2$ serie, our proposal is that $H_C$ is quite decoupled from $H_K$ . For $P \rightarrow P_C$ , $H_C \rightarrow 0$ while $H_K$ is finite. The same phenomena seems to occur for $YbRh_2Si_2$ (Tokiwa et al 2004).

To summarize, there is evidence that the SF approach and its excitation spectrum describes experiments at first approximation as reported here for $CeRu_2Si_2$ or $CeNi_2Ge_2$ above $P_C$. However a zoom at a given ω or T window show even here the need for theoretical and experimental improvements. The focus just at $P_C$ or just on its PM side leads to neglect the AF boundary as $P \rightarrow P_C$ . In the reports on $CeCu_{6-x}Au_x$ (von Löhneysen 1994-1966-2000), $Ce_{1-x}La_xRu_2Si_2$ (Fisher 1991) or $Ce_{1-x}Pd_xNiGe_2$ (Knebel et al 1999), a maximum of γ at $P_C$ rarely exists. The dogma of a unique singularity at $P_C$ must be challenged. It is obvious that we prefer to break the popular consensus of a QCP with a universal second order transition.

## *2.5 – On the electron symmetry between Ce and Yb Kondo lattice : $YbRh_2Si_2$*

The recent fashionable material is $YbRh_2Si_2$ considered to be the hole Kondo lattice analog of $CeRh_2Si_2$. A strong similarity is expected between Yb and Ce intermetallic compounds with the difference that in Ce HFC the pressure induces a transition from AF to PM while, in Yb HFC, the pressure induces a reverse effect with a transition from PM to AF. The stable trivalent state corresponds to low pressure for the Ce center and high pressure for the Yb center. In the trivalent $Yb^{3+}$ configuration, the Kondo effect is produced by an absorption of an extraelectron on the 4f shell leading to its full saturation (14 electrons on the 4f shell) while for $Ce^{3+}$ the Kondo effect is created by the release of the electron from the 4f shell leaving an empty 4f shell.

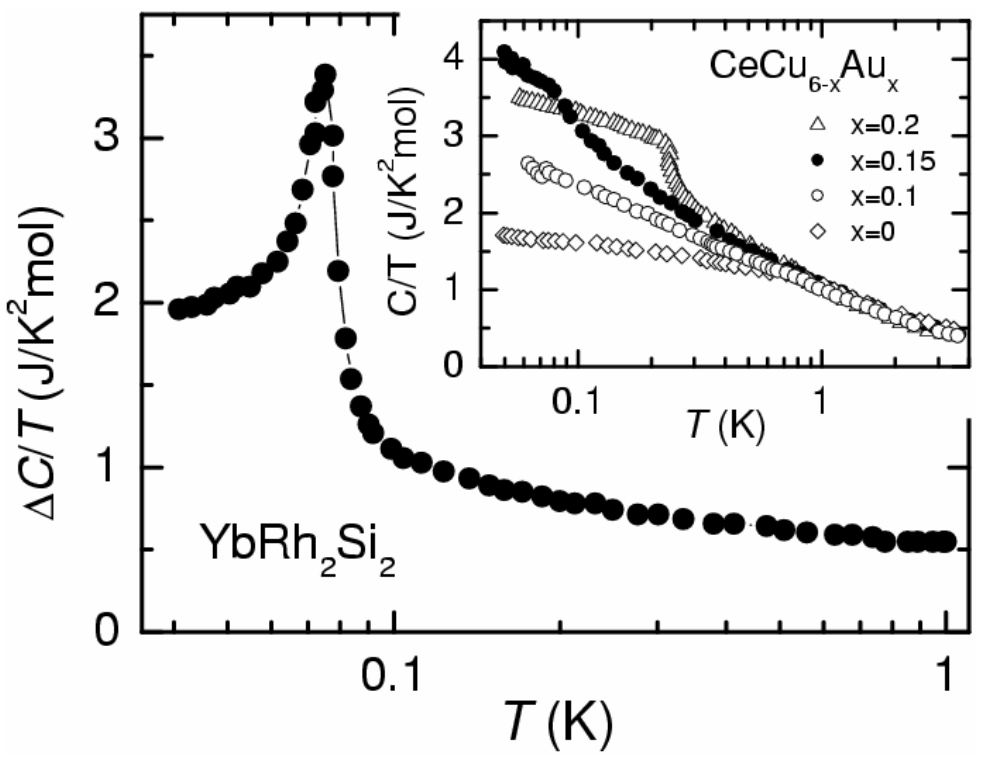

Figure 23 : Specific heat as ΔC/T vs T (on a logarithmic scale) for $YbRh_2Si_2$ . (from Trovarelli et al 2000 and Gegenwart et al 2002). Comparison with the specific C/T of $CeCu_{1-x}Au_x$ in the border of the QCP ($x_c$ = 0.1).

Excellent crystals of $YbRh_2Si_2$ have been obtained by the flux technic but on the form of thin tablets. The novelty in $YbRh_2Si_2$ is that nice specific heat and resistivity anomalies (figure 23) occur at $T_N$ = 70 mK for P = 0 even for a tiny ordered moment ($M_o = 10^{-3}\ \mu_B$) (Trovarelli et al 2000, Ishida et al 2003). In Ce compounds, the corresponding anomalies become highly difficult to follow below $M_o \sim 0{,}1\ \mu_B$ i.e x < 0.2 in the insert of figure 23 (see Löhneysen et al 1996). That strongly suggests that the Ytterbium case is not equivalent to the cerium case at least concerning their Kondo lattices close to $P_C$ . Evidence for this is given by the observation of an electron spin resonance at low temperature (Sichelschmidt et al 2003). Obviously, internal structures stabilize the existence of $Yb^{3+}$ moments below $T_K$ suspected to be near 25 K and even favored a large based magnetic anisotropy. In the Yb Kondo lattice, the spin coherence appears to be preserved during a long period and thus the electronic spin may be sensitive to any fine structure. Even the bare hyperfine interaction A between the nuclear spin I and the localized spin S is not a small perturbation (AIS ~ 1K for some Yb nuclei). Four different stable isotopes exist for Yb. Two $^{171}$Yb and $^{173}$Yb with the isotopic abundance of 14 and 16 percent have nuclear spin of I = 1/2 and I = 5/2 (see Flouquet 1978), a large variety of phenomena must be considered with different spin and orbital channels. The hyperfine coupling AIS may play the role of a cut off which must be compared with the characteristic energies $k_BT_K$ for the single Kondo impurity (see Frossati et al 1976), or $k_B\ T_{KL}$ or $k_B\ T_N$ for the Kondo lattice. It is amazing that when $T_N$ = 20 mK in the studies of $YbRh_2Si_2$ doped with 5% of Ge (Custers et al 2003), the critical field $H_C$ ~ 200 Oe for restoring the PM state is roughly the field where the electronic and nuclear spin of $^{171}$Yb and $^{173}$Yb will become decoupled. In order to test if the hyperfine coupling play a role, a crystal was measured with monoisotopic $^{174}$Yb which, as even neutron-proton nuclei, has no nuclear moment. As for the natural Yb case, the specific heat shows a very pronounced peak at $T_N$ = 80 mK and also the linear temperature dependence of the resistivity just above $T_N$ persists. Thus the particular effects of this Yb compound is not due to a fancy hyperfine dynamic but must be linked to the microscopic description of the 4f Ytterbium electrons.

To our opinion, the symmetry electron-hole between Ce and Yb may be not so valid as their respective coupling with the d electrons are not equivalent. Of course, there is similarity between $YbRh_2Si_2$ and $CeRh_2Si_2$ for a similar strengh of $T_K$ : amplitude of the maxima of the resistivity, T linearity of ρ just above $T_N$ , shape of the resistivity anomaly at $T_N$ but there are differences such as the sharpnes of the 4f density of states in $YbRh_2Si_2$ by comparison to $CeRh_2Si_2$ found in band calculations with local density approximation (Harima 2004). The 4 f orbital of Yb are far more localized (0.25 Å) that of Ce (0.37 Å) (Waler and Cromer 1965). Thus the degree of hybridisation must be higher in Ce HFC than in Yb HFC. In the picture of virtual bound state, the width $\Delta_{Yb}$ will be an order of magnitude smaller than that of Ce in the same non magnetic lattice. Thus for a given $T_K^{3+}$, the departure $(1 - n_f)$ of the occupation number $n_f$ unity will be ten times bigger for Yb than for Ce case. That may lead to drastic differences in the appearance of long range magnetism. Let us point out that a supplementary factor to preserve the local character of the Yb atoms is that the spin orbit coupling $\lambda'_{SO}$ of each 4f electron in the $j = \ell - s$ and $j = \ell + s$ individual configuration (see Abragam and Bleaney) is one order of magnitude higher than in the cerium case. i.e in the Yb case, $\lambda'_{SO} >> \Delta \sim C_{CF}$, in the Ce case $\lambda'_{SO} \sim \Delta > C_{CF}$. The hidden problem is the role of the strength of the hybridisation ($\Delta$, $T_K$ and $n_f$ ) on the crystal field and thus the relation between $C_{CF}$ and $T_K$ which will govern Ising, planar or Heisenberg spin dynamics.

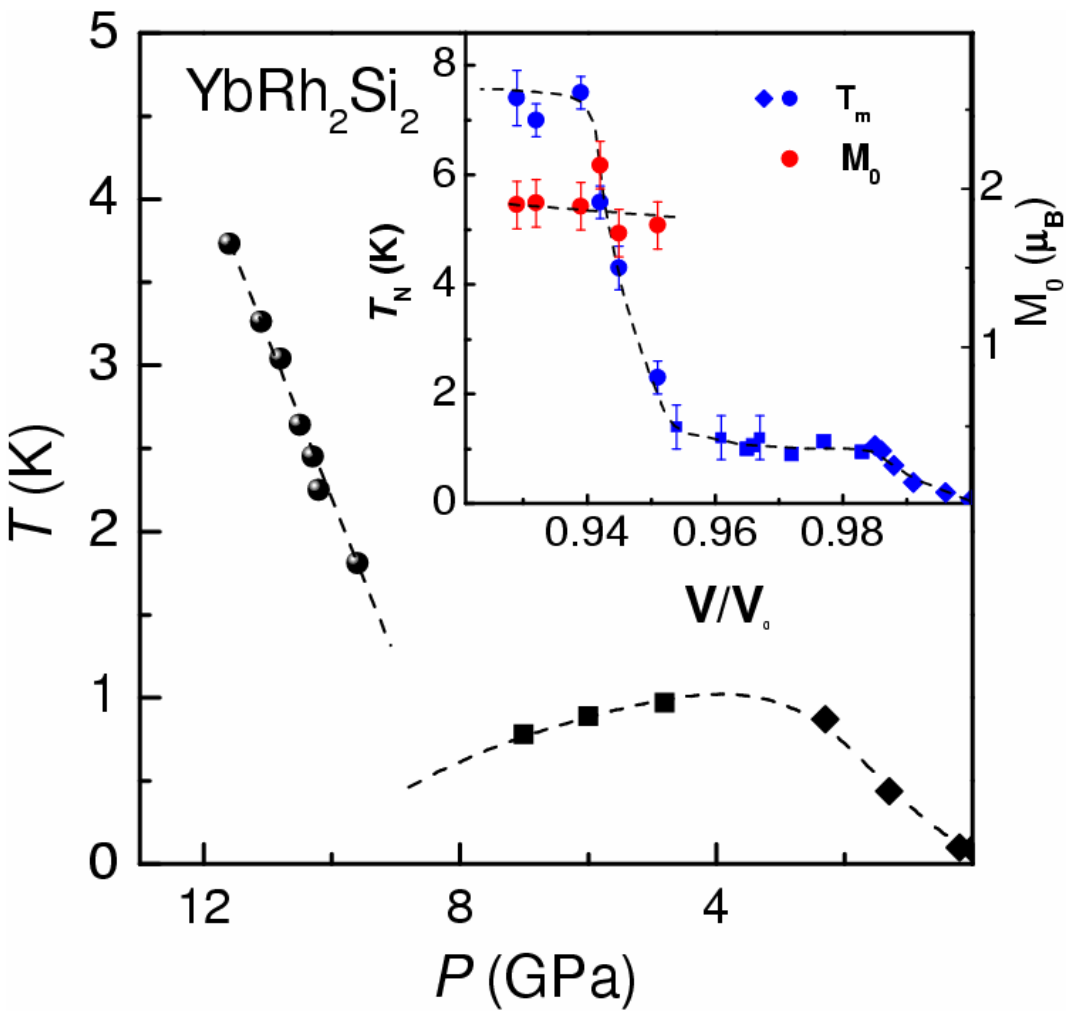

Figure 24 : The insert in the upper part shows the pressure –temperature magnetic phase diagram for $YbRh_2Si_2$ based on the experimental data for the ordering temperature $T_N$ determined by ME (●) and electrical resistance (■) and (◆) (Plessel et al 2003). The inner part show the recent result obtained with resistivity and microcalorimetric experiment under pressure (Knebel 2005).

A manifestation of the difference between $YbRh_2Si_2$ and $CeRh_2Si_2$ (see chapter 4, figure 29) is the pressure dependence of their Néel temperature. In $CeRh_2Si_2$ basically a collapse of $T_N$ occurs at $P_C$ = 1 GPa with a corresponding broadening and collapse of the specific heat anomaly. In $YbRh_2Si_2$ , (Plessel et al 2003, Knebel et al 2005, Dionicio et al 2005 - figure 24), increasing the pressure leads to a first regime where $T_N$ increases at the rate of $\frac{\partial T_N}{\partial P} \sim$ 0.4 K/GPa up to $P_1$ = 2 GPa ; then $T_N$ reaches a smooth pressure variation with a maxima of $T_N \sim$ 1K between 2 GPa and 7 GPa. Suddendly above P* ~ 9 GPa, $T_N$ strongly increases up to 12 GPa with a slope $\frac{\partial T_N}{\partial P} \sim$ 0.8K/GPa. Finally above 12 GPa, $T_N$ seems to saturate to $T_N \sim$ 8K with a sublattice magnetization $M_o \sim 2\mu_B$ (Plessel et al 2003).

Our feeling is that due to the already reported differences in valence mixing between Ce and Yb, the $Yb^{3+}$ configuration can live longer than the $Ce^{3+}$ one . Similar phenomena has been recently observed for IVC of Sm where also the Kondo effect on $Sm^{3+}$ corresponds to an absorption of a 5d electron on its 4f shell (Barla et al 2004 and 2005). Even if the d electrons from the Rh ions form a narrow band, locally they will be trapped on the Yb site. Increasing the pressure will delocalize strongly the d electron i.e decrease the f-d correlation and thus lead to recover a situation rather antisymmetric from the Ce case. Following the idea that for the same $T_K^{3+}$, $(1 - n_{Yb}) \sim 10\ (1-n_{Ce})$, the new phenomena in the Yb case is that AF can occur for a relative large departure of $n_f$ from unity. In this domain, the magnetic interaction will not be given by the simple RKKY dependence of $E_{ij}$ in $\Gamma^2$ N ($E_F$) but will contain a drastic dependence on $n_f$ . The competition between $T_K$ and Eij ($n_f$ ) can lead to different pressure regimes with the recovery of the Doniach picture only when $n_f \rightarrow 1$ i.e for P > P*.

The unusual temperature variations of the specific heat and of the resistivity just above $T_N$ and of their field dependences in $YbRh_2Si_2$ and $YbRh_2Si_2$ –x$Ge_x$ at P = 0 was a supplementary boost towards the possibility of local criticality. In the spirit of the breakdown of the heavy Fermi surface, proposed for $CeCu_6$ , a divergence of m* at the transition field $H_C$ from AF ($T_N$ = 80 mK, $H_C$ = 600 oe) was recently suggested (Gegenwart et al 2002). The claims of the divergence of m* by complementary studies on $YbRh_2Si_{1.95}G_{0.05}$ ($T_N$ ~ 20 mK, Hc = 200 Oe) leads to the statement that : "all ballistic motion of electron vanishes at the magnetic quantum critical point $H_C$ forming a new class of conductor in which electrons decay into collective current carrying motions of the electron fluid" (Custers et al 2003)".

In $YbRh_2Si_2$ $(Si_{0.95}Ge_{0.05})_2$, $\gamma(H)$ has mainly a log H decrease above $H^* = 10\ H_C$. Below 10 $H_C$ , in the large field region 10 $H_C$ > H > $H_C$ , $\gamma$ (H) seems to vary like $((H - H_C))^{-0.33}$ however there is no convincing evidence that m* diverges at $H_C$ since for the closest value reported of H to $H_C$, $\gamma(H_C + \varepsilon)$ is quite similar to the zero field value $\gamma_0 = 1.7$ J mole$^{-1}$K$^{-2}$ measured for the pure lattice of $YbRh_2Si_2$ . The large field quantum regime $\Delta H_C \sim 2000$ Oe where $\gamma$ (H) increases strongly on approaching $\gamma(H_C)$ is quite comparable to that found in $CeRu_2Si_2$ at $H_M$ . As in this field window, AF and F compete (Ishida 2002) it is not surprising that the Kadowaki – Woods ratio is not observed and that $A/\gamma^2$ increases at $H \rightarrow H_C$ . The relevance of 2d critically with strong local fluctuations (Si 2001, 2003 a and b) may be favored by the large anisotropy between the susceptibility $\chi_a$ and $\chi_c$ respectively $\perp$ and // to the c axis ($\chi_a/\chi_c = 200$). Thus the local ion has a clear planar anisotropy. Preliminary dHvA experiments with the detection of only few orbits does not show a 2d dimensional character which will reinforce the hypothesis of 2d fluctuations (Sheikin et al 2004). The results on $YbRh_2Si_2$ continue to boost new theoretical developments such as fractionalization of Fermi surface (Pépin 2004) and underscreened Kondo model (Coleman and Pépin 2003).

To summarize, $YbRh_2Si_2$ as $CeRu_2Si_2$ is a clean system with a simple axial symmetry. At P = 0, its position right on the AF side of $P_C$ opens a view which are quite complementary to that achieved in $CeRu_2Si_2$ or $CeNi_2Ge_2$ where studies concerned already at P = 0 the PM phase. Furthermore the planar local anisotropy of the spin in $YbRh_2Si_2$ is quite different from the respective Ising and Heisenberg character found in $CeRu_2Si_2$ and $CeNi_2Ge_2$. More systematic measurements on Yb HFC need to be realized to specify its microscopic description. That includes low energy experiments as well as high energy spectroscopy.

# 3/ Unconventional superconductivity

*Outlook :*

- Conventional superconductivity : only gauge symmetry is broken.
- Unconventional superconductivity : additional symmetry is broken.
- In HFS, interplay between the orbital and Pauli limit of the superconducting upper critical field $H_{c2}$ .
- In conventional superconductivity, antiferromagnetism and superconductivity may live peacefully while ferromagnetism usually destroys superconductivity.
- Unconventional superconductivity can be found in F and AF spin fluctuation approach. Magnetic and electronic anisotropies are favourable factors to increase the pairing strength.
- For a Kondo lattice, an efficient Cooper mechanism is also the density fluctuation

## *3.1 - Generalities*

Usual conventional superconductors are well described by the Bardeen Cooper Schrieffer theory (BCS) first established for a s wave singlet state with pairing of electrons with opposite spin and zero angular momentum. The order parameter $\Delta$ (k) is mainly isotropic. When the superconducting transition occurs in conventional superconductors, the gauge symmetry is the only symmetry broken at the superconducting transition. Due to strong coulomb repulsion among the f electrons, the existence of conventional s wave Cooper pairs with finite amplitude on a given site is precluded. This prohibition can be overcome with an anisotropic pairing like the triplet p wave (case of $^3He$) or the spin singlet d wave channel. We will see that ferromagnetic fluctuations or antiferromagnetic fluctuations can lead to the two situations.

In unconventional superconductivity, another symmetry is broken : point group, odd parity or time reversal. The latter occurs when the superconducting state has orbital, spin moments or odd frequency pairing. $\Delta$ (k) can be written for even and odd parity pairing respectively as :

$$\Delta (k) = \Psi(k) i \sigma_y$$

$$\Delta (k) = (\sigma . \hat{d}(k)) i\sigma_y$$

where $\Psi(k)$ and $\hat{d}$ (k) are respectively the even scalar and odd vector dependent of the momentum k ; $\sigma$ is the Pauli spin matrix. Often due to the additional broken symmetry, $\Delta$ (k) vanishes on point nodes or lines of nodes on the FS. The occurrence of zeroes will allow low energy excitations and produce temperature power law dependences of transport and thermodynamic quantities. That contrasts with the exponential collapse of the number of low energy thermal excitations for s wave superconductors. As gapless superconductors can also lead to power law temperature dependence, the proof of unconventional superconductivity requires careful studies as a function of the purity (i.e. of residual resistivity for example). Important support can be given by the discovery of multiple superconducting phases (case of $UPt_3$ ), triplet spin pairing (via NMR or other related techniques), of an anisotropy in the thermal conductivity, in the penetration depth and in ultrasound or of a violation of time reversal symmetry. Classification of the different order parameters based on group theory arguments has been given in reference (Gorkov 1987). We will later discuss in more detail the case of $UPt_3$ and $UPd_2Al_3$ .

In unconventional superconductors non magnetic impurities are effective pair breakers since the impurity scattering destroys the anisotropic Cooper pair wave function. In conventional superconductors, only magnetic impurities are efficient pair breakers as explained by Abrikosov and Gorkov (1961). Furthermore, it was stressed rapidly by (Pethik and Pines 1986) that in order to explain the results of ultrasound or thermal conductivity in $UBe_{13}$ or in $UPt_3$ , a large phase shift $\delta = \pi/2$ (in the strong scattering unitary limit) must be assumed. This assumption seems a "rule" applied now to all strongly correlated electronic systems. For unconventional superconductivity, it is generally admitted that the clean limit must be achieved i.e an electronic mean free path $\ell >> \xi_0$ where $\xi_0$ is the superconducting coherence length. Generally as $\ell \sim \rho_0^{-1}$ increases, $T_C$ increases as do the values of the upper critical field. We will discuss later how the impurity and pressure gradient can modify the (P, T) contour of superconductivity. For s wave dirty superconductors, the addition of non magnetic centers is a wellknown process to increase $H_{C2}$ (T) without changing drastically $T_C$ . In heavy fermion superconductors (HFS), any impurity will be pair breaking thus both $H_{C2}$ (T) and $T_C$ depend strongly on $\ell$. Improving the mean free path leads to obtain the optima values of $T_C$ and $H_{C_2}$ (0).

Before focusing on the mechanism of unconventional superconductivity driven by spin fluctuations, let us stress the particular situation of heavy fermion systems with regard to the field restoration of the normal phase at $H_{C2}$(T). The consequence of the huge effective mass m* is that the orbital limitation of the $H_{C2}$(T) given by :

$$H_{c_2}^{orb}(T) = \frac{\Phi_0}{2\pi\xi^2(T)} \; with \; H_{c_2}^{orb}(T=0) \approx m^{*^2} T_C^2$$

is large since the coherence length goes as $(m^*)^{-1}$

$$\xi_0 = 0.18 \frac{\hbar k_F}{m^* k_B T_C}$$

($k_F$ and $k_B$ being respectively the Fermi wavevector and the Boltzman constant).

At T ~ $T_C$ , the dominance of the orbital limitation leads to an initial linear temperature variation of $H_{C2}$. As the orbital contribution can be large, the breaking of the Cooper pair by the Zeeman effect (the so called Pauli limit $H_P$) can be efficient on cooling. For s wave superconductors, at T $\rightarrow$ 0K, $H_P(0)$ is equal to

$$H_P\ (O) = \frac{\sqrt{2}}{g\mu_B} \Delta_0 = 1.85\ T_C \ in\ Tesla$$

assuming g = 2 for the g factor of the conduction electrons. It will be effective for unconventional superconductors when the spin susceptibility decreases below $T_C$ . For triplet pairing, when H is perpendicular to d (direction where the spin components $S_z$ is zero), no Pauli limit occurs. For s wave superconductors, the reference article for $H_{C_2}$ (T) is from Wertharmer et al (1966). Scharnberg and Klemm (1988), has derived $H_{C_2}$ (T) for p wave triplet superconductor assuming no effective mass anisotropy.

When the Pauli limit dominates, for s wave superconductors in the case of the clean limit (generally required), Fulde Ferrel (1964) and Larkin Ovchinnikov (1965) (FFLO) predict that the entrance in the superconducting state below $T_{FFLO} \sim 0.56\ T_C$ is not the isotropic state but a spatially modulated structure. When the orbital limitation becomes comparable to the Pauli limit, $T_{FFLO}$ is lower and the occurence of the FFLO state is governed by the strength of the Maki parameter defined by :

$$\alpha = \sqrt{2}\ \frac{H_{C_2}^{orb}\ (0)}{H_p\ (0)} = 0.27\ g\left(\frac{\partial H_{C_2}}{\partial T}\right)_{T_C}$$

Favorable conditions seems to exist in $CeCoIn_5$, $UPd_2Al_3$ and $URu_2Si_2$ (H//c). When the orbital limit $H^{orb}$ (0) < 1.27 $H_P$ (0), the FFLO state disappears. The analysis of the superconducting properties of $UBe_{13}$ will lead to discuss the interplay between the different mechanisms as well as new effects due to strong coupling. Evidence of a FFLO state has been recently invoked in the new HFC $CeCoIn_5$ (see chapter 4). It was proposed for $UPd_2Al_3$ (Gloos and al 1993) and then rejected (Norman 1993) and suggested in $UBe_{13}$.(see chapter 6) but also with still no further confirmation.

As in heavy fermion systems, $T_C$ may be comparable to the effective Fermi temperature $T^*_F \sim T_K$, strong coupling can be considered. For $T_C \sim T_K$, the thermal disorder leads to a decrease of $T_C$ ; however on cooling, the superconducting properties will be reinforced. For s wave superconductors in an Einstein phonon model, it can be shown that $\Delta(0) \sim \sqrt{\lambda}\ k_B\ T_C > 1.76\ k_B\ T_C$ for $\lambda >> 1$. The extreme strong coupling phenomena are rare in HFC. Examples may be found in $UBe_{13}$ and $CeCoIn_5$ according to the criterium based on the specific heat jump ($\Delta$C) at Tc since for strong coupling superconductors $\Delta C/C$ can overpass the value (1.43) predicted for a BCS superconductor ($T_C < T_F$) . That is however an art effect of the fact that at $T_C$ the Kondo lattice may not have achieved its zero temperature value thus leading to a larger $\Delta C/\gamma T_C$ ratio if $\gamma$ is identified to C/T at $T_C$ .

## *3.2 - Magnetism and conventional superconductivity*

The pair breaking of s wave superconductivity by paramagnetic impurities was formulated by Abrikosov and Gorkov (1961). Experimental and theoretical discussions can be found in the volume V of the series Magnetism (1973) (Maple, Fischer and Peter Müller-Hartmann). The theory was extended to Kondo impurities (Müller-Hartmann and Zittartz, 1971). We will focus here on the interplay between long range ordered magnetism and superconductivity.

The coexistence of antiferromagnetism and superconductivity is well established in conventional s wave superconductivity. Basically on the scale of the superconducting coherence length, the Cooper pairs feel a zero exchange interaction. The field was very active after the discovery of the ternary compounds of rare earth (RE) elements and molybdenum sulfide (RE $Mo_6S_8$) in 1975 and of a serie of rhodium boride alloys (RE $Rh_4B_4$) in 1977 (see Fischer 1978-1990). It has been revived with the appearance of superconductivity in the new borocarbide family (RE $Ni_2B_2C$) in 1994. The major improvement brought by this family was the possibility to grow large crystals of high quality which allow a large diversity of experiments (Canfield et al 1998). The interesting features are the initial value of $T_C$ (16 K for $LuNi_2B_2C$) and the persistence of superconductivity even when $T_N$ becomes larger than $T_C$ ($T_N$ ~10K, $T_C$ ~5 K for $DyNi_2B_2C$). For further reading on the interplay between antiferromagnetism and superconductivity in borocarbide systems, the review article (Müller 2001) is recommended as well as the recent review of Thalmeier and Zwicknagl (2004-a).

The problems of the interplay between s wave superconductivity and ferromagnetism is more intriguing. It was continuously discussed in the last decades starting with the first paper by Ginzburg (1957) who points out that "the probability of finding superconductivity of ferromagnets in ordinary measurements is as small as that of finding non ferromagnetic superconductors placed in an external field with a magnetization of several thousand oersted". At this time, before the discovery of type II superconductivity, the first consideration was on the internal magnetic induction $B_0 = 4\pi M_0$ created by the magnetization density. Typical values are reported in the following table (6).

Table 6

| Material | $B_0$ in gauss |
|---|---|
| Fe | 22,000 |
| Co | 18,500 |
| Ni | 6,400 |
| Gd | 24,800 |
| $UGe_2$ | 2400 |
| $ZrZn_2$ | 400 |

Ferromagnetism places stringent limits on the existence of superconductivity. It can easily break the Cooper pair. In the experiments on erbium rhodium boride ($ErRh_4B_4$) and holmium molybdenum sulphide ($HoMoS_8$), it is now well established that superconductivity is destroyed by the onset of a first order ferromagnetic phase transition. For example, $ErRh_4B_4$ is a superconductor below 8.7 K. When it is cooled to a temperature Tm ~1 K a modulated magnetic structure appears rather than ferromagnetic ordering. Neighbouring magnetic moments are aligned in the same direction but a magnetic modulation occurs on a period $d < \xi_0$. The material is not ferromagnetic but presents domain like structures of period d/2 which have been detected by neutron diffraction ($d \sim 80$ Å, $\xi_0 = 200$ Å) (see Fischer 1978-1990 for reference). From the point of view of superconductivity, this magnetic structure is like an antiferromagnetic one but almost ferromagnetic on atomic scales. In these compounds, the energy gained by the atoms through the magnetic ordering $T_N \sim \Gamma^2/T_F$ exceeds the gain related to the superconducting transition as the Cooper pair modifies the electronic spectrum in a very small energy window $k_BT_C << k_BT_F$ . As the number of Cooper pairs is small ($T_C /T_F <<1$), the energy gain per atom due to the Cooper condensation is low $k_B T_C^2/T_F < k_BT_N$. At least for static magnetic centers, the magnetism is a much more robust phenomenon than superconductivity. Superconductivity cannot prevent the magnetic transition and is able only to modify it slightly. (In SCES this argument must be revisited as the magnetic ordering is driven by $k_BT_K$ and the pairing energy per atom becomes comparable to $k_BT_K$ if $\Delta \sim k_B T_K$ since $T_F \sim T_K$ : superfluidity of fermions may look like boson superfluidity). However such a phase is not stable at the lowest temperatures as the creation of microdomains costs energy. Cooling to $T_{Curie}$ = 0.8 K brings $ErRh_4B_4$ via a first order transition into a ferromagnetic phase with the disappearance of superconductivity (full restoration of the resistivity).The ferromagnetic phase destroys superconductivity as the exchange field acting on the light conduction electrons exceeds $T_C$ (see Flouquet and Buzdin 2003).

Superconductivity and weak ferromagnetism were found to coexist in $ErNi_2B_2C$ below $T_{Curie}$ = 2.3 K (Gammel et al 2000). The ferromagnetic component (0.33 $\mu_B$ /Er atom) is weak as only one of the 20 Er atoms contributes to the ferromagnetic order (periodicity near 35 Å) (Choi et al 2001, Kawano et al 2001, Deflets et al 2003). Up to now, a spontaneous flux lattice has not been observed at H = 0 as suggested by Ng and Varma (1997) but enhanced critical currents characterized the entrance into the weakly ferromagnetic cases (Gammel et al 2000). Ferromagnetism and superconductivity have been claimed to coexist in $Eu_{1.5}Ce_{0.5}RuSr_2Cu_2O_{10}$ (Felner et al 1997) and $RuSr_2GdCu_2O_8$ (Bernhard et al 1999). In these cases, the two states might occur in different structural layers. An exotic case is that of nuclear ferromagnetism ($T_{Curie}$ = 37 µK) on the superconductivity of $AuIn_2$ ($T_C$ = 207 mK, $H_C$ = 1.45 mT) (Rehmann et al 1997). However it was suggested that the contact hyperfine exchange interaction may also lead to a domain structure as described for $HoMo_6S_8$ and $ErRh_4B_4$ (Kulic et al 1997).

In these cases, different electrons are involved in magnetic and Cooper pairing . Switching to the heavy fermion case we will now assume that the same electrons are involved in the magnetism and superconductivity. Of course, unconventional superconductivity and ferromagnetism can occur as emphasized previously for triplet superconductivity and notably for equal spin pairing (EPS) between parallel (up up or down down) spins. In such a case, the exchange field cannot break the Cooper pair via the Pauli limit and the upper critical field limitation will be limited only by the orbital limit which can be very high if the effective mass is huge. The discovery of superconductivity in the ferromagnet $UGe_2$ with a Curie temperature $T_{Curie}$ ~ 30K far higher than $T_C$ ~ 0.7 K at its optimum $P_{opt}$ = 1.3 GPa opens new perspectives (chapter 5).

### *3.3 - Spin fluctuations and superconductivity*

The relevance of nearly ferromagnetic spin fluctuations for anisotropic BCS states was illustrated by the p wave superfluidity of liquid $^3He$ (Anderson and Brinkman 1973, Nakajma 1973). The p wave superconducting transition temperature for paramagnon induced pairing in nearly ferromagnetic itinerant systems was first calculated by Layzer and Fay in 1971. The vanishing of $T_C$ at $P_C$ is correlated with the vanishing of $T_I$ ie the divergence of the effective mass as the pairing is a strong function of frequency. In the paramagnetic state, $T_C$ is the same for parallel and antiparallel spin pairs. Of course, a different situation (fig. 25) will occur in the ferromagnetic state calculated by Fay and Appel (1980). The ESP interaction between ↑↑ or ↓↓ component of the triplet with an angular momentum's transfer q is related to the non interacting Lindhard response of the spin susceptibility $\chi_0^{\uparrow}$ or $\chi_0^{\downarrow}$ and the onsite Coulomb repulsion U by the relation :

$$V^{\uparrow\uparrow} = \left( \frac{U^2 \chi_0^{\downarrow\downarrow}}{1 - U^2 \chi_0^{\uparrow\uparrow}(q)\, \chi_0^{\downarrow\downarrow}(q)} \right)$$

Due to the spin conservation, the exchange of spin waves or transverse fluctuations does not contribute to $V^{\uparrow\uparrow}$. As for the PM phase, $T_C$ reaches a maximum before collapsing at $P_C$ . Figure (25) shows the variation of $T_C$ versus I or P – $P_C$. In the ferromagnetic domain, $T_C$ differ for the majority (↑) and minority (↓) spin.

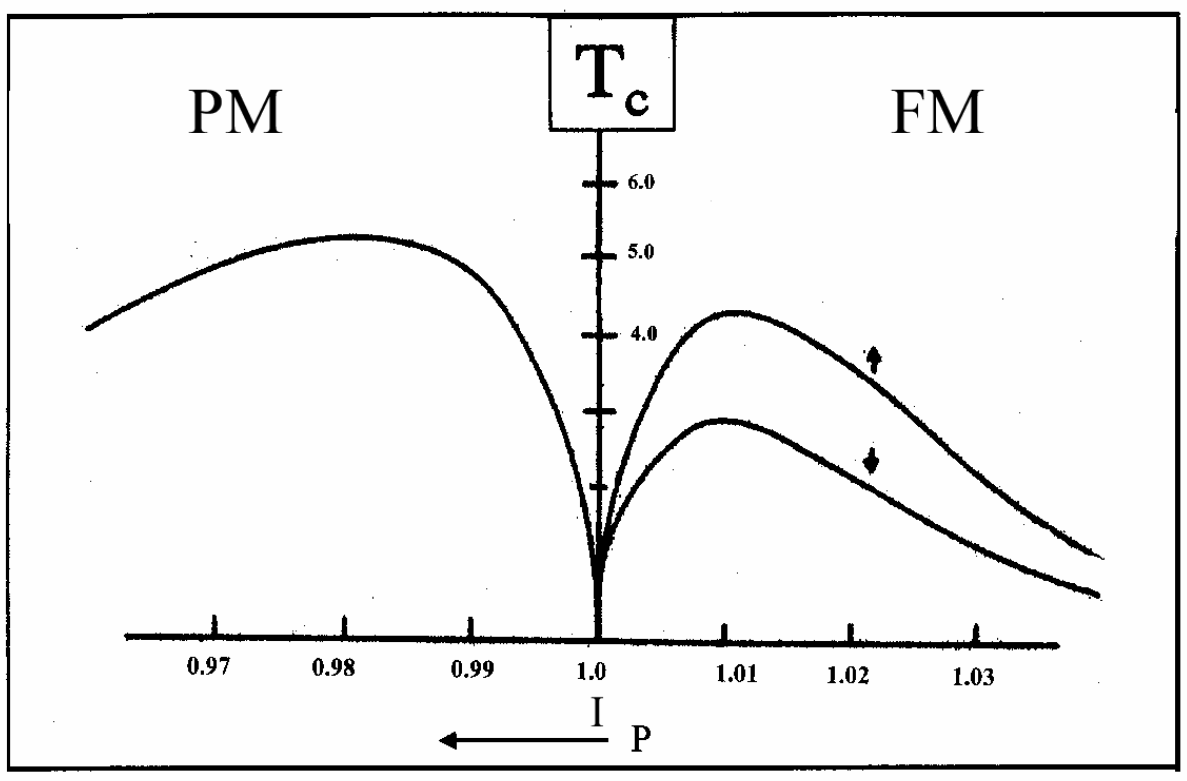

Figure 25 : First prediction for superconducting transition $T_C$ of a ferromagnet under pressure P (Fay and Appel 1980).

A more complete treatement (Roussev and Millis 2001) shows that $T_C$ does not vanish at $P_C$ but reaches a minimum in the vicinity of $P_C$. Furthermore for small values of ($P – P_C$ ), the $T_C$ behavior is predicted to be universal. The coexistence of ferromagnetism and superconductivity was revisited recently by Kirpatrick and Belitz (2003). We have already pointed out that an enhanced impurity scattering occurs at $P_C$ (Miyake and Narikiyo 2002).

The discussion of superconductivity induced by antiferromagnetic fluctuations has been boosted by the discovery of superconductivity in high $T_C$ oxydes and the rapid demonstration that the spin dynamics play an important role in the normal phase properties and react on the superconductivity onset (see Rossat Mignod et al 1992). Early considerations on superconductivity mediated by antiferomagnetic fluctuations can be found in ref (Emery 1983, Hirsch 1985, Miyake et al 1986). Reviews have been written on the AF spin fluctuation model and d wave superconductivity ($d_{x2}$-$_{y2}$ for high $T_C$ ) (Moriya and Ueda 2000, 2003-b) (Chubukov et al 2002) with discussions of the relevance to the different SCES. It was stressed that a key point is the occurrence of the so-called hot spots in the Fermi surface i.e. peaked magnetic coupling at the wavevector $k_o$, and that d wave pairing is favoured more by AF spin fluctuation than magnon like excitations. Two distinct frequencies have strong impact on the pairing : the characteristic spin fluctuation energy $\hbar\omega_{sf} \sim k_BT_I$ and the energy $\hbar\varpi_0 \sim k_BT_0$ related to the effective Fermi temperature.

The d wave spin pairing in nearly AF systems is generally stronger than a triplet pairing in the nearly F case as both longitudinal and transverse fluctuations can mediate superconductivity (Monthoux and Lonzarich 2001). For comparable parameters, the strength of the pairing increases with the magnetic $\alpha_m$ and electronic $\alpha_t$ anisotropy (Monthoux and Lonzarich 2003) independently of the type of the magnetic instability (ferromagnetic or antiferromagnetic). Few theoretical studies concern the respective boundary of $P_C$ and the pressure $P_{-S}$, $P_{+S}$ where superconductivity emerges and disappears ($T_C$ = 0 K at $P_{-S}$ and $P_S$) . A recent discussion can be found in reference (Takimoto and Moriya 2002) where different cascades of ground states are pointed out : for example a first order transition from commensurate spin wave (SDW) to superconductivity (S), cascade of SDW to incommensurate (I) SDW followed by the coexistence of ISDW and S, then of S and PM and finally PM phases. Due to the occurrence of soft modes near electronic, magnetic or structural instabilities, very often $T_C$ reaches its maximum at the instability point in agreement with the famous MacMillan optimisation between coupling and energy cut off. A good example is given by the disappearance of the charge density wave of α uranium at $P_{CDW}$ where $T_C$ has its maximum (see Smith and Fisher 1973). Let us indicate the new approach where the orbital splitting energy ε plays a key role of control parameter for the AF and S quantum phase transition in HFC (Takimoto et al 2003). As ε increases ie for example $C_{CF} / k_B$ versus $T_K$, the system transits from PM, and AF with a superconducting dome right at $P_C$ (see also Hotta and Ueda 2003).

### *3.4 - Atomic motion and retarded effect*

An important point in HFC is their huge Grüneisen parameter which points out huge anharmonicity or strong mixing between differents modes. Neglecting the electron, in a harmonic crystal, the coefficient of the thermal expansion (α) must be zero as the pressure required to maintain a given volume does not vary with temperature. Anharmonic terms play a big role and are responsible of the phonon thermal expansion $\alpha \sim T^3$ . A typical phonon Grüneisen parameter is $\Omega$ ($\theta_D$) ~ 2 for the associated phonon contribution characterized by their Debye temperature $\theta_D$ (Aschcroft and Mermin 1976). The electronic contribution for normal metal as copper will lead to a linear temperature thermal expansion ($\alpha \sim T$) as its Fermi temperature varies as $V^{-2/3}$ and effective mass near the free electron mass. Again the corresponding Grüneisen parameter is weak $\Omega(T_F) = +0.66$. In HFC, as underlined, Ω* can reach 100 or 1000. The electronic thermal expansion $\frac{\partial V}{\partial T}$ goes basically as $m^{*2}$ (i.e $\frac{\partial V}{\partial T}$ ~ $\Omega^* m^*$ and $\Omega^* \sim m^*$). An amplification of four orders of magnitude of $\frac{\partial V}{\partial T}$ can be achieved by comparison to a normal metal at low temperature. Thus the displacement of the atom can reach a value at 1K only achieved in ordinary metal above 100 K through the volume variation of the Fermi temperature or through the phonon anharmonicity. The density fluctuation of HFC is very large. In conventional superconductors, although the direct electrostatic interaction is repulsive, the ion motion overscreens the coulomb interaction and leads to the attraction (see Aschcroft and Mermin 1976). As the interaction spreads over an energy interval $k_B\theta_D$, it operates over a finite interval.

In HFC the two ingredients of pairing as for the electron phonon coupling may exist : atomic motion linked to large density fluctuation and retarded effect linked to the long lifetime $\tau_{KL}$ of the Kondo cloud or the spin fluctuation near $P_C$ . Thus the image of its superconductivity may be rather similar than that developed for the electron phonon interaction. In the early time of $CeCu_2Si_2$ discovery it has been proposed that the electron phonon mecanism can explain the superconductivity of $CeCu_2Si_2$ (see Ambrumenil and Fulde 1985, Razafimanchy et al 1984, Tachiki and Maekawa 1984). As pointed out in chapter 4, the Kondo coupling favors longitudinal fluctuations and thus the difference in pairing between AF and F may be not so high. An interesting idea is the possible difference due to large retarded effects close to $P_C$ (long lifetime of excitations) and instantaneous coupling far from $P_C$ (see Fuseya et al 2003). Near $P_C$ , the possible achievement of p wave spin singlet superconductivity with a gap function $\Delta$ (k, i$\omega$) odd in momentum and frequency has been found to be more likely than d wave singlet superconductivity. This phase will have no gap in the quasiparticle spectrum anywhere on the FS due to the odd frequency. It will exist on both sides of $P_C$ if $P_C$ is a second order QCP. The conditions for this stabilization in a narrow pressure range around $P_C$ are that : 1/ FS is not nested i.e basically the origin of AF is the exchange interaction between the local spin component, 2/ strongly retarded effects which occur at P ~ $P_C$ where the effective interaction is strongly frequency dependent. A crucial theoretical point is to clarify the Meissner effect since at simple level of calculations a negative Meissner effect has been found in odd frequency pairing (Abrahams et al 1995). Evidence of gapless superconductivity close to $P_C$ in HFC has been found in the $CeCu_2(Si_{1-x}Gex)_2$ series (Kitaoka et al 2001) and will be reported here for $CeRhIn_5$ (Kawasaki et al 2003-b, Knebel et al 2004). In this last example the associated diamagnetic response is broad and shifted to high temperature with respect to the temperature $T_C$. The tiny specific heat or NMR superconducting anomalies at $T_C$ can be due to extrinsic properties. The debate on the intrinsic gapless properties is open.

# 4/ Superconductivity and antiferromagnetic instability in cerium compounds

*The outlooks are :*

- An universal second order singularity at $P_C$ may not be the correct vision.
- Observation of magnetic phase separation or first order transition in $CeIn_3$, $CeRh_2Si_2$ and $CeRhIn_5$ .
- Two distinct superconducting phases in $CeCu_2Si_2$ : two mechanisms ?
- The new 115 serie : importance of quasibidimensional fluctuations
- A new field induced superconducting phase in $CeCoIn_5$.
- $T_C$ dependence on the anisotropic ratio c/a between c and a lattice parameters ; the record with $PuGaIn_5$
- Recent exotic superconductors : $CePt_3Si/PrOs_4Sb_{12}$

## *4.1 – Superconductivity near magnetic quantum critical point $CeIn_3$, $CePd_2Si_2$ and $CeRh_2Si_2$*

For clarity, we will not use the chronological order of the discovery of superconductivity near $P_C$ but select successive systems where the magnetic instability can be tuned under pressure. In the chosen examples of $CeIn_3$, $CePd_2Si_2$ and $CeRh_2Si_2$, $P_C$ is respectively 2.6, 2.8 and 1.0 GPa and the superconductivity domain is centered around $P_C$ . In the cases of $CeCu_2Si_2$ and $CeCu_2Ge_2$ superconductivity appears just below $P_C$ but its temperature maximum $T_C^{max}$ is shifted to higher pressures. We will discuss this situation later. The main difference seems to be that for the three first examples, $P_C$ and $P_V$ coincides or at least are very near while in the $CeCu_2(Si_{1-x}Ge_x)_2$ familly, $P_C$ and $P_V$ are quite distinct.

**4.1.1 - $CeIn_3$ : phase separation.** At P = 0, $CeIn_3$ is a AF HFC with $T_N$ = 10 K, $M_o$ = 0.5 $\mu_B$ and ½ ½ ½ propagation vector (Lawrence and Shapiro 1980, Benoit et al 1980). The spin dynamics at zero pressure are now very well documented by neutron scattering experiments with the determination of the crystal field splitting ($C_{CF}$ = 10 meV), the observation of a quasielastic line and damped spin wave (Knafo et al 2003). Neutron diffraction experiments under pressure have already suggested that AF in the cubic lattice of $CeIn_3$ will collapse near $P_C$ = 2.6 GPa (Morin et al 1988).

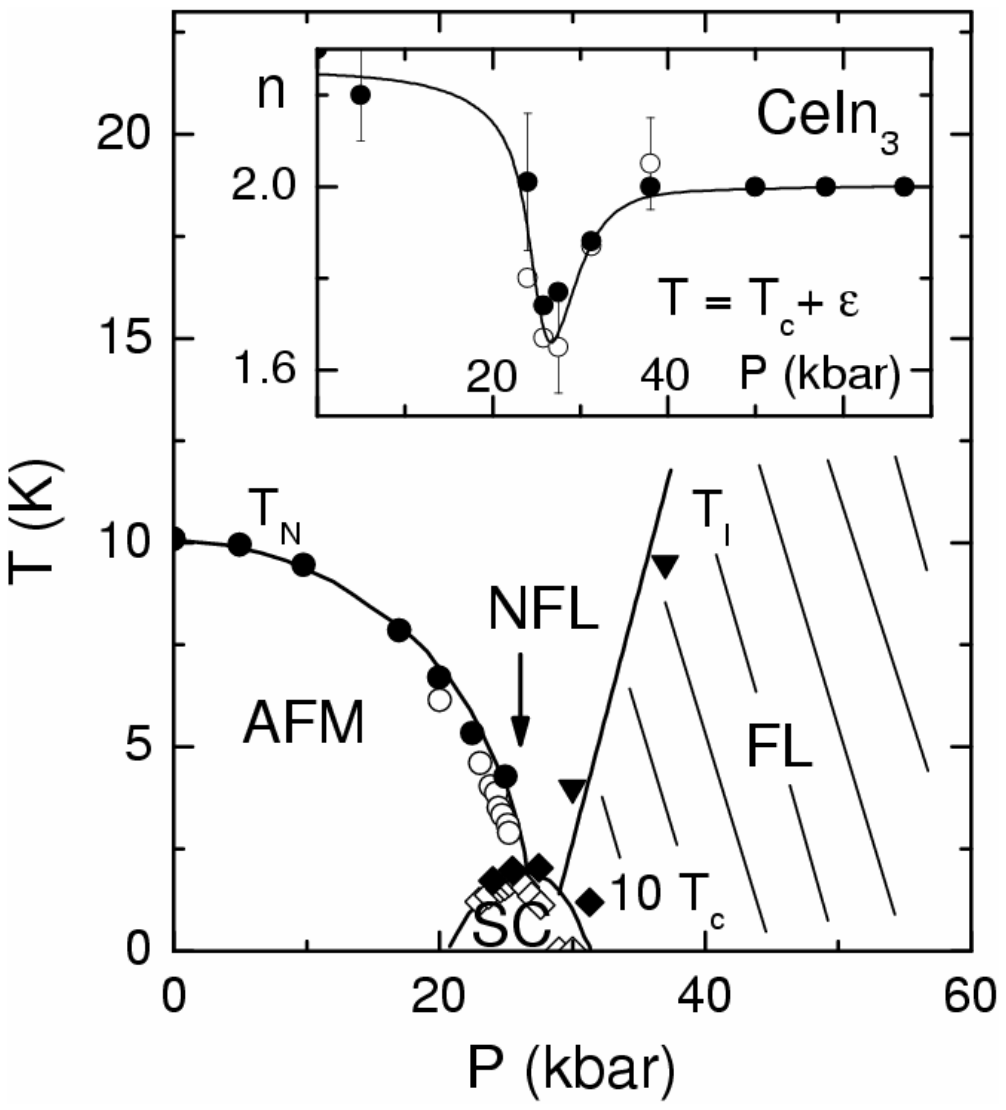

Figure 26 : Phase diagram of $CeIn_3$. $T_N$ indicates the Néel temperature, $T_I$ the crossover temperature to the Fermi liquid regime. The superconducting transition temperature $T_C$ is scaled by a factor 10 (◊) after Mathur et al (1998). The pressure dependence of the resistivity exponent n is shown in the insert. The minimum of the exponent n in the temperature dependence $T^n$ of the resistivity corresponds to the critical pressure $P_C$ (Knebel 2002).

$P_C$ is marked by a deep minima of the exponent n of the temperature dependence of the resistivity $\rho = A_nT^n$. Fermi liquid properties are recovered on both sides of $P_C$ ; n reaches 2 for P > 3.7 GPa above $P_C$ and a value n > 2 for P = 2.4 GPa below $P_C$ as the spin wave contribution will add an extra scattering contribution to ρ (figure 26) . It is worthwhile to reemphasize that $P_C$ can be more easily defined via the deep minimum of n than by its resistivity anomaly at $T_N$ which becomes difficult to detect in the vicinity of $P_C$. A strong broadening of the resistivity anomaly at $T_N$ is already pointed out 1 GPa below $P_C$ (Knebel et al 2002). The analysis of the resistivity show that $P_C$ and $P_V$ coincide. dHvA experiments under pressure have just been performed through $P_C$ (Settai et al 2003). In the experiments, only a slight change occurs for the spherical FS (referred as d, d') with hump along (1,1,1). At least the effective mass increases smoothly on approaching $P_C$ (factor 1.3 and 2 respectively for d and d'). The recent publication of T Elihara et al (2004), Endo et al (2004) and Settai et al (2005) report new features in magnetic field and pressure not discussed in the previous works of Settai et al (2003).

Experiments on a high quality crystal in Cambridge ($\rho_0$ = 1μΩcm) (Mathur et al 1998, Grosche et al 2000) show that superconductivity occurs ($T_C^{max}$ = 400 mK) in a narrow P range around $P_C$ . The large initial slope of the upper critical field $H_{c2}\ (T) \approx (m^*)^2$ demonstrates that the heavy quasiparticles themselves condensate in Cooper pairs. A full superconducting resistive transition was confirmed by further experiments made in Grenoble (Knebel et al 2002) and Osaka (Kobayashi et al 2001).

The upper critical field can be analysed in a single band superconducting model assuming a g factor equal to 1.4 and a strong coupling parameter $\lambda$ = 1.3. This points out that the mass enhancement coming from non local fluctuation m* is not large by comparison to the band mass renormalization $m_K$ driven by local Kondo fluctuation : $m^*/m_K = \lambda + 1 = 2.3$. Nuclear quadrupolar resonance NQR on the In site were very successful to study the spin dynamics notably in the (AF and S) coexisting regime (Kohori et al 2000-a, Kawasaki et al 2001). The second order nature of the magnetic collapse at $P_C$ must be questioned as two NQR signals (AF and Pa) appears at $P_C - \varepsilon$ (figure 27). The coexistence of both phases points to a phase separation in a pressure interval $P_{KL} - P_C \sim 0.3$ GPa (Kawasaki et al 2004). Evidence for the unconventional nature of the superconductivity in both phases is mainly given by the temperature variation of the nuclear relaxation time $T_1$ which follows the $1/T_1 \sim T^3$ law reported in many unconventional exotic superconductors. It is taken as a proof of a line node (Asayama 2002, Kohori et al 2000-a). The superconductivity of $CeIn_3$ has been studied theoretically on the basis of a three dimensional Hubbard model (Fukawaza and Yamada 2003). The suggested d wave pairing is induced by AF spin fluctuations. In agreement with other theoretical studies, $T_C$ is lower by one order than that for the 2d case.

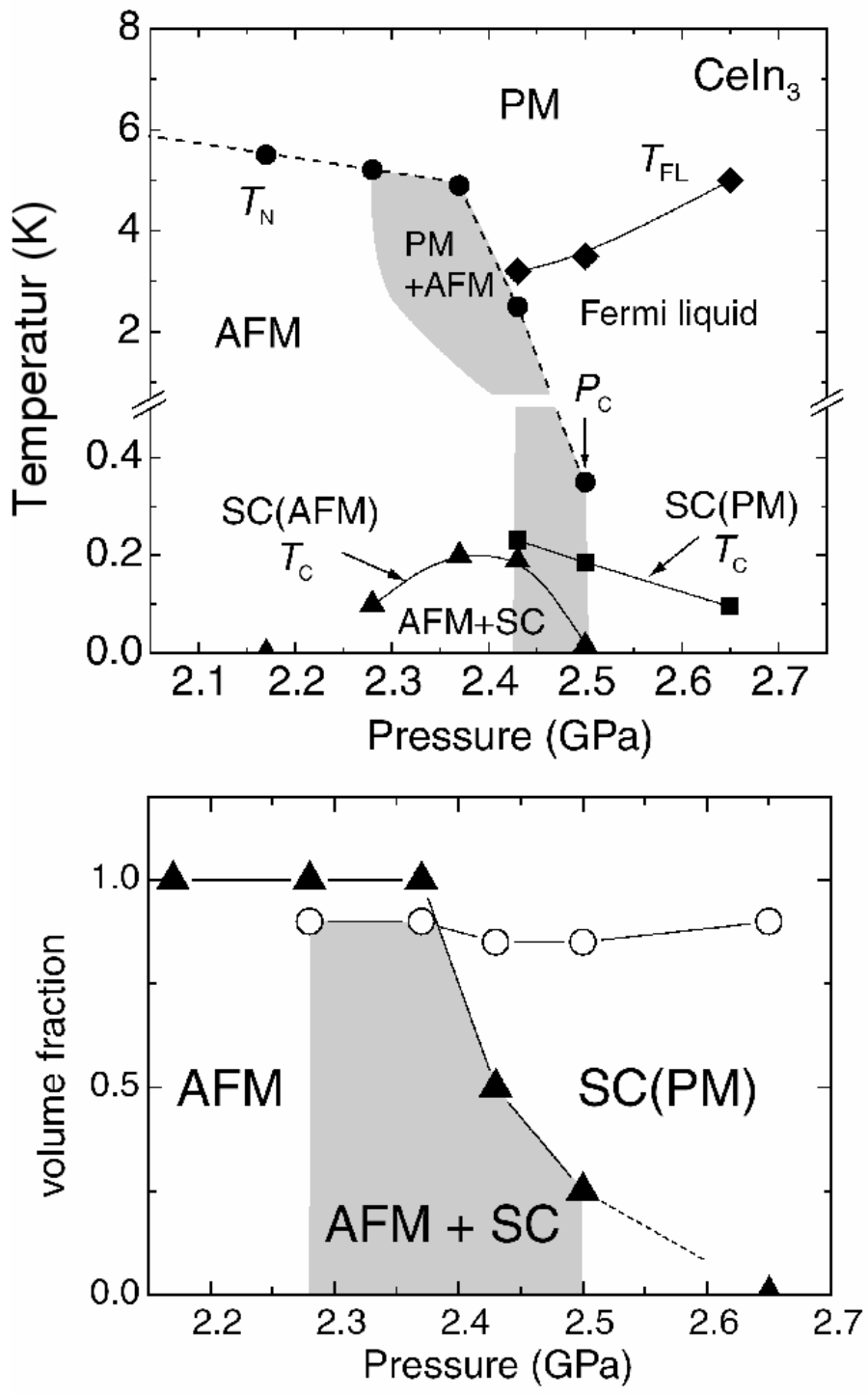

Figure 27 : NMR data of Kawasaki et al : $T_N$ , $T_C$ , $T_I$ of $CeIn_3$ and domain of phase separation (grey area). Volume fraction of AF and S states as a function of P (Kawasaki 2004).

### 4.1.2 - $CePd_2Si_2$. : questions on the range of the coexistence.

At P = 0, $CePd_2Si_2$ is a AF with $T_N$ ~ 10 K and a (½, ½, 0) propagation vector. As for $CeIn_3$, the P = 0 situation is very well documented by macroscopic and microscopic measurements. By contrast to the cubic $CeIn_3$ compound, in this tetragonal crystal (1, 2, 2) a strong axial anisotropy occurs. The easy direction of the magnetization is found in the basal plane. Below $T_N$ both spin waves and quasielastic excitations can be detected (van Dijk et al 2000). Above $T_N$ , the magnetic fluctuations are dominated by Kondo fluctuations (Fak et al 2004). The exchange constants extracted from the spin wave excitations have only a weak anisotropy. The Fermi surface starts to be determined by the dHvA torque technique on tiny crystals. At least parts of Fermi surface show partial 4f itineracy (Sheikin et al 2003) ($P_{KL}$ < $P_V$) . As in $CeIn_3$, a minimum in the resistivity exponent n appears at $P_C$ with a lower value 1.3 instead of 1.6 for $CeIn_3$. Reasons for this difference can be disorder, dimensionality or proximity to a valence transition as we will see for $CeCu_2Si_2$ or $CeCu_2Ge_2$. The first one is very unlikely as n = 1.3 appears a robust limit whatever is the residual resistivity $\rho_0$ (Demuer et al 2001). The temperature analysis of the resistivity shows also that $P_C$ ~ $P_V$. However no P experiments either by resistivity, specific heat or neutron scattering (Demuer et al 2001-2002,

Kernavanois et al 2003) succeed to follow the magnetic ordering closed to $P_C$ and thus cannot answer the question of the order of the magnetic transition at $P_C$. At least, in recent neutron scattering experiments up to P = 2.45 GPa, the magnetic transition in T appears second order and any first order transition at $P_C$ will be weak (Kernavanois et al 2005).

Our aim was to clarify the process of the magnetic collapse at the so called QCP by simultaneous ac calorimetry and resistivity experiments with a zoom on their respective anomalies at $T_N$ (Demuer 2000). To avoid any parasitic effect due to non hydrostaticity, the measurements were realized on diamond anvil cell with helium transmitting medium. In the orthodox vision of QCP, for $T \rightarrow 0K$, the molecular field description must become more and more valid for a second order phase transition (Zülicke and Millis 1995). Contrary to these classical statements, both normalized anomalies are broadened gradually. For example, for P = 2.3 GPa, $T_N$ = 3K, the specific heat broadening reaches already 15% (we will see that similar effects are observed for $CeRh_2Si_2$ and $CeRhIn_5$). However, explanation of the gradual broadening of the specific heat anomaly at $T_N$ as $T_N \rightarrow 0$ may come from the weakness of a first order transition with large fluctuations.

The discovery of superconductivity in $CePd_2Si_2$ was also made in Cambridge (Mathur et al 1998). Further experiments have been completed in Geneva (Raymond and Jaccard 2000, Demuer et al 2002) and Grenoble (Sheikin et al 2001, Demuer et al 2001). The analysis of $H_{c2}$ (T) requires an anisotropy in the g factor for the weight of the Pauli limit in qualitative agreement with the magnetic basal plane anisotropy and again a moderate $\lambda$ = 1.5 coefficient (Sheikin et al 2001). The new experimental feature is the observation of the superconducting specific heat anomaly at 2.7 GPa (Demuer et al 2002). However, this superconducting calorimetric signature is observed only very close to $P_C$ . Bulk homogeneous gapped superconductivity may occur in a narrower region than that claimed from resistitivy measurements. This statement is reinforced by the difficulty to reach zero resistivity on each side of $P_C$ . The realization of the clean limit is crucial for unconventional superconductivity ($\ell > \xi_0$). When $T_C$ collapses, $\xi_0$ increases as $T_C^{-1}$. So the impurities are a severe cut off.

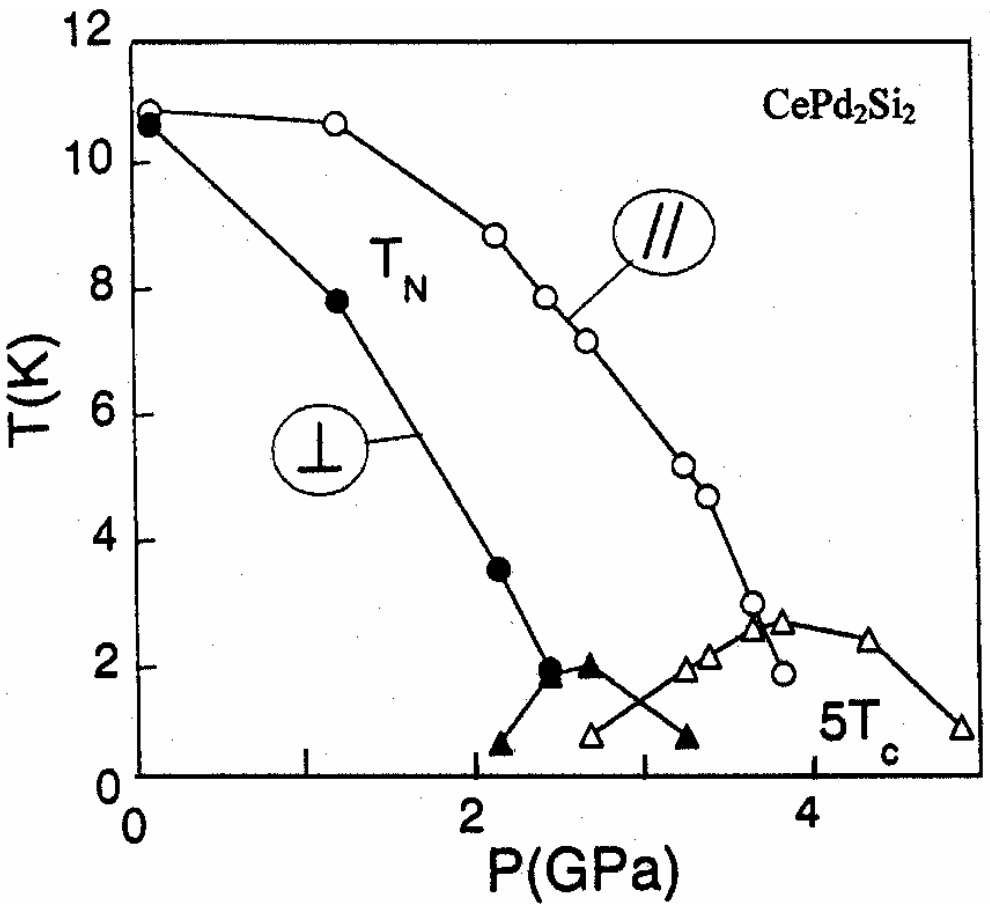

Figure 28 : Phase diagram of the two samples (filled and open symbols for samples ⊥ and // respectively) pressurised as described in reference (Demuer et al 2002). (● or ○ $T_N$ ▲ or △ $T_C$ ).

Simultaneous P measurements, on two crystals oriented with c axis parallel or perpendicular to the force load direction in a Bridgman anvil cell with a quasi hydrostatic steatite P medium, show a large shift in the P phase diagram as indicated figure (28). Correlatively a good correspondance is demonstrated between the magnetic instability, the position of $P_V$ and the optimum of superconductivity (Demuer et al 2002). In this axial crystal, the application of a strain along the c and a axis leads to opposite shifts of $T_N$ (van Dijk et al 2000). Thus the high magnetic sensitivity of the $CePd_2Si_2$ tetragonal crystal to a non uniform pressure (even a weak percent of $P_C$ ) was suspected. The nice result is that a non ideal P set up but a careful designed experiment prove that AF and S boundaries remain bounded at $P_C$.

The fast erasing of the superconducting specific heat anomaly when P is not near $P_C$ seems to be consequence that the intrinsic superconducting dome $T_C^0$(P) is associated with a high sensitivity to impurities as in the pair breaking mechanism represented by the wellknown Abrikosov-Gorkov formula (Abrikosov – Gorkov 1961). The important parameter is $x = \xi_0/\ell$. As $T_C^0$ (P) decreases, $\xi_0$ will increase and thus x. That leads to a supplementary decrease of the pressure range $P_{-S}$ – $P_S$ where superconductivity can be detected. Indeed, numerical simulation (Brison 2004) shows that the specific heat anomaly at $T_C$ is rapidly smeared out. These considerations can be applied to any quantum phase transition at $P_C$ or $P_S$ where a steep P collapse of the critical temperature may occur. Our message is that the determination of the contour of S and AF extrapolated to ideal conditions is a "tour de force". (see $CeRhIn_5$).

The additional experimental pressure anisotropy will lead to minor effects for a cubic material like $CeIn_3$. It becomes a major perturbation in anisotropic materials like the 122 cerium family (like $CePd_2Si_2$) and even more the 115 series. Even at P = 0 (see later $CeIrIn_5$ ) pressure gradients of (0.1 GPa) exist near imperfections : dislocations, stacking faults. These may produce superconducting nanostructures inside the material. Superconductivity can also occur by proximity on these objects. (For an array of stacking fault, Abrikosov and Buzdin (1988) have even proposed the possibility of a tiny splitting for the superconductivity transition of conventional superconductors).

### 4.1.3 - $CeRh_2Si_2$ : First order and superconductivity.

By contrast to the two first examples where, at P = 0, $T_K$ is around 10K, in $CeRh_2Si_2$ the Kondo temperature $T_K$ = 30 K is higher. However the magnetic exchange among the Ce ions is strong enough for AF to occur at $T_N$ = 36 K the Rh ions may play an important role in the strength of $T_K$ and Eij. Magnetism disappears at $P_C$ ~ 1.0 GPa (Kawarazaki et al 2000). For P = 0, just above $T_N$ , the anisotropy of the magnetic susceptibility between the c axis and the a axis is near 4 but it drops to an almost isotropic situation for the Pauli susceptibility at T << $T_N$ (Mori et al 1999). Just below $P_C$ , the propagation vector $k_0$ of AF is again (½, ½, 0) and the magnetic moment is transverse by respect to $k_0$ and oriented along the (0, 0, 1) direction. Evidence that the transition may be first order at $P_C$ comes from the steep pressure variation of $T_N$ on approaching $P_C$, from a rapid wipe out of the specific heat anomaly on approaching $P_C$ (Haga et al 2005) (figures 29, 30) and also from drastic modifications of the FS as detected by the change of dHvA frequencies which can be analysed with a localized 4f picture below $P_C$ and an itinerant above $P_C$ (Araki et al 2002) ($P_{KL}$ ~ $P_C$ ). Another indirect evidence of a discontinuity (i.e volume) is given by the persistence of $T^2$ resistivity law on each side of $P_C$ and the absence of NFL behaviors down to very low temperature as $P \to P_C$ (Araki et al 2002). A recent confirmation can be found in Ohashi et al (2003).

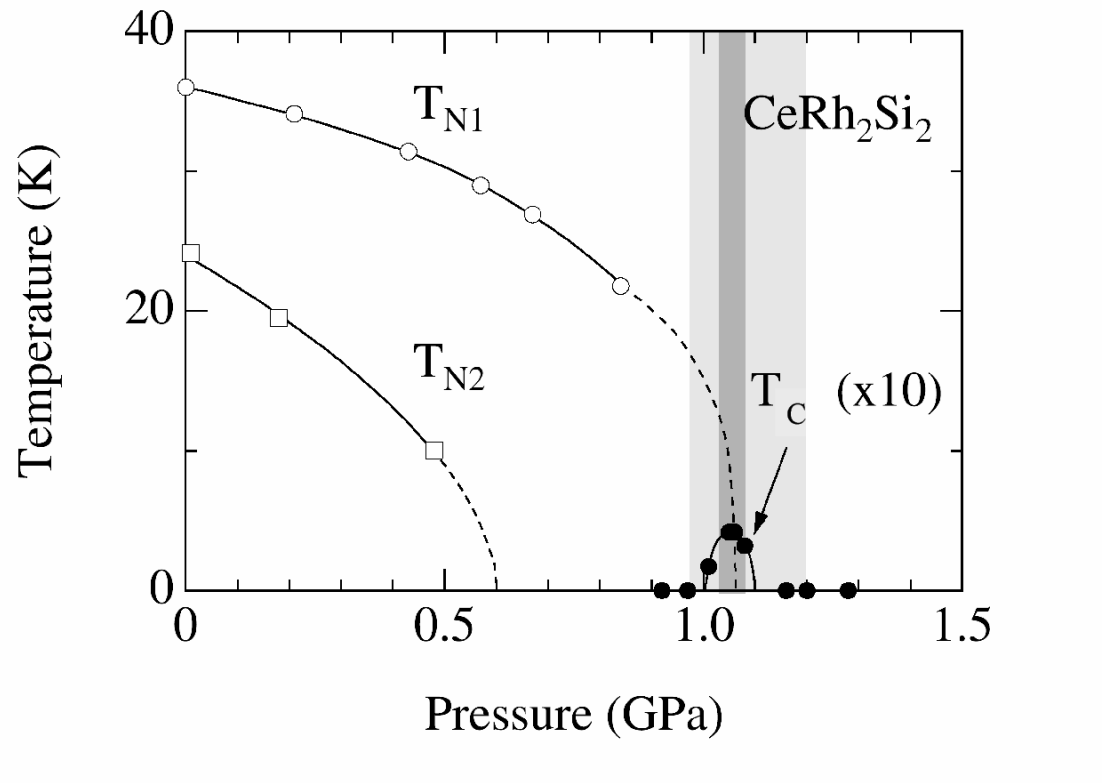

Figure 29 : $CeRh_2Si_2$ (T, P) phase diagram drawn by resistivity (○ or ●) ac calorimetry (■) and neutron scattering (□) (see Araki et al 2000, Haga et al 2004, Kawarazaki et al 2000). The superconducting border (x 10) in dark correspond to the full resistivity drop and in grey to the onset of the resistivity (Araki et al). We don't discuss the low pressure AF phase at $T_{N2}$.

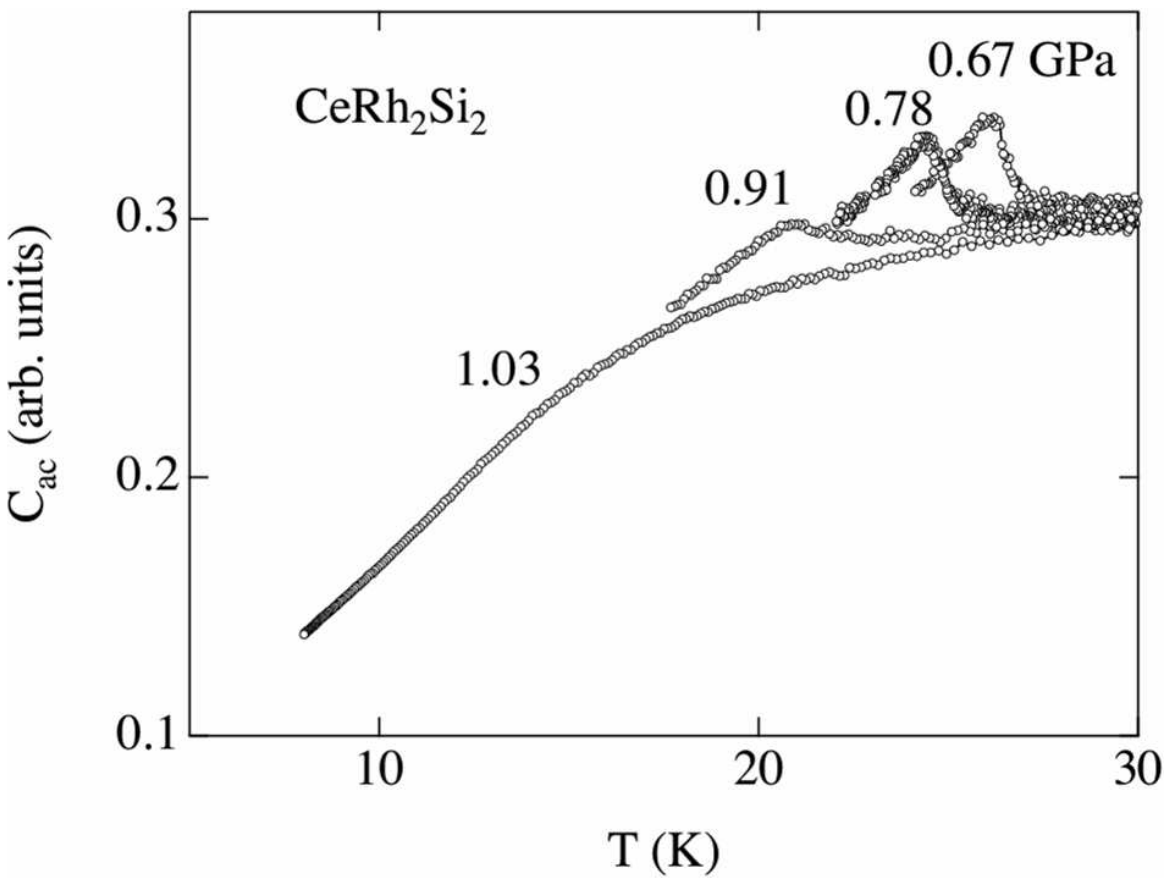

Figure 30 : ac specific heat of $CeRh_2Si_2$ with the broadening and collapse of the AF anomaly on approaching $P_C$ . A quite common phenomena in HFC (Haga et al 2004).

Kawarazaki et al (2000) have followed carefully the pressure evolution of $T_N$ (P) and $M_o$ (P). Up to 0.85 Gpa, $M_o$ is proportional to $T_N$ . The magnetic moment is a longitudinal variable. The reason lies in the fact that it is created through an induced magnetism in accordance to the Doniach picture. In the model of Benoit et al (1979) $M_o \sim T_N$ . As for the linear polarized light which is the sum of equal right and left circular polarization, the induced electronic moment is the combination of the electron and hole component. The proportionality $T_N \sim M_o$ (quite general in magnetic ordered HFC) deserves special attention. It is not predicted by SF theory where $T_N \sim M_o^{4/3}$ of course nor for local Heisenberg magnetism $T_N \sim$ $Mo^2$ with a P invariant exchange. A linear relation of $T_N$ with $M_o$ has been also observed for the antiferromagnetism in chromium alloys (Koehler et al 1996). It has been explained for this spin density wave structure by the feedback of the Fermi surface due to nesting. With the dual character of the magnetism here, one may expect rather similar phenomena even for commensurate structure with missing Fermi surface related to AF gap (see Fuseya et al 2003).

The occurrence of two time scales is manifest by the discrepancy between $M_o$ detected by neutrons ($M_o \sim 1.38\ \mu_B$ Ce) and by NMR ($M_o \sim 0.3\ \mu_B$ ) (Kawasaki et al 2002). "There may be a longitudinal fluctuation of the f electron moment which has a lifetime longer than the characteristic time of observations for neutrons and shorter than for NMR" (Kawarazaki et al 2000). This switch between two frequencies may be the result of the slow motion of the heavy fermion condensate in the ordered magnetic medium. In $CeRh_2Si_2$ at $P_C$ , the localized moment becomes delocalised and some itinerant magnetism picture occurs above $P_C$ . The specific heat anomaly at $T_N$ has almost collapsed at 0.9 GPa (figure 30). Above $P_C \sim 1$ GPa, a drastic difference appears between the pressure evolution of $T_N$ and $M_o$ . A slow P variation of $T_N$ contrasts with a continuous drop of $M_o$ . This exotic SMAF signature was pointed out for $CeRu_2Si_2$ and will appear later for $UPt_3$ and $URu_2Si_2$ .

Superconductivity in $CeRh_2Si_2$ was first reported by Movshovich et al 1996. Recently it was stressed even for a high quality crystal ($\rho_0 \sim 0.8\ \mu\Omega$cm) that the achievement of zero resistivity or a sharp resistive transition in the superconducting phase is only realized close to $P_C$ in a narrow pressure region (1.04 – 1.07 GPa) (Araki et al 2002). Precursor effects of superconductivity (resistivity onset) will give a wider pressure window (~ 0.2 GPa) (figure 29). The situation is rather similar to that found in $CePd_2Si_2$ . Again in $CeRh_2Si_2$ the thermodynamic boundary of superconductivity and long range magnetism is not yet defined.

## *4.2 - $CeCu_2Si_2$ and $CeCu_2Ge_2$ : spin and valence pairing*

For $CeCu_2Si_2$ and $CeCu_2Ge_2$ the new phenomenon is that the maximum of $T_C$ is located far above $P_C$ (a few GPa). For $CeCu_2Si_2$, the precise location for $P_C$ was still open up to achievement of recent successful neutron scattering experiments (Stockert et al 2004). We will not enter into the details of all systematic studies realized on this material but roughly $P_C$ may be a few kbar below P = 0. Tiny differences in composition can induce AF. Recently, the $CeCu_2(Si_{1-x}Ge_x)_2$ series was explored intensively by elastic neutron scattering experiments. The same incommensurate wave vector was observed for all concentrations suggesting a spin density wave instability given by a nesting property of the Fermi surface (Stockert et al 2004). The difficulty to grow large crystals of $CeCu_2Si_2$ has precluded up to now to collect inelastic information however deep extensive NMR works have been realized (see Kawasaki et al 1998 and Ishida et al 1999). An extensive discussion on the superconductivity of $CeCu_2Si_2$ can be found in the recent article of Thalmeier et al (2004) with the new highlights that :

i/ below T = $T_A$ the mysterious A phase which was found to envelop superconductivity in (H, T) phase diagram is the reported spin density wave in good agreement with the nesting wavevector predicted from the heavy fermion Fermi Surface sheets (Zwicknagl 1992 and Zwicknagl and Pulst 1993). When $T_C$ and $T_A$ are closed, superconductivity and long range magnetism seems to repel ; similar effects are observed in $CeRhIn_5$ for T ~ $P_C$ .

ii/ for a proper doping of Ge the large P stability of superconductivity can be broken in two domes associated with magnetic and valence transition (Yuan et al 2003). This last observation precises and confirms a serie of P experiments realized by Jaccard and co workers on $CeCu_2Si_2$ and $CeCu_2Ge_2$.

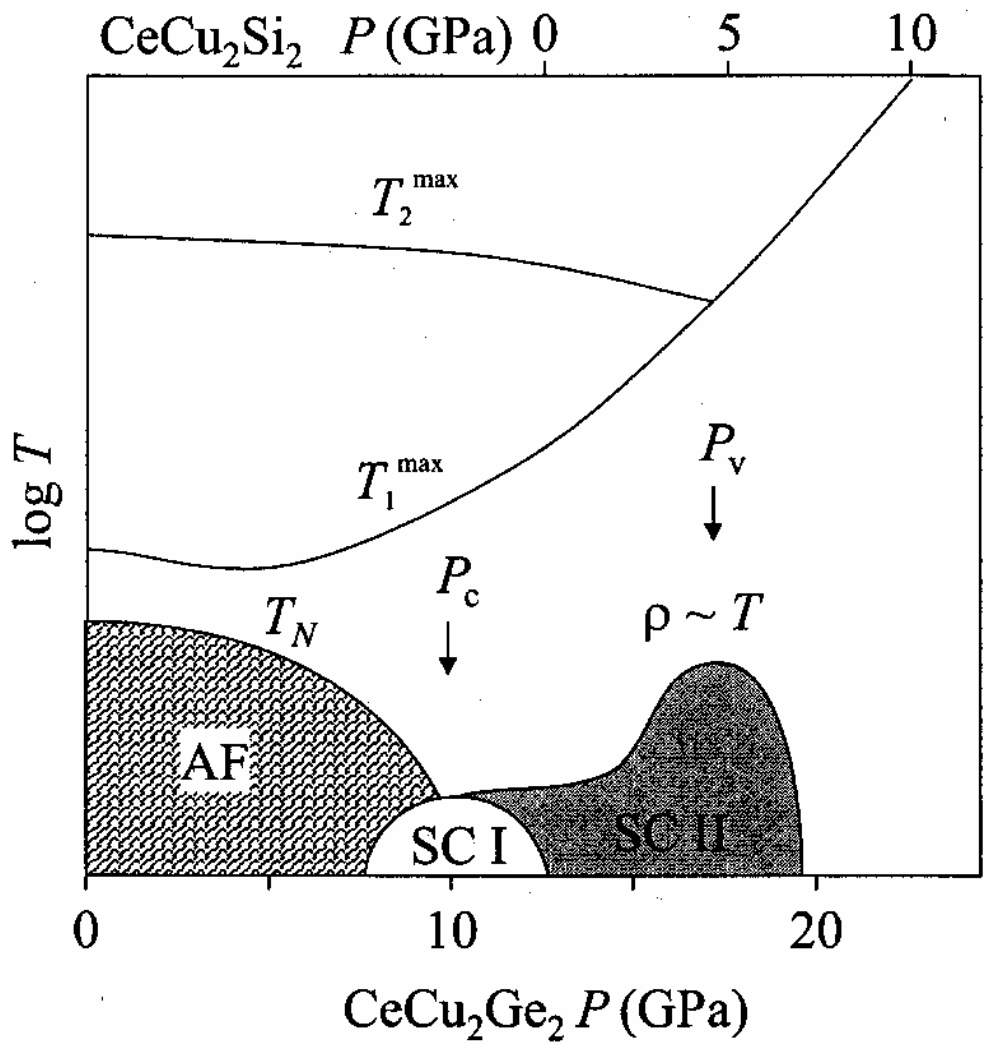

Figure 31 : Schematic P-T phase diagram for $CeCu_2(Si/Ge)_2$ showing the two critical pressures $P_C$ and $P_V$. At $P_C$ , where the antiferromagnetic ordering temperature $T_N \rightarrow 0$, superconductivity in region SC is governed by antiferromagnetic spin fluctuations. Around $P_V$, in the region SC II, valence fluctuations provide the pairing mechanism and the resistivity is linear in temperature. The temperatures $T_1^{MAX}$ , and $T_2^{MAX}$ of the temperature of the maxima of the resistivity merge at a pressure coinciding with $P_V$. (Holmes et al 2004).

$CeCu_2Ge_2$ was the first case (Jaccard et al 1992) where it was found that superconductivity occurs near $P_C$ = 6 GPa with a first rather flat pressure variation of $T_C$ = 0.6K followed by a bump to $T_C^{max}$ = 2. K at P = 17 GPa. The relevance of such a structure of $T_C$ was reinforced by similar observations already made in $CeCu_2Si_2$ (Bellarbi et al 1984, Jaccard et al 1985) where $P_C$ is assumed to be shifted almost to zero. It is amazing to notice that this key result obtained two decades ago has reached a real impact only last years. Recent simultaneous resistivity and ac specific heat measurements under pressure on $CeCu_2Si_2$ over 6 GPa (Holmes et al 2004-a) clarify the correlation between the collapses of the two maxima $T_1^{max}$, $T_2^{max}$ of the temperature variation of the resistivity on warming at $P_V$ and the optima of $T_C$ (figure 31). In agreement with the previous cases, a first superconducting domain will be centered on $P_C$ and a second one at $P_V$ suggesting that new superconductivity is mediated by valence fluctuations. The particular P variation of $T_C$ (P) was already interpreted via two contributions (Thomas et al 1996) : a smooth one assumed to be due to the pressure increase of $T_K$ and sharper additional features reflecting topological changes in the renormalized heavy bands. However the monotony ot the scaling of $H_{C_2}$ (o) by $m^{*2}T$ in reduced temperature $T/T_C$ leaves questionable two different sources of pairing (Vargoz et al 1998)

Valence fluctuations near $P_V$ ((Holmes et al 2004-a, Onishi and Miyake 2000) seem a favourable factor in the increase of $T_C$. The departure from $P_C$ will decrease the spin pairing potential. So a new attractive potential is needed. The importance of valence fluctuations is stressed by the fact that near $P_C$ the usual NFL 3d AF behavior is recovered ($\rho \sim T^n$ with ~ 3/2) while near $P_V$ a linear temperature crossover in the resistivity is observed. This dependence is well explained by soft valence fluctuations which will produce large angle scattering of the quasiparticles. They are efficient in a wide region of the Brillouin zone and strongly coupled to Umklapp process of the quasiparticle scattering. At least, the shift of $P_V$ from $P_C$ seems a convincing explanation for the large shift of $T_C^{max}$ from $P_C$ . The direct mark of a valence change in $CeCu_2Si_2$ comes from $L_{III}$-Xray absorption experiment (Roehler et al 1988). Other indirect evidences are the decrease of the Kadowaki Wood ratio, a small hump in the $\gamma$ term and strong enhancement of the residual resistivity. In $CeIn_3$ , $CePd_2Si_2$ and $CeRh_2Si_2$ Pc and $P_V$ cannot be separated.

## *4.3 - From 3d to quasi 2d systems : the new 115 family : $CeRhIn_5$ and $CeCoIn_5$*

The link between superconductivity and magnetism was recently boosted with the Los Alamos discovery of superconductivity (see Thompson 2001) in the so called 115 cerium compounds like $CeRhIn_5$, $CeIrIn_5$ or $CeCoIn_5$. A planar anisotropy is induced by inserting in $CeIn_3$ a single layer of $MIn_2$. Single layers of $CeIn_3$ are stacked sequentially along the c axis according to the relation $Ce_nM_mIn_{3n+2m}$, in 115 n = 1, m = 1. For n = 2, there will be two adjacent layers of $CeIn_3$ separated by a single layer m =1 of $MIn_2$ (M transition metal). A planar anisotropy is produced. The Fermi surface is dominated by a slighty warped cylindrical sheet even in $LaRhIn_5$. That contrasts with the previous cases of 3d complex Fermi surfaces.

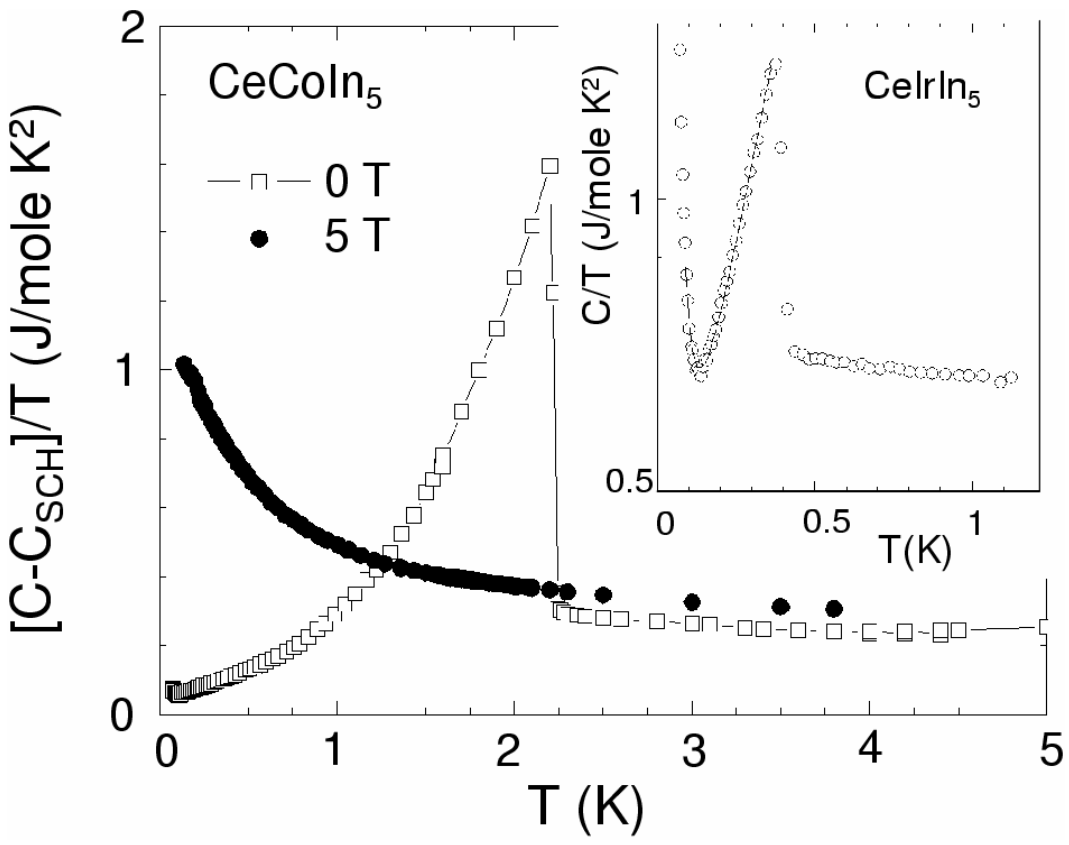

Figure 32 : Superconducting specific heat anomaly of $CeCoIn_5$ and $CeIrIn_5$ from Petrovic et al (2001) and (Movshovich et al 2001). The nuclear Schottky contribution of In has been substracted fo $CeCoIn_5$ not for $CeRhIn_5$.

The gold "mine" of these compounds are that 3 examples (where the crystal growth by flux is easy) cover all possibilities. $CeRhIn_5$ (AF at P = 0 with $T_N$ = 3.8 K) has $P_C$ ~ 2 GPa. The two others ($CeIrIn_5$ and $CeCoIn_5$) are already on the PM side of the magnetic instability and superconductors at $T_C$ = 0.4 K and $T_C$ = 2.3 K at P = 0 (Petrovic et al 2001, Thompson et al 2001) (figure 32). As for the other cerium heavy fermion compounds, in $CeRhIn_5$, superconductivity emerges near $P_C$ . There is already a large diversity of studies on the 115 serie notably for the FS determinations of the normal phases at P = 0. In $CeRhIn_5$, the 4f electrons are localized and itinerant in $CeIrIn_5$ and $CeCoIn_5$ (Shishido et al 2002). A new generation of P through $P_C$ for $CeRhIn_5$ (Shishido et al 2005) show that the FS changes ; the 4f electrons become itinerant above P ~ 2.4 GPa. It is slightly higher than $P_C$ but magnetostriction corrections must be made. Indeed, the (H, T) domain of the AF boundary must be precised close to $P_C$ in order to extrapolate the data of quantum oscillations at finite H to the FS topology at H = 0.

Extensive NQR experiments on the In site demonstrate the 3d behavior for the spin fluctuations of $CeRhIn_5$ above $T_N$ (Mito et al 2001). The quasi 2d behavior of the PM phase compound in $CeCoIn_5$ has been studied on In and Co sites (Kohori et al 2002, Kawasaki et al 2003-a). Unconventional singlet d wave superconductivity has been tested by a large panoply of techniques : for $CeCoIn_5$, NMR (Kohori et al 2000-b, 2001, 2002) with the proof of d wave pairing (from the drop of the Knight shift below $T_C$) and of a line of zero (with the $T^3$ decrease of $T_1^{-1}$), the observation of power laws in the T dependence of the specific heat thermal conductivity (Movskovich et al 2001), and microwave conductivity (Ormeno et al 2002). Anisotropy have been detected in the angular variation of the thermal conductivity (Izawa et al 2001) and of the specific heat (Aoki et al 2004) in magnetic fields.

### 4.3.1 - $CeRhIn_5$ : Coexistence and exclusion.

We will focus here on the coexistence of superconductivity and magnetism near $P_C$ in $CeRhIn_5$ which can be studied under pressure but also by alloying. Extensive works can be found for the $CeRh_{1-x}Co_xIn_5$ and $CeRh_{1-x}Ir_xIn_5$ (Zapf et al 2001, Pagliuso et al 2001). In the first resistivity report on $CeRhIn_5$ under pressure (Hegger et al 2000), magnetism and superconductivity seem to repeal each other i.e. a drastic decrease of $T_N$ coincides with the sudden appearance of superconductivity at $P_C$ ~1.5 GPa. Careful specific heat measurements in a quasi hydrostatic solid pressure medium indicate nice magnetic specific heat anomalies below $P_C$ and superconducting jumps above $P_C$ (Fisher et al 2002). Just right near $P_C$, the interpretation is complex as even above $P_C$ a maximum appears in the temperature variation of C/T i.e. in its PM state. The differentiation between the PM and AF states, i.e. $P_C$, has been made through the location of the maximum of $\gamma$. With such a procedure, superconductivity appears only on the PM side.

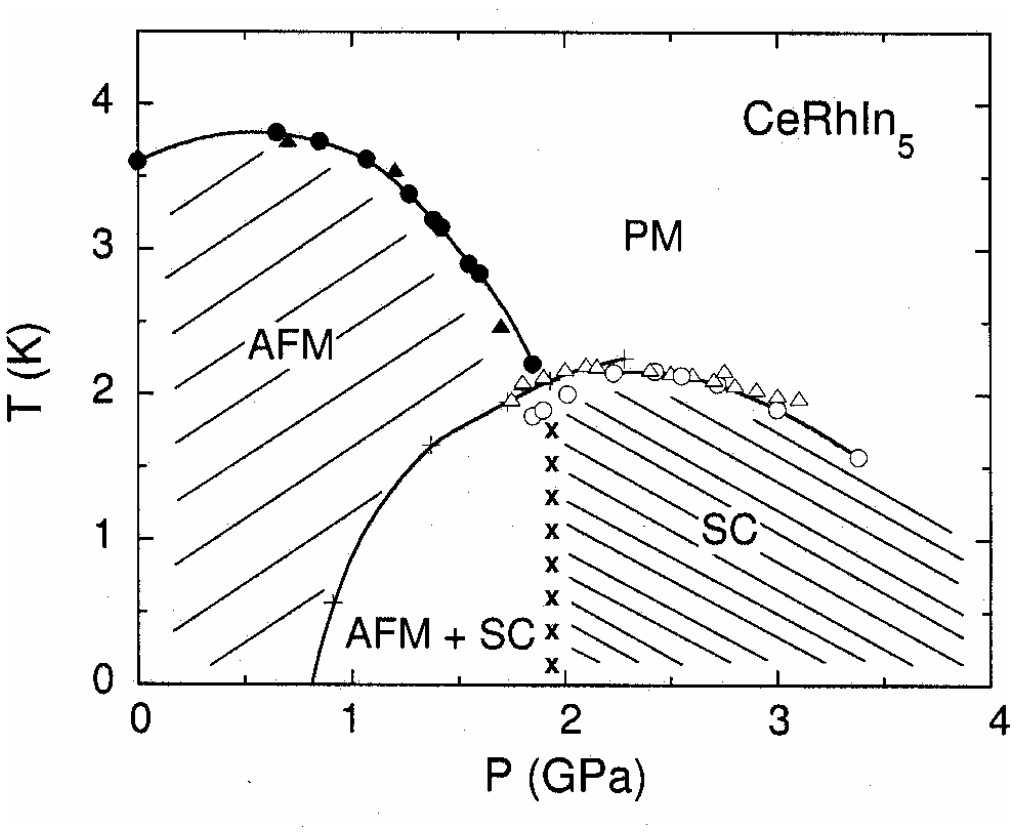

Figure 33 : (T, P) phase diagram of $CeRhIn_5$ (●, ○) specific heat anomalies at $T_N$ , $T_C$ and corresponding susceptibility (△). The AF anomaly disappears suddently at $P_C$ ~ 1.9 GPa. Gapped superconductivity is observed above $P_C$ . Inhomogeneous ungapped superconductivity may occur below $P_C$ (white domain) (Knebel et al 2004).

By improving the sample quality and the accuracy in the resistivity measurements (Llobet et al 2003), it was found that a complete resistive superconducting transition at $T_C$ (ρ) occurs far below $P_C$ ($T_C$ = 0.7 K at P ~ 1.6 GPa). Simultaneous measurements of ac susceptibility, NQR spectrum and nuclear spin relaxation time ($T_1$) (Kawasaki et al 2003-b) show that the temperature derivative of $\chi_{ac}$ has its maxima at $T_C$ lower than the previous determination of $T_C$ (ρ) by resistivity. Furthermore a strong broadening of the diamagnetic signal occurs below $P_C$ which is located now near 1.8 GPa.

To clarify the situation, a recent attempt has been made (Knebel et al 2004) by ac calorimetry to detect $T_N$ and $T_C$ in more hydrostatic conditions (now Argon pressure transmitting medium) than in the previous calorimetric experiment of Fisher et al (2002). The critical pressure as shown on figure (33) seems located near 1.9 GPa. Qualitatively the important feature is that clear AF specific heat anomalies are observed below $P_C$ and a clear superconducting specific heat one just above $P_{C.}$ Tiny AF or superconducting (S) anomalies are detected in the vicinity of $P_C$ but it can be due to residual internal stress (figure 34). The domain of homogeneous coexistence of AF and gapped superconducting phases may not exist. Furthermore, ac susceptibility experiments on a sample coming from the same batch show again only a broadened diamagnetism below $P_{C,}$ furthermore at higher temperature than the superconducting specific heat anomaly. A sharp diamagnetic transition occurs only for P > $P_{C,}$ now in good agreement with the specific anomaly. It was proposed that the observation of the inhomogeneity may be not a parasitic effect but an intrinsic property of a new gapless superconducting phase of odd parity pairing directly linked with strong retarded effects which may occur near $P_C$ (see chapter 3). An interesting observation is that only the cascade AF → S have been observed on cooling (see also $CeIn_3$) but, as here, to our knowledge there is no example of HFC where a AF or F phase appears after the entrance in a superconducting state at high temperature. That strongly suggests a clear demarcation between AF and PM phase i.e a first order transition. In conventional magnetic superconductors as Chevrel phase usually $T_C > T_N$ .

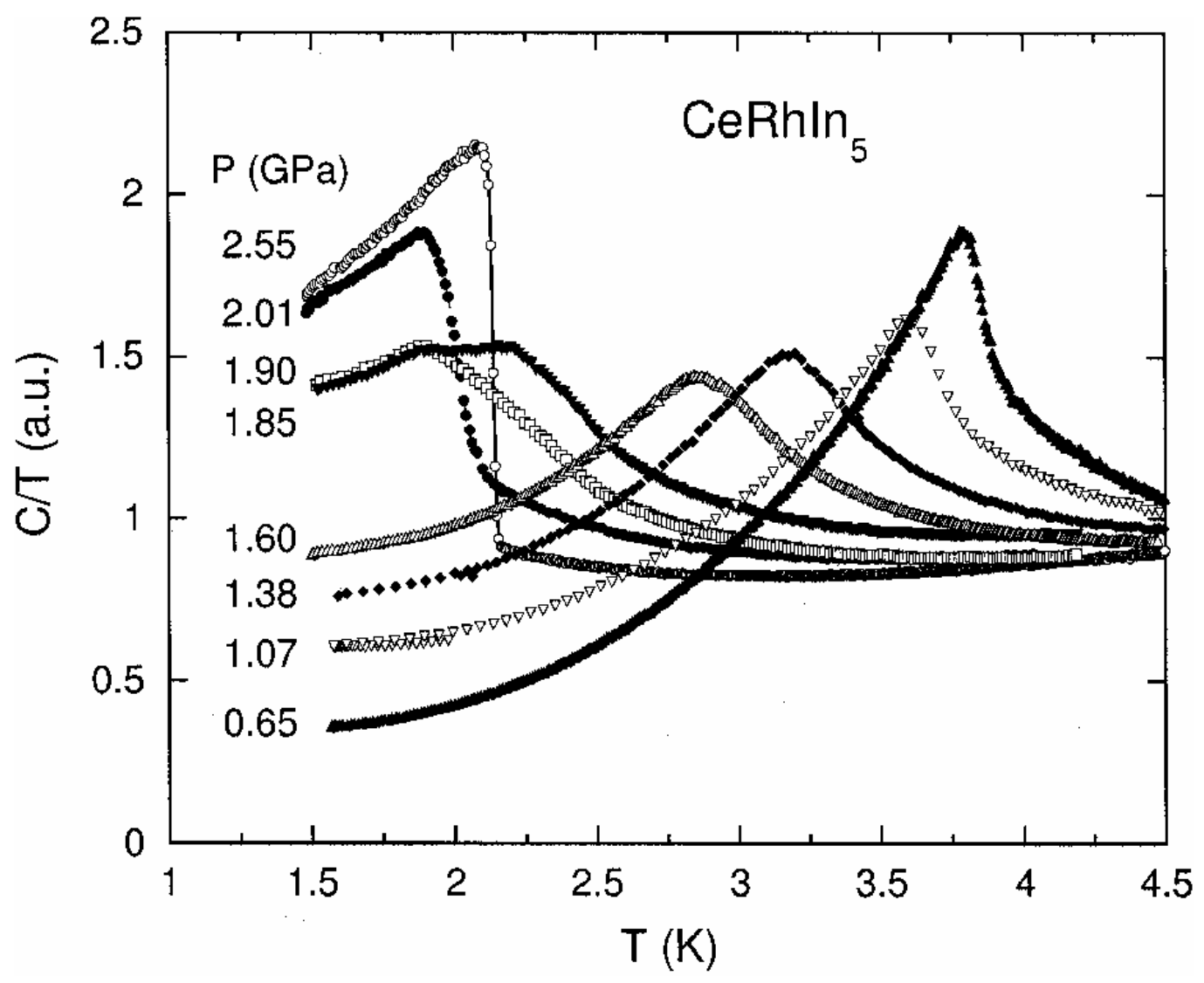

Figure 34 : Specific heat anomalies of $CeRhIn_5$ at different pressure. At 1.9 GPa, there is a superposition of tiny superconducting and magnetic anomalies (Knebel et al 2004).

In the $CeRh_{1-x}Co_xIn_5$ (Zapf et al 2001) and in $CeRh_{1-x}Ir_xIn_5$ (Pagluiso et al 2001), the coexistence of superconductivity and antiferromagnetism is claimed over a large doping range respectively 0.4 < x < 0.6 for Co substitution and 0.3 < x < 0.6 for Ir substitution. However the questions on the respective amplitude of the specific heat anomalies and their broadening at $T_N$ and $T_C$ as well as on their low temperature behaviors must be revisited before claiming that the derived phase diagrams are evidences of AF and S coexistence. Microscopically, one must worry about the effects of internal strain, mismatch between the a, c lattice parameters and their ratio c/a on doping (see $URu_2Si_2$ discussion in chapter 6). Even for the pure compound $CeIrIn_5$ , the puzzle still remains of the large difference between the value of $T_C(\rho) = 1.3$K and $T_C(C) = 0.38$ K measured respectively by resistivity and specific heat (Thompson 2001). At least it is obvious that the domain of AF and S coexistence deserves systematic studies, careful analysis and excellent hydrostatic conditions. The disorder by washing out the first order transition seems to restore a homogeneous situation or at least the simultaneous observation of AF and S anomalies over a large doping region. Let us note that as for $CeRhIn_5$, discrepancies appear between NMR (collapse of $M_o$ at $P_C$ ) (Mito et al 2001) and magnetic neutron diffraction (discontinuity of $M_o$ at $P_C$ ) (Llobet et al 2003).

**4.3.2 – $CeCoIn_5$ a new field induced superconducting phase :** A large variety of superconducting studies have been performed on $CeCoIn_5$. If the d singlet state is well established, the previous claim of a $d_{x2-y2}$ order parameter mainly from the anisotropy of the thermal conductivity in a rotating magnetic field (Izawa et al 2001) has been questioned in a field-angular study of the specific heat. The superconducting gap may be most probably of $d_{xy}$ type (H. Aoki et al 2004). This $d_{xy}$ order parameter may be driven by valence fluctuations since AF fluctuations mainly in the (Π, Π) directions of the basal plane may induce a $d_{x2-y2}$ order parameter (Miyake 2004).

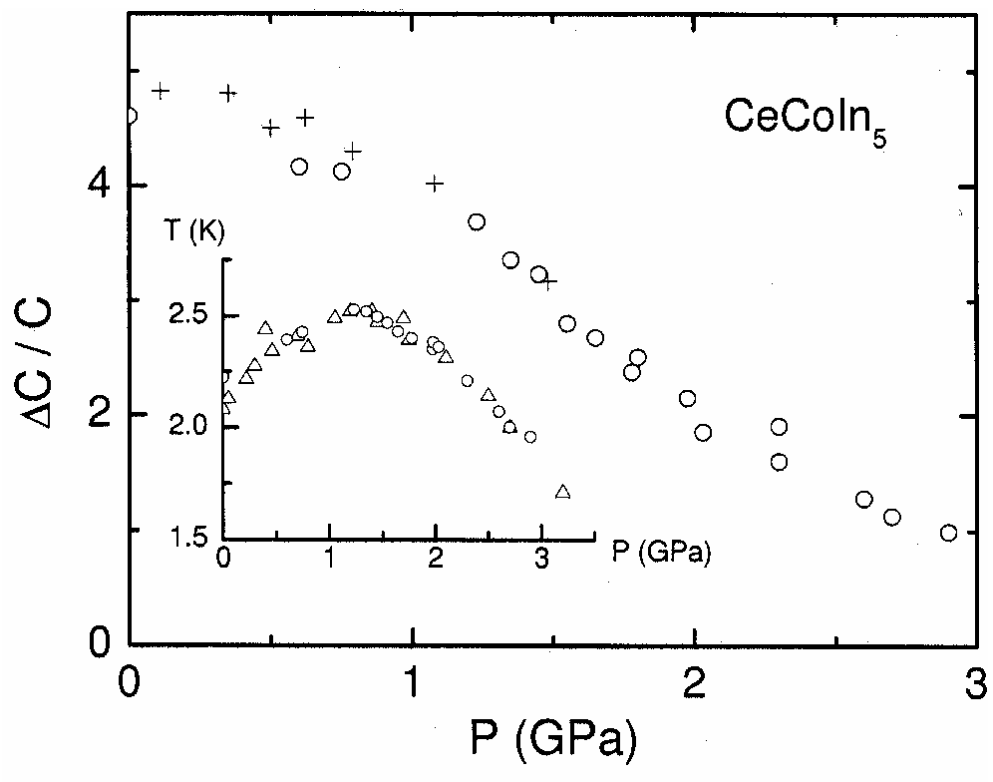

Figure 35 : $CeCoIn_5$. Jump of the superconducting specific anomaly normalized to the value just above $T_C$ (○) from Knebel et al (2004), + (Sparn 2001). Insert variation of $T_C$ (P) measured by ac calorimetry ○ and ac susceptibility △.

Special attention has also be given to strong coupling effects as superconductivity appears at $T_C$ higher than the Fermi liquid temperature $T_I$ or $T_{KL}$ which can be attributed to FL low energy excitations. As discussed in chapter (3), the large specific heat jump $\frac{\Delta C}{C(T_C+\varepsilon)}$ may not reflect a large strong coupling (λ>> 1) but a delay in the achievement of the normal phase Fermi liquid value which will be reached only at very low temperature (see Kos et al 2003). The appearance of superconductivity drives in a coherent regime and consequently to fill already at $T_C$ the large heavy fermion γ value of C/T at T → 0K. Recent experiments made by Knebel et al (2004) have completed the P calorimetric studies of Sparn et al (2001) up to 1.5 GPa. The pressure variation of $T_C$ (P) and of the specific jump $\Delta C/C(T_C)$ is now established up to 3 GPa (figure 35). $T_C$ (P) reaches its maximum for P = 1.5 GPa while the specific heat jump at $T_C$ continuously decreases under pressure. Neglecting strong coupling effects (see Kos et al 2003), the jump normalized to the value of m* at T = 0K must be universal $\frac{\Delta C}{m^* T_C} = cte$. With this hypothesis the effective mass decreases gradually under pressure (a factor 4 between 0 and 3 GPa).

Two new features appear in magnetic field which have been predicted three decades ago (Saint James et al 1969) for the specific case where the upper magnetic field is governed by the Pauli limitation at 0K :

1/ a crossover from second order to first order in $H_{C2}$ (T) at To (Izawa et al 2001, Bianchi et al 2002, Tatayama et al 2002),

2/ a new high field phase reminiscent of the FFLO phase

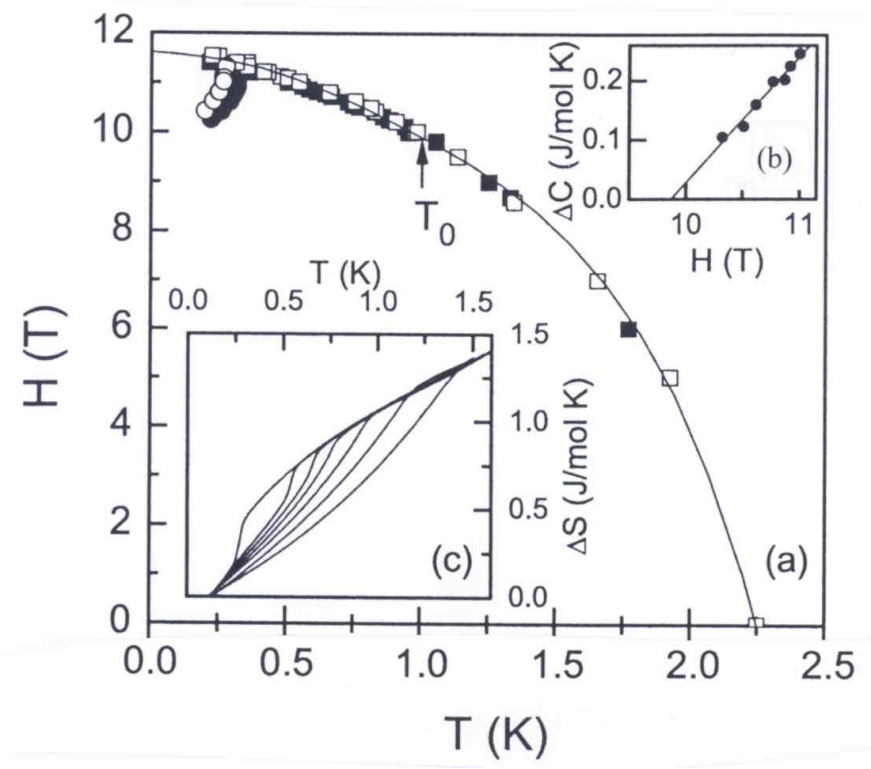

Figure 36 : (a) : H-T phase diagram of $CeCoIn_5$ with both H || [110] (filled symbols) H || [100] (open symbols). (○) and (●) indicate the $T_{FFLO}$ anomaly for H || [100] and H || [110], respectively. Inset (c) : entropy gain from T = 0.13 K for fields of 11.4 T, 11 T, 10.8 T, 10.6 T, 10.22 T, 9.5 T and 8.6 T from left to right. Inset (b) : specific heat jump at the $T_{FFLO}$ (from Bianchi et al 2003-a)

We will discuss mainly this last discovery. Figure (36) representes the domain of the new high magnetic field superconducting phase which appears below a temperature $T_C^* < T_0$ in specific heat measurements (Bianchi et al 2003-a), other thermodynamic experiments (Radovan et al 2003, Watanabe et al 2004) and recently in thermal conductivity experiments (Capan 2004). The FFLO state of $CeCoIn_5$ has been discussed theoretically by Won et al (2003). In FFLO, the Maki parameter depends only on the pressure variation of $Hc_2$ (orb) ~ $(m^*T_C)^2$. The initial pressure dependence of $T_{FFLO}/T_C$ is weak (see chapter 3). Only above 1.5

GPa, $T_{FFLO}$ is predicted to collapse (using the previous relative pressure variation of m* and $T_C$ ).

As discussed in chapter 2 for $CeRu_2Si_2$ , the supplementary interesting features will be if the magnetic interaction is modified for H < $H_{C2}$ (0). In $CeCoIn_5$, field induced quantum critical point at $H_M$ is suspected for H = 5T and 8.5 T respectively for H//c and H//a axis (Paglione et al 2003, Radovan et al 2003, Bianchi et al 2003b). Extrapolating from $CeRu_2Si_2$ , one can guess that : 1/ a strong dilatation at $H_M$ will be a supplementary favourable factor to drive the phase transition at $H_{C2}$ from second order below $H_M$ to first order above $H_M$ at $T_0$, 2/ the new field induced superconducting phase may be boosted by the enhancement of the magnetization and also can appear when a strong coherence due to Cooper pairing exists in the polarized phase i.e below $T_\propto$ (H). The role of the magnetic correlations in the field behavior of the superconducting phase of $CeCoIn_5$ may open other perspectives than the classical FFLO frame.

Finally, a surprising result on superconductivity was found by extending the studies on 115 cerium compounds to 115 plutonium systems. The high $T_C$ record in HFC was beaten recently with the discovery of superconductivity in $PuCoGa_5$ : $T_C$ = 18.5 K (Sarrao et al 2002) again in Los Alamos. The heavy quasiparticule leads to $\gamma$ = 77 mJmole$^{-1}$K$^{-2}$ in their normal phase and a corresponding huge value of the initial slope of $H_{c2}$ (T) ~- 59KOe/K leading to an orbital limit of 740 KOoe at T = 0 K roughly twice the estimated Pauli limit of 340 KOe. The superconducting parameters $\xi_0$ = 21 Å, $\lambda_L$ = 1240 Å and $\kappa$ = 32 are rather similar to those found in SCES superconductors. The proximity of spin fluctuations seems reflected by the low value of the Curie Weiss temperature $\theta$ (2K) of the susceptibility which obeys roughly a Curie Weiss law. The high $T_C$ of $PuCoGa_5$ may be due to an increase in the effective Fermi temperature. A spectacular observation (Pagliuso et al 2002 and Kumar et al 2004) is that the variation of $T_C$ versus the ratio c/a of the c by a lattice parameters is a beautiful unique straight line for the Ce and Pu 115 compounds. This huge variation of Tc versus the separation of the basal plane is a nice evidence of the importance of the magnetic and electronic anisotropies. It points out the key role of lattice deformations.

## *4.4 – Recent exotic superconductors $CePt_3Si/PrOs_4Sb_{12}$*

A new unexpected result (Bauer et al 2004) was the discovery of superconductivity in the non centrosymetric compound of $CePt_3Si$ as it was stressed that the lack of an inversion center may not be favourable for superconductivity. It was even accepted that a material without inversion center would be a bad candidate for spin triplet pairing (Anderson 1984). Gorkov and Rashba 2001 clarifies that in fact the order parameter will be a mixture of spin singlet and spin triplet components. The supplementary interest of this new material is the coexistence of long range magnetism at $T_N$ = 2.2 K (Metoki et al 2004) and of superconductivity at $T_C$ = 0.78 K with clear heavy fermion properties ($\gamma \sim 400$ mJmole$^{-1}$k$^{-2}$). The occurrence of an hybrid situation with both singlet and triplet components is manifested in NMR properties which appear as a mixture of conventional and unconventional behaviors (Yogi et al 2004) : The nuclear relaxation rate $T_1^{-1}$ shows a kind of Helbel Slichter anomaly usually absent in unconventional superconductors and a temperature variation which cannot be described in either hypothesis. As in $UPd_2Al_3$ , superconductivity appears to coexist with the AF state. The link with AF QCP is not obvious even the origin of the pairing mechanism as, due to the lack of inversion center, ferromagnetic and antiferromagnetic fluctuations will be inevitably mixed. The discovery of superconductivity in $CePt_3Si$ appears during the revision of the review so I will not go further on the recent activity in this subject. An excellent report on experimental status can be found in Bauer et al (2005). Recent theoretical papers are Frigeri et la (2004), Samokhin et al (2004) and Mineev (2004).

Another interesting case has been provided by the discovery of the superconductivity in the skutterudite heavy fermion compound $PrOs_4Sb_{12}$ with $T_C$ ~ 1.8 K (Bauer et al 2002). The new insight of the Pr case by comparison to the Ce or Yb ones (which are Kramer's ions with even magnetic degenerate crystal field levels) is that now the crystal field state can be a singlet. Thus, if an extra hybridisation occurs, there are two competing mechanism for the formation of a singlet state : the crystal field and the Kondo effect. Furthermore, depending on the crystal field scheme, fancy couplings and ordered phase can occur with dipole, quadrupole, octopole order parameter. For general considerations, the reader can look at Maple et al (2003), Aoki et al (2005) or Sakakibara et al (2005). Let us also point out that an extra source of the mass enhancement can also occur due to the dressing of the itinerant electron by the inelastic scattering from the singlet ground state $\Gamma_1$ to the excited triplet $\Gamma_5$ (Fulde and Jansen 1983). The feedback ingredient in this new toy (Frederik et al 2004) is the strong dependence of $\gamma$ on $C_{CF}$. It is of course directly related to the weakness of $C_{CF}$ = 8 K and the large value of $\gamma \sim 350$ mJmole$^{-1}$K$^{-2}$ (Bauer et al 2002, Maple et al 2002). Tiny variations of $C_{CF}$ can drive the material to a multipolar instability.

After the discovery of superconductivity, the two surprises concerning superconductivity are :

i/ the observation of a double transition at $T_C^A$ and $T_C^B$ (Vollmer et al 2003) and

ii/ the unusual weak sensitivity to doping as demonstrated studies on $Pr_{1-x}La_xOs_4Sb_{12}$ and $PrOs_{4(1-x)}Ru_{4x}Sb_{12}$ ($T_C$ of $LaOs_4Sb_{12}$ and $PrRu_4Sb_{12}$ are respectively equal to 0.74 K and 1.04 K) (see Rotundu et al 2004, Frederik et al 2004). It is quite different from the previous doping sensitivity of the unconventional superconductors of chapter 4.

That pushes to search for arguments on the double apparent superconducting transition which may be not related with an exotic multicomponent order parameters as described on chapter 6 for $UPt_3$ . Two macroscopic observations lead to our proposal that the lowest temperature $T_C^B = 1.75\,K < T_C^A = 1.85$ K at H = 0K may be the evidence of either a sharp crossover or a real condensation of the previous overdamped dispersive crystal field excitations (see Kuwahara 2004, 2005 and Raymond et al 2005) referred often as magnetic excitons :

1/ in mainly magnetic field, $H_{C_2}$ ($T_A$) and $H_{C_2}$ ($T_B$) are mainly parallel (Tayama et al 2003, Méasson et al 2004),

2/ in pressure, the splitting is roughly preserved (Measson 2005).

The strong support that $T_C^B$ is associated with a feedback effect of superconductivity on the magnetic excitons was given by inelastic neutron scattering experiments on single crystals at $T_C^A$ (Raymond et al 2005, Kuwahara et al 2005). The coherence of the Cooper pair drives to a shift of $C_{CF}$ to lower energy and also to the narrowing of its width $\Gamma$. The strong interplay of $C_{CF}$ with $\Gamma$ and $\gamma$ may produce the emergence of the second anomaly at $T_C^B$. Tiny change of $C_{CF}$ at $T_C^A$ can affect strongly the specific heat and may lead to a sharp crossover at $T_C^B$. The underlining question is if the overdamped magnetic excitons above $T_C^A$ becomes real excitations at $T_C^B$ associated with very weak induced multipolar component. Tiny magnetic component was observed in $\mu$SR experiments below $T_C$ and interpreted as evidence of unconventional superconductivity with broken time reversal symmetry (Aoki et al 2003). Experiment on thermal conductivity were analysed by multiple phases depending on temperature and external magnetic field (Izawa et al 2003). We will argue that the reported 4 fold anisotropy observed may be due to a weak change in $C_{CF}$ as the field is rotated in the (0, 0, 1) plane.

The exotic situation of $PrOs_4Sb_{12}$ and the unusual properties of its superconductivity has boosted many theoretical developments (see Miyake et al 2003, Goryo 2003, Maki et al 2003, Ichioka et al 2003, Sergienko and Curnoe 2004) with perspective of unconventional superconductivity. An interesting proposal is that in $PrOs_4Sb_{12}$ may be mediated by the magnetic excitons as it will be discussed later for $UPd_2Al_3$ (Matsumoto and Koga 2004).

By contrast to the debate on superconductivity, the physics of the field induced ordered phase in $PrOs_4Sb_{12}$ appears very well established by neutron scattering experiments (Khogi et al 2003) and well described by a theoretical model assuming an interaction of quadrupolar moments of Pr 4f electrons in the single – triplet crystal field levels (Shina et al 2004).

# 5/ Ferromagnetism and superconductivity

*The Outlooks are :*

- For $UGe_2$ : Dual aspect of ferromagnetism. Pressure switch from $FM_2$ to $FM_1$ (two successive ferromagnetic phases) at a first order transition pressure $P_X$. Metamagnetism of $FM_1$ toward $FM_2$ and of the paramagnetic phase PM via two successive metamagnetic transition to $FM_1$ and $FM_2$ . Appearance of superconductivity centered on $P_X$ with consequence on the superconducting phase S2 coupled to $FM_2$ and S1 coupled to $FM_1$ .
- For URhGe as in $UGe_2$, the superconductivity is not directly associated to a ferromagnetic quantum critical point.
- For $ZrZn_2$ : extra proofs are required.
- For εFe : The superconductivity appears unconventional induced by FM spin fluctuation.

## *5.1 – The Ferromagnetism of $UGe_2$*

### 5.1.1 - Ferromagnetism

$UGe_2$ is an orthorombic crystal which was already studied for its ferromagnetic phase one decade ago (Onuki et al 1992). The uranium atoms are arranged as zig zag chains of nearest neighbours that run along the crystallographic a axis which is the easy magnetization direction. The chains are stacked to form corrugated sheets as in α uranium but with Ge atoms inserted along the b axis. The difference between $UGe_2$ and α uranium is that in $UGe_2$, the nearest neighbour separation $d_{u-u}$ = 3.85 Å is larger. That leads to a greater localisation of 5f electrons and much larger entropy at low temperature. At P = 0, $UGe_2$ is a ferromagnet below $T_{Curie}$ = 54 K. It has a moderate residual $\gamma$ = 35 mJ mole$^{-1}$K$^{-2}$ term at T $\rightarrow$ 0K. The value of the ordered moment $M_o$ = 1.48 $\mu_B$/uranium is far lower than the full moment 3.5 $\mu_B$ of a free uranium ion and smaller than the Curie Weiss moment (2.8 $\mu_B$) measured above $T_{Curie}$ (Huxley et al 2001). On cooling the resistivity first decreases slowly to $\rho$ ($T_{Curie}$ ) ~ 100 µΩcm and then drops rapidly below $T_{Curie}$. This variation is reminiscent of properties found in the magnetically ordered Kondo lattice. Under pressure, $T_{Curie}$ decreases ; the critical pressure $P_C$ = 1.6 GPa appears as a first order transition. The ($T_{Curie}$ , P) contour has been drawn by different technics and notably recently by neutron diffraction. For ferromagnetism, magnetization measurements give also direct access to the order parameter (Pfleiderer and Huxley 2002, and Huxley et al 2003a).

The assertion that even at P = 0, $UGe_2$ is itinerant was based on the already reported large P dependence of $T_{Curie}$ and $M_o$ but also on band calculations with their qualitative success to explain quantum oscillation frequencies found in dHvA experiments. Band structure analysis on $UGe_2$ may suggest some tendencies to quasi two dimensional majority carrier Fermi surface sheets below $T_{Curie}$ with even sections which may be parallel i.e. giving thus some 1D character. However the nesting predictions for the wavevector Q differ, either Q $\perp$ a for Yamagami (2003) or Q // a for Shick and Picket 2001. The densities of states of spin up and spin down electrons have a large 5f contribution at the Fermi level.

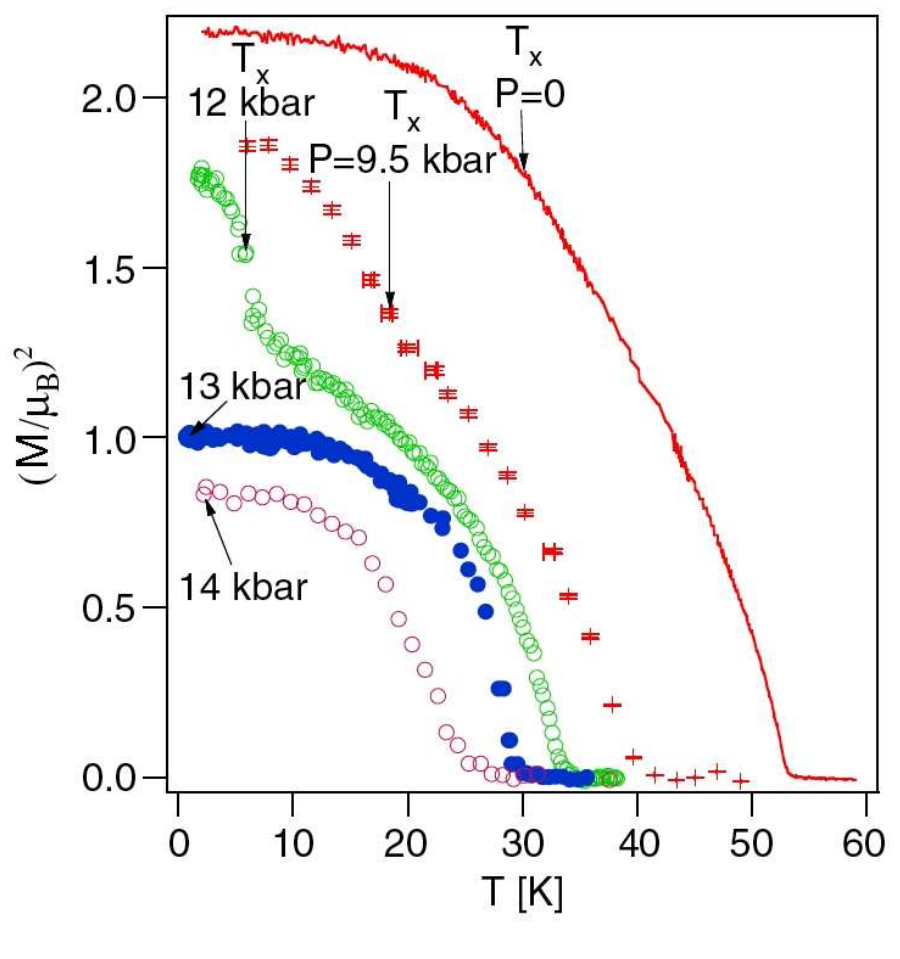

Figure 37 : The temperature dependence of the ordered moment squared at different pressure deduced from neutron scattering measurements (Huxley et al 2003).

Figure (37) shows the temperature dependence of the ordered moment M (T) squared at different pressures as detected on the nuclear Bragg reflections by neutron diffraction. At a critical pressure $P_X$ = 1.20 GPa, a change occurs in the temperature dependence of $M^2(T)$ at $T_X$. The emergence of another temperature $T_X < T_{Curie}$ below $P_X$ is also marked by an extra drop of the resistivity (figure 38). Careful neutron diffraction measurements have failed to detect any additional reflections. Through $P_X$ , ferromagnetism persists. Magnetization experiments indicate that the extrapolation of the ordered moment $M_o$ at T = 0K leads to a discontinuity of $M_o$ at $P_X$ (Pfleiderer and Huxley 2002). Thus the transition from a high moment ferromagnetic phase $FM_2$ to the other low moment ferromagnetic phase $FM_1$ is first order at low temperature. As the signature of the $FM_1 \rightarrow FM_2$ transition disappears rapidly when P decreases from $P_X$ , the critical end point at $P_{cr}$ must be few tenths of GPa below $P_X$ . However a crossover temperature $T_X$ can be drawn down to P = 0 (where $T_X$ = 28 K) in the temperature derivative of ρ(T) as well as in thermal expansion experiments (Oomi et al 1993).

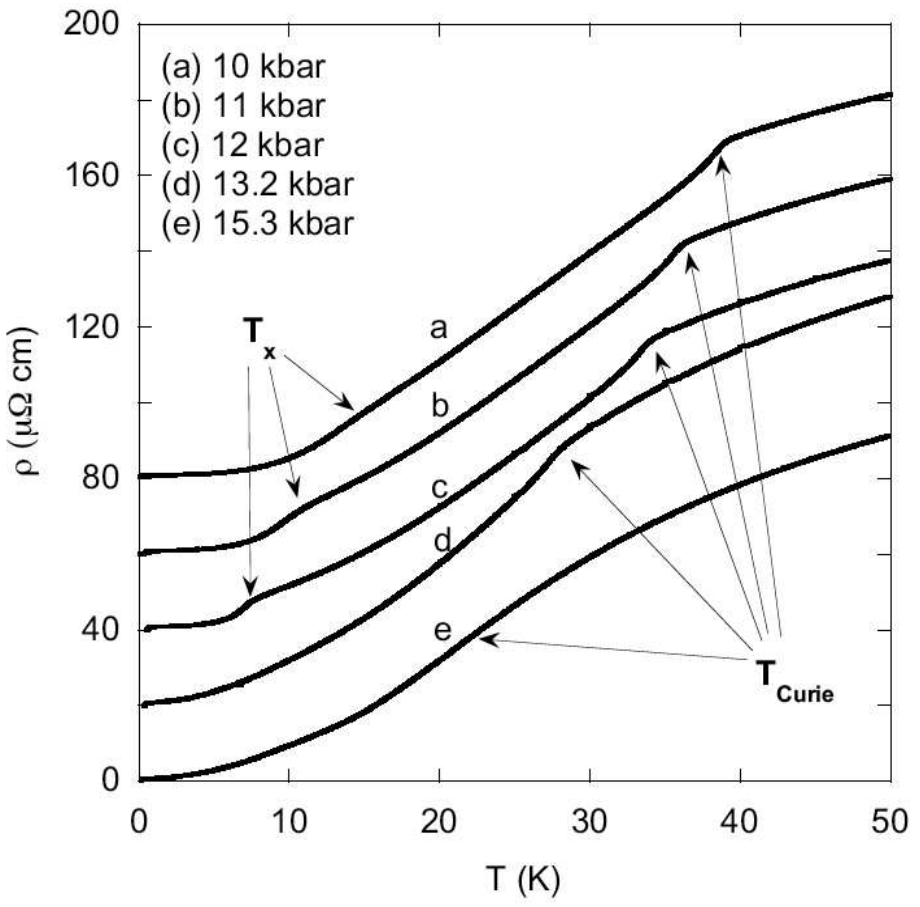

Figure 38 : Temperature variation of the resistivity of $UGe_2$ at different pressure with the emergence of the two anomalies at $T_{Curie}$ and $T_X$ below $P_X$ = 12.2 kbar (Sheikin 2000 and Huxley et al 2001).

Magnetic form factor experiments (Kernavanois et al 2001) have been performed to check if there is a change in the orientation of the magnetic moment by comparing data on quite different wavevectors (040 and 001) for P = 1.2 GPa. The excellent scaling through $T_X$ supports the persistence of the same type of magnetic structure. The agreement between magnetic form factors and the square of the magnetization suggests a homogeneous scenario of the $FM_2 \rightarrow FM_1$ transition. However the weak difference between the (001) and (111) reflection points out a slight modification in the origin of the magnetism i.e. a difference in orbital component related to a change in electronic structure (tiny modification of the valence may lead to a drastic change in the electronic structure). This statement is supported by recent band theory picture : over a range of volumes two nearly degenerate FM states exist which differ most strikingly in their orbital character on the uranium site (Shick et al 2004).

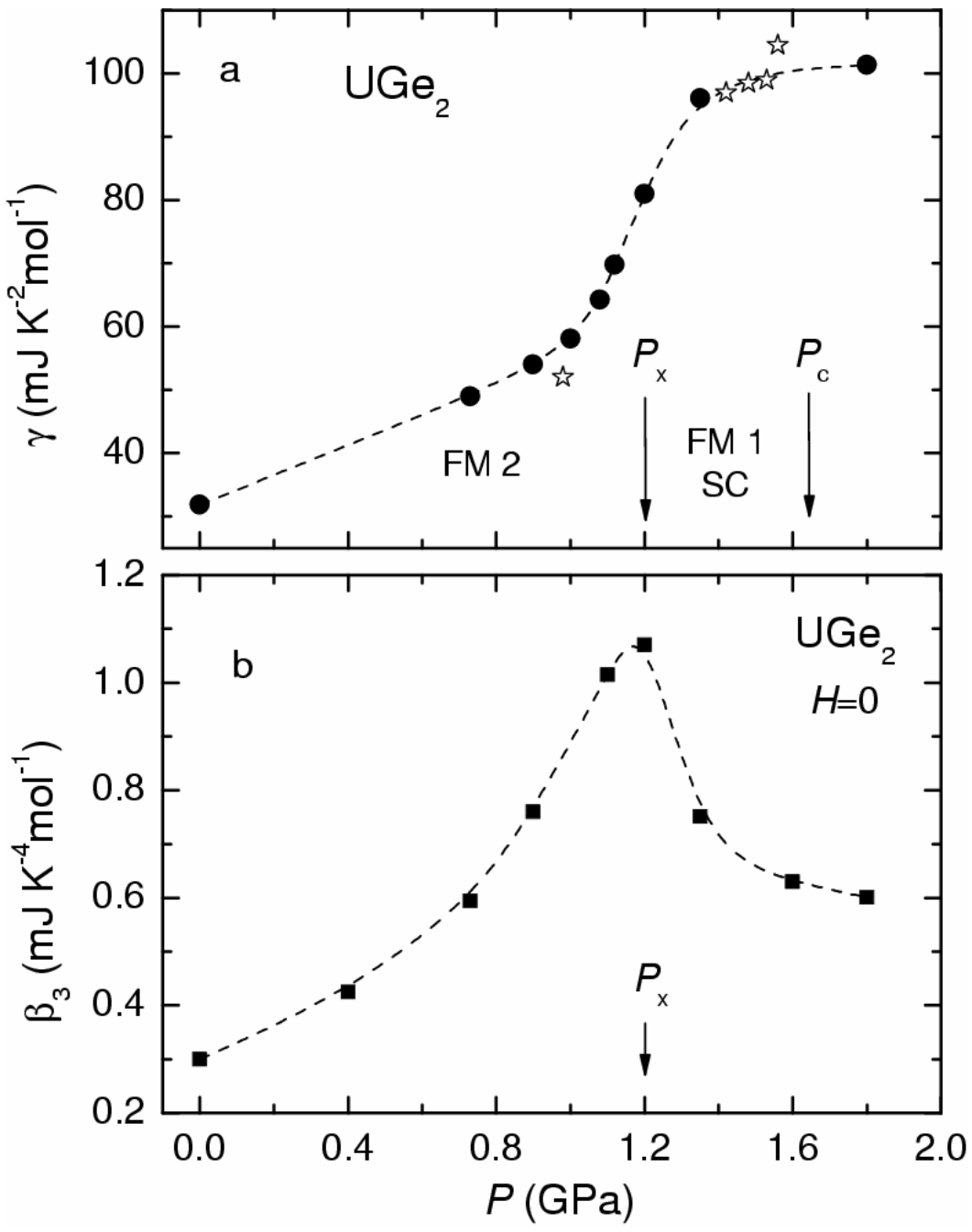

Figure 39 : In (a) (●) Plot of $\gamma$ vs P for $UGe_2$. Note the rapid rise in $\gamma$ for $P_X$ in the superconducting region of the phase diagram (Fisher et al 2005). In (b) $\beta_3$ vs P from fits to the low temperature specific heat of $UGe_2$ in H = 0. It is unlikely that $\beta_3$ for the lattice would increase with pressure, or that it has any significant P dependence to P = 1.80 GPa. Therefore, $\beta_3$ must be a composite of the weak $T^3$ lattice term plus an additional strong term. A low excitation mode can be represented by a $T^3$ term in the limited temperature range (15 K) of the fit. Furthermore, at 1.80 GPa, in the paramagnetic region of the phase diagram, a larger $T^3$ term than that at P = 0 indicates some additional contribution to the specific heat. (Fisher et al 2004).

Resistivity measurements show that the A coefficient of the $AT^2$ inelastic contribution jumps by a factor 4 above $P_X$. The related increase in the renormalized density of states is directly observed in P specific heat experiments reported figure (39-a) (Tateiwa et al 2003-a, Fisher et al 2005). At T = 0, $\gamma$ (P = 1.15 GPa = $P_X$ - $\varepsilon$) is roughly half the value of $\gamma$ at P = 1.28 GPa = $P_X$ + $\varepsilon$. At P = 1.15 GPa, the transition at $T_X$ ~ 5K is characterized by a deep drop of C/T (Tateiwa et al 2003). Analysis of the P low temperature specific heat data (Fisher et al 2005) in $\gamma T + \beta T^3$ contributions shows a maximum of $\beta$ and $P_X$ (figure 39-b) i.e where $T_C$ reaches also its maximum. Quantum oscillation experiments show drastic changes between the $FM_2$ phase below $P_X$ and the PM phase above $P_C$ in agreement with band structure calculations (figure 40) (Settai et al 2002, Terashima et al 2001). The dHvA signals between $P_X$ and $P_C$ are still controversial so we will not enter in this debate. Intrinsic inhomogeneity may wipe out the dHvA signal between $P_X$ and $P_C$ . However, we keep the message that superconductivity will be strongly influenced by FS reconstruction which occur at $P_X$ and $P_C$ .

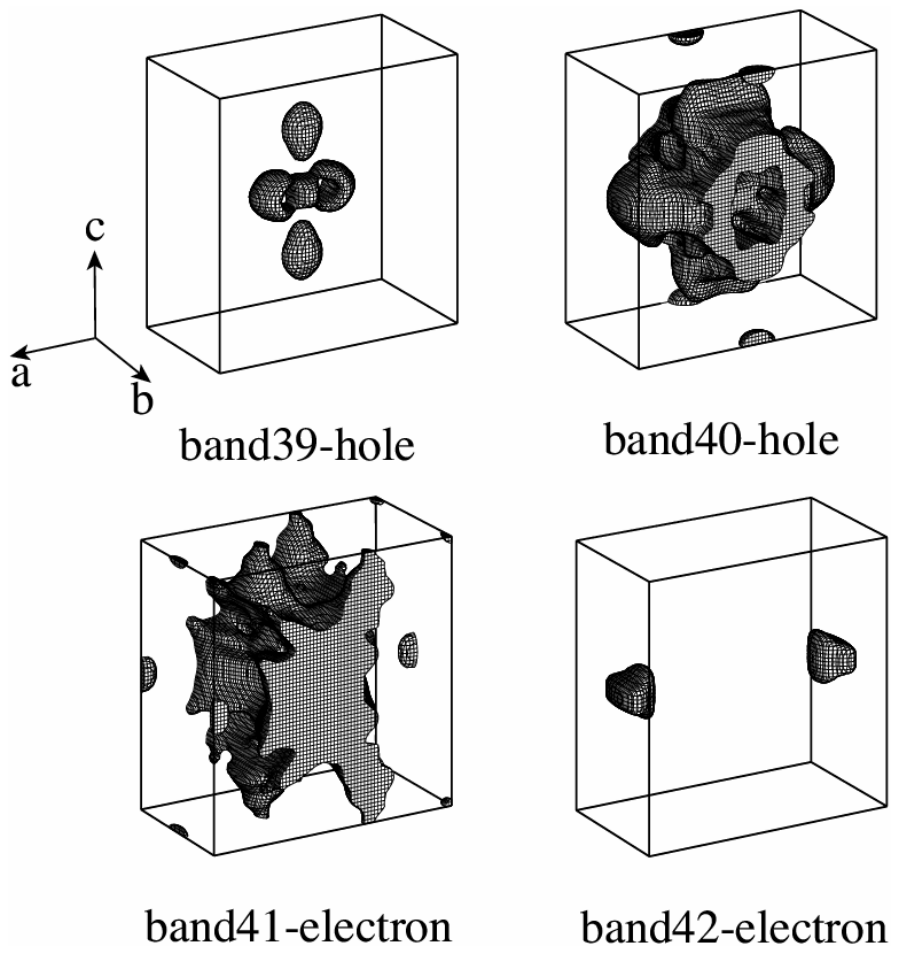

Figure 40 : a) Fermi surfaces in the ferromagnetic state of $UGe_2$ ($FM_2$ ) (Settai et al 2002)

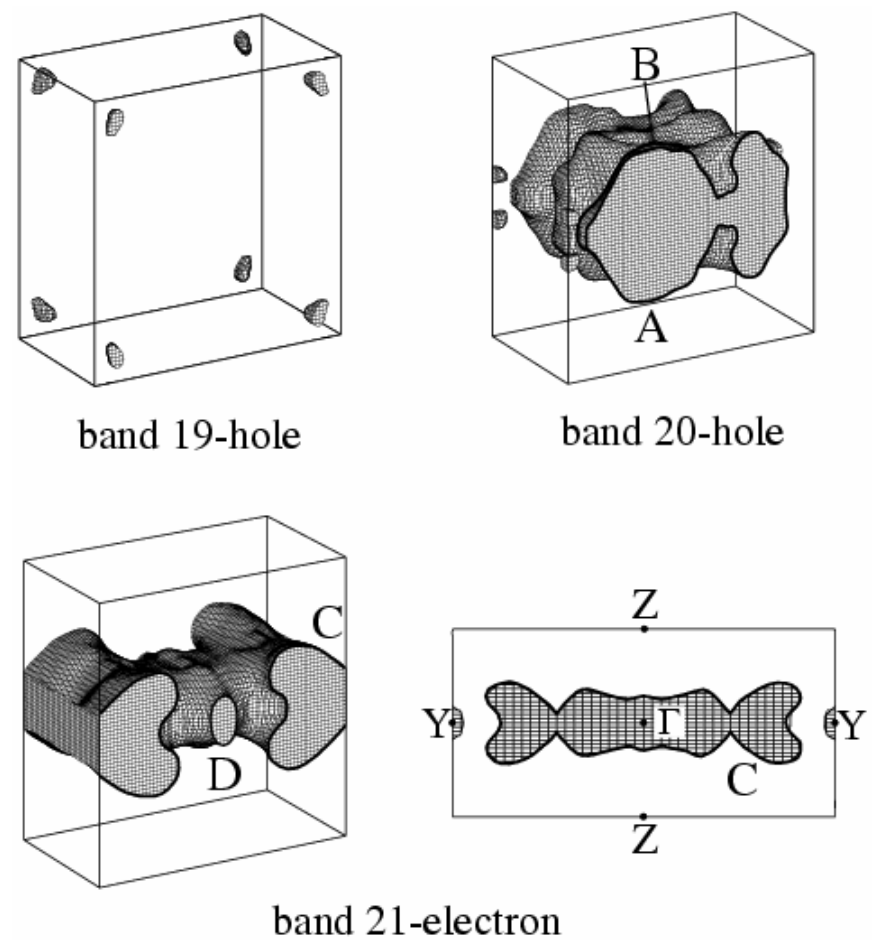

Figure 40 b : b) Fermi surfaces in the paramagnetic state of $UGe_2$ (Settai et al 2002)

Correlatively to the transition from the $FM_2$ to the $FM_1$ ground state in zero field under pressure, the magnetic field can restore the $FM_2$ above $P_X$ via a metamagnetic transition at $H_X$ and even lead to the cascade PM $\rightarrow$ $FM_1$ $\rightarrow$ $FM_2$ , through 2 successive metamagnetic transitions at $H_C$ and $H_X$ (figure 41). As observed for $CeRu_2Si_2$ , the transition at $H_X$ corresponds to a critical value of the magnetization independent of pressure. This suggests that metamagnetism takes place when the Fermi energy crosses a sharp maximum in the density of states for one spin polarization. Analysis of the magnetization along the a and c axis show that, below $P_X$ in the $FM_2$ state, the Pauli susceptibility is quite isotropic while above $P_X$ , $\chi_a \sim 4\,\chi_c$. This ratio increases slightly above $P_C$ (Huxley et al 2003-a).

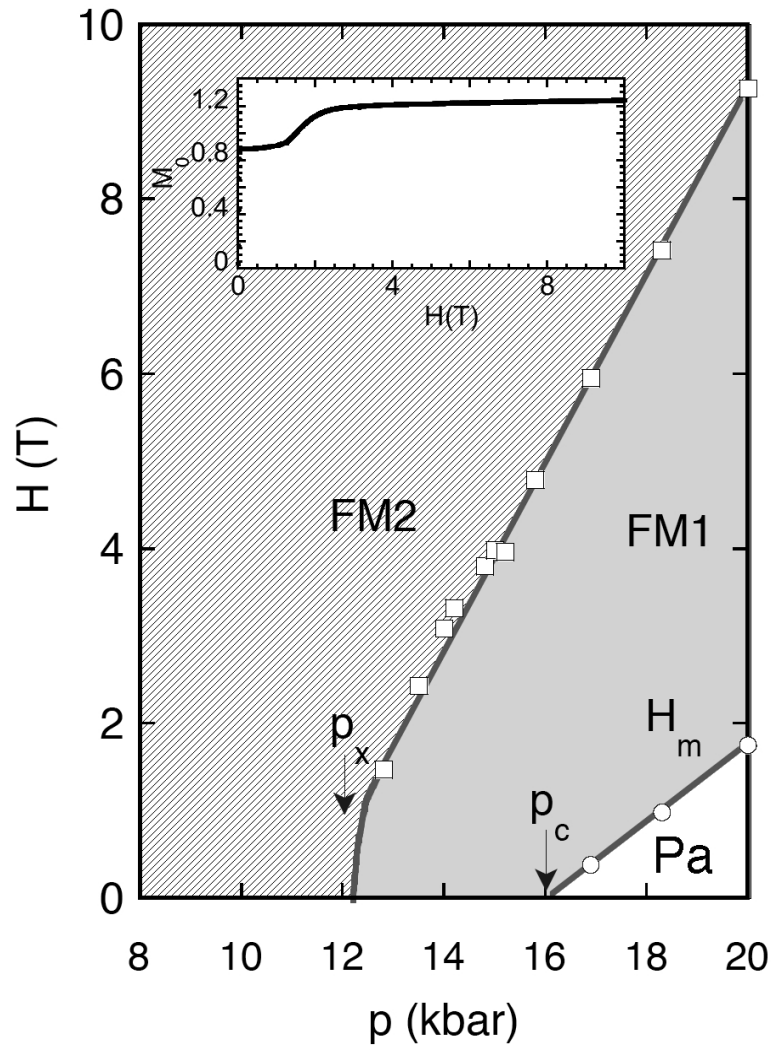

Figure 41 : (H, P) phase diagramme of $UGe_2$ the insert show the jump of $M_o$ in $\mu_B$ at the transition $FM_1 \rightarrow FM_2$ at $P_X$ for T = 2.3 K (Pfleiderer and Huxley 2002)

Spin dynamics have been studied recently by inelastic neutron scattering (Huxley et al 2003-b). In contrast to the case found for d metal ferromagnets, the magnetic excitations at small q extend to higher energies in $UGe_2$. The relaxation rate $\Gamma_q$ of the magnetization density does not vanish linearly as $q \rightarrow 0$ but has strong independent q component. The product $\chi_q\Gamma_q$ is weakly q dependent above $T_{Curie}$ (figure 42). For a clean metal not too close to $T_C$ , in 3d intermetallic compounds, it depends linearly on q (Lonzarich and Taillefer 1985). Here $\chi_q\Gamma_q \sim$ 0.70 meV as $q \rightarrow 0$. This large strength of $\chi_q\Gamma_q$ implies that long wave length fluctuations of magnetization relax rapidly. Thus the average magnetization density will be no longer a conserved quantity (similar phenomena is wellknown for liquid $^3He$ via the dipolar interaction). As for nearly ferromagnetic d compounds, $\chi_q\Gamma_q$ is T independent above $T_C$ . Below $T_C$ , it decreases strongly. Temperature analysis of $\chi_q$ gives a finite magnetic coherent length $\xi_m^o$ ($T \rightarrow 0$) ~ 24 Å larger than the typical values found in localized systems (~ 6 Å). Since the relaxation $\Gamma_q$ is related with the energy scale for magnetically mediated superconductivity, the finite quasi uniform damping of $\Gamma_q$ may enhance $T_C$. It is worthwhile to notice that positive muon spin relaxation measurements show magnetic correlations with tiny magnetic moments (Yaouanc et al 2002).

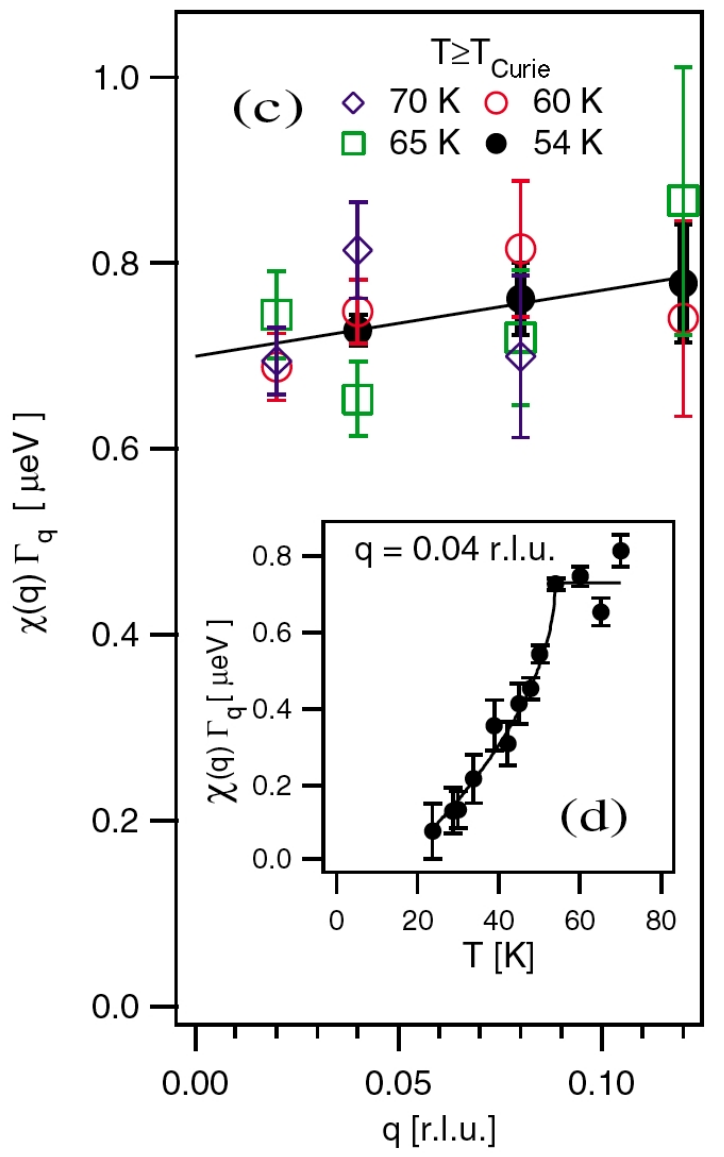

Figure 42 : The q dependence of the product $\chi(q)\Gamma_q$ at different temperatures above $T_C$ . Insert : the temperatures dependences of $\chi(q)\Gamma q$ at q = (0, 0, 0.04 r.l.u) above and below $T_C$ . The lines serve to guide the eye (Huxley et al 2003-b).

## 5.2 – $UGe_2$ a ferromagnetic superconductor

The discovery of superconductivity (Saxena et al 2000) just near 1.0 GPa and below $P_C$ i.e. inside the ferromagnetic domain (figure 43) is remarkable as superconductivity occurs when the Curie temperature is still high. The maximum of $T_C$ at $T_C^{max}$ ~ 700 mK at P ~ 12.5 kbar appears far below the Curie temperature $T_{Curie}$ ~ 35 K and in a high moment $M_o$ ~ 1.2 $\mu_B$ ferromagnetic state. As in a simple one electron picture, the exchange magnetic field (near 100 T assuming a simple formula $M_o$ x $H_{ex} = k_B T_C$ or $M_o = H_{ex}\ \chi_{pauli}$) exceeds the Pauli limit ($H_p$ ~ 1T at $T_C^{max}$ = 0.7 K), triplet superconductivity seems likely. The bulk nature of the superconductivity was suggested in flux flow resistivity experiments (Huxley et al 2001) and established without ambiguity by the observation of a nearly 30% specific heat jump at $T_C^{max}$ for P = 1.22 GPa ~ $P_X$ (Tateiwa et al 2001) and in the calorimetric experiment in Berkeley (Fisher et al 2004). The estimate of the electronic mean free path $\ell$ ~ 1000 Å from $\rho_0$, specific heat and $Hc_2$ shows that the clean condition ($\ell > \xi_0$~ 100 Å) is achieved in the Grenoble-Cambridge or Osaka experiments performed on high quality single crystals ($\rho_0$ < 1 $\mu\Omega$ cm). $T_C^{max}$ appears to coincide with the collapse of $P_X$ so is probably strongly related with the $FM_2$ - $FM_1$ transition at $P_X$ .

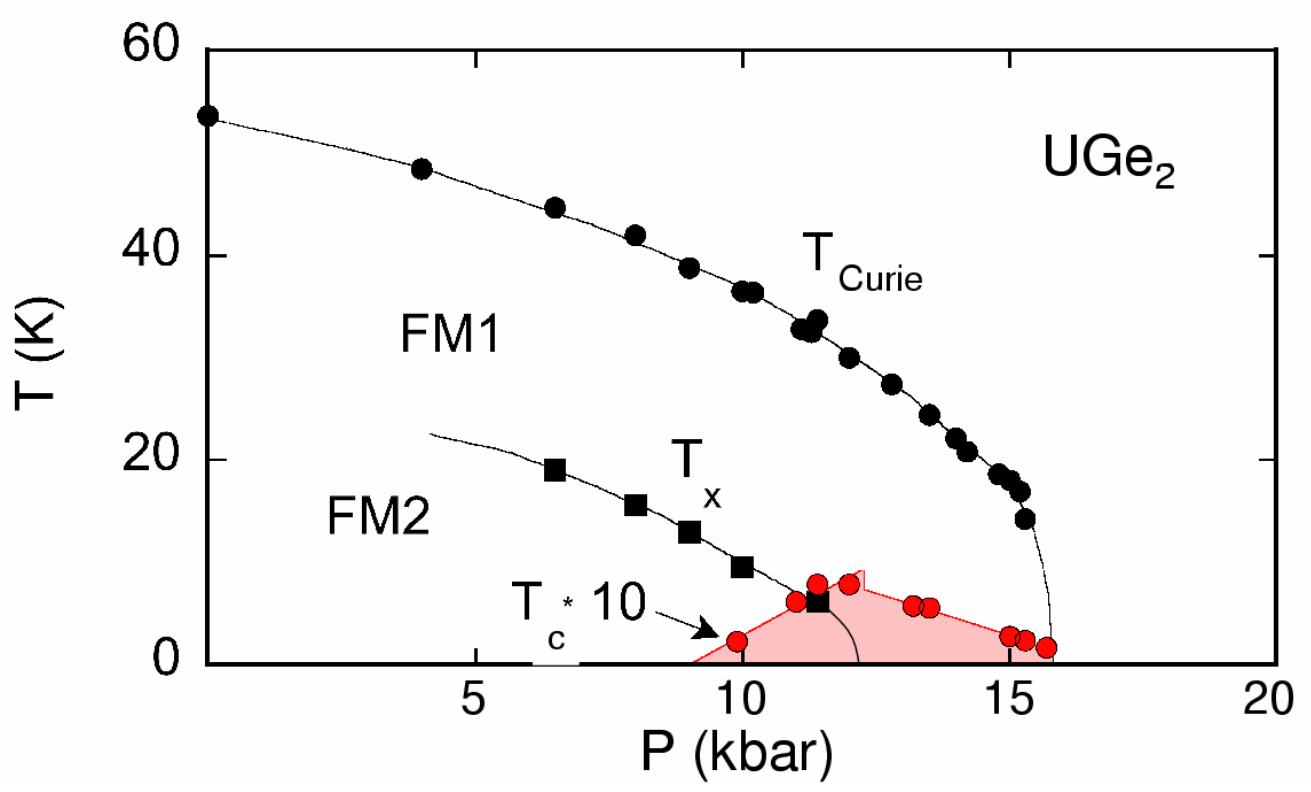

Figure 43 : The (T, P) phase diagram of $UGe_2$. The Curie temperature $T_{Curie}$ (●) the supplementary characteristic temperature $T_X$ (■) which leads to first order transition at T → 0K and the superconducting temperature $T_C$ (●) are shown.

After subtraction of a small temperature contribution due to nuclear scattering, the temperature dependence of the intensity of the (0, 0, 1) Bragg reflection measured by neutrons shows no reduction below $T_C$ = 0.7 K due to superconductivity. The persistence of the same magnetic polarisation is strong evidence of triplet pairing with the coexistence of ferromagnetism and superconductivity (Huxley et al 2003-a).

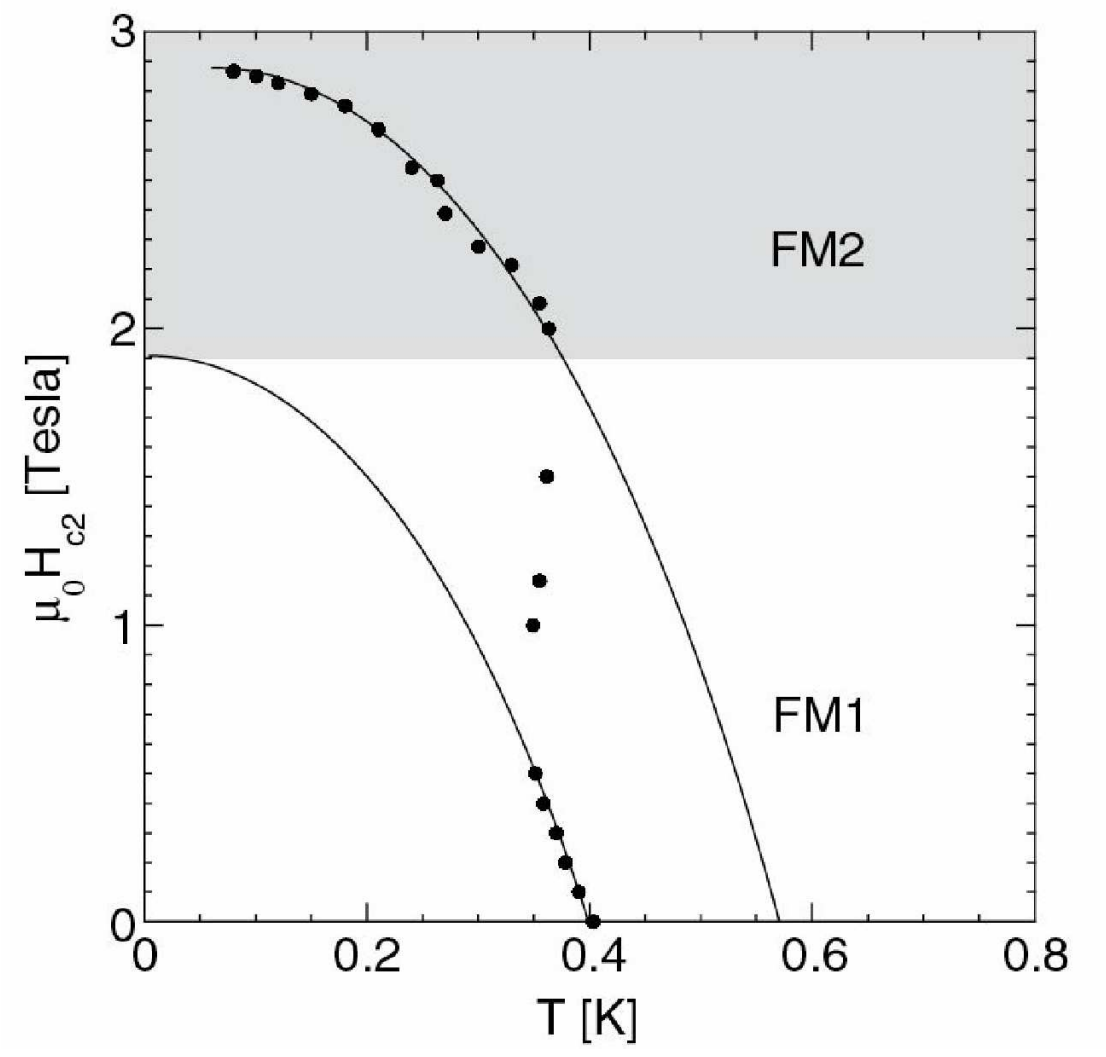

Figure 44 : The temperature field coordinates of the mid-points of the superconducting transition measured by resistivity are shown at a pressure of 13.5 kbar. The position of the onset of the bulk $FM_1$ → $FM_2$ transition seen in the resistivity in the normal phase at 1K is also indicated as the lower limit of the shaded region. (Huxley et al 2003)

The transition from $FM_1$ to $FM_2$ at $H_X$ is directly felt in the field variation of $Hc_2$ as shown at 13.5 kbar just above $P_X$ in figure (44) where the $Hc_2$(T) data at low fields below $H_X$ in the $FM_1$ state appears to belong to a different phase than the points in the $FM_2$ high field state. The $FM_1$ phase seems to have a lower maximum in the critical temperature than the $FM_2$ phase (Sheikin et al 2001) but also a lower upper critical field at T $\rightarrow$ OK. No direct accurate experiments on the pressure dependence of $T_C$ near $P_X$ are yet been performed. The puzzling result is that the specific jump $\frac{\Delta C}{\gamma T_C}$ firstly at $P_X$ is far below the BCS value, secondly drops rapidly on either side of $P_X$ (figure 45) (Tateiwa et al 2003-b and 2004). Of course, departure from $T_C^{max}$ leads to a huge sensitivity to pressure gradients and impurities. A marked jump of $\frac{\Delta C}{\gamma T_C}$ only near $P_X$ deserves confirmations and explanations.

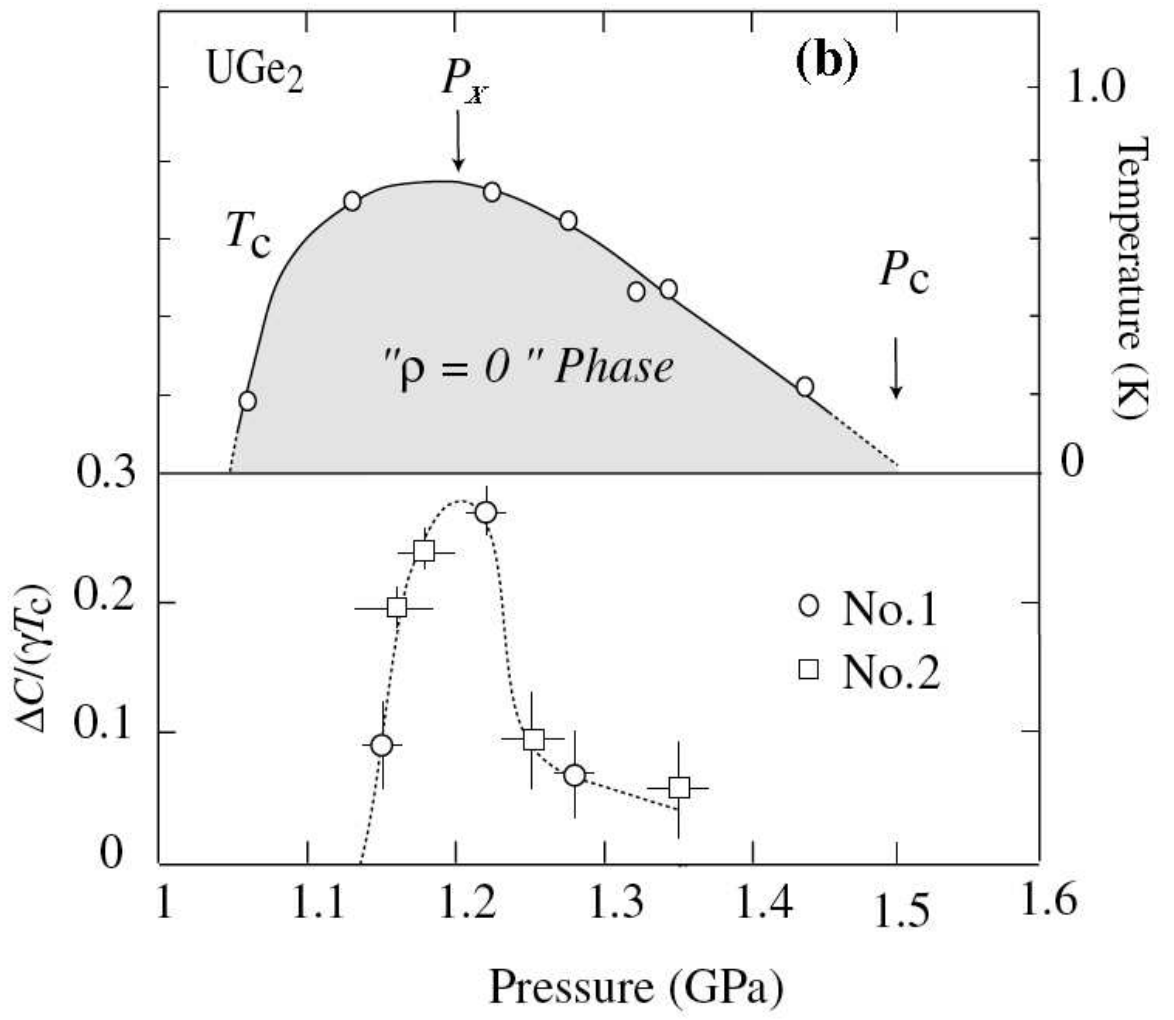

Figure 45 : Pressure dependence of $T_C$ and the value $\frac{\Delta C}{\gamma T_C}$ (from Tateiwa et al 2003-b and 2004)

Evidences for microscope coexistence of ferromagnetism and unconventional superconductivity seem now well established by $^{73}$Ge NQR study under pressure (Kotegawa et al 2003-2005, Harada et al 2005). The NMR data are summarized on figure 46. The nuclear spin lattice relaxation rate $1/T_1$ has probed the ferromagnetic transition exhibiting a peak at $T_{Curie}$ as well as a decrease without a coherence peak below $T_C$ . The coexistence of ferromagnetism and unconventional superconductivity with presumably line-node gap is clear. By comparison to previous Tateiwa results, the P range of gapped superconductivity is larger : the residual density of states at T = 0K is respectively 65%, 37% and 30% of the normal phase density of states at P = 1.15, 1.2 and 1.3 GPa. By comparison to our conclusion (from $H_{C_2}$ (T) jump at P ~ 1.3 GPa), it is found that, just below $P_X$ in the $FM_2$ phase, $T_C$ ~ 0.35 K. Just above $P_X$, in the $FM_1$ phase, $T_C$ ~ 0.7 K. This opposite conclusion shows that careful measurements are needed in the vicinity of $P_X$. A theoretical treatment on the motion of the problem.

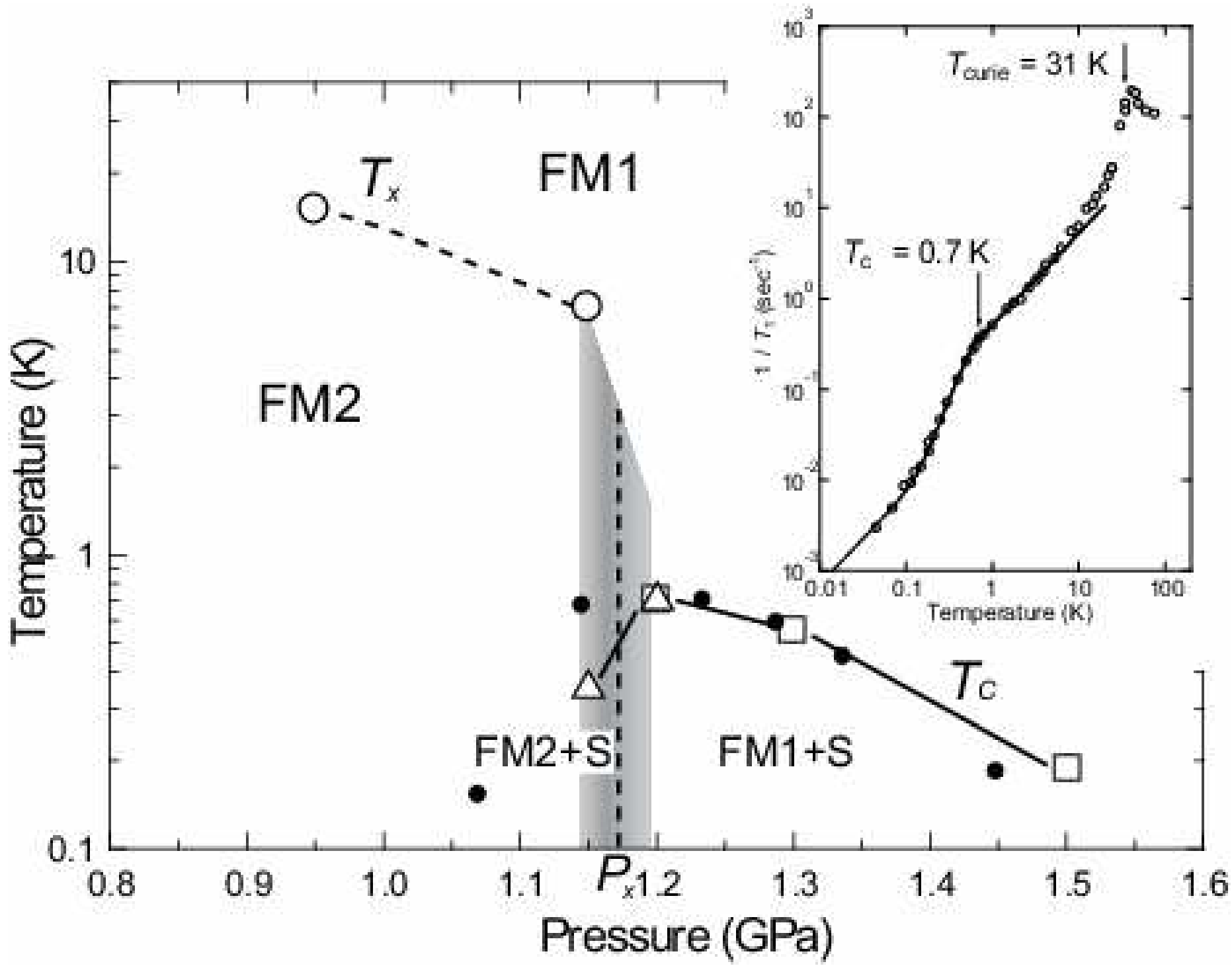

Figure 46 : (T, P) phase diagram of $UGe_2$ detected in NQR experiment (Kotegawa 2004-2005) : $P_X$ is marked by the dashed line and the shaded region corresponds to phase separation. The solid wide referred to previous work, the NQR determination of $T_X$ and $T_C$ are respectively (o) and (Δ and •). In insert, temperature variation of $1/T_1$ with the clear signature of ferromagnetism and superconductivity. The solid line is a calculation assuming unconventional line node gap.

The reproducibility of superconductivity was first verified inside the Cambridge/Grenoble collaboration and then rapidly confirmed in Osaka (Tateiwa et al 2001), Nagoya (Motoyama et al 2001) and La Jolla (Bauer et al 2001). The occurrence of superconductivity in $UGe_2$ is well established. However the homogeneity of the effect as well as the conditions of its observation (clean or dirty limit) are still controversial. Experiments on $UGe_2$ polycrystals can lead to the conclusion that the clean limit requirement ($\ell > \xi_0$) may not be a necessary condition so implicitly opens the possibility of conventional pairing. To precise the unconventional nature of superconductivity a new generation of systematic measurements is clearly needed on different crystals.

It was claimed by G. Motoyama et al (2003) that for P < $P_X$ a clear sharp anomaly can be observed in the ac magnetic susceptibility at $T_{Curie}$ while an imperfect superconducting shielding effect occurs at $T_C$. As P increases away from $P_X$ , the reverse is observed i.e. the peak anomaly at $T_{Curie}$ becomes broad while the diamagnetic susceptibility approaches a perfect superconducting shielding. We have already stressed that the selfconsistency between the square of the magnetization and the ferromagnetic intensity detected on Bragg reflection is an important proof of ferromagnetic homogeneity on both sides of $P_X$ . Crude conclusions on the homogeneity of the superconductivity and ferromagnetism on each side of $P_X$ and $P_C$ may lead to abusive statements as the respective broadening of the transitions are sensitive to the P dependence of the characteristic temperatures which are strong below $P_X$ for $T_C$ and above $P_X$ for $T_{Curie}$ . As described, the recent accurate NMR data support a homogeneous coexistence of FM and S.

## *5.2 – Ferromagnetism and superconductivity in URhGe and $ZrZn_2$*

URhGe appears a new promising material as here the superconductivity at $T_C$ ~ 300 mK is achieved already at P = 0. This favourable situation seems correlated with a smaller U – U distance, $d_{u-u}$ = 3.5 Å, than in $UGe_2$ corresponding to the set in of ferromagnetism below $T_{Curie}$ = 10 K with a residual $\gamma$ = 160 $mJmole^{-1}k^{-2}$ and a sublattice magnetization of 0.4 $\mu_B$ /U atom (Aoki D. et al 2001). The bulk nature of the superconductivity is demonstrated by the specific heat jump $\frac{\Delta C}{\gamma T_C} \approx 0.45$ at $T_C$ . Of course, with this result made on a polycristal, one may speculate that the relative weak jump of $\frac{\Delta C}{\gamma T_C}$ and the high value of C/T at T = 0K even in the superconducting state are indications for a zero gap for the minority spin while the majority spins may lead to a temperature dependence of C/T characteristic of a polar state (line of zeroes).

However further developments require to succeed in the growth of excellent crystals and then to perform careful measurements as will be reported for $UPt_3$ or $UPd_2Al_3$ . The growth of excellent crystals with convincing superconducting properties has been unsuccessful in different laboratories. Systematic measurements on polycristals underline that the clean limit must be fulfilled for the occurrence of superconductivity. Very recently superconducting single crystals have been produced in Grenoble (Huxley and Hardy 2004). The upper critical field on single crystals can be well fitted with a triplet pairing (Hardy 2004, Hardy et al 2005).

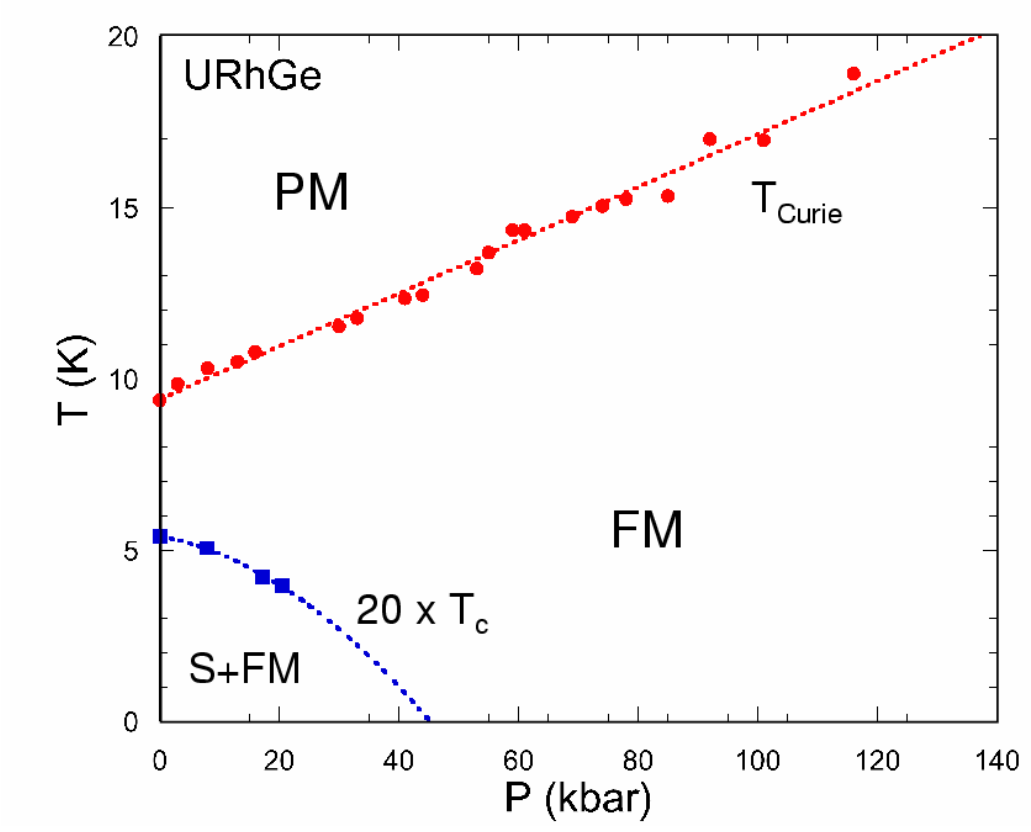

Figure 47 : Pressure dependence of $T_{Curie}$ , $T_C$ of URhGe (Hardy 2004)

Of course pressure experiments have been performed ; contrary to $UGe_2$, $T_{Curie}$ increases under pressure at least up to 10 GPa while a linear extrapolation of the P decrease of $T_C$ suggest $P_S$ = 3 GPa. The URhGe (T, P) phase diagram is different to the $UGe_2$ one, $P_S << P_C > 10$ GPa (figure 47). The driving mechanism for superconductivity may be not ferromagnetic spin fluctuations even if the superconductivity must adjust its pairing with the spin arrangement. In the easy c axis of magnetization, in low magnetic field (H < 500 oe) up to create a single ferromagnetic domain, $T_C$ is field invariant.

A supplementary interest of URhGe is the field switch of the easy axis of magnetization from c to b axis at $H_C \sim 12$ T. Of course, this transition are associated with changes of the Fermi surface and related variations of the spin and charge dynamics. The big surprise is the H re-entrance of superconductivity. (Levy et al 2005)..

Just before the report of superconductivity in URhGe, it was claimed that the wellknown weak Heisenberg itinerant ferromagnet $ZrZn_2$ ($T_{Curie}$ = 30 K, $M_o$ = 0.17 $\mu_B$ at P = 0) may become a superconductor below $T_C$ = 0.2K (Pfleiderer et al 2001). The non convincing points were the nonvanishing value of the resistivity below $T_C$ , the lack of a maximum in the imaginary part $\chi''(q = 0, \omega \to 0)$ of the susceptibility at $T_C$ and furthermore the absence of any specific heat jump at $T_C$ . The main evidence of intrinsic phenomena was the apparent collapse of $T_{Curie}$ and $T_C$ at $P_C$ = 1.5 GPa. After this first report, a full superconductivity resistivity drop was observed below $T_C$ as well as a smooth maximum in $\chi''$ at $T_C$ (see Pfleiderer et al 2003, Hayden 2003). However still no specific heat anomaly has been found at $T_C$ and contradictory links are reported between the occurrence of superconductivity and sample purity. Recently in Grenoble (Boursier 2005), for single crystals of $ZrZn_2$ with RRR ~ 30 a complet resistive superconducting transition was detected with $T_C$ = 80 mK and $H_{C2}$ (0) ~ 1 KOe. In Sendai (Kimura et al 2003), no trace of superconductivity can be found even for RRR ~ 60 and in Tokyo with RRR ~ 140 (Takashima et al 2004). Here new materials and careful tests are needed to understand the superconductivity of $ZrZn_2$ and even to assign its intrinsic character. A recent publication by Uhlarz et al (2004) show that the collapse of ferromagnetism occur at P = 1.65 GPa through a first order transition.

Unsuccessful attempts to discover superconductivity have been made in different cerium ferromagnetic HFC ($CeRu_2Ge_2$ , $CeRh_3B_2$, $CeAg_2Sb$) as well as in uranium ferromagnets UP and $U_3P_4$ and 3d systems such as $Ni_3Al$ (Niklowitz et al 2003) or $Y_2Ni_3$. The problem of unconventional superconductivity and ferromagnetism is far less documented than the previous one of unconventional superconductivity and antiferromagnetism.

## *5.4 – Ferromagnetic fluctuation and superconductivity in ε Fe ?*

The other major breakthrough is the discovery of superconductivity in the high pressure (P > 13 GPa) crystallographic ε hexagonal compact phase of Fe : $T_C^{max}$ = 3K at P = 22 GPa (Shimizu et al 2001 (figure 48). In this ε phase, the ground state is paramagnetic but with an enhanced density of state. Band calculations suggest a proximity to antiferromagnetism (Saxena and Littlewood 2001). Nevertheless, it has been predicted that in ε Fe, the ferromagnetic fluctuation will lead to higher $T_C$ than the AF fluctuation (Jarlborg 2002). Of course, the martinsitic process of the α to ε conversion favours the persistence of a "ferromagnetic" memory over short distancies.

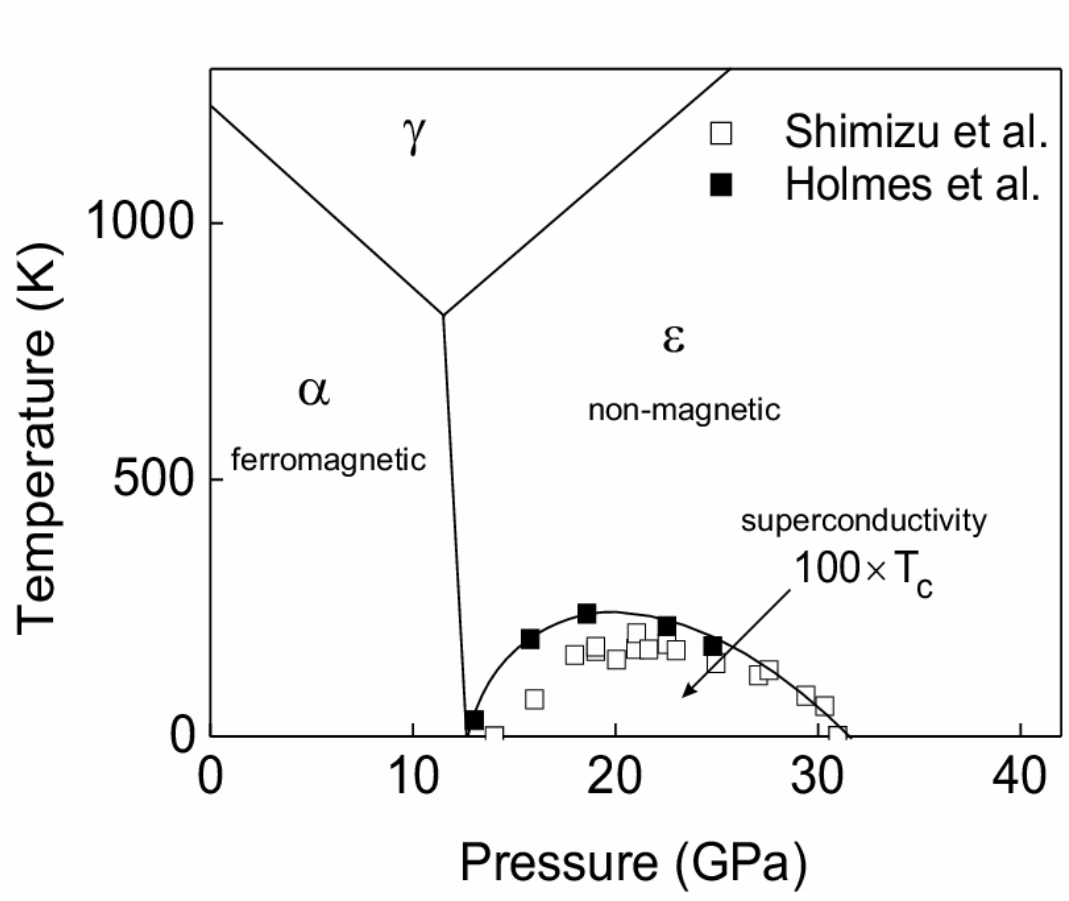

Figure 48 : The body-centered-cubic α phase is the well known ferromagnet, while if the high pressure and high temperature γ phase was stable at low temperatures it would form an antiferromagnet at 100 K at ambient pressure. The hexagonal-close-packed ε phase is non-magnetic and becomes superconducting at low temperatures. (from Shimizu et al (□) and Holmes et al (■) 2001, 2004-b).

Careful absolute resistivity measurements have been recently performed on different crystals of Fe. In good agreement with the previous remarks, the temperature variation of the resistivity just above $T_C$ follows the $T^{5/3}$ law predicted for ferromagnetic fluctuation. The large slope $\frac{\partial Hc_2}{\partial T}$ at $T_C$ confirms that heavy particles are involved in the pairing ($m^* \sim 10\ m_o$) (Jaccard et al 2002, Holmes et al 2004-b, Jaccard et al 2005). The necessity to respect the clean limit condition ($\ell > \xi_0$) points to an unconventional pairing. By contrast to the previous examples of $UGe_2$, URhGe and $ZrZn_2$ , here the superconductivity occurs in the PM state. So εFe, the inner constituant of the earth appears to have concomitant NFL properties and unconventional superconductivity near P = 20 GPa. Furthermore the superconductivity may keep memory of the spin and charge dynamics of the low pressure α cubic phase. The discovery of this new class of unconventional superconductors pushes for new theoretical developments.

## *5.5 - Theory on ferromagnetic superconductors*

The different superconducting states in ferromagnetic phases for crystal with cubic (Samokhin and Walker 2002) and orthorhombic structure (Mineev 2002) have been classified on general symmetry arguments. For the $ZrZn_2$ cubic case (Walker and Samokhin 2002), it was predicted that the gap nodes change when the magnetization is rotated by the magnetic field. Tests can be easily made in ultrasound attenuation and thermal conductivity experiments. For an orthorhombic point group, only one dimensional representations are possible. This can lead to a magnetic superconducting phase with spontaneous magnetization when superconductivity occurs inside the ferromagnetic region ($T_{Curie} > T_C$ ) for the case of a strong spin-orbit coupling (Fomin 2001).

In general, no symmetry nodes exist, if the pairing amplitude with the zero projection of the Cooper pair spin is not vanishing. If for some reason the pairing amplitude with zero projection of the Cooper pair spin (the $\hat{z}$ component of the order parameter) is absent then the gap for the superconducting gap Δ can have point nodes parallel to the magnetic axis at $k_x = k_y = 0$ (Fomin 2002). The possible absence of the z-component of the order parameter also practically eliminates any possibility of the occurrence of a Fulde-Ferrell-Larkin-Ovchinikov type state (no paramagnetic limitation).

One question is the reason for the stabilization of superconductivity in the ferromagnetic domain. It was suggested by Walker and Samokin (2002) that a positive feedback occurs due to an exchange interaction between the magnetic moments of Cooper pair and the magnetization density. For the previous orthorhombic case, the possible superconducting states are non-unitary and Cooper pairs have a spin momentum, which is proportional to the difference in the density of populations of pairs with spin up and spin down. The interaction of the spontaneous magnetic moment of the Cooper pair with the exchange field due to the ferromagnetism $H_{ex}$ stimulates the superconducting state (Mineev 2002). This effect exists only due to the difference in the density of states on the Fermi surfaces for the spin-up and spin-down quasiparticles (Ambegaokar and Mermin 1973) and leads to an enhancement of the critical temperatures $\frac{T_C\ (H_{ex}) - T_C}{T_C} \approx \frac{\mu_B H_{ex}}{\varepsilon_F}$.

In parallel, the ferromagnetic magnetization creates a magnetic field $H_{em}$ that acts on the orbital electron motion to suppress the superconducting state. The reduction of the critical temperature due to the orbital effects is :

$$\frac{T_C\ (H_{em}) - T_C}{T_C} \approx \frac{d^2 H_{em}}{\Phi_0}$$

where d is the characteristic length scale over which the order parameter changes near the upper critical field. For a pure superconductor d simply coincides with the usual coherence length $\xi_0$. In the vicinity of the critical pressure $P_S$ the coherence length $\xi_0 \sim v_F/T_C$ will eventually exceed the mean free path $\ell$. Thus close to $P_S$ in this unconventional clean superconductor, d is given by $\ell$. The comparison of the previous two expressions shows that triplet superconductivity can be stimulated by ferromagnetism.

An alternative idea for a weak Heisenberg ferromagnet such as $ZrZn_2$ is linked to the disappearance of the transverse magnetic fluctuation for coherent magnons below $T_{Curie}$ with an enhancement of $T_C$ on the ferromagnetic side because the coupling of magnons to the longitudinal magnetic susceptibility enhances strongly $T_C$ in the FM state respectively to the paramagnetic state (Kirkpatrick et al 2001). This possibility seems to be ruled out in the $UGe_2$ Ising ferromagnet as transverse modes like magnons will be missing.

Further, explanations have been proposed for $UGe_2$ on the remarkable coincidence that $T_C$ is optimum just at $P_X$ . The drop of the resistivity at $T_X$ as well as the coincidence of the maximum in $T_C$ when $T_X$ collapses are reminiscent of the paramagnet $\alpha$ Uranium where $T_X$ is identified as the charge density wave temperature $T_{CDW}$. Furthermore the common point between $UGe_2$ and $\alpha$ Uranium is their zig zag Uranium chain (Huxley et al 2001). These anologies plus the unusual temperature variation of the neutron intensity of ferromagnetic Bragg reflections (at $T_X$) and a bump in C/T near $T_X$ push to propose a model where a charge density wave CDW may occur below $T_X$ (Watanabe and Miyake 2002-a and b). This model is able to explain the field instability at $H_X$ and the unusual shape of $Hc_2$ (T). However up to now, as indicated above, no lattice and magnetic superstructure has been detected. In the same spirit, a zero temperature Stoner model (Sandeman et al 2003) is proposed on a system which has a twin peak structure in the electronic density of states(DOS) i.e. the necessary ingredients for two metamagnetic instabilities at $H_C$ and $H_X$ . Triplet superconductivity is driven in the ferromagnetic phase by tuning the majority spin Fermi level through one of the two peaks. The maximum of $T_C$ is found at $P_X$ i.e. at the magnetization jump.

Of course, another possibility is to consider the possible occurrence of s wave superconductivity by bypassing the argument on the strength of the exchange field seen by the conduction electrons. Such a possibility is considered mainly for $UGe_2$ since the ferromagnetism may come from the localized 5f part. It was shown that the coupling of two electrons via a localized spin can be attractive (Suhl 2001) and demonstrated that this s wave attraction holds for the whole FS (Abrikosov 2002). The supplementary condition for the occurrence of superconductivity is a large density of states at Fermi level i.e. the occurrence of heavy fermions. The applicability to $UGe_2$ is an open question as underlined previously its ferromagnetism has both localized and itinerant characters.

There are particular features associated with unconventional superconductivity in ferromagnets. It was recently underlined that triplet ferromagnetic superconductors with up up spin pairs and down down spin pairs with scattering at a finite spin orbit coupling are two band superconductors. The consequence is that the dependence of $T_C$ on the impurity content is non universal but determined by two independent dimensionless parameters linked to the respective scattering time and to the pairing. The departure from the universal Abrikosov Gorkov law will be a qualitative proof of the two band character (Mineev and Champel 2004).

An interesting new feature is that the superconducting order parameter is coupled to the magnetization (Fomin 2001, Machida and Ohmi 2001). The spin direction of the dominant pairing amplitude is fixed by the magnetization direction. As this direction changes from one domain to another, the properties of this layered structure can differ from those of a domain structure with singlet pairing. One should expect here that domain walls play the role of weak links.

Another supplementary consideration (Buzdin and Mel'nikhov 2003) at least in resistivity experiments, is that of domain wall superconductivity. In each domain a finite average magnetic induction $4\pi M_o$ exists (near 2000 Oe for $UGe_2$). Assuming a thin domain wall ($\ll \xi$) and modelling the domain interface by a step like function $\pm M_o$ on each side of the wall, the orbital effect is cancelled. On cooling, the superconductivity will first appear locally at a domain wall not inside a magnetic domain. Furthermore depending on the relative orientation of M with respect to H, different critical temperatures will occur between two opposite domains. That must lead to an unusual broadening at $H = 0$ near $T_C$ which will disappear rapidly in magnetic field.

# 6/ The four uranium heavy fermion superconductors

*Outlooks :*

- Diversity of the (P, T) superconducting phase with respect of the universality which may happen for hypothetic quantum critical point.
- $UPt_3$ : an example of the role of the Kondo lattice condensate to unconventional superconductivity. Careful experiments which demonstrate the unusual low temperature excitations of unconventional superconductors.
- $UPd_2Al_3$ : evidence of d wave pairing in inelastic neutron scattering and tunnelling experiments.
- $URu_2Si_2$ : Switch from SMAF or hidden order to LMAF magnetism at $P_X$ . Disappearance of superconductivity at $P_{cr}$ the critical pressure where the first order transition line ($T_X$, $P_X$ ) meets or approches the magnetic line ($T_N$ , P).
- $UBe_{13}$ : A superdense Kondo lattice due to the low carrier density ?

## *6.1 – Generalities*

The Kondo picture cannot be applied straightforward to actinide compounds. In U systems, evidences of $f^n \Leftrightarrow f^{n\pm1}$ transition is lost for example in photoemission spectra (Allen 1992). Already even without renormalization, a rather broad f band exists. The valence is not near an integer number and furthermore the fluctuations will often occur between 2 magnetic configurations of $U^{3+}$ and $U^{4+}$. For Sm and Yb the fluctuations happens between a trivalent configuration which is a Kramer's ion and a divalent non magnetic configuration ; for Tm as for U the valence fluctuation involves 2 magnetic configurations. The description of U compounds require consideration and knowledges on their bandstructure while, for the Ce, HFC qualitative previsions can be made from its local behavior ($n_f$ , $T_K$, $\Delta_{CF}$). For a view on the magnetism of U intermetallic systems the reader can look to the review article of Sechovsky and Havela (1998). We will focus mainly on the interplay between superconductivity and magnetism.

In the previous works, either the requirement to apply a pressure or the difficulty to grow large single crystals have restricted experimental studies. Extensive data have been obtained on four different uranium heavy fermion superconductors $UPt_3$ , $UPd_2Al_3$ , $URu_2Si_2$ and $UBe_{13}$. The growth of large crystals has allowed to perform combined studies notably neutron scattering experiments, quantum oscillation (de Haas van Alphen) NMR and various macroscopic probes. Our main aim is to stress important features of these unconventional superconductors :

- the necessity to treat the impurity scattering in the unitary limit,

- the consequence of lines of zeros or point nodes in the angular variation of the gap $\Delta_k$ (low energy excitations, sensitivity to the Doppler shift even at low magnetic field, appearance of a normal fluid component at very low temperature directly linked with the change of sign of $\Delta_k$)

- the relation between the spin pairing of the order parameter and $Hc_2$ (T).

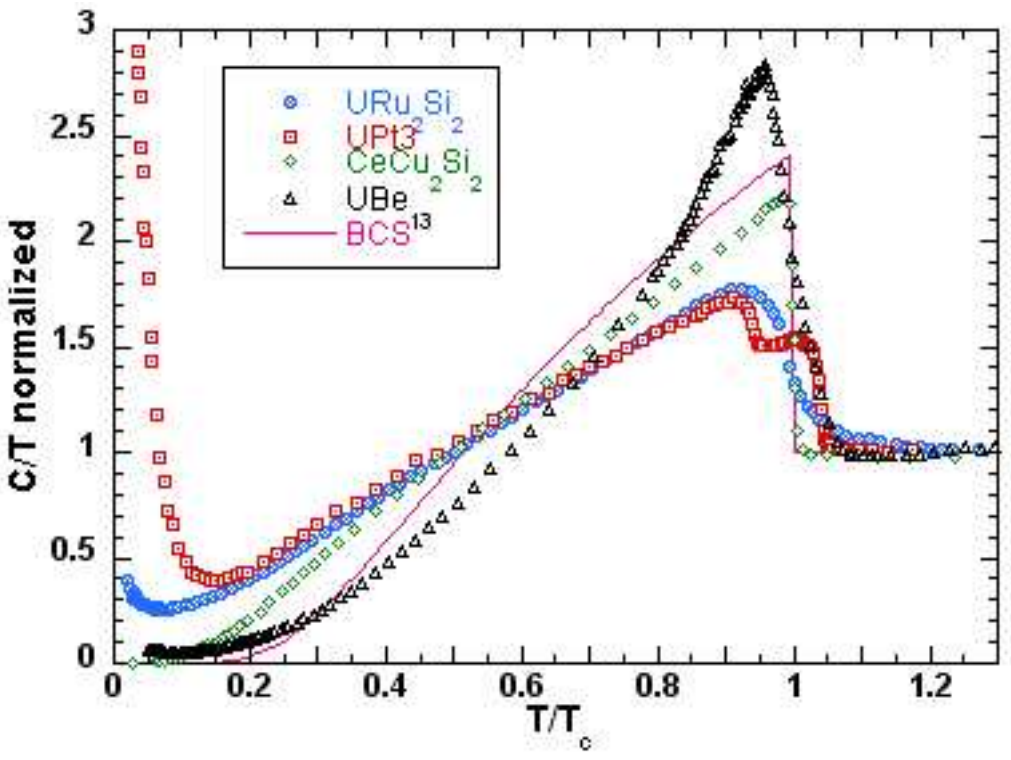

Figure 49 : Specific heat of various HFC normalized to the value of C/T in the normal phase just above $T_C$ (Brison et al 1994).

Let us summarize briefly the main phenomena at zero pressure. The occurrence of bulk superconductivity is proved with the observation of marked specific heat anomalies at $T_C$ (figure 49) . Outside $UBe_{13}$ , the three other compounds have specific heat jumps at $T_C \left( \frac{\Delta C}{\gamma T_C} \approx 1 \right)$ often smaller than the usual BCS value of 1.43 observed for s wave superconductors. The decrease of C below $T_C$ is not exponential but follows mainly a $T^2$ power law for $UPt_3$ , $URu_2Si_2$ and $UPd_2Al_3$ and $T^3$ for $UBe_{13}$ (Brison et al 1994-a),. That leads to suspect a line of zeros for the three first cases and point nodes for $UBe_{13}$. $UPt_3$ is an intriguing superconductor as a double superconducting transition occurs. This phenomenon appears highly reproducible and boosts the focus on $UPt_3$ with a large diversity of experiments (Joynt and Taillefer 2002). Such a multi-component phase diagram cannot be explained in the framework of conventional superconductivity. We will see later that it needs also a new concept for SMAF at least for its dynamics. As we will see, the main trend is to assume a bidimensional $E_{2u}$ order parameter split by a symmetry breaking field (SBF) (see figure 50). The possibility of a multi-component phase diagram was already stressed in the study of $U_{1-x}Th_xBe_{13}$ (0.02 < x < 0.047) (Ott et al 1985) but the requirement of doping has prevented careful studies in the low temperature phase. The controversy still exist if the second low temperature transition is linked or not with the onset of large range magnetism.

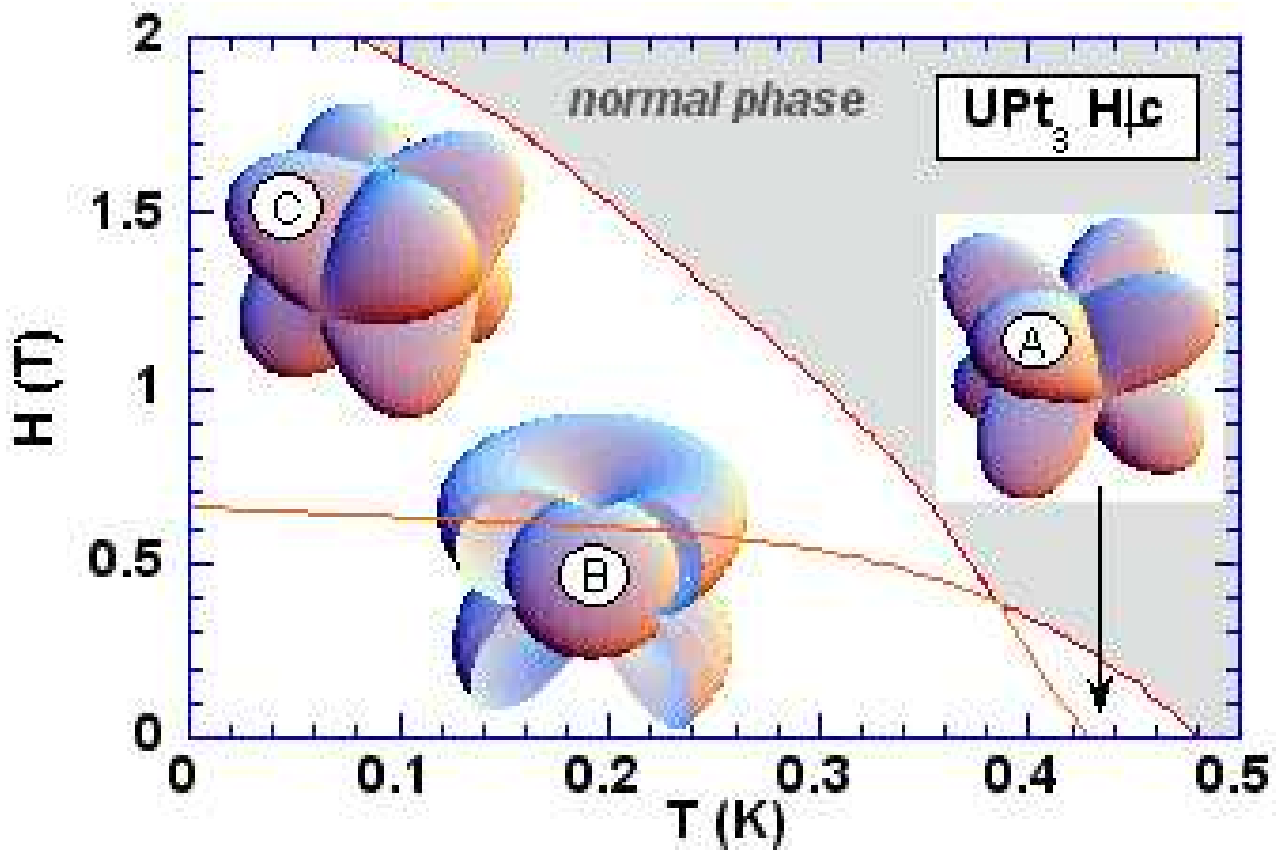

Figure 50 : Schematic phase diagram of the three superconducting phases of $UPt_3$ , together with the hypothetical corresponding gap structures in the $E_{2u}$ model for a spherical Fermi surface (apparent on the B phase gap scheme). The A and C phases differ by a rotation of the azimuthal line of nodes. Only a second order point node on the c-axis remains in the B phase (Brison et al 2000).

In these four cases, the superconductivity is not obviously related with the proximity to a magnetic instability. In $UPt_3$ , below $T_N$ ~ 6K neutron scattering experiments as well as X ray scattering (Aeppli et al 1988-a and b, Isaacs et al 1995) show the appearance of tiny ordered moment ($M_o$ = 0.02$\mu_B$/U-atom) along the a* axis of the basal plane. No sign of static magnetic order has been found in macroscopic experiments or low energy spectroscopy (NMR - µSR). In $UPd_2Al_3$ , a clear AF ordering occurs below $T_N$ = 14 K with a large sublattice magnetization $M_o$ ~ 1 $\mu_B$ (large moment antiferromagnetism LMAF) located in the basal plane and a propagation vector $Q_0$ = (0, 0, 1/2) (Krimmel et al 1992) along the c axis of the hexagonal crystals. In $URu_2Si_2$ below $T_N$ ~ 17.2 K, a large specific heat anomaly characterises the onset of long range ordering however only a small sublattice magnetization $M_o$~ 0.03 $\mu_B$ is detected along the c axis with a propagation vector $Q_0$ = (0, 0, 1) (Broholm et al 1987 and 1991). In this so called SMAF phase, a hidden order (H.O) must exist which has not yet been identified (Mydosh 2003, Coleman 2002). In $UBe_{13}$, no long range magnetism has been detected at least above $T_C$. The interesting feature is that when superconductivity appears, no simple Fermi liquid regime is yet established (as for $CeCoIn_5$). For example the resistivity has reached its maximum only at $T_M$ ~ 2.5 K. At $T_C$ , large inelastic scattering still occurs leading to $\rho(T_C)$ ~ 100 µΩcm.

Under pressure, the tiny ordered moment $M_o$ in $UPt_3$ vanishes linearly with P at $P_X$ ~ 0.4 GPa however the drop of $M_o$ (Hayden et al 1992) is not correlated with a concomitant variation of $T_N$ as observed in the previous cerium case of $CeIn_3$ or $CePd_2Si_2$. The important point is that the collapse of $M_o$ appears linked to the P disappearance of the superconducting splitting. The antiferromagnetism may be at the origin of the SBF (Behnia et al 1990, Trappmann et al 1991, de Visser et al 2002). $T_C$ may vanish at $P_S$ ~ 5GPa far above 0.4 GPa.

In $UPd_2Al_3$ , $T_N$ decreases initially strongly under pressure with a slope $\frac{\delta T_N}{\delta p} = 900\, mK/GPa$. No magnetic signal can be detected in resistivity experiments above 7.5 GPa. A linear P extrapolation suggest $P_C$ ~ 15 GPa. $T_C$ by contrast is first quite insensitive to pressure up to 7 GPa and then decreases at a rate of 90 mK/GPa (Link et al 1993). Extrapolating to zero will give $P_S$ ~ 30 GPa quite different from $P_C$ . Up to now no data exists in the critical regime $P_C$ or $P_S$. A high pressure structural study points out a pressure induced electronic transition at P = 25 GPa from hpc to orthorombic phase (Krimmel et al 2000).

In $URu_2Si_2$ , recent experiments show that above $P_X$ = 0.5 GPa at low temperature, the H.O phase with its weak antiferromagnetism and superimposed hidden order is replaced by a large moment antiferromagnetic phase (LMAF) with the same Q wavevector but a sublattice magnetization (jump of $M_o$ from 0.15 $\mu_B$ to 0.35 $\mu_B$ ) (Amitsuka et al 1999). The superconductivity of $URu_2Si_2$ disappears at a pressure $P_S$ = 1.2 GPa with a large pressure rate of 1000 mK/GPa (Mc Elfresh et al 1987, Schmidt et al 1993) while $T_N$ increases initially under pressure with a variation $\frac{\partial T_N}{\partial p} = 1900\, mK/GPa$. $P_S$ =1.2 GPa seems to be near the critical pressure where the first order $T_X$ line may end up or meets the two second order H.O and LMAF ($T_N$) lines.

In $UBe_{13}$, Tc decreases with a slope of 130 mK/GPa suggesting $P_S$ = 8 GPa (Chen et al 1984). Surprisingly, thermoelectric power experiments suggest that magnetic ordering may occur above 6.7 GPa (Mao et al 1988).

It is interesting to compare the experimental and theoretical evaluation of the London penetration depth $\lambda_L$ :

$$\lambda_L = \left( \frac{m*}{\mu_0 n e^2} \right)^{1/2}$$

as it is a sensible quantity of the ratio of the effective mass to the carrier number. On table 7, the average measured values $(\lambda_L^{exp})$ by muon experiments are compared with the results obtained in a free electron assumption assuming either 3 electron carrier per unit formula $(\lambda_L^{3+})$ or a number of carrier derived from band calculations $(\lambda_L^{BC})$ (see $URu_2Si_2$ , $UBe_{13}$ paragraph) :

| table 7 | Vm ($cm^3$/mold'U) | $\gamma$ $mJmole^{-1}K^{-2}$ | $(\lambda_L^{3+})$ in Å | $(\lambda_L^{BC})$ in Å | $(\lambda_L^{exp})$ in Å | Ref. |
|---|---|---|---|---|---|---|
| $UPt_3$ | 42 | 457 | 3600 | 3600 | 5200 | Yaouanc et al 1998 |
| $UPd_2Al_3$ | 62 | 140 | 2100 | 2100 | 4510 | Knetsch et al 1993 |
| $URu_2Si_2$ | 49 | 70 | 700 | 17000 | 9000 | Feyerherm et al 1994 |
| $UBe_{13}$ | 81 | 1000 | 6000 | 90000 | > 121000 | Dalmas 2000 |

The large values of $\lambda_L^{exp}$ for $URu_2Si_2$ and $UBe_{13}$ agree well with the low carrier case of these two materials. Their magnitude deserves further considerations as the dependence between the effective mass m* and the number of carrier $n_e$ may not be given by a free electron hypothesis (see chapter 1). All these compounds are strong type II superconductors with a large parameter kappa ($\kappa$ ~ 50 or 100).

Of course in these complex materials, a large diversity of cases may happen. The relevance of spin dynamics to superconductivity is not obvious but at least inelastic neutron scattering measurements point out the occurrence of large magnetic responses. Subtle games between parameters may lead to unique situations well suitable for a given property. That is the case for magnetic dynamical susceptibility in $UPd_2Al_3$ where in the AF state, the normal phase is caracterized by an important quasielastic contribution peaked around the AF wave vector $Q_0$ and by a spin wave or magnetic exciton dispersion. In contrast for the previous cerium HFC, there is no such strong peaked dynamic component. This intense quasielastic signal is completely modified below $T_C$ . The energy redistribution is a direct consequence of a gap opening and characteristic of a d wave even pairing. (Bernhoeft 2000, Sato et al 2001).

## *6.2- $UPt_3$ : multicomponent superconductivity and slow fluctuating magnetism*

The magnetism of U atoms is first dominated on cooling by a fluctuating local moment (~ $2\mu_B$ ) at a characteristic energy of 10 meV (Aeppli et al 1988-b) ; paramagnetic moments on nearest neighbour sites in adjacent planes start to be antiferromagnetically coupled below T = 20 K at the wavevector Q = (0, 0, 1) with a typical energy of 5 meV. At lower energies (~ 0.3 meV) another antiferromagnetic correlation appears at $Q_0$ = (1/2, 0, 1) with a small moment 0.1 $\mu_B$ ; it corresponds to AF coupling of neighbouring sites on the a* or b axis inside the basal plane (Aeppli et al 1988-a). A slow magnetic component of the itinerant quasiparticles was detected in a time of flight neutron scattering experiments (Bernhoeft and Lonzarich 1995).

Below $T_N$ ~ 6K, a part of the AF fluctuations at $Q_0$ appears static in a neutron scattering experiment with a tiny ordered moment $M_o$ = 0.02 $\mu_B$ /U atom at T $\rightarrow$ 0K. Comparable signals are absent in probes having a longer time scale (NMR – µSR) as well as in macroscopic properties (C, $\chi$, $\rho$). The magnetic coherence length is finite (300 Å). A linear temperature dependence of the intensity of the magnetic Bragg peak in neutron scattering pushes for an extrinsic origin for this SMAF or to precursor effects linked to the slowdown of the correlation while the real phase transition may appear at far lower temperature. Evidences for this last possibility are the observation of a specific heat anomaly at $T_N$ ~ 18 mK (Schuberth et al 1992, Brison et al 1994-a) and the concomitant increase of the correlation length with a divergence near 20 mK (Metoki et al 2000). The tiny moment antiferromagnetic correlation can be qualitatively interpreted with the crystal field singlet – doublet scheme as the thermal fluctuations can play a main role practically in the whole temperature region where correlations set in (Fomin and Flouquet 1996). This framework can also explain that under pressure $M_o$ collapses linearly with P up to 0.5 GPa with no concomitant P dependence of $T_N$ . The observation of uncorrelated P dependences of $M_o$ anf $T_N$ can also be explained in a inhomogeneous scheme with the P disappearance of a fraction of a magnetic phase caracterized by LMAF. However as the same tiny ordered moment $M_o$ ~ 0.02 $\mu_B$ occurs in high quality crystals, a self consistent intrinsic mechanism should control SMAF. Attempts have been made to discover crystallographic structural modulations or trigonal distortions but deeper experiments (Dalmas et al 2005) rule out the proposals (Midgley et al 1992, Walker et al 2001). The small angle neutron scattering shows intense diffusion along the a* and c* axis with "rod" shapes suggesting planar defects (Huxley et al 2000, van Dijk et al 2002). Their implication in the magnetism of $UPt_3$ is open.

The claim of a double intrinsic superconducting transition with our colleagues in Berkeley (Fisher et al 1989) was the end of a long track which :

- firstly starts with the observation of a split specific heat transition (Sulpice et al 1986) on a polycristal but unfortunately which is not associated with a unique diamagnetic shielding at $T_C^A$,

- secondly continues with the confirmation of the same features on another polycristal (Ravex et al 1987),

- thirdly leads to assign the phenomena to intrinsic properties as it survives in the specific heat measurement realized in Berkeley on a single crystal now available in Grenoble with the arrival of L. Taillefer from Cambridge, (Fisher et al 1989),

- fourthly suggests the collapse of the splitting in magnetic field at a tetracritical point (Hasselbach et al 1989)

Accurate determinations of the (H, T) superconducting phase diagram with the three phases A, B, C were achieved rapidly by ultrasound measurements (Adenwalla et al 1990, Bruls et al 1990) by magnetocaloric effect (Bobenberger et al 1993) and by magnetostriction (Van Dijk et al 1993 - 1994). Despite the reproductibility of the phenomena in different laboratories, I was not completely confident during a decade of its intrinsic nature keeping in mind that for an experimentalist it is best to find by his own error. However, the intrinsic nature of the phenomena seems established.

Systematic studies were performed on $U(Pt_{1-x}Pd_x)_3$ alloys (de Visser 2002). The striking results is that the superconductivity collapses at the concentration x ~ 0.6% above which the SMAF phase is replaced by a LMAF phase. In this last state, the link between $T_N$ and $M_o$ is recovered starting from zero at $x_c$ = 0.6% reaching $T_N$ = 6K and $M_o$ = 0.6 $\mu_B$ for x = 5%. The coincidence between the collapses of both $T_C$ and $T_N$ at $x_c$ at least proves that $T_C$ is not correlated with an AF QCP as in chapter 4. It has been suggested that doping may be associated with a shift in the spectral weight from ferromagnetism to antiferromagnetism and thus with the disappearance of the odd parity superconducting state. However a strong decrease of $T_C$ will occur on doping in this unconventional superconductor. Substitutions on the U sites whatever is the doping, (non magnetic (La, Th) or magnetic (Gd)), destroy superconductivity above $x_c$. The absence of differences between paramagnetic substituants or non magnetic impurities have suggested already an odd parity pairing (Dalichaouch et et 1995).

We have emphasized that in unconventional superconductors, any type of impurities have pair breaking effects. The reason is that in the scattering processes on impurities, the wavevector of the incident electrons may be changed to a position on the FS where the order parameter has a different sign. It was rapidly stressed that in order to explain the thermal transport in these exotic superconductors, a maximum of scattering must be realized corresponding to the unitary limit with a phase shift $\delta = \pi/2$ (Pethick and Pines 1986). Now it is taken as a general hypothesis for all non conventional superconductors. For this pair breaking, magnetic and non magnetic impurities give similar effects. An important consequence of the unitary limit is the creation of virtual bound states which are revealed by a sharp resonance at the Fermi level. In the gapless regime of the unconventional superconductors the normal fluid component leads to residual linear T terms in specific heat and thermal conductivity (Hirschfeld et al 1988-1986, Schmidt Rink et al 1986).

Several experiments(see Sauls 1994) favour the choice of the $E_{2u}$ bidimensional order parameter reported in figure (50). The first possibility of a $E_{2u}$ state appears for the description of the upper critical field $Hc_2$ (T) measured along the a and c axis. The absence of a Pauli limit for H// a can be explained if the d vector of an odd parity order parameter is parallel to c. Indeed, no Pauli limit will occur when H is perpendicular to c as in this equal spin pairing scheme (↑↑, ↓↓), the Pauli susceptibility is invariant through $T_C$ (Shivaram et al 1986, Piquemal et al 1987). At least, in strong spin orbit limit (SO) a Pauli limit is predicted for H//c as observed experimentally on $H_{C_2}(T)$. The $E_{2u}$ state of the B phase is called a hybrid state as lines of zeros exist in the basal plane and point nodes in the c axis with an energy gap $\Delta_k$ vanishing as $\sin^2\theta$, θ being the azimutal angle between the k and c vectors. The other bidimensional $E_{1g}$ even state in the B phase has a line of zeros again in the basal plane but $\Delta_k$ vanishes as sinθ as $\vec{k}$ reaches $\vec{c}$.

Thermal conductivity (κ) experiments are a powerful technique to test the hybrid state as a $T^3$ dependence of K is predicted for both $E_{1g}$ and $E_{2u}$ states at least above the characteristic temperature T* of the normal fluid component induced by the impurities. In the low temperature regime 20 mK < T < 70 mK the thermal conductivity in both directions ($\kappa_a$ and $\kappa_c$) show the hybrid state $T^3$ variation (Suderow et al 1997-1998). Below 20 mk, down to the lowest measured temperature of 16 mk, the linear T variation emerges. This component stems from the band of impurities present in the case of symmetry enforced gap nodes. A vanishing gap without a change of sign of the order parameter would be in the contrary smeared out by impurities. It was stressed that, in the gapless regime for a $E_{2u}$ state, κ/T may reach the same universal value along c as that along a (Lee 1993, Graf et al 1996) while for $E_{1g}$ no unique extrapolation of κ/T will occur. With the present unique results on the thermal conductivity in the gapless regime of $UPt_3$ , the data does not appear consistent with the $E_{2u}$ choice (universal constant of κ/T along a). But, the validity of the present theoretical hypothesis (phase shift, isotropy of the relaxation state) has to be re-examined.

The low energy excitations lead to a strong reactivity to the magnetic field. The quasiparticle energies are Doppler shifted by the superfluid flow induced around each vortex. Unlike in conventional superconductors, quasiparticles can be now excited even in small magnetic field. Volovik first points out that for lines of nodes a $\sqrt{H}$ contribution will dominate the usual linear H dependence of C/T produced by the creation of the normal core component (Volovik 1993). Furthermore, by considering combined H and thermal variations, a scaling relation with a single variable $x = \frac{T}{T_C}\sqrt{\frac{H_{c2}}{H}}$ (Kopnin and Volovik 1996, Simon and Lee 1997) must be respected in the specific heat and the thermal transport. The scaling is observed for both, a and c, directions in $UPt_3$ (Suderow et al 1998). Thus field and thermal response agree with a hybrid gap at least in the B phase.

Another interesting test is to follow the thermal anisotropy $\kappa_c/\kappa_{a \text{ or } b}$ normalized to their normal phase values. A first attempt (Flouquet et al 1991) on quite different crystals failed to detect any anisotropy. However new generations of experiments in better conditions (high quality crystal ($T_C$ ~ 530 mK $\rho_0$ ~ 0.5 µΩcm) and on adjacent pieces from the same batch) show a large temperature variation in the ratio $\kappa_c/\kappa_a$ (Lussier et al 1994, Huxley et al 1995) The difficulty is that the comparison of the experiment with theoretical models depends on the parametrization of the order parameter close to the nodes. The initial weak dependence of $\kappa_c/\kappa_b$ close to $T_C$ and the still large value of $\kappa_c/\kappa_b$ at low temperature seems to favour the $E_{2u}$ choice. Ultrasonic attenuation experiments have been used also to probe the quasiparticle spectrum. Combined measurements of transverse and longitudinal ultrasonic attenuation confirm the hybrid gap in the phase B and gap structure with additional nodal planes in the A phase as predicted for both $E_{1g}$ and $E_{2u}$ scenarios (Ellman et al 1996).

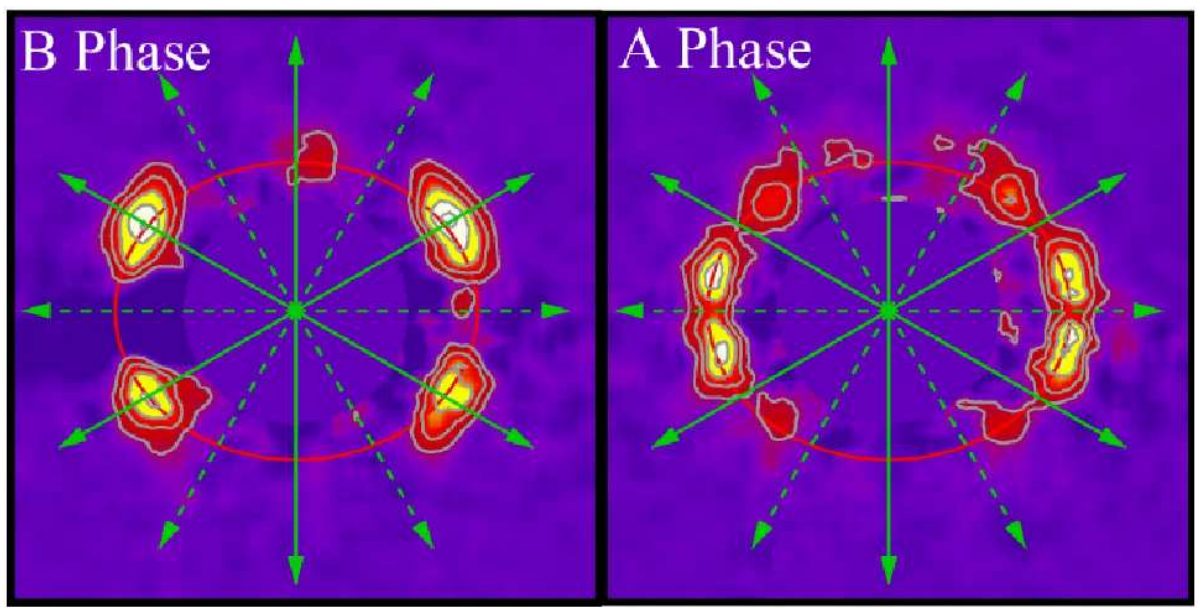

Figure 51 : Flux line lattice in the B and A phases of $UPt_3$ observed by small angle neutron scattering experiments (Huxley et al 2000).

Macroscopic evidence of the difference between the B and A order parameter was given in the observation of the flux lattice in small angle neutron scattering experiments (figure 51) (Huxley et al 2000). In the high temperature phase A, the flux lattice arrangement is governed by the strong gap anisotropy in the basal plane. It gives rise to different domains for the orientation of the vortex lattice. At lower temperature in the B phase, the gap with its line of zeros is isotropic in the basal plane. In the B state, the weak anisotropy of the Fermi surface will fix the vortex orientation. A theoretical analysis has been proposed by Champel and Mineev (2001). The non local correction in the Landau free energy may not play a big role in the orientation of the vortex lattice by contrast to classical low dimensional superconductors. Independently of the models used for the order parameter, the results in the A phase can be explained only if the order parameter belongs to a 2D irreducible representation (Rodière 2001).

It was hoped that the confirmation of a $E_{2u}$ odd parity order parameter could be easily found in a NMR experiment. In the strong spin orbit limit, assuming the Knight shift is dominated by the Pauli susceptibility, it is predicted that there would be no change through $T_C$ for H//a but that a drastic decrease for H//c may occur. As tiny variations of the Knight shift (Tou et al 1996-1998) were found in each direction, different proposals have been given to understand the experiments. One of them suggests a drastic change in the hypothesis on the spin orbit coupling : switching from strong to weak limit will allow to explain a field orientation of the d vector. The proposed orbital order parameter is quite similar to that of the $E_{2u}$ state (Ohmi and Machida 1996). Another remark starts with the consideration that the Van Vleck susceptibility may be large and thus also the contribution to the Knight shift. Furthermore it is shown that new type of Fermi liquid state can be realized in the specific situation of $5f^2$ configuration for a singlet crystal field ground state with no enhancement of the Pauli susceptibility but enhancement of the density of states (Ikeda and Miyake 1997). Tiny temperature decreases of the Knight shift through $T_C$ are of course in excellent agreement with this picture. Assuming an opposite hypothesis of Kramers ions with a doublet ground state, it was recently pointed out, that a selfconsistent analysis of the Korringa relaxation, the γ Sommerfeld coefficient of C/T, and the Knight shift gives, in $UPt_3$, large enough Pauli contribution in the Knight shift to be probed experimentally. The NMR results of different HFC leads to classify $UPt_3$ , $UNi_2Al_3$ with odd parity pairing and $CeCu_2Si_2$, $CeCoIn_5$, $UPd_2Al_3$ with even parity pairing (Tou et al 2003). The figure (52) compares the Knight shift variation of $UPt_3$ and $UPd_2Al_3$ along the main orientation axis. The solid lines in both figures are calculations using the Sommerfeld coefficient and d wave singlet model.

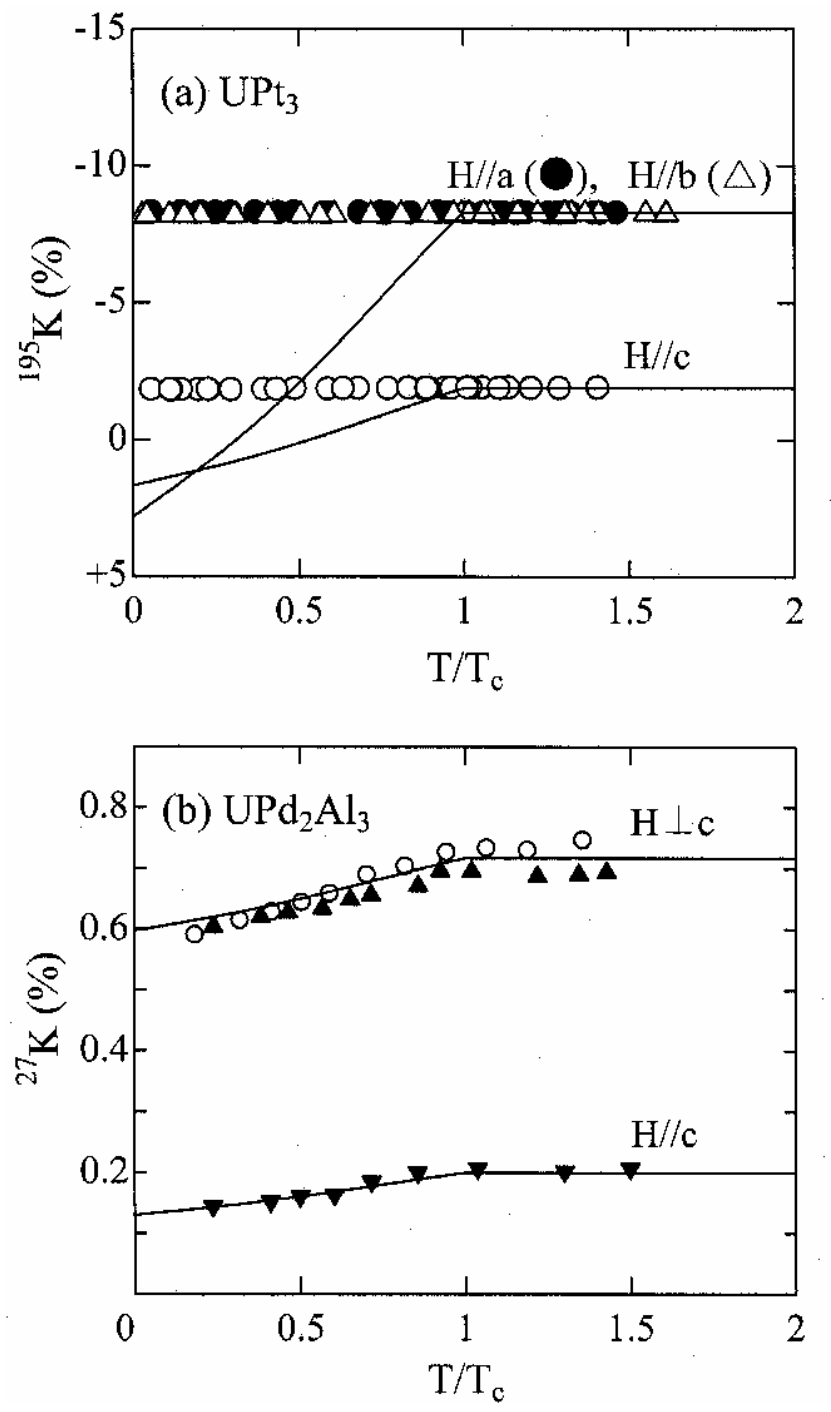

Figure 52 : T dependence of the Knight shift of $UPt_3$ and $UPd_2Al_3$ measured along the (a) and (c) axis. The solid lines in both Fig.s are calculations using the Sommerfeld coefficient γ for a d wave singlet model (Tou et al 2003).

In the complex $UPt_3$ material, a complete understanding of the different phases has not been achieved. The main puzzles are the link between SMAF which looks mainly like a slow fluctuating cluster above $T_C$ and the underlining SBF, the apparent necessity to revisit the strong spin orbit hypothesis or enhancement between magnetic responses and density of states. If the magnetic fluctuations are slow $(\approx 10^{-6}\,\text{sec})$ enough by comparison to the superconducting lifetime of a Cooper pair $\tau \sim \xi_0/v_F \sim 2.10^{-13}$ sec they can be efficient SBF (van Dijk et al 2002). However no calculation exists on their strength on $\delta T_C$ . NMR and neutron scattering time scales are $10^{-6}$ sec to $10^{-12}$ sec. In the Kondo cloud image, it is quite probable that the anisotropy linked to the slow dynamic is completely different from that suspected from instantaneous pictures. Traveling along a loop path $\ell_{KL}$ the quasiparticle feels the different atomic properties of U ant Pt atoms and thus during this orbital motion the initial strong spin orbit condition can be smeared out. An interesting observation is that two distinct and isotropic Knight shifts have been found by muon experiments for the field in the basal plane with drastic and opposite temperature dependence around $T_N$ . Unfortunately the threshold field for this two component magnetic response is yet not determined (Yaouanc et al 2000).

## *6.3 - $UPd_2Al_3$ : localized and itinerant f electrons. A magnetic exciton pairing.*

Extensive discussion can be found in the recent review of Thalmeier et al (2004) with emphasis on the dual model for U based systems (Yotsubashi et al 2001, Sato et al 2001, Zwicknagl and Fulde 2003) and on nodal superconductivity mediated by magnetic excitons which originates from crystal field transitions (see $PrOs_4Sb_{12}$ on chapter 4). The main difference with spin fluctuation mechanism is that they require a localized electron component.

In $UPd_2Al_3$ , a well defined AF ordering occurs at $T_N$ = 14 K with $M_o$ = 0.85 $\mu_B$ (Krimmel et al 1992) and a moderate Sommerfeld coefficient $\gamma$ = 145 $mJmole^{-1}K^{-2}$. At $T_C$ = 2K, the large specific heat jump $\frac{\Delta C}{\gamma T_c} \sim 1.2$ proves that the heavy fermion quasiparticules themselves condense in Cooper pairs (Geibel et al 1991). As high quality single crystals have been obtained, excellent macroscopic and microscopic experiments have been performed. De Haas van Alphen, experiments succeed to detect eight kinds of dHvA branches which have been obtained in band calculations based on different technics (Inada et al 1994, Knöpfle et al 1996). In a so called dual model, good agreement has been found by keeping two of the 2 electron localized on the U sites and another delocalised. The same approach seems also capable of explaining the $UPt_3$ Fermi surface (Zwicknagl and Fulde 2003).

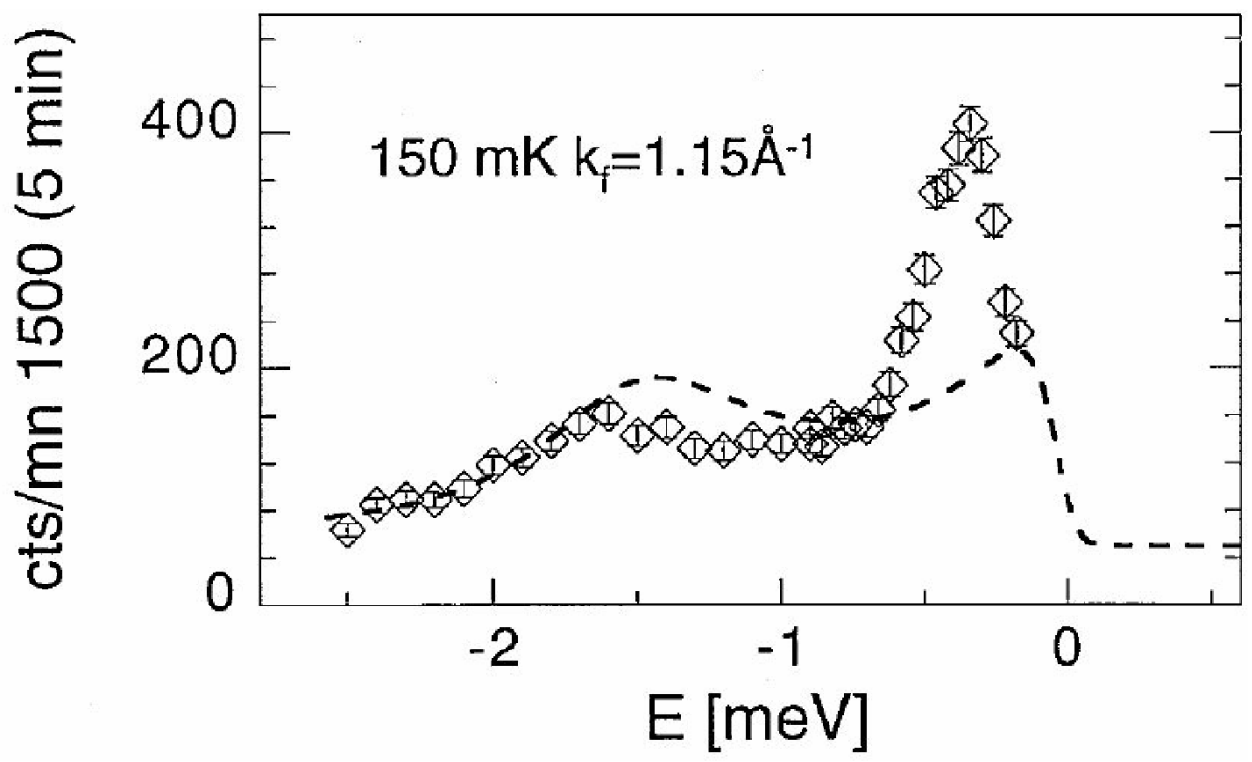

Figure 53 : Inelastic neutron data (◊) from $UPd_2Al_3$ at the AF wave vector at 150 mK. The dashed line is the extrapolated response for the normal antiferromagnetic state at this temperature (Bernhoeft et al 2000)

Concerning superconductivity, AF ordering may play a minor role as emphasized for conventional superconductivity. Observations of power laws notably in the nuclear relaxation time $T_1^{-1} \sim T^3$ suggests here also unconventional superconductivity with nodes (Kyogaku et al 1993). The even nature of the pairing is well established from the drop of the Knight shift (see figure 52) below $T_C$ as well as from the Pauli limitation of $Hc_2$ (T) for both directions.

As already pointed out, the unique feature of $UPd_2Al_3$ among HFC is its sharp signal in inelastic neutron experiments. Just above $T_C$ , two contributions appear in the dynamical susceptibility $\chi''(q, \omega)$ (Metoki et al 1998, Bernhoeft et al 1998). The first one can be regarded as damped spin waves or magnetic excitons due to the action of the intersite exchange on the CEF excitations. It softens and becomes overdamped as T approaches $T_N$. These modes can extend up to 10meV. But, for a wavevector along the c axis, they became sharp and well defined (see Hiess et al 2004). Of course heavy fermions exist with a quasielastic component but it is strongly peaked around the ordered wavevector. Crossing through $T_C$ leads to a drastic change with a resonance like structure of the low energy response (simultaneous inelasticity and enhancement of the signal) (figure 53). The feedback between magnetic excitations and the low energy component illustrates the coupling between the modes. It has been argued that the magnetic excitations may produce the attractive interaction of the itinerant electrons. Quantitative analysis of the spectrum by two different approaches (Bernhoeft 2000, Sato et al 2001) lead to the conclusion that an even parity unconventional gap must occur along the c axis :

$\Delta_{SC}$ (k) = $\Delta_0$ cosk$_z$c

where $\Delta_0$ is the amplitude of the gap function and c the lattice constant (A1g representation). A line of nodes will happen at the intersection of k = ± 0.5 $Q_0$ with the Fermi surface. Furthermore, the gap function as well as the inelastic spectrum can explain selfconsistently tunneling experiment results obtained along the c axis in a N-I-S junction performed on high quality $UPd_2Al_3$ film (figure 54) (Jourdan et al 1999, Huth and Jourdan 1999 - 2000). An anomaly in the tunnelling spectrum occurs at the energy of the mediated boson (magnetic excitations) which is comparable to the gap. The conductivity modulation at about 1.2 meV is roughly the energy (1.5 meV) of the magnetic excitation at the magnetic Bragg point (0, 0, 1/2) which is found in the normal phase. In a weak BCS limit, such a coincidence is not expected. In strong coupling, its reflets strong retardation effects which may be caused by a slow velocity of the magnetic excitons by comparison to the Fermi velocity of the heavy fermion quasiparticles (Sato et al 2001). In agreement with a A1g representation for the order parameter, there is no node of the gap function along the crystallographic c axis normal here to the surface of the $UPd_2Al_3$ film.

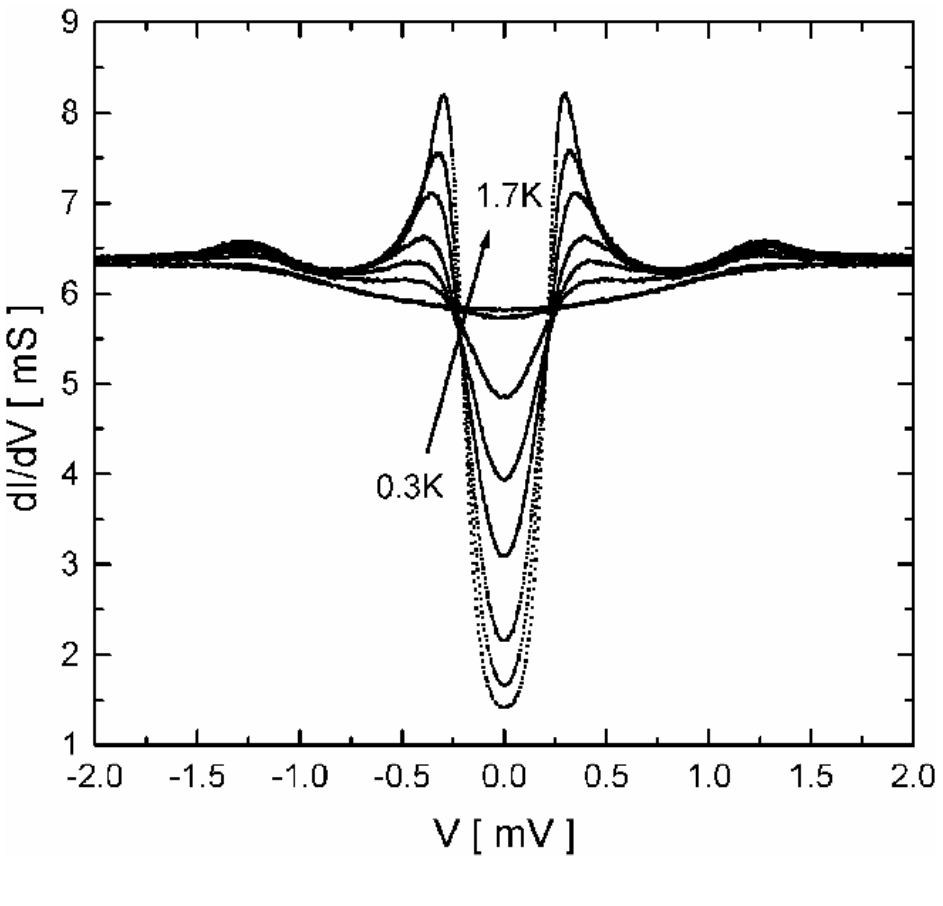

Figure 54 : Differential conductivity of a $UPd_2Al_3$ , $AlO_x$, Pb junction as a function of the temperature (0.3/0.5/0.7/0.9/1.1/1.3/1.5, 1.7) (Jourdan et al 1999)

$UPd_2Al_3$ appears as a model case where both the order parameter and the mechanism of superconductivity can be established. It is worthwhile to mention that a resonance spectrum below $T_C$ has been first observed for $YBa_2Cu_3O_7$ and is now considered as part of the evidence of $d_{x2-y2}$ neutron singlet pairing (Rossat Mignod et al 1992). By comparison to $UPd_2Al_3$ , the complete inelastic spectrum is far more complex. So the discussion on high $T_C$ mechanism is still controversial.

### 6.4 - $URu_2Si_2$ : from hidden order to large moment.

The new interest in the superconductivity of $URu_2Si_2$ is that its collapse at $P_S \sim 1.2$ GPa may be related to recent features discovered on the pressure stability of its magnetic order. The nature of the phase transition at $T_N = 17.5$ K was enigmatic (see Mydosh et al 2003) as the huge $\lambda$ anomaly in the specific heat suggests the onset of a magnetic transition associated with a large sublattice magnetization ($M_o \sim 1\ \mu_B$ / U atom) (Schlabitz et al 1986). However the tiny AF sublattice magnetization ($M_o \sim 0.03\ \mu_B$ / U atom) (Broholm et al 1987 – 1991) at the wave vector $Q_0 = (0, 0, 1)$ is too small to explain the amplitude of the specific heat anomaly (see chapter 2) in a static picture. A hidden order (HO) must be coupled to this weak AF component. Below $T_N$ , clear features of (HO) in inelastic neutron scattering experiments are two minima at $Q_0$ and $Q_1$ characterized by the respective low energy gaps $\omega_{Q_0} = 1.6$ meV and $\omega_{Q_1} = 4.5$ meV. Above $T_N = 17.5$ K, the excitation at $Q_0$ becomes quasielastic ($\Gamma = 1$ meV) at $T_N$ and vanishes already above 25 K while at $Q_1$ it stays inelastic but with a very strong damping ($\Gamma = 3$ meV) (see also Bourdarot). Taking into account this spin dynamic through $T_N$ , there is no more mystery on the specific heat anomaly but the nature of the order parameter is open.

Theoretical proposal (Chandra et al 2002) is that the hidden order corresponds to incommensurate orbital antiferromagnetism which may be due to circulating current between the uranium ions. Recently a specific search for hidden orbital currents by neutron scattering (Wiebe et al 2004) was unsuccessful. The microscopic signatures of the hidden order are the previous gap $\omega_0$ which explains basically the size of the specific heat anomaly at $T_N$ and an observation by NMR on Si of a field independent contribution to the linewidth (Bernal et al 2001). As we will see there are controversies on the intrinsic character of the tiny ordered moment $M_o \sim 0.02\ \mu_B$ . Restricting the problem to local order parameters of $5f^2$ shells, Kiss and Fazekas (2004) propose that the best candidate for HO may be a staggered order of octupoles. As we will see later, the duality between itinerant and localized states may lead to another issue.

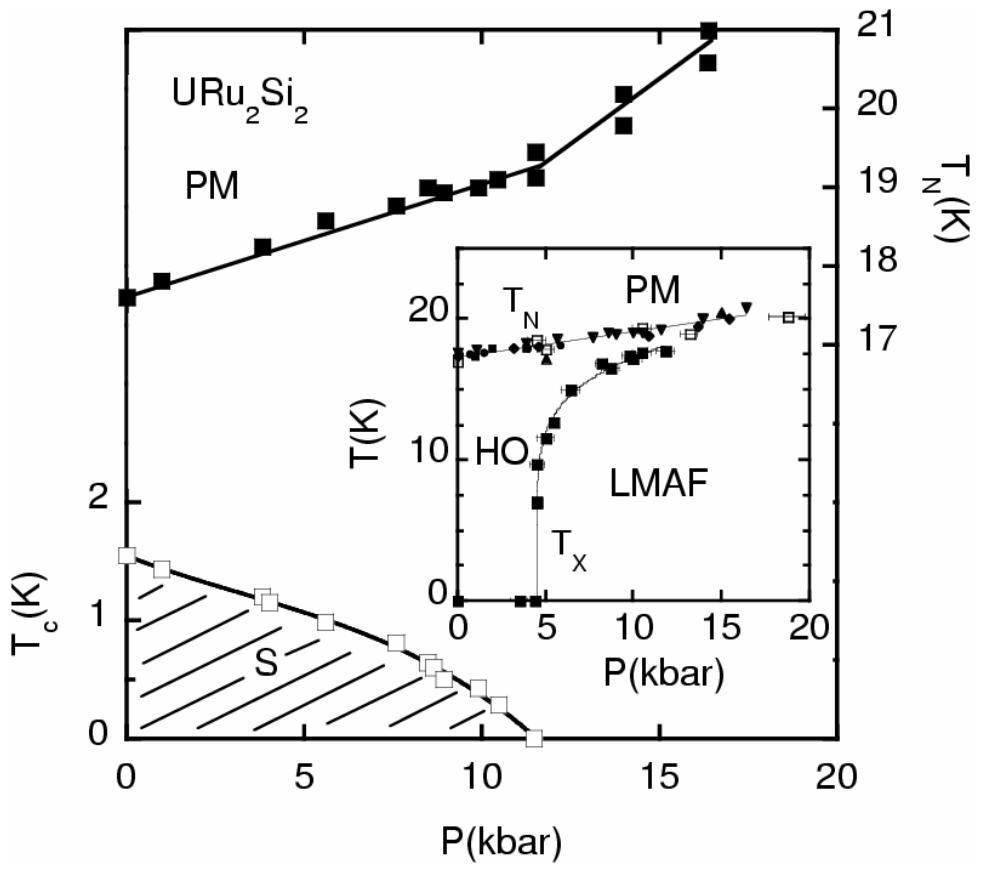

Figure 55 : (T, P) phase diagram of $URu_2Si_2$ with pressure variation of $T_N$ and $T_C$ (Schmidt 1993) measured by resistivity. $P_X$ is the first order transition between HO and LMAF phase at T = 0K (Motoyama et al 2003). In the insert, domain of existence ($T_X$) of HO state and LMAF detected by neutron scattering below 1.2 GPa (Bourdarot et al 2004)

The new highlight was the observation that the HO phase becomes unstable under pressure. At P = $P_X$ ~ 0.5 GPa, the long range order switches from "mysterious" HO to a "classical" LMAF (Amitsuka et al 1999). Furthermore using strain gauges to detect a volume discontinuity, Motoyama et al (2003) found that a low temperature the transition is first order. Later it was precised in neutron scattering experiment (Bourdarot et al 2003-b) that the line ($T_X$, $P_X$ ) of the phase transition approaches or ends up at ($T_N$ , P) for P = $P_{cr}$ ~ 1.2 GPa (figure 55). The change from HO to LMAF is accompanied by important modifications of the inelastic neutron spectrum : disappearance of the excitation $\omega_{Q_0}$ at $Q_0$, and the pressure increase of $\omega_{Q_1}$ at $Q_1$ which reaches 9.2 meV at 11 GPa. The issue on the dipole magnetic contribution at low pressure below $P_X$ is open.

The first order nature of the transition can be invoked to keep a few percentages of the LMAF phase even below $P_X$ . The main support of this possibility is that the NMR frequency of the Si site characteristic of a large ordered moment is constant whatever the pressure range but the fraction (f) of LMAF (i.e the intensity of the NMR signal) increases continuously with pressure before reaching 1 at $P_{cr}$ (figure 56) (Mastuda et al 2001, 2003). If the fraction f is a function of P and H, the system contains enough variables to explain NMR and neutron scattering.

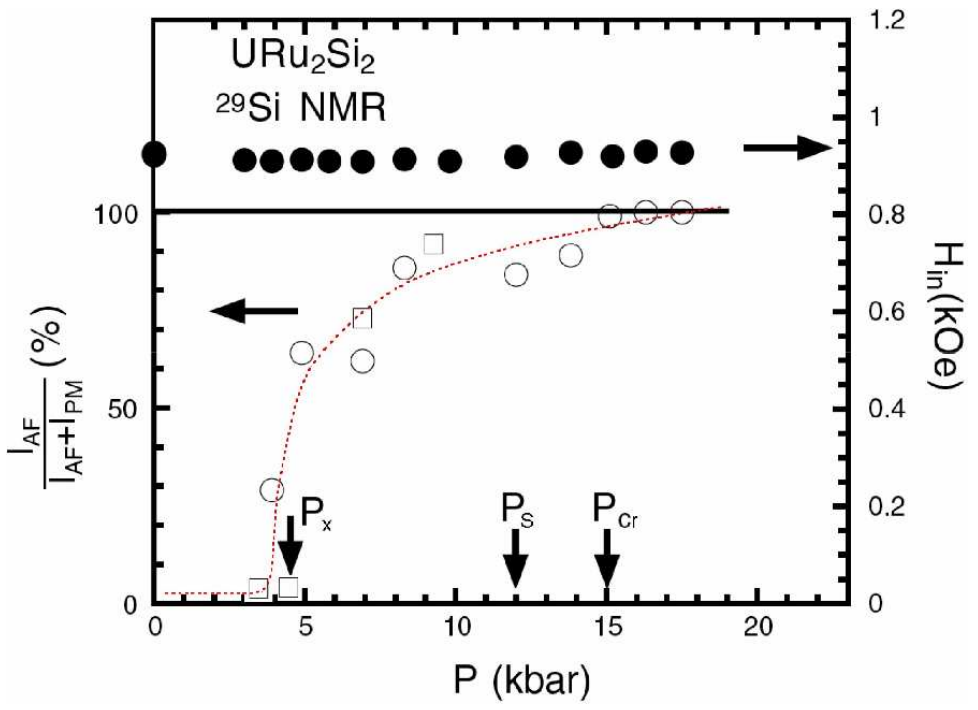

Figure 56 : The pressure dependence of internal field $H_{in}$ and the normalized intensity (O) of the $H_{in}$-split line $\frac{I_{AF}}{I_{AF}+I_{PM}}$ with phase separation between AF and Pa signal (Matsuda et al 2003). The full point represents the invariance of the NMR frequency on Si sites. The open squares the AF fraction estimated from muon experiments (Amato et al 2004). The full line is a guide for eyes showing the 100% LMAF. 100 % is achieved at $P_{cr}$ ~ 1.5 GPa not too far $P_S$. A deep increase in the fraction f of LMAF state occurs at $P_X$ .

In a dual model, Okuno and Miyake (1998), following the Ising Kondo lattice model of Sikkema (1996) have proposed that a spin density wave (SDW) occurs in the itinerant system due to partial nesting of the Fermi surface with the feedback to induce an extra weak ordered moment $M_o$ on the localized singlet levels added to the tiny moment m created by the SDW. The SDW is also associated with a charge gap ($\Delta_G$) in the electronic spectrum ($\Delta_G \sim k_B T_N$) (see recent development : Fomin 2004, Mineev and Zhitomirsky 2004). The hidden order can be defined by m and M or the charge gap $\Delta_G$ and the spin gap $\omega_0$. Unlike the other models, the SDW scenario is based on the coexistence of two orderings with the same AF dipole symmetry. It has been shown (Mineev and Zhitomirsky 2004) that the field dependences of the staggered magnetization and of the spin gap is in quantitative agreement with the experiments. In magnetic field, $\Delta_G$ decreases while $\omega_{Q_0}$ increases.

Neutron scattering experiments in magnetic field on a single crystal (Bourdarot et al 2003-a) suggest that the weak magnetic moment of the H.O phase is intrinsic. The reproductibility of the weak magnetic moment at P = 0 on single crystals pushes also for its intrinsic origin. An explanation for weak magnetism may be linked with the microscopic nature of the magnetism itself. The spin and orbital contribution can change with H and P. Band calculations point out that the tiny ordered moment at the uranium site may come from a cancellation between spin and orbital 5f moment respectively –0.75 $\mu_B$ and + 0.86 $\mu_B$ while the d component has a weak ( - 0.01 $\mu_B$ ) equal weight of spin and orbital part (Yamagami and Hamada 2000). It is worthwhile to remark that this d component has the magnitude of $M_o$ . In HFC, there is the underlining possibility that a singular behavior may be due to d-f correlations. The possibility of SDW was underlined first by Maple et al (1986) and Schoenes et al (1987) and confirmed recently by thermal transport (Bel et al 2004, Behnia et al 2005).

Phenomenological solution in a full homogeneous frame for the stability of two phases, is to assume the hidden order parameter $\psi$ ~ m coupled to the ordered moment $M_o$ with $M_o$ and $\psi$ of the same symmetry. The two states above and below $T_X$(P) are phases with reversed size of $\psi$ and M i.e large $\psi$ coupled with small $M_o$ or small $\psi$ coupled with large $M_o$ . The predicted first order line will end up at $P_{cr}$ , $T_{cr}$ (Bourdarot et al 2003-b, Mineev and Zhitomirsky 2004). In this scheme, the inhomogeneous effects detected in NMR or muon experiments may be due to supplementary specific experimental conditions (for example powdering for NMR, supercooling or superheating at the first order transition $T_X$). A careful verification must be performed now to demonstrate if $T_{cr}$ ends up on the $T_N$ line.

Let us stress that $URu_2Si_2$ has rich multiple magnetic field phases which may be the combination of a quantum critical field (basically like found in $CeRu_2Si_2$ at $H_M$ ) and magnetic phase diagrams (Kim et al 2003, Harrison et al 2003). It was proposed that a reentrant HO phase is created in the vicinity of a quantum critical end point for H = 35 T. We will not go further in the complex interplay. The pressure evolution of these rich (H, T) phases will give some keys in the understanding of the H.O phase.

As already emphasized, $T_C$ drops rapidly under pressure as it vanishes near 12 kbar. Thus $T_C$ collapses at the pressure $P_S$ ~ $P_{cr}$ where the magnetic order switches from H.O to LMAF (figure 55). The apparent survival of superconductivity observed by resistivity up to P ~ $P_C$ may be a consequence of the slow disappearance of the H.O fraction and the smooth increase of the LMAF fraction. So one may suspect that superconductivity is associated only with the H.O phase. The bulk nature of superconductivity may disappear at $P_X$ .New experiments on superconductivity will certainly clarify the nature of the normal phase and notably will infirm or not an intrinsic phase separation between H.O and LMAF order. By comparison to $UPt_3$ and $UPd_2Al_3$ , the superconductivity of $URu_2Si_2$ has been far less studied due to the difficulty to obtain high quality crystals. For example, the superconducting transition measured by specific heat does not coincide with the resistive transition and a strong broadening occurs in magnetic fields for the determination of $Hc_2$ (T) (Brison et al 1995). The striking result is the quasi absence of Pauli limitation along the a axis which suggests an odd parity pairing with d vector along the c axis. Careful studies of the superconducting transition by specific heat rule out the occurrence of a double transition (Hasselbach et al 1991, Ramirez et al 1991). At least, the SMAF, with the propagation vector ($Q_0$ = (0, 0, 1)) which preserves the lattice symmetry, cannot play the role of SBF.

Quantum oscillations (dHvA, Haas Shunikov) (Bergmann et al 1997, Ohkuni et al 1997, 1999) were successfully observed in the normal phase. In agreement with a gap opening at $T_N$, few orbits have been detected. One may expect that the Fermi surface will react to the change of magnetic ground state at $P_X$ . But only slight increases in the dHvA frequency occur up to 2 GPa (Nakashima et al 2003). For the spherical ∝ branch, the effective mass changes drastically going from $m^* \sim 17\ m_o$ at P = 0 kbar to $m^* = 8\ m_o$ at P = 18 kbar in agreement with a drop of the A coefficient ($A \sim m^{*2}$) roughly from 0.12 µΩcm $k^{-2}$ at P = 0 to 0.03 µΩcm $k^{-2}$ at P = 20 kbar (Schmidt 1993). This dependence may be related with the collapse of $T_C$ .

The interesting new ingredient is that $URu_2Si_2$, in its AF phase, can be classified as a compensated semi metal with a total carrier content near 0.07 per formula unit (Yamagami and Hamada 2000). Thus per carrier, the effective mass may be comparable to that of $UPt_3$ (Maple et al 1986, Schoenes et al 1987, Bel et al 2004). Another example of semi metal will be $UBe_{13}$. the drastic difference between the two cases is that, in $URu_2Si_2$, the Fermi liquid regime seems well established above $T_C$ since the onset of long range ordering occur at high temperature ($\frac{T_N}{T_C}$ ~10), while, in $UBe_{13}$, the effective Fermi temperature is comparable to $T_C$ and no long range ordering is detected above $T_C$ . However in high magnetic field, the feedback of the magnetostriction on the carrier number may be an important parameter.

$URu_2Si_2$ is an illustrating example on the variation of $T_N$ or $T_C$ by the application of a uniaxial strain $\sigma$ along a and c axis or hydrostatic pressure P. According to the Ehrenfest relation, the $\sigma$ variation of the critical temperature is linked to the corresponding jump of the longitudinal thermal expansion ($\Delta\alpha_i$) and of the specific heat jump ($\Delta C$) by the relation :

$$\frac{\partial T_C}{\partial \sigma_i} = \frac{V_m \Delta\alpha}{\Delta C / T_c}$$

For $URu_2Si_2$ the corresponding variations are (van Dijk et al 1995, de Visser et al 1986 and Guillaume 1996)

$$\frac{\partial T_N}{\partial \sigma_a} = +\ 900 mK/GPa \qquad \frac{\partial T_N}{\partial \sigma_c} = -\ 410 mK/GPa$$

$$\frac{\partial T_C}{\partial \sigma_a} = -620 mK/Gpa \quad \frac{\partial T_C}{\partial \sigma_c} = +\ 430 mK/GPa$$

An uniaxial strain along a increases $T_N$ and decreases $T_C$ . Along c, $\sigma_c$ will decrease $T_N$ and increase $T_C$ . The high σ sensitivity of HFC puts of course experimental constraints on the hydrostatic P conditions and emphasizes the possibility of filamentary or induced phenomena near dislocations or stacking faults where pressure gradient near kbar can occur over few atomic distances. The relevance of an excellent hydrostaticity was already reported for $CePd_2Si_2$ where already it was known from thermal expansion experiments that $\sigma_a$ and $\sigma_c$ have opposite effect on $T_N$ (van Dijk et al 1995). The possibility of a mismatch inside $CeIrIn_5$ was also underlined ; recent experiments show that strong opposite uniaxial effects occur on $T_C$ (Oeschler et al 2003-a).

## *6.5 – The $UBe_{13}$ enigm : the low density carrier ?*

As pointed out , at P = 0 superconductivity appears in $UBe_{13}$ at $T_C$ from a paramagnetic metallic phase far above the temperature of the Fermi liquid regime (Ott et al 1983). Similar situations may occur in $CeCoIn_5$, $PuCoGa_5$ or $CeCu_2Si_2$ at P = 0. The link of superconductivity of $UBe_{13}$ with its heavy quasiparticles (suggested early on that of $CeCu_2Si_2$) may not be a strange exotic case but another example of a new general class of unconventional superconductors.

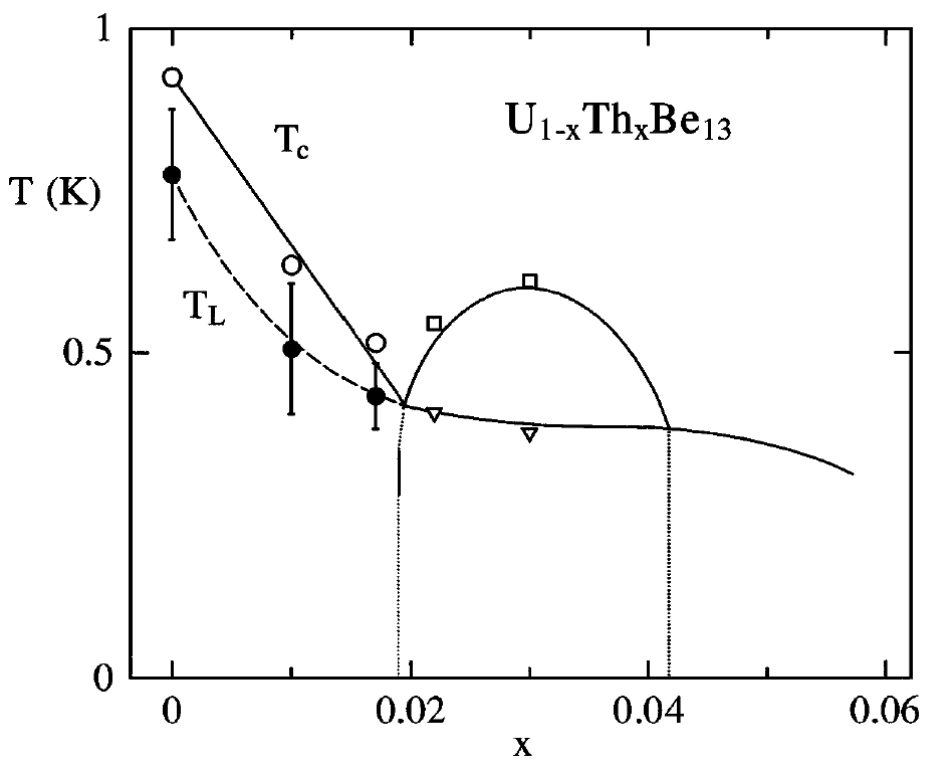

Figure 57 : Schematic T-x phase diagram of $U_{1-x}Th_xBe_{13}$ . Full lines represent phase transitions after the recent work of Schreiner el al 1998, while the broken line $T_L$ (x) corresponds to a crossover reported in Kromer et al 2000.

The interest on $UBe_{13}$ was reinforced when it was shown that under doping with Th ($U_{1-x}Th_xBe_{13}$ alloys) (Ott et al 1985), two successive phase transitions occur at $T_{C_1}$ and $T_{C_2}$ in the concentration range $0.020 = x_1 < x < 0.042 = x_2$ as shown in figure (57). The transition at higher temperature ($T_{C_1}$ for $x_1 < x < x_2$) corresponds to a superconducting transition. It was questionned if the second transition truly coincides with a phase transition to another superconducting order parameter since a magnetic component ($M_o = 0.01\ \mu_B$ ) has been discovered by µSR below $T_{C_2}$ (Heffner et al 1987). The second phase transition may indicate a long range magnetic ordering or the entrance in a new superconducting phase. Most of the proposals assume a change in the superconducting order parameter. For example, an initial even d wave A phase ($x < x_1$) strongly suppressed by doping would be supplemented by a s wave at $T_{C_1}$ and then a s + id state at $T_{C_2}$ (Kumar and Wölfle 1987). The cascade may also be from a single odd component to a multicomponent odd order parameter on cooling (Sigrist and Rice 1989). Recently a theory called of "ferrisuperconductivity" (Martisovits et al 2000) has been proposed with coherent pair motion and incoherent quasiparticles. In this approach a yet unobserved charge density wave (CDW) is predicted below $T_{C_2}$.

Experimentally, the $U_{1-x}Th_xBe_{13}$ phase diagram has been re-examined by macroscopic measurements (Oeschler et al 2003-b). At low temperature, the study of residual $\gamma T$ term in the specific heat (which is sensitive to the resonant defect scattering) shows structures at $x_1$ and $x_2$ .Thus it gives evidence of changes in unconventional pairing at these boundaries (Schreiner et al 1999). Simultaneous specific heat and thermal expansion measurements on pure $UBe_{13}$ point out a crossover temperature $T_L$ below $T_C$ which is a precursor of $T_{C_2}$. As its field response at $H_L$ differs from $Hc_2$ , it has been suggested that $T_L$ is linked to magnetic correlations.

Among the four archetype superconductors, $UBe_{13}$ was the only one where AF correlations were not detected rapidly. Experiments on polycrystals show two quasielastic q independent responses, one broad and one narrow, with respective Γ/2 half linewidth of 13 meV and 1.5 meV (Goldman et al 1986, Lander et al 1992). New measurements on $UBe_{13}$ ($T_C$ = 0.9 K) and $U_{1-x}Th_xBe_{13}$ (x = 3.5%, $T_{C_1}$ = 0.55 K, $T_{C_2}$ = 0.4K) single crystals allow to detect longitudinal AF fluctuations with the wave vector (1/2, 1/2, 0) ; for both cases, the energy window extends near 1-2 meV while the correlation length is quite short. The coupling is restricted to the next nearest neighbour uranium ions (Coad et al 2000, Hiess et al 2002). Above 20 K, this AF magnetic fluctuation disappears. The search for CDW for x = .35% below $T_{C_2}$ was unsuccessful despite the great neutron diffraction sensitivity to any movement of the Be cages. The experimental limit of 0.003 Å for the displacement is smaller than that (0.10 Å) proposed in the theory of Martisovits et al (2000). Furthermore, no elastic AF magnetic diffraction line was detected below $T_{C_2}$ i.e any ordered moment will be lower than 0.025 $\mu_B$ .

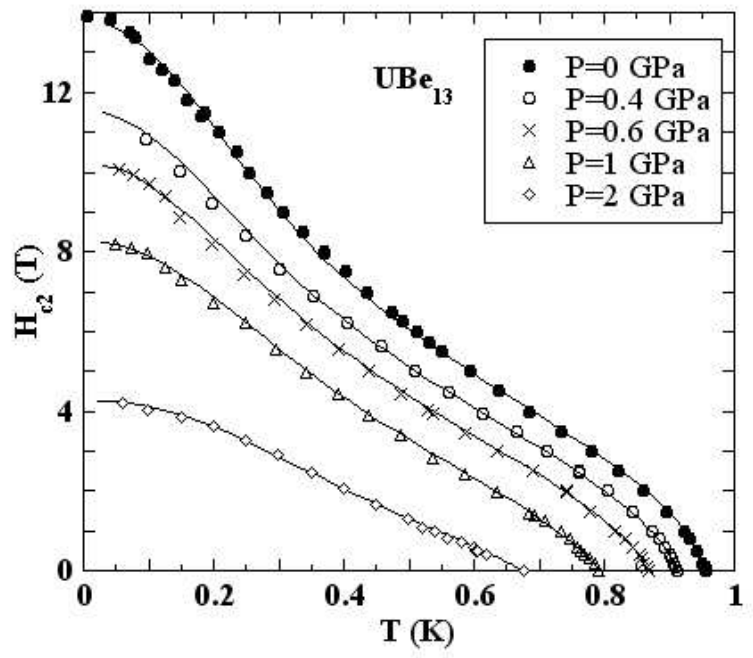

Figure 58 : $UBe_{13}$ : the upper critical field at different pressure P from Glémot et al 1999

Another striking effect in $UBe_{13}$ at P = 0 is the unusual temperature dependence of $Hc_2(T)$ with a huge initial slope $\frac{\partial Hc_2}{\partial T} \approx -\ 55\ T/K$, with a strong negative curvature and a convex shape at intermediate temperature (figure 58) (Thomas et al 1995). The temperature and pressure dependence of $Hc_2$ has been described in a simple strong coupling model for even superconductivity (Glemot et al 1999, Brison et al 2000). The conflict with the paramagnetic limit at T = 0 seems resolved by its enhancement due both to direct strong coupling effect (increase of the ratio of the gap energy by $k_B T_C$ ) and to the formation of a spatially modulated superconducting FFLO state favoured by the dominance of the paramagnetic limit. The values of $\lambda \sim 15$ at P = 0 are difficult to justify and far above the usual value ($\lambda \sim 1.4$) found in other HFC where similar fits have been performed ($CeCu_2Si_2$, $CeIn_3$ and $CePd_2Si_2$). At least the derived relative pressure dependence of $\lambda \sim m^*/m$ agrees well with the pressure dependence of the effective mass obtained from the specific heat or the slope $\frac{\partial Hc_2}{\partial T}$ . The possibility of a FFLO state comes only from the unusual temperature dependence of $Hc_2$ (T). Contrary to $CeCoIn_5$ there is no confirmation of FFLO states by other technics.

Table 7

Pressure dependence of the parameters used in the fit of $_{HC2}$ (T) for $UBe_{13}$.

(Glémot et al 1999)

| P GPa | $\lambda$ | $(\partial H_{C2}/\partial T)_{Tc}$ in T/K | $T_{FFLO}/T_C$ |
|---|---|---|---|
| 0 | 15 | - 55 | 0.45 |
| 0.4 | 13 | - 42 | 0.42 |
| 0.6 | 12 | - 32 | 0.37 |
| 1.0 | 11 | - 21 | 0.26 |
| 2.0 | 6.5 | - 8.5 | 0.10 |

It has been proposed that an alternative route is a model with a field induced mixture of two odd parity irreducible representations (a mixture of $A_{1u}$ and $E_u$) (Fomin et Brison 2002). The agreement is less satisfactory than in the even pairing case with strong coupling but inclusion of other effects can correct the discrepancies : mixing with other odd representations, mixture between odd and even representations and introduction of strong coupling. In this model, the second phase ($E_u$ ) does not appear in zero field at zero pressure ; the magnetic field introduces the mixtures of the representation. Here one may expect a pressure change in the pressure variation of $T_C$ depending on wether a $A_{1u}$ or $E_u$ phase is achieved. Extrapolation suggests that the $E_u$ representation will appear first in zero field above 30 kbar. As the $UBe_{13}$ situation appears unique, a singular point may occur for this system.

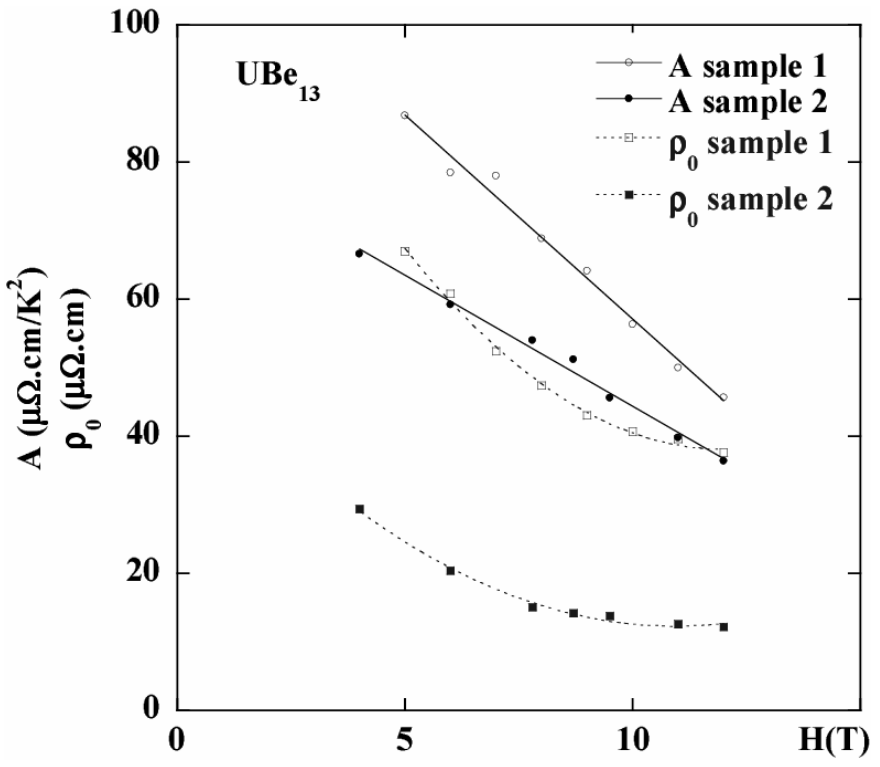

Figure 59 : Field variation of $\rho_0(H)$ and $A_H T^2$ of $UBe_{13}$ at P = 0 (Brison et al 1989).

One paradox is that γ and $\chi = \dfrac{\partial M}{\partial H}$ both have a weak H dependent. So one expects that the A coefficient of the $T^2$ resistivity law should also have a weak H variation (see below $CeRu_2Si_2$ and $UPt_3$ ). As shown figure 59, it is not the case (Brison 1989). Both A and $\rho_0$ decrease strongly with field, however in a first approximation $\rho_0/A$ are weakly H dependent. The main H effect is a change in the carrier number. It looks as though the carrier is released by H.

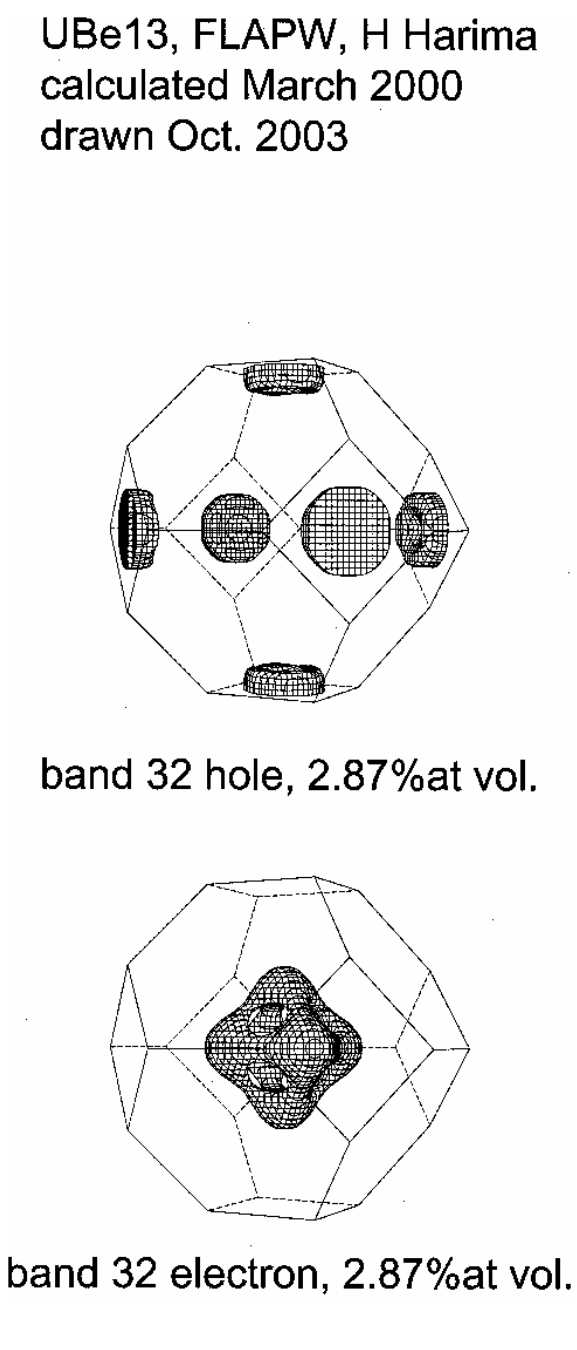

Figure 60 : Fermi surface of $UBe_{13}$ (Harima 2004, private communication)

It was already suggested that, in $UBe_{13}$, the carrier number may be low ( Takegahara et al 1986, Norman et al 1987, Brison et al 1989). Recent band calculations (Takegahara and Harima et al 2000) assuming the 5f electron itinerant indicates that the FS is remarkably simple (figure 60) with only $n_h$ = 2.87% holes and $n_e$ = 2.87 % electrons per atomic volume in this compensated metal. So $UBe_{13}$ fulfills the condition of super dense Kondo lattice with a deficiency of charge carriers per atomic site. In this superdense Kondo lattice, simple arguments with rigid normalized density of states will never work. As $\rho_0 \sim (n_e + n_h)^{-1}$ decreases with H for a given mean free path, the magnetic field seems to act as a carrier pump. This pumping may stop near the field called $H_L$ in the Dresden experiment (Oeschler et al 2003-b).

The magnetostriction induces a crystal change from cubic to tetragonal (van Dijk et al 1994). The deformation parallel to the field is opposite to that in the plane perpendicular to the field. Again here weak effects in ordinary metals will be magnified by the huge Grüneisen parameter. Tiny effects with a change of symmetry are reinforced by the exceptional condition of a semimetal. In this picture, at each field H corresponds a carrier concentration n (H) and thus an extrapolated value $T_C$ (n(H), H = 0). The field variation of carriers occurs mainly up to $H_L$ ~ 6T above which $UBe_{13}$ becomes a normal metal. In doping with $ThBe_{13}$ the striking point in this "composite" system is that the double transition occurs basically near 3% i.e roughly for the concentration where the number of charge doubles by comparison to the pure compound $UBe_{13}$ . $ThBe_{13}$ is a nice metal with a large number of carriers $n_e$ ~ 1 per atomic volume (see Harisson et al 1986). Under pressure, as the wavefunctions overlap more, the number of carrier will increase. The H and P effects are nicely evident in the 3d plot of $\rho_0$ (P, H) (by Aronson et al 1990). For example for P > 20 kbar, the monotonous variation of $\rho_0$ (H) suggests that the field pump of the carrier is no longer efficient, P electron pump is a better tool.

By contrast to $UBe_{13}$, in $CeRu_2Si_2$ and $UPt_3$ $\gamma$, $\chi$ and $\sqrt{A}$ are field insensitive below their pseudometamagnetic field $H_M$ respectively equal to 7.8 and 21 T while $\rho_0$ (H) has a strong positive linear residual magnetoresistivity :

$$\rho_0 (H) = \rho_0 + cte\ H$$

at least extrapolated from T > $T_C$ (0) ~ 0.55K for $UPt_3$ (Taillefer et al 1988). Below $T_C$ (0), in $UPt_3$, a different regime occurs as the quasiparticle enters in a collisionless regime ($\omega_c \tau \sim 1$ $\omega_c$ = cyclotron frequency, $\tau$ relaxation time). The appealing idea is that the field behavior in the collision regime ($T > T_C$ (0)) may be a specific reaction of the lattice with the creation of a longitudinal orbital voltage antagonist to the normal current flow. If we assume as for a normal metal that the magnetoresistivity is a function of the product of the cyclotron frequency ($\omega_C$ ~ H) by the collision time $\tau$ : $\Delta\rho_0(H)/\rho_0 \approx \omega_C \tau$. $\Delta\rho$ becomes of course independent of $\rho_0$ : $\Delta\rho_0$ (H) ~ $R_L H$. The question of a collective field response (here $R_L H$) of the heavy particle is open. A classical interpretation of the H linear residual magnetoresistivity was given by Ohkawa (1990) ; the field reveals the local disorder of the ligand which induces a distribution in the Kondo temperature. At H = 0 all Kondo centers will be equivalent (unitary limit). Our message is that a deep look on simple transport data may lead to unexpected insights.

# 7/ Conclusion and perspectives

The main trends are :

**Chapter 1** : Important progresses have been made in the self consistent treatment of the correlations including the feedback on the crystal field. At least, basic questions emerge notably the consequences on the localized-itinerant duality of 4f or 5f electrons. At T = 0K, our attempt to classify HFC by three pressures $P_{KL}$ , $P_C$ and $P_V$ (respective P switch from localized to itinerant 4f description, from AF to PM and to valence or large orbital fluctuation) is of course a simplification extrapolating the phase diagram of Ce metal to a first order transition down to 0 K. It is also possible to consider that another critical point ($P_{KL}$ , $T_{KL}$) will appear at low temperature.

**Chapter 2** : Careful experiments have been realized on normal phase properties ($CeRu_2Si_2$ , $CeCu_6$ , $CeNi_2Ge_2$, $YbRh_2Si_2$ ). An interesting point is when the itinerant picture of the 4f Yb electron will be recovered. It may happen that very low temperatures need to be achieved as the Fermi liquid regime below $T_I$ may depend on $P – P_C$ but also on a renormalization parameter as $\Delta/k_BT_K$ . The key question is the location of $P_{KL}$ by respect to $P_C$ and the coincidence or not of $P_C$ from $P_V$. The increase in competitive studies by quantum oscillations will boost soon the understanding. The effects of the magnetic field are rich as it acts on the intersite correlations, on the Kondo effect but also on the mixing or decoupling of the crystal field level.

**Chapter 3** : The interplay of spin dynamics and density fluctuations are strong in HFS due to the huge electronic Grüneisen parameter. The weight of each contribution in the Cooper pairing depends on the relative position of $P_{+S}$, $P_{-S}$ for the onset and disappearance of superconductivity by respect to $P_C$ and $P_V$. Of course, other sources of pairing as magnetic excitons can occur.

**Chapter 4** : There are already excellent basis of HFS where AF fluctuations play a key role : $CeIn_3$, $CePd_2Si_2$, $CeRh_2Si_2$. The canonical case of $CeCu_2Si_2$ seems now fully understood with the direct observation of the magnetic structure at P = 0 and also with a clear appearance under pressure of two different superconducting domains. The vitality of HFC thema is continuous reactivated by the discovery of new materials : few years ago the 115 Ce HFC, recently $CePt_3Si_2$ and $PrOs_4Sb_{12}$. That pushes to elucidate the relevance of the dimensionality, of the crystal symmetry and of combined effects of nesting and multipolar orderings.

**Chapter 5** : The growth of excellent crystals of $UGe_2$ and recently of URhGe has led to the discovery of ferromagnetic superconductors. Triplet superconductivity is suspected to be crucial for the coexistence as it breaks the conventional wisdom from s wave superconductors that ferromagnetic and superconductivity are antagonist. In this recent subject, new windows appear as a careful study of superconductivity through the first order pressure $P_X$ from $FM_2$ to $FM_1$ in $UGe_2$, a fine analysis of reentrant superconductivity in URhGe including the magnetostriction effect , the resolution of $ZrZn_2$ mysterious superconductivity and the hope for new examples.

**Chapitre 6** : Despite a large activity on $UPt_3$ there is still mysteries concerning the symmetry breaking field of the multidimensional order parameter i.e basically what is the origin of the weak antiferromagnetic component. For $URu_2Si_2$ , the link between the low pressure hidden order phase and superconductivity will be certainly confirmed soon. For $UPd_2Al_3$ after the nice results obtained by tunnelling and neutron scattering experiments, a new generation of experiments will be designed to precise the interplay between magnetic nodes (spin wave or exiton) and the Cooper pair. Finaly for $UBe_{13}$ progresses in the high pressure technics will allow to follow the change in strong coupling conditions for superconductivity, the restoration of Fermi liquid properties and the possible emergence of long range antiferromagnetism.

Here we have focus mainly on paramagnetic normal phase properties and their link with superconductivity. A missing important domain concerns a careful analysis of the magnetic or multipolar order parameter as well as their excitations. A highly documented case is CeB6 with its two successive quadrupolar and dipolar ordering (see Shina et al 1977). The $URu_2Si_2$ discussion on hidden order parameter gives at least an idea on the large variety of different possibilities. The new skutterudite family opens a wide domain.

To precise the characteristic lengths involved in HFC physics is important. Up to now, almost no direct derivation has been given on the magnetic correlation length close to $P_C$ . Indirect observations point out that weakly first order transition may be often achieved close to $P_C$ . The generality of HFC in condensed matter is that the forces are atomic (Å) the correlation are often nanometric (NA) (Kondo length, superconducting coherence length, magnetic correlation). A mean electronic free path of a tenth of micron can be achieved, the nice achievement of the clean limit condition i.e physics governed by the correlations is realized. Furthermore, the experimentalists have made major progresses in the handling of micrometric (MI) crystals. In this A.NA.MI process, low cost investment have revealed fascinating and unexpected horizons.

Developments coincide often with the discovery of new materials, the mastery in the microhandling, the increase in the sensitivity but also in the fiability of the measurements (appearance of cantilever for quantum oscillation, P tuning in situ) and also the combination of different probes (see the realization of excellent P or H experiments by macroscopic technics but also microscopic ones : quantum oscillations, NMR, neutron scattering and synchrotron radiation). Of course direct imaging by tunnelling technics will be a tremendous progress. Young physicist will certainly discover unexpected facets of heavy fermion matter.

I like to express my gratitude to J. Friedel, R. Tournier and F. Holtzberg. K. Behnia and K. Miyake suggest many improvements of the manuscript. J.P. Brison, P. Haen and G. Knebel help me by their comments but also by using different results even yet unpublished. I thank F. Bourdarot, D. Braithwaite, M. Continentino, B. Fak, D. Jaccard, J. Thompson, A. Huxley, Ph. Niklowitz, F. Lapierre, P. Lejay, G. Lapertot, S. Raymond, L.P. Regnault, H. Suderow, J.L. Tholence, J.P. Sanchez for their suggestions. I have greatly enjoyed the hospitality of Berkeley (R.A. Fischer, N. Phillips), Cambridge (G. Lonzarich, S.S. Saxena), Osaka (K. Amaya, K. Asayama, K. Miyake, Y. Miyako, Y. Kitaoka), Tokkai (S. Kambe, Y. Haga, D. Aoki), Toyama (J. Sakurai), Geneva (J. Sierro, D. Jaccard), Moscow (A. Buzdin, I. Fomin, Y. Gaidukov, V. Mineev, G. Volovik) and IBM Yorktown Heights (F. Holtzberg and S. von Molnar). The new PhD students R. Bel, R. Boursier, J. Derr, F. Hardy, A. Holmes, S. Kawasaki, W. Knafo, M.A. Measson, give a strong stimulation to break often the usual consensus. I thank also the already young researchers and previous PhD students A. Demuer, P. Rodière and I. Sheikin. My participation to the recent workshops organized in the Lorentz Center of Leiden by G. Stewart, A. de Visser and Q. Si, and in Santa Fe by D. Pines, J. Thompson and J. Sarrao (ICAM) was very stimulating as well as meetings from Center of Excellence in Osaka University and Tokyo Metropolitan University. Inside Europe, heavy fermion matter has profited from the FERLIN network of the European Science Fundation (ESF). M. Perrier has transformed my old hand fashion manuscript to the present document. J.P. Brison and G. Knebel has played a major role for the improvement of the figures.

A first version of the article was send to the editor in March 2004. After the referee report in July 2004, the new version was corrected during my stay in Tokyo Metropolitan University in November 2004. I thank my host, Professor H. Sato and his collaborators, (Pr. Y. Aoki, M. Khogi, O. Sakai and Dr. K. Kuwahara) for stimulating discussions. Prof. H. Harima was an excellent teacher in the new subject of skutterudite. This review will be part of the new issue of the serie "Progress in low temperature physics" edited by W.P. Halperin for Elsevier (2005).

| | | | | |
|---|---|---|---|---|
| Abragam A. and Bleaney . | Electron paramagnetic resonance of transition ions. Clarendon Press, Oxford | | | (1970) |
| Abrahams E. et al | Phys. Rev. B | 52 | 1271 | (1995) |
| Abrikosov A.A. | J. Phys. Condens. Matter | 13 | L943 | (2002) |
| Abrikosov A.A. | Fundamentals of the theory of metals (North Holland) | | | (1988) |
| Abrikosov A.A. and Buzdin A.I | Pis'mev Zh. Eksp.& Teor Fiz | 47 | 204 | (1988) |
| Abrikosov A.A. and Gorkov L.P. | Sov. Phys. JETP | 12 | 1243 | (1961) |
| Adenwalla S. et al | Phys. Rev. Lett. | 65 | 2298 | (1990) |
| Aeppli G and Fisk Z. | Comments Condens. Matter Phys. | 16 | 155 | (1992) |
| Aeppli G. et al | Phys. Rev. Lett. | 57 | 122 | (1986) |
| Aeppli G. et al | Phys. Rev. Lett. | 60 | 615 | (1988-a) |
| Aeppli G. et al | J. Magn. Magn. Mater | 76-77 | 385 | (1988 b) |
| Aliev F.G. et al | Some modern aspects of the physics of strongly correlated electron systems. Institute Universitario de Cienca de Materials Nicolas Cabrera | | | (2001) |
| Allen J.W. | Solid State Commun. | 123 | 469 | (2002) |
| Allen J.W. and Martin R.M. | Phys. Rev. Lett. | 49 | 1106 | (1982) |
| Amadou B. et al | To be published | | | (2005) |
| Amato A. | J. Phys. Cond. Mat. | 16 | S4403 | (2004) |
| Amato A. et al | J. Low Temp. Phys. | 68 | 371 | (1987) |
| Amato A. et al | J. Low Temp. Phys. | 77 | 195 | (1989) |
| Amato A. et al | Physica B | 186-188 | 276 | (1993) |
| Amato et al | Rev. Mod. Phys. | 69 | 1119 | (1997) |
| Ambegaokar V. and Mermin N.D. | Phys. Rev. Lett | 30 | 81 | (1973) |
| Ambrumenil N. and Fulde P. | J. Magn. Magn. Mat. | 47-48 | 1 | (1985) |
| Amitsuka H. et al | Phys. Rev. Lett. | 83 | 5114 | (1999) |
| Anderson P.W. and Brinkman W.F. | Phys. Rev. Lett. | 30 | 170 | (1973) |
| Anderson P.W. | Problème à N corps – Many body Physics. Ed. C. Hewitt and R. Balian, Gordon and Breach | | | (1967) |
| Anderson P.W. and Brinkman W.F. | The helium liquids, Academic Press edited by G.M. Armitage and I.E. Farguhar, Academic Press | | 315-416 | (1975) |
| Anderson P.W. et al | Phys. Rev. | B30 | 4000 | (1984) |
| Andres K et al | Phys. Rev. Lett. | 43 | 1779 | (1975) |
| Aoki D. et al | Nature | 413 | 613 | (2001) |
| Aoki H. et al | Physica B | 206-207 | 26 | (1995) |
| Aoki H. et al | J. Phys. Soc. Japan | 62 | 3157 | (1993) |
| Aoki H. et al | J. Phys. Condens. Mat. | 36 | L13 | (2004) |
| Aoki Y. et al | J. Magn. Magn. Mat. | 177-181 | 271 | (1998) |
| Aoki Y. et al | J. Phys. Soc. Jpn | 74 | 209 | (2005) |
| Aoki Y. et al | Phys. Rev. Lett. | 91 | 067003 | (2003) |
| Araki S. et al | J. Phys. Condens. Matter | 14 | 377 | (2002) |
| Aronson M.C. et al | Physica B | 163 | 515 | (1990) |
| Asayama K. | Nuclear magnetic resonance in electron system, Shokabo, Tokyo | | | (2002) |
| Aschcroft N.W. and Mermin N.D. | Solid State Physics, Harcourt College Publishers | | | (1976) |
| Baber W.G. | Proc. Roy. Soc. A | 158 | 3838 | (1937) |
| Barbara J. et al | J. Appl. Phys | 50 | 2300 | (1979) |
| Barla A. et al | Phys. Rev. Lett. | 92 | 066401 | (2004) |
| Barla A. et al | To be published J. Phys. Condens. Matter | | 0406166 | (2005) |
| Bauer E. et al | Phys. Rev. Lett. | 92 | 027003 | (2004) |
| Bauer E. et al | Phys. Rev. Lett. | 92 | 027003 | (2004 |
| Bauer E. et al | Physica B to be published | | | (2005) |
| Bauer E.D. et al | J. Phys. Condens. Matter | 13 | L759 | (2001) |
| Bauer E.D. et al | Phys. Rev | B65 | 100506R | (2002) |
| Beal Monod M.T. and Lawrence J.M. | Phys. Rev. | B21 | 5400 | (1980) |
| Behnia K. et al | J. Phys. Condens. Matter | 16 | 5187 | (2004) |
| Behnia K. et al | J. Appl. Phys. | 67 | 5200 | (1990) |
| Behnia K. et al | J. Phys. Condens. Matt. | | 0412619 | (2005) |
| Bel R. et al | Phys. Rev. Letters | 92 | 217002 | (2004-a) |
| Bel R. et al | Phys. Rev. | B70 | 220501 (R) | (2004-b) |
| Belitz TR. et al | Phys. Rev. Letters | 82 | 4707 | (1999) |
| Bellarli B. et al | Phys. Rev. B | 30 | 1182 | (1984) |
| Benoit A. et al | Journal de Physique Lettres | 39 | 94 | (1978) |
| Benoit A. et al | Valence fluctuations in solids. L.M. Falicov, W. Hanke and M.P. Maple editors, North Holland publishing Company | | 283 | (1981) |
| Benoit A. et al | Journal de Physique C5 | 40 | 328 | (1979) |
| Benoit A. et al | Solid State Commun | 34 | 293 | (1980) |
| Benoit A. et al | J. Physique Lett. | 46 | L923 | (1985) |
| Bergmann C. et al | Physica B | 230-232 | 348 | (1997) |
| Bergmann G. | Phys. Rev. Lett. | 67 | 2545 | (1991) |
| Bernal O.O. et al | Phys. Rev. Lett. | 87 | 196402 | (2001) |
| Bernhard C. et al | Phys. Rev. B | 59 | 14099 | (1999) |
| Bernhoeft N. | Eur. Phys. J. B | 13 | 685 | (2000) |
| Bernhoeft N. | J. Phys. Condens. Matter | 13 | R771 | (2001) |
| Bernhoeft N. et al | Phys. Rev. Letters | 81 | 4244 | (1998) |
| Bernhoeft N.R and Lonzarich G.G. | J. Phys. Condens. Matter | 7 | 7325 | (1995) |
| Besnus M.J. et al | Solid State Commun. | 55 | 779 | (1985) |
| Bianchi A. et al | Phys. Rev. Lett. | 91 | 187004 | (2003-a) |
| Bianchi A. et al | Phys. Rev. Letters | 89 | 137002 | (2002) |
| Bianchi A. et al | Phys. Rev. Lett. | 91 | 257001 | (2003-b) |
| Bickers et al | Phys. Rev. Letters | 54 | 230 | (1985) |
| Blandin A. | In Magnetism V edited by H. Suhl, Academic Press | | 57 | (1973) |
| Bobenberger B. et al | Physica B | 186-188 | 248 | (1993) |
| Bourdarot F. et al | Cond-Mat. 0312206 | | | (2003-b) |
| Bourdarot F. et al | Phys. Rev. Lett. | 90 | 067203 | (2003-a) |
| Boursier R. et al | Thèse Université de Grenoble | | | (2005) |
| Braithwaite D. et al | J. Phys. Condens. Matter | 12 | 1339 | (2000) |
| Brandt N.B. and Moschalkov V.V. | Advances in Physics | 33 | 373 | (1984) |
| Bredl C.D. et al | Z. Phys. B. | 29 | 327 | (1978) |
| Brison J.P. | Private communication | | | (2004) |
| Brison J.P. et al | Physica B | 199-200 | 70 | (1994-a) |
| Brison J.P. et al | Physica C | 256 | 128 | (1995) |
| Brison J.P. et al | J. Low Temp. Phys. | 95 | 145 | (1994-b) |
| Brison J.P. et al | Physica B | 280 | 165 | (2000) |
| Brison J.P. et al | J. Physique | 50 | 2795 | (1989) |
| Brison J.P. et al | Physica | 230-232 | 406 | (1997) |
| Brodale G.E. et al | Phys. Rev. lett. | 56 | 390 | (1985) |
| Broholm C. et al | Phys. Rev. B | 43 | 12809 | (1991) |
| Broholm C. et al | Phys. Rev. Letters | 58 | 1467 | (1987) |
| Bruls G. et al | Phys. Rev. Lett. | 65 | 2294 | (1990) |
| Burdin S. et al | Phys. Rev. Lett. | 85 | 1048 | (2000) |
| Buzdin A.I. and Mel'nikov | Phys. Rev. | 67 | 020503R | (2003) |
| Canfied P.C., et al | Phys. Today | 51 | 40 | (1998) |
| Capan C. et al | Phys. Rev. | 70 | 134513 | (2004) |
| Castro Neto A.H. and Jones B.A. | Phys. Rev. B | 62 | 14975 | (2000) |
| Champel T. and Mineev V. | Phys. Rev. Lett. | 86 | 4903 | (2001) |
| Chandra P., et al | Nature | 417 | 831 | (2002) |
| Chen J.W. et al | Physica | | | (1984) |
| Chikorek T. et al | Acta Physica Polonica B | 34 | 371 | (2003) |
| Choi S.M. et al | Phys. Rev. Lett. | 87 | 107001 | (2001) |
| Chubukov A.V. et al | ArXiv : Cond. Mat/0201140 | | | (2002) |
| Chubukov A.V. et al | ArXiv:cond-mat 0311420v1 | | | (2003) |
| Cibin R. | Thesis, University of Geneve | | | (1990) |
| Coad S. et al | Physica B | 276-278 | 764 | (2000) |
| Coleman P and Pépin C. | Phys. Rev. B | 68 | 220405(R) | (2003) |
| Coleman P. | Local moment physics in heavy fermion electron systems, editor F. Mancini, American Institute of Physics | | 79-160 | (2002) |
| Coleman P. | Physica B | 259-261 | 353 | (1999) |
| Coleman P. et al | J. Phys. Condens. Matter | 13 | R723 | (2001) |
| Continentino M.A. | To be published Brazilian Journal of Physics | | | (2004-a) |
| Continentino M.A. | Quantum scaling in many body systems. World scientific, Lecture notes in physics (World scientific New York) | 67 | | (2001) |
| Continentino M.A. and Ferreira A.S. | Phys. Rev. | B69 | 233104 | (2004) |
| Continentino M.A. et al | Physica B to be published | | | (2005) |
| Coqblin B. and Schrieffer J.R. | Phys. Rev. | 185 | 847 | (1969) |
| Coqblin B. et al | Phys. Rev. | B67 | 064417 | (2003) |
| Cornut B., Coqblin B. | Phys. Rev. | B5 | 4541 | (1972) |
| Cox D. and Maple M. | Physics Today | 48 | 232 | (1995) |
| Curro N.J. et al | Cond-Mat/0402179V2 | | | (2004) |
| Custers C.V. et al | Nature (London) | 424 | 524 | (2003) |
| Dalichouch V. et al | Phys. Rev. Lett. | 75 | 3938 | (1975) |
| Dalmas de Réotier P. et al | To be published | | | (2005) |
| Dalmas de Réotier P. et al | Phys. Rev. B | | 6377 | (2000) |
| De Visser A. | Physica B | 319 | 233 | (2002) |
| De Visser A. | PhD Amterdam | | | (1986) |
| Deflets C. et al | Cond-mat/0306742 | | | (2003) |
| Degiorgi L. | Rev. of Modern Physics | 71 | 687 | (1999) |
| Demuer A. | PhD Grenoble | | | (2000) |
| Demuer A. et al | J. Phys. Condens. Matter | 14 | L529 | (2002) |
| Demuer A. et al | J. Phys. Condens. Matter | 13 | 9335 | (2001) |
| Denlinger J.D. et al | Electron spectroscopy phenomena | 117-118 | 347 | (2000) |
| Doiraud N. et al | Nature | 425 | 595 | (2003) |
| Doniach S. | Physica B | 91 | 231 | (1977) |
| Ebihara T. | Phys. Rev. Lett. | 93 | 246401 | (2004) |
| Edwards D.O. | Proceedings of 1970 ultralow temperature symposium, Naval research laboratory, Washington DC NRL report | (7133) | 27 | (1970) |
| Elliot R.J. | Magnetic properties of rare earth metal, Plenum press | | | (1972) |
| Ellman B. et al | Phys. Rev. B | 54 | 5043 | (1996) |
| Emery W.J. | J. Phys. (Paris) Colloq 44 | C3 | 977 | (1983) |
| Endo M. | Phys. Rev. Lett. | 93 | 247003 | (2004) |
| Eremets M. et al | High pressure experimental methods. Oxford University Press | | | (1996) |
| Evans S.M.M. | J. Magn. Magn. Mat. | 108 | 135 | (1992) |
| Fak B. et al | J. Low Temp. Phys. | 110 | 417 | (1998) |
| Fak B. et al | J. Magn. Magn. Mat. | 272-271 S1 | E13 | (2004) |
| Fak B. et al | J. Phys. Condens. Matter | 12 | 5423 | (2000) |
| Fay D. and Appel J. | Phys. Rev. B | 22 | 3173 | (1980) |
| Fazekas P. | Lectures notes on electron correlation and magnetism, World Scientific, Singapore | | | (1999) |
| Felner I. et al | Phys. Rev. B | 55 | 3374 | (1997) |
| Feyerherm A. et al | Phys. Rev. Lett. | 73 | 1849 | (1994) |
| Fischer Ø. | Appl. Phys. | 16 | 1 | (1978) |
| Fischer Ø. | Magnetic superconductors in ferromagnetic materials. KH.A. Buschow and E.P. Wohlforth ed. (Science publishers BV, Amsterdam) | | | (1990) |
| Fischer Ø. and Peter M. | In Magnetism V edited by H. Suhl, Academic Press | | 327 | (1973) |
| Fisher R.A. el al | Phys. Rev. Lett. | 62 | 1411 | (1989) |
| Fisher R.A. et al | J. Low Temp. Phys. | 84 | 49 | (1991) |
| Fisher R.A. et al | Phys. Rev. | 65 | 224509 | (2002) |
| Fisher R.E. et al | To be published | | | (2005) |
| Fisk Z. et al | Science | 239 | 33 | (1998) |
| Flouquet J. | Prog. Low Temp. Physics VII. Eds. Brewver North Holland | | | (1978) |
| Flouquet J. and Buzdin A. | Physics World | 41 | January | (2002) |
| Flouquet J. et al | Physica B | 319 | 251 | (2002) |
| Flouquet J. et al | J. Magn. Magn. Mat. | 76-77 | 285 | (1988) |
| Flouquet J. et al | J. Magn. Magn. Mat. | 272-271 | 27 | (2004) |
| Flouquet J. et al | J. Appl. Phys. | 53 | 2126 | (1982) |
| Flouquet J. et al | Physica C | 185-189 | 372 | (1991) |
| Flouquet J. et al | J. Phys. Soc. Jpn | 74 | 178 | (2005) |
| Fomin I. A and Flouquet J. | Solid State Commun | 99 | 795 | (1996) |
| Fomin I.A. | JETP Letters | 74 | 116 | (2001) |
| Fomin I.A. | Sov. Phys. JETP | 95 | | (2002) |
| Fomin I.A. | Private communication | | | (2004) |
| Fomin I.A. and Brison J.P. | Eur. Physics Lett. | | | (2002) |
| Forgan E.M. et al | J. Phys. Condens. Matter | 2 | 10211 | (1990) |
| Frederick N.A. et al | Phys. Rev. | 69 | 024523 | (2004) |
| Frederick N.A. et al | Phys. Rev. | B69 | 024523 | (2004) |
| Frigeri P.A. et al | Phys. Rev. Lett. | 92 | 097001 | (2004) |
| Frossati G. et al | Phys. Rev. Lett. | 36 | 203 | (1976) |
| Fukazawa H. and Yamada K. | J. Phys. Soc. Jpn | 72 | 2449 | (2003) |
| Fukuhara T. et al | J. Phys. Soc. Jpn | 65 | 1559 | (1996) |
| Fulde P. | J. Low Temp. Phys. | 95 | 45 | (1994) |
| Fulde P. and Ferrel R.A | Phys Rev. A | 135 | 550 | (1964) |
| Fulde P. and Jensen J. | Phys. Rev. B | 27 | 4085 | (1983) |
| Fulde P. et al | Solid State Physics. (Academic press) H. Ehreureich and Turnbull editors. | 41 | | (1988) |
| Fuseya Y. et al | Journal Phys. Soc. Jpn | 72 | 2914 | (2003) |
| Gammel P.L. et al | Phys. Rev. Lett. | 84 | 2497 | (2000) |
| Gegenwart P. et al | Phys. Rev. Lett. | 89 | 056402 | (2002) |
| Gegenwart P. et al | Phys. Rev. Lett. | 82 | 1293 | (1999) |
| Geibel G. et al | Z. Phys. B | 84 | 1 | (1991) |
| Georges A. et al | Rev. Mod. Phys. | 68 | 13 | (1996) |
| Georges A. et al | Lectures on the physics of highly correlated electron systems VIII. American Institute; Physics Conference Proceedings | Vol 75 | | (2004) |
| Ginzburg V. | Sov. Phys. JETP | 4 | 153 | (1957) |
| Glemot L. et al | Phys. Rev. Letters | 82 | 169 | (1999) |
| Gloos K. et al | Phys. Rev. Lett. | 70 | 501 | (1993) |
| Glyde H.R. et al | Phys. Rev. B | 61 | 1421 | (2000) |
| Goldman A. et al | Phys. Rev. B | 33 | 1626 | (1986) |
| Gorkov L. | Sov. Sci. Rev. Sect | A9 | 1 | (1987) |
| Gorkov L.P. and Rashba E.I. | Phys. Rev. Lett. | 87 | 037004 | (2001) |
| Goryo J. | Phys. Rev. | B67 | 184511 | (2003) |
| Graf M.J. et al | J. Low Temp. Phys. | 102 | 367 | (1996) |
| Grempel D.R. and Si Q. | Phys. Rev. Lett. | 91 | 156404 | (2003) |
| Grewe M. and Steglich F. | Handbook on the Physics and the Chemistry of Rare Earth. Elsevier Science publishers KA Gschneidner and L. Eyring editors. | 14 | 343 | (1991) |
| Grosche F.M. et al | J. Phys. Condens. Matter | 13 | 2845 | (2001) |
| Grosche F.M. et al | J. Phys. Cond. Matter | 12 | L533 | (2000) |
| Grüneisen E. | Ann. Phys. Leipzig | 39 | 257 | (1912) |
| Guillaume A. | Thesis, Université de Grenoble | | | (1999) |
| Haen P. et al | J. Phys. Soc. Jpn | 65, Suppl. B | 27 | (1996) |
| Haen P. et al | J. Low Temp. Phys. | 67 | 391 | (1987) |
| Haga Y. et al | To be published | | | (2005) |
| Hanzawa K. et al | J. Phys. Soc. Jpn | 56 | 678 | (1987) |
| Harada A. et al | To be published | | | (2005) |
| Hardy F.et al | To be published | | | (2005) |
| Hardy F. | Thesis, Université de Grenoble | | | (2004) |
| Harima H. | Private Communication | | | (2004) |
| Harrison N. et al | Phys. Rev. B | 61 | 1779 | (2000) |
| Harrison N. et al | Phys. Rev. Lett. | 90 | 096402 | (2003) |
| Hasegawa H. and Moriya T. | J. Phys. Soc. Jpn | 36 | 1542 | (1974) |
| Hasselbach K. et al | Phys. Rev. Lett. | 63 | 93 | (1989) |
| Hasselbach K. et al | Phys. Lett. A | 156 | 313 | (1991) |
| Hayden S. | Private Communication | | | (2002) |
| Hayden S. et al | Phys. Rev. B | 46 | 8675 | (1992) |
| Heffner R.H and Norman M.R. | Comments Condens. Matter Phys. | 17 | 361 | (1996) |
| Heffner R.H. et al | Europhys. Lett. | 3 | 751 | (1987) |
| Hegger H. et al | Phys. Rev. Letters | 84 | 4986 | (2000) |
| Held K. et al | Phys. Rev. Lett. | 87 | 276404 | (2001) |
| Helfrich R. | PhD Thesis Darmstadt | | | (1995) |
| Hertz J.A. | Phys.Rev. | B14 | 1165 | (1976) |
| Hess D.W. et al | Encyclopedia of applied physics (VCH Weinheim) | 7 | 435 | (1993) |
| Hewson A.C. | The Kondo problem to heavy fermions (Cambridge University Press) | | | (1992) |
| Hiess A. et al | Phys. Rev. B | 66 | 064531 | (2002) |
| Hiess A. et al | | | Cond-Mat. 0411041 | (2004) |
| Hirsch J.E. | Phys. Rev. Lett. | 54 | 1317 | (1985) |
| Hirschfeld P.J. et al | Phys. Rev. B | 37 | 83 | (1988) |
| Hirschfeld P.J. et al | Solid State Commun | 59 | 111 | (1986) |
| Holmes A.T. et al | Phys. Rev. | B69 | 024508 | (2004-a) |
| Holmes A.T. et al | J. Phys. Cond. Mat. | 16 | S1121 | (2004-b) |
| Holtmeier S. | PhD thesis Grenoble | | | (1994) |
| Hotta T. and Ueda K. | Phys. Rev. B | 67 | 104518 | (2003) |
| Houghton A. et al | Phys. Rev. B | 35 | 5123 | (1987) |
| Huth et al | Physica B | 281-282 | 882 | (2000) |
| Huth M and Jourdan M. | Advances in Solid state Physics | 39 | 351 | (1999) |
| Huxley A et al | Phys. Rev. letters | 91 | 207201 | (2003-b) |
| Huxley A. and Hardy F. | Private Communication | | | (2004) |
| Huxley A. et al | Phys. Rev. B | 63 | 144519 | (2001) |
| Huxley A. et al | J. Phys. Condens. Matter | 5 | S1945 | (2003-a) |
| Huxley A. et al | Nature | 406 | 160 | (2000) |
| Huxley A. et al | Phys. Letters A | 209 | 365 | (1995) |
| Ichioka M. et al | J. Phys. Soc. Jpn | 72 | 1322 | (2003) |
| Ikeda H. | PhD Thesis, University of Osaka | | | (1997) |
| Ikeda H. and Miyake K. | J. Phys. Soc. Jpn | 66 | 3714 | (1997) |
| Ikezawa I. et al | Physica B | 237-238 | 210 | (1997) |
| Inada Y. et al | Physica B | 199-200 | 119 | (1994) |
| Isaacs E.D. et al | Phys. Rev. Lett. | 75 | 1178 | (1995) |
| Ishida K et al | Phys. Rev. | B68 | 184401 | (2003) |
| Ishida K. et al | Phys. Rev. Letters | 89 | 167202-1 | (2002) |
| Ishida K. et al | Phys. Rev. Lett. | 82 | 5353 | (1999) |
| Izawa K. et al | Phys. Rev. lett. | 87 | 057002-1 | (2001) |
| Izawa K. et al | Phys. Rev. Lett. | 90 | 117001 | (2003) |
| Jaccard D et al | J. Appl. Phys. | 26 | 517 | (1987) |
| Jaccard D. and Flouquet J. | J. Magn. Magn. Mat. | 47-48 | 45 | (1985) |
| Jaccard D. et al | Phys. Letters A | 299 | 282 | (2002) |
| Jaccard D. et al | J. Magn. Magn. Mat. | 76-77 | | (1988) |
| Jaccard D. et al | Phys. Lett. A | 163 | 475 | (1992) |
| Jaccard D. et al | Physica B to be published | | | (2005) |
| Jaccard D. et al | J. Magn. Magn. Mat. | 47-48 | 23 | (1985) |
| Jacoud | Thèse Univsersité de Grenoble | | | (1989) |
| Jarlborg T. et al | Phys. Letters A | 300 | 518 | (2002) |
| Jayaraman A. | Phys. Rev. A | 137 | 179 | (1965) |
| Jones B.A. et al | Phys. Rev. Lett. | 61 | 125 | (1988) |
| Jourdan M. et al | Nature | 398 | 47 | (1999) |
| Joynt R. and Taillefer L. | Rev. Moderns Physics | 74 | 235 | (2002) |
| Julian S.R. et al | Physica B | 199, 200 | 36 | (1994) |
| Jullien R. et al | J. Appl. Phys. | 50 | 7555 | (1979) |
| Kadowaki H. et al | Phys. Rev. | B68 | 140402 | (2003) |
| Kadowaki H. et al | Phys. Rev. Lett. | 092 | 097204 | (2004) |
| Kadowaki K. and Woods S.B. | Solid State Commun. | 58 | 507 | (1986) |
| Kambe S. and Flouquet J. | Sol State Commun | 103 | 551 | (1997) |
| Kambe S. et al | J. Phys. Low. Temp. | 117 | 101 | (1999) |
| Kambe S. et al | Solid State Commun | 95 | 449 | (1995) |
| Kambe S. et al | J. Low Temp. Phys. | 102 | 477 | (1996) |
| Kambe S. et al | J.  Low Temp. Phys. | 108 | 383 | (1997) |
| Kawano-Furukawa H. et | Phys. Rev. B | 65 | R180508 | (2002) |
| Kawarasaki et al | Phys. Rev. B | 61 | 4167 | (2000) |
| Kawasaki S. et al | J. Phys. Soc. Jpn | 73 | 1647 | (2004) |
| Kawasaki S. et al | Phys. Rev. B | 65 | 020504R | (2001) |
| Kawasaki S. et al | Phys. Rev. Letters | 91 | 137001-1 | (2003-b) |
| Kawasaki Y. et al | J. Phys. Soc. Jpn | 72 | 2308 | (2003-a) |
| Kawasaki Y. et al | Phys. Rev. B | 58 | 8634 | (1998) |
| Kawasaki Y. et al . | Phys. Rev. | B66 | 224502 | (2002) |
| Kernavanois N. et al | Acta Physica Polonica | 34 | 721 | (2003) |
| Kernavanois N. et al | To be published | | | (2004) |
| Kernavanois N. et al | Phys. Rev. B | 64 | 174500 | (2001) |
| Kim K.H. et al | Phys. Rev. Lett. | 91 | 256401 | (2003) |
| Kimura N. | Private communication | | | (2004) |
| King C.A. and Lonzarich G.G. | Physica B | 171 | 161 | (1991) |
| Kirkpatrick T.R. and Belitz D. | Phys. Rev. B | 67 | 024515 | (2003) |
| Kirkpatrick T.R. et al | Phys. Rev. Lett. | 87 | 127003 | (2001) |
| Kiss A. and Fazekas P | | | Cond-Mat 0411029 | (2004) |
| Kitaoka Y. et al | J. Phys. Condens. Matter | 13 | L79 | (2001) |
| Knafo W. | Phys. Rev. | B70 | 174401 | (2004) |
| Knafo W. | J. Phys. Condens. Matter | 15 | 3741 | (2003) |
| Knebel G. et al | Physica B to be published | | | (2005) |
| Knebel G. et al | Phys. Rev. B | 59 | 12390 | (1999) |
| Knebel G. et al | Phys. Rev. B | 65 | 024425 | (2002) |
| Knetsch A. et al | Physica B | 186-188 | 300 | (1993) |
| Knöpfle K. et al | J. Phys. Condens. Matter | 8 | 901 | (1996) |
| Kobayashi T.C. | Private communication | | | (2001) |
| Koehler WC. et al | Phys. Rev. | 151 | 405 | (1966) |
| Kohgi M. et al | J. Phys. Soc. Jpn | 73 | 1002 | (2003) |
| Kohori Y. et al | Physica B | 281-282 | 12 | (2000-a) |
| Kohori Y. et al | Phys. Rev. | B64 | 134526 | (2001) |
| Kohori Y. et al | Physica B | 312-313 | 126 | (2002) |
| Kohori Y. et al. | Eur. Phys. B | 18 | 601 | (2000-b) |
| Kondo J. | J. Phys. Soc. Jpn Effect 40 years after the discovery | 74 | 1-238 | (2005) |
| Kopnin N.B and Volovik G.E. | JETP Lett. | 64 | 690 | (1996) |
| Kos S. | Cond-mat 0302089 | | | (2003) |
| Kotegawa H. | J. Phys. Soc. Jpn. | 74 | To be published | (2005) |
| Kotegawa H. et al | J. Phys. Condens. Matter | 15 | S2043 | (2003) |
| Kouroudis I. et al | Phys. Rev. Lett. | 58 | 820 | (1987) |
| Krimmel A. et al | Z. Phys. B | 86 | 161 | (1992) |
| Krimmel A. et al | J. Phys. Condens. Matter | 12 | 8801 | (2000) |
| Kromer F. et al | Phys. Rev. B | 62 | 12477 | (2000) |
| Küchler et al | Phys. Rev. lett | 91 | 066405 | (2003) |
| Kulic M. et al | Phys. Rev. B | 56 | R11415 | (1997) |
| Kumar P. and Wölfle P. | Phys. Rev. Lett. | 59 | 1954 | (1987) |
| Kumar R.S. et al | Cond. Mat. 0405043 | | | (2004) |
| Kuramoto Y. and Kitaoka Y. | Dynamics of heavy electrons. Clarendon Press, Oxford | | | (2000) |
| Kuramoto Y. and Miyake K. | J. Phys. Soc. Jpn | 59 | 2831 | (1990 |
| Kuwahara K. et al | J. Phys. Soc. Jpn | 73 | 1438 | (2004) |
| Kuwahara K. et al | To be published | | | (2005) |
| Kyogaku M. et al | J. Phys. Soc. Jpn | 62 | 4016 | (1993) |
| Lacerda A. et al | Phys. Rev.B | 40 | 11429 | (1989) |
| Lander G. et al | Phys. Rev. B | 46 | 5387 | (1992) |
| Lapertot G. et al | Physica B | 186-188 | 454 | (1993) |
| Larkin A.I. and Ovchinnikov Y.N. | Sov. Phys. JETP | 20 | 762 | (1965) |
| Laughlin R.D. and Pines D. | PNAS | 97 | 27 | (2000) |
| Lavagna M. et al | Phys. Lett. | 90A | 210 | (1982) |
| Lawrence J. and Shapiro S. | Phys. Rev. B | 22 | 4379 | (1980) |
| Layzer A. and Fay D. | Int. J. Magn. | 1 | 135 | (1971) |
| Lee P.A. | Phys. Rev. Lett. | 71 | 1887 | (1993) |
| Leggett A.J. | Rev. Mod. Phys. | 47 | 331 | (1975) |
| Leggett A.J. | J. Magn. Magn. Mat. | 63-64 | 406 | (1987) |
| Levanyuk A.P. | Some modern aspects of the physics of strongly correlated systems : Instituto Universitario de Ciencia de materials, Nicolas Cabrera UAM ed F.G Aliev, J.C. Gomez-sal, H. Suderow and R. Villar (Eds) | | 15 | (2001) |
| Levy F. et al | To be published | | | (2005) |
| Lin CL. et al | Phys. Rev. Lett. | 54 | 2541 | (1985) |
| Link P. et al | J. Phys. Condens. Matter | 7 | 373 | (1993) |
| Lloblet A. et al | Phys. Rev. B | 69 | 024403 | (2003) |
| Lonzarich G.G. | J. Magn. Magn. Mat. | 76,77 | 1 | (1988) |
| Lonzarich G.G. and Taillefer L. | J. Phys. C | 18 | 4339 | (1985) |
| Lousnamaa O.V. | Experimental Principles and Methods below 1K. Academic Press. | | | (1974) |
| Lussier B. et al | Phys. Rev. Lett. | 73 | 3294 | (1994) |
| Machida K. and Ohmi T. | Phys. Rev. Lett. | 86 | 850 | (2001) |
| Maki K. et al | Europhys. Lett. | 64 | 496 | (2003) |
| Makoshi K. and Moriya T. | J. Phys. Soc. Jpn | 38 | 10 | (1975) |
| Malterre D. et al | Adv. Phys. | 45 | 299 | (1996) |
| Mao S.Y. et al | J. Magn. Magn. Mat | 76-77 | 241 | (1988) |
| Maple B. | In Magnetism V edited by H. Suhl, Academic Press | | 289 | (1973) |
| Maple M.B. et al | Phys. Rev. Lett. | 56 | 185 | (1986) |
| Maple M.B. et al | Acta Physica Polonica B | 34 | 919 | (2003) |
| Marcenat C. et al | J. Magn. Magn. Mat. | 76-77 | 115 | (1988) |
| Martisovitz V. et al | Phys. Rev. Lett. | 84 | 5872 | (2000) |
| Mathur N.D. et al | Nature | 394 | 39 | (1998) |
| Matsuda K. et al | Phys. Rev. Lett. | 87 | 087203 | (2001) |
| Matsuda K. et al | J. Phys. Condens. Matter | 15 | 2363 | (2003) |
| Matsumoto M. and Koga M. | J. Phys. Soc. Jpn | 73 | 1135 | (2004) |
| Mc Elfresh M.W. et al | Phys. Rev. B | 35 | 43 | (1987) |
| Measson M.A. | Private Comumnication | | | (2005) |
| Measson M.A. et al | Phys. Rev. B | 70 | 064516 | (2004) |
| Metoki N. et al | Physica B | 280 | 362 | (2000) |
| Metoki N. et al | Phys. Rev. Letters | 80 | 5417 | (1998) |
| Metoki N. et al | J. Phys. Cond. Mat. | 16 | L207 | (2004) |
| Midgley P.A. et al | Phys. Rev. Lett. | 70 | 678 | (1993) |
| Millis A. | Phys. Rev. | B48 | 7183 | (1993) |
| Mineev V. | International Journal of Modern Physic B | 18 | 2963 | (2004-a) |
| Mineev V. | J. Phys. Cond. Mat. | | 0405672 | (2004-b) |
| Mineev V. and Champel T. | Phys. Rev. | B69 | 054508 | (2004) |
| Mineev V. and Samokhin K.V. | Introduction to unconventional superconductivity. (Gordon and Breach SV Publ.) | | | (1999) |
| Mineev V.P. | Phys. Rev. B | 66 | 134504 | (2002) |
| Mineev V.P. and Zhitomirsky M. | Cond-Mat 0412055 | | | (2004) |
| Miranda E. et al | Phys. Rev. Lett. | 78 | 290 | (1997) |
| Mito T. et al | Phys. Rev. | B63 | 220507 (R) | (2001) |
| Miyake K. | Private communication | | | (2004) |
| Miyake K. and Kohno H. | J. Phys. Soc. Jpn. | 75 | 254 | (2005) |
| Miyake K. and Kuramoto Y | J. Magn. Magn. Mat. | 90-91 | 438 | (1990) |
| Miyake K. and Maebashi H. | J. Phys. Soc. Jpn | 71 | 1007 | (2002) |
| Miyake K. and Narikiyo O. | J. Phys. Soc. Jpn | 71 | 867 | (2002) |
| Miyake K. et al | Solid State Commun. | 71 | 1149 | (1989) |
| Miyake K. et al | Phys. Rev. | 34 | 6554 | (1986) |
| Miyake K. et al | J. Phys. Condens. Matter | 15 | L275 | (2003) |
| Monthoux P. and Lonzarich G. | Phys. Rev. B | 63 | 054529 | (2001) |
| Monthoux P. and Lonzarich G. | Phys. Rev. B | 66 | 224 | (2003) |
| Mori H. et al | Physica B | 259-261 | 58 | (1999) |
| Morin P. et al | J. Low Temp. Phys. | 70 | 377 | (1988) |
| Moriya T. | Spin fluctuation in itinerant electron magnetism. Springer Berlin | | | (1985) |
| Moriya T. | Acta Physica Polonica | B34 | 287 | (2003-a) |
| Moriya T. and Takimoto T. | J. Phys. Soc. Jpn | 64 | 960 | (1995) |
| Moriya T. and Ueda K. | Advances in Physics | 49 | 555 | (2000) |
| Moriya T. and Ueda K. | Rep. Prog. Phys. | 66 | 1299 | (2003-b) |
| Motoyama G. et al | Phys. Rev. B | 65 | 020510R | (2001) |
| Motoyama G. et al | Phys. Rev. Lett. | 90 | 166402 | (2003) |
| Mott N.F. | Pro. Phys. Soc. | 47 | 571 | (1935) |
| Movshovich R. et al | Physica B | 223-224 | 126 | (1996) |
| Movshovich V. et al | Phys. Rev. Letters | 86 | 5152 | (2001) |
| Müller K.H. | Rep. Prog. Phys. | 64 | 943 | (2001) |
| Müller-Hartmann E. | In Magnetism V edited by H. Suhl, Academic Press | | 353 | (1973) |
| Müller-Hartmann E. and Zittartz J. | Phys. Rev. Lett. | 26 | 428 | (1971) |
| Mydosh J.A. et al | Acta Physica Polonica B | 34 | 659 | (2003) |
| Naidyuk Yu G. and Yanson I.K. | J. Phys. Cond. Matter | 10 | 8905 | (1998) |
| Nakajma S. | Progress of theoretical physics | 50 | 1101 | (1973) |
| Nakashima N. et al | J. Phys. Condens. Matter | 13 | L569-576 | (2001) |
| Nakashima N. et al | J. Phys. Condens. Matter | 15 | S2011 | (2003) |
| Nakatsuji et al | Phys; Rev. Letters | 92 | 016401/1 | (2004) |
| Newns D.C. and Read N. | Advances in Physics | 36 | 799 | (1987) |
| Ng N.K. and Varma C.M. | Phys. Rev. Lett. | 78 | 330 | (1997) |
| Niklowitz Ph. | PhD Thesis, Cambridge | | | (2003) |
| Norman M.R. | Phys. Rev. Lett. | 71 | 3391 | (1993) |
| Norman M.R. et al | Phys. Rev. | B36 | 4058 | (1987) |
| Norman M.R. et al | Phys. Rev. Lett. | 90 | 16601 | (2003) |
| Nozières P. | Cours, Collège de France, Magnetisme et localisation dans les liquides de Fermi. | | | (1986) |
| Nozières P. | Eur. Phys. | B6 | 447 | (1998) |
| Nozières P. | J. Low Temp. Phys. | 17 | 31 | (1974) |
| Nozières P. | Ann. Phys. Fr | 10 | 19 | (1985) |
| Oeschler et al | Phys. Rev. Lett. | 91 | 076402 | (2003-a) |
| Oeschler N. et al | Acta Physica Polonica | 34 | 255 | (2003-b) |
| Ohashi M. et al | Phys. Rev. | 68 | 144.428 | (2003) |
| Ohkawa F. | Phys. Rev. Lett. | 64 | 2300 | (1990) |
| Ohkuni H. et al | J. Phys. Soc. Jpn | 66 | 945 | (1997) |
| Ohmi T. and Machida K. | J. Phys. Soc. Jpn | 65 | 4018 | (1996) |
| Okhuni et al | Philosophical Mag. | 79 | 1045 | (1999) |
| Okuno Y. et Miyake K. | J. Phys. Soc. Japan | 67 | 2469 | (1998) |
| Ong N.P. and Wang Y. | Physica C | 408-410 | 11 | (2004) |
| Onishi Y. and Miyake K. | J. Phys. Soc. Jpn | 69 | 3955 | (2000) |
| Onuki Y. and Komatsubara T. | J. Magn. Magn. Mat. | 63, 64 | 281 | (1987) |
| Onuki Y. et al | J. Phys. Soc. Jpn. | 61 | 293 | (1992) |
| Onuki Y. et al | Acta Physica Polonica B | 34 | 667 | (2003) |
| Oomi G. et al | Physica B | 186-188 | 758 | (1993) |
| Ormeno R.J. et al | Phys. Rev. Letters | 88 | 047005 | (2002) |
| Osheroff D. et al | Phys. Rev. Lett. | 44 | 792 | (1980) |
| Ott H.R. | Prog. Low Temp Phys XI ed. D.F. Brewer, Elsevier Science Publishers | | 215-289 | (1987) |
| Ott H.R. et al | Phys.Rev. B | 11 | 1651 | (1985) |
| Ott H.R. et al | Phys. Rev. Lett. | 74 | 4734 | (1983) |
| Paglione et al | Phys. Rev. Lett. | 91 | 246405 | (2003) |
| Pagliuso P.G. et al | Phys. Rev. | B64 | 100503 R | (2001) |
| Pagluiso et al | Physica B | 312 | 129 | (2002) |
| Pankov S. et al | Phys. Rev. B | 69 | 054426 | (2004) |
| Paschen S. et al | Nature | 432 | 881 | (2004) |
| Paulsen et al | J. Low Temp. Phys. | 81 | 317 | (1990) |
| Payer K. et al | Physica B | 186-188 | 503 | (1993) |
| Pépin C. | Cond-mat/0407155 | | | (2004) |
| Pethick C.J. and Pines D. | Phys. Rev. Lett | 57 | 118 | (1986) |
| Petrovic C. et al . | J. Phys. Condens. Matter | 13 | 337 | (2001) |
| Peyrard J. | PhD dissertation, Grenoble | | | (1980) |
| Pfleiderer C. | Acta Physica Polonica B | 34 | 679 | (2003) |
| Pfleiderer C. and Huxley A. | Phys. Rev. Lett. | 89 | 147005 | (2002) |
| Pfleiderer C. et al | Nature | 412 | 58 | (2001) |
| Pfleiderer C. et al | Nature | 427 | 227 | (2004) |
| Pietri R. and Andraka | Phys. Rev. B | 62 | 8619 | (2000) |
| Piquemal F. et al | J. of Magn. Magn. Mat. | 63-64 | 469 | (1987) |
| Plessel J. et al | Phys. Rev. B | 67 | 180403R | (2003) |
| Radovan H.A. et al | Nature | 425 | 51 | (2003) |
| Ramakrishan T.V. and Sur K. | Phys. Rev. B | 26 | | |
| Ramirez A.P. et al | Phys. Rev. B | 44 | 5392 | (1991) |
| Ravex A. et al | J. Magn. Magn. Mat. | 63-64 | 400 | (1987) |
| Raymond S. and Jaccard D. | Phys Rev. | B61 | 8679 | (2000) |
| Raymond S. et al | Physica B | 259-261 | 48 | (1999-a) |
| Raymond S. et al | J. Low. Temp. Phys. | 109 | 205 | (1997) |
| Raymond S. et al | J. Phys. Condens. Matter | 11 | 5547 | (1999-b) |
| Raymond S. et al | J. Phys. Condens. Matter | 13 | 8303 | (2001) |
| Raymond S. et al | Physica B, to be published | | | (2005) |
| Razafimanchy et al | Z. Phys. | B 54 | 111 | (1984) |
| Regnault L.P. et al | Phys. Rev. | B38 | 4481 | (1988) |
| Rehmann S. et al | Phys. Rev. Lett. | 78 | 1122 | (1997) |
| Reinders P.H.P. | J. Magn. Magn. Mat. | 63-64 | 297 | (1987) |
| Ribault M. et al | Journal de physique – Lettres | 40 | 413 | (1979) |
| Rodière P. | Thèse, Université de Grenoble | | | (2001) |
| Roehler J. et al | J. Magn. Magn. Mat. | 76-77 | 340 | (1988) |
| Rosch A. | Phys; Rev. Lett. | 82 | 4280 | (1999) |
| Rosch A. | Phys. Rev. B | 62 | 4945 | (2000) |
| Rosch A. et al | Phys. Rev. Lett. | 79 | 159 | (1997) |
| Rossat Mignod J. | Neutron scattering in condensed matter research. Ed. K.L. and D.L. Price) Academic Press, chapter 20 | | | (1986) |
| Rossat Mignod J. et al | J. Magn. Magn. Mat. | 76-77 | 376 | (1988) |
| Rossat Mignod J. et al | Physica B | 180-181 | 383 | (1992) |
| Rotondu C.R. et al | Cond-Mat 042599 | | | (2004) |
| Roussev R. and Millis A.J. | Phys. Rev. B | 63 | 140504(R) | (2001) |
| Sachdev S. | Quantum phase trantitions. Cambridge University Press | | | (1999) |
| Saint James D. et al | Type II superconductivity. Pergamon Press, Oxford | | | (1969) |
| Sakai O. et al | To be published | | | (2005) |
| Sakakibara T. et al | Physica B, to be published | | | (2005) |
| Sakurai J. | J. Low Temp. | 95 | 165 | (1994) |
| Sakurai J. | Encyclopedia Science and Technology ISPN 0.0804311526 8018, Elsevier science Ltd | | | (2001) |
| Salce B. et al | Rev. of Scientific Instruments | 71 | 2461 | (2000) |
| Samokhin K.V. and Walker M.B. | Phys. Rev. B | 66 | 024512 | (2002) |
| Samokhin K.V. et al | Phys. Rev. | 69 | 094514 | (2004) |
| Sandeman K. et al | Phys. Rev. Lett. | 90 | 167005 | (2003) |
| Sarrao J.L. et al | Nature | 420 | 297 | (2002) |
| Sato M. et al | J. Phys. Soc. Jpn. | 73 | 3418 | (2004) |
| Sato N.K. et al | Nature | 410 | 340 | (2001) |
| Satoh H. and Okawa F.J. | Phys. Rev. B | 63 | 184401 | (2001) |
| Sauls J.A. | Ad. Phys. | 43 | 113 | (1994) |
| Saxena S. et al | Nature | 406 | 587 | (2000) |
| Saxena S.S. and Littlewood P.B | Nature | 412 | 290 | (2001) |
| Scharnberg K. and Klemm R.A. | Phys. Rev. | B22 | 11290 | (1988) |
| Schick A.B.S. et al | Phys. Rev. | B70 | 134506 | (2004) |
| Schilling J.S. | Adv. Phys. | 28 | 657 | (1979) |
| Schlabitz W. et al | Z. fur Physik B | 62 | 171 | (1986) |
| Schmidt L. | PhD thesis, UJF Grenoble | | | (1993) |
| Schmidt Rink S. et al | Phys. Rev. Lett. | 57 | 2575 | (1986) |
| Schoenes et al | Phys. Rev. B | 35 | 5375 | (1987) |
| Schreiner T. et al | Europhys. Lett. 48 | 568 | 568 | (1999) |
| Schröder A. et al | Phys. Rev. Lett. | 80 | 5623 | (1998) |
| Schröder O. et al | Nature, London | 407 | 351 | (2000) |
| Schuberth E.A. et al | Phys; Rev. Lett. | 82 | 2378 | (1999) |
| Sechovsky V. and Havela | Handbook of magnetic materials, edited by Wohfarth and Breschow, North Holland (Amsterdam) | Vol. 11 | (1998 1 and vol. 4 (1988) 309) | |
| Senthil T. et al | Phys. Rev. B | 69 | 035111 | (2004-a) |
| Senthil T. et al | Cond-mat/0409033 | | | (2004-b) |
| Sergienko I.A. and Curnoe S.H. | Cond-Mat 0309382 | | | |
| Settai R. et al | J. Phys. Condens. Matter | 14 | L29 | (2002) |
| Settai R. et al | J. Magn. Magn. Mat. | 272276 | 223 | (2003) |
| Settai R. et al | Physica B, to be published | | | (2005) |
| Shaginyan V.R. | JETP Lett. | 77 | 99 and 178 | (2003) |
| Sheikin I. | Private Communication | | | (2004) |
| Sheikin I. et al | Phys. Rev. B | 64 | 220504R | (2001) |
| Sheikin I. et al | Phys. Rev. B | 67 | 094420 | (2003) |
| Sheikin I. et al | J. Low Temp. Phys | 122 | 591 | (2001) |
| Shick A.B.S. and Picket W.E. | Phys. Rev. Lett. | 86 | 300 | (2001) |
| Shiina R. et al | J. Phys. Soc. Jpn | 67 | 941 | (1998) |
| Shimizu K. et al | Nature (London) | 406 | 316 | (2001) |
| Shina R. et al | J. Phys. Soc. Jpn | 73 | 3453 | (2004) |
| Shishido H. et al | Journal of the Physical Society of Japan | 71 | 162 | (2002) |
| Shishido H. Et al | J. Phys. Soc. Jpn. To be published | | | (2005) |
| Shivaram B.S | Phys. Rev. Lett. | 57 | 1259 | (1986) |
| Si Q. | Cond. Mat 0302110V1 | | | (2003-a) |
| Si Q. et al | Phys. Rev. | B68 | 115103 | (2003-b) |
| Si Q. et al | Nature | 413 | 804 | (2001) |
| Sichelschmidt J. et al | Phys. Rev. Lett. | 91 | 156401 | (2003) |
| Sigrist M. and Rice T.M. | Phys. Rev. B | 39 | 2200 | (1989) |
| Sikkema et al | Phys. Rev. B | 54 | 9322 | (1996) |
| Simon H and Lee P.A. | Phys. Rev. Lett. | 78 | 1548 | (1997) |
| Smith T.F. and Fisher E.S | J. Low Temp. Phys. | 12 | 631 | (1973) |
| Sparn G. et al | H.D. Hochkeimer et al eds. Frontier of high pressure research II. Kluwer Academic Publishers, Netherlands | | 413-Y22 | (2001) |
| Springford M. | Electron Cambridge University Press | | | (1997) |
| Steglich F. | Festkörperproblem Advance in Solid State Physics (Vieneg, Braunschweig 1977) ed. Treusch, Vo. XVII | | 319 | (1977) |
| Steglich F. et al | Phys. Rev. Lett. | 43 | 1892 | (1979) |
| Steglich F. et al | Proceedings of Physical Phenomena at High Magnetic Fields II, World Scientific Singapore, editor Z. Fisk, L. Gorkov, D. Metzer, R. Schieffer | | 125 | (1996) |
| Stewart G. | Rev. Modern Physics | 73 | 797 | (2001) |
| Stewart G. et al | Phys. Rev. Lett | 52 | 679 | (1984) |
| Stockert O. et al | Phys. Rev. Lett. | 92 | 136401 | (2004) |
| Stockert O. et al | Phys. Rev. Lett. | 80 | 5627 | (1998) |
| Suderow H. et al | J. Low Temp. Phys. | 108 | 11 | (1997) |
| Suderow H. et al | Phys. Rev. Lett. | 80 | 165 | (1998) |
| Suhl H. | Phys. Rev. lett | 87 | 167007 | (2001) |
| Sulpice A. et al | J. Low Temp. Phys. | 62 | 39 | (1986) |
| Sun P. and Kotliar G. | Phys. Rev. Lett. | 91 | 037209 | (2003) |
| Tachiki M. and Maekawa S. | Phys. Rev. | B 29 | 2497 | (1984) |
| Taillefer L. And Lonzarich G. | Phys. Rev. Lett. | 60 | 1570 | (1988) |
| Taillefer L. et al | Physica C | 153-155 | 451 | (1988) |
| Takabatake T. et al | J. Magn. Magn. Mat. | 177-181 | 277 | (1998) |
| Takashima S. et al | Private communication | | | (2004) |
| Takashita M. et al | J. Magn. Magn. Mat. | 177-181 | 417 | (1998) |
| Takegahara K. and Harima H. | Physica B | 281-282 | 764 | (2000) |
| Takegahara K. et al | J. Phys. F Metal Phys. | 16 | 1961 | (1986) |
| Takegahara M. and Harima H. | Physica B | 281-282 | 764 | (2000) |
| Takimoto T. and Moriya T. | Solid Stat Commun. | 99 | 457 | (1996) |
| Takimoto T. and Moriya T. | Phys. Rev. B | 66 | 134516 | (2002) |
| Takimoto T. et al | Cond. Mat/0212467v1 | | | (2003) |
| Takke R. et al | Z. Phys. B | 44 | 33 | (1981) |
| Tatayama T. et al | Phys. Rev. Letters | B65 | 180504 | (2002) |
| Tateiwa N. et al | J. Phys. Condens. Matter | 13 | L17 | (2001) |
| Tateiwa N | Phys. Rev. | B69 | 180513 | (2004) |
| Tateiwa N. et al | Acta Physica Polonica B | 34 | 515 | (2003-a) |
| Tateiwa N. et al | Physica C | 388-389 | 527 | (2003-b) |
| Tayama T. et al | J. Phys. Soc. Jpn | 72 | 1516 | (2003) |
| Terashima T. et al | Phys. Rev. lett. | 87 | 166401 | (2001) |
| Thalmeier P. and Lüthi B. | Handbook on the Physics and Chemistry of Rare Earths, edited by K.A. Gschneider Jr and L. Eyring, Elsevier Science Publishers BV | 14 | 225 | (1991) |
| Thalmeier P. and Zwicknagl G. | Handbook on the physics and chemistry of Rare Earth | | | (2004-a) |
| Thalmeier P. et al | Frontiers in superconducting material, eds. A. Narlikaa, Springer Verlag Berlin, to be published | | | (2004-b) |
| Thessieu C. et al | Solid State Commun. | 95 | 707 | (1995) |
| Thessieu C. et al | J. Magn. Magn. Mat. | 177-181 | 609 | (1998) |
| Thessieu C. et al | Physica B | 259-261 | 847 | 1999) |
| Thomas F. et al | J. Phys. Condens. Matter | 8 | 51 | (1996) |
| Thomas F. et al | J. Low Temp. Phys. | 102 | 117 | (1995) |
| Thompson J.D. and Lawrence J.M. | In Handbook on the Physics and Chemistry of Rare Earths and Actinides, eds K.A. Gschneider et al, Elsevier Amsterdam | 19 | | (1994) |
| Thompson J.D. et al | J. Magn. Magn. Mat. | 226-230 | 5 | (2001) |
| Tokiwa W. et al | J. Magn. Magn. Mat | 272-276 | 87 | (2004) |
| Tou H et al | Cond-Mat. 0308562 | | | (2003) |
| Tou H. et al | Phys. Rev. Letters | 80 | 3129 | (1998) |
| Tou H. et al | Phys. Rev. Letters | 77 | 1374 | (1996) |
| Trappmann T. et al | Phys. Rev. B | 43 | 13174 | (1991) |
| Trovarelli O. et al | Phys. Rev. Lett. | 85 | 626 | (2000) |
| Tsujii N. et al | J. Phys. Condens. Matter | 15 | 1993 | (2003) |
| Tsunetsugu H. et al | Review of modern Physics | 69 | 809 | (1988) |
| Ueda K., Onuki Y. | Physics of heavy electron systems. Shokadu Tokyo | | | (1998) |
| Uhlarz M. et al | Phys. Rev. Lett. | 93 | 256404 | (2004) |
| Van Dijk et al | Physica B | 319 | 220 | (2002) |
| Van Dijk N. et al | Phys Rev | B61 | 8922 | (2000) |
| Van Dijk N.H. et al | Physica B | 186/188 | 267 | (1993) |
| Van Dijk N.H. et al | Phys. Rev. B | 51 | 12665 | (1995) |
| Van Dijk N.H. et al | PdH thesis, Amsterdam | | | (1994) |
| Vargoz E. et al | Solid State Commun | 106 | 631 | (1998) |
| Varma C.M. et al | Physics Reports | 361 | 267 | (2002) |
| Vettier C. et al | Phys. Rev. Lett. | 56 | 1980 | (1986) |
| Vojta M. | Report Porg. Phys. | 66 | 2069 | (2003) |
| Vollhardt D. and Wölfle P. | The superfluid phases of helium 3, Taylor and Francis, London | | | (1990) |
| Vollmer Y. et al | Phys. Rev. Lett. | 90 | 05707 | (2003) |
| Volovik G.E. | JETP Lett. | 58 | 469 | (1993) |
| von Löhneysen H. et al | Phys. Rev. Letters | 72 | 3262 | (1994) |
| von Löhneysen H. et al | J. Phys. Soc Jpn | 69 | Suppl. A 63 | (2000) |
| von Löhneysen H. et al | J. Phys. Condens. Matter | 8 | 9689 | (1996) |
| Wachter P. | Handbook of physics and chemistry of rare earths. Ed by K.A Gscheidner et al North Holland, Amsterdam (1994). | 19 | 177 | (1993) |
| Waler J.T. and Cromer T. | J. Chem. Phys. | 42 | 4112 | (1965) |
| Walker D.A. et al | Phys. Rev. B | 63 | 054522 | (2001) |
| Walker M.B. and Samokhin K.V. | Phys. Rev. Lett. | 88 | 207001 | (2002) |
| Wang Y. And Cooper B.R. | Phys. Rev | 172 | 539 | (1968) |
| Wang Y. Et al | Phys. Rev. Lett. | 88 | 257-003 | (2002) |
| Wang Y.L. and Cooper B.R. | Phys. Rev. | 185 | 696 | (1969) |
| Watanabe S. and Miyake K. | Physica B | 312-313 | 115 | (2002-a) |
| Watanabe S. and Miyake K. | J. Phys. Chem. Solids | 63 | 1465 | (2002-b) |
| Watanabe T et al | Phys. Rev. | B70 | 020506R | (2004) |
| Welp U. | PhD Thesis | | | (1988) |
| Welp U. et al | J. Magn. Magn. Mat. | 63-64 | 28 | (1987) |
| Wertharmer N.R. et al | Phys. Rev. | 147 | 295 | (1996) |
| Wiebe CR. Et al | Phys. Rev. B | 69 | 132418 | (2004) |
| Wihelm H. et al | J. Phys. Cond. Matter | 13 | L329 | (2001) |
| Wilhelm H. and Jaccard D. | Phys. Rev. | B69 | 214408 | (2004) |
| Wilson A.H. | Proc. Roy. Soc. A | 67 | 580 | (1938) |
| Wilson K. G. | Rev. Mod. Physics | 47 | 773 | (1975) |
| Won H. et al | Cond-mat/0306548 | | | (2003) |
| Yamada K. | Prog. Theor. Phys. | 53 | 970 | (1975) |
| Yamagami H. | J. Phys. Condens. Matter | 15 | S2271 | (2003) |
| Yamagami H. and Hamada N. | Physica B | 384 | 1295 | (2000) |
| Yaouanc A. et al | Phys. Rev. Lett. | 89 | 147001 | (2002) |
| Yaouanc A. et al | Phys. Rev. Lett. | 84 | 2702 | (2000) |
| Yaouanc A. et al | J. Phys. Condens. Matter | 10 | 9791 | (1998) |
| Yeh A. et al | Nature (London) | 419 | 459 | (2002) |
| Yogi M. et al | Phys. Rev. Lett. | | | |
| Yotsubashi S. et al | J. Phys. Soc. Japan | 70 | 186 | (2001) |
| Yu W. et al | Phys. Rev. Lett. | 92 | 086403 | (2004) |
| Yuan H.Q. et al | Science | 302 | 2104 | (2003) |
| Zaanen J. | Phil. Mag. | 81 | 1485 | (2001) |
| Zapf V.S. et al | Phys. Rev. | B65 | 014506-1 | (2001) |
| Zhu J.X et al | Phys.Rev. Lettt. | 91 | 156404 | (2003-a) |
| Zhu L. et al | Phys. Rev. Lett. | 91 | 066404 | (2003-b) |
| Zlatic V. et al | Phys. Rev. | 68 | 104432 | (2003) |
| Zülicke U. and Millis H.J. | Phys. Rev. B | 51 | 8996 | (1995) |
| Zwicknagl G. | Advances in Physics | 41 | 203 | (1992 |
| Zwicknagl G. and Fulde P. | J. Phys. Condens. Matter | 15 | S1911 | (2003) |
| Zwicknagl G. and Pulst | Physica B | 186 | 895 | (1993) |